%% file: polar-harmonics-siggraph-main.tex
\PassOptionsToPackage{dvipsnames,table}{xcolor}%

\documentclass[acmtog,screen,authorversion]{acmart} %

\acmSubmissionID{0213} %

\input{mypackages}

\input{mymacros}

\input{szpackage} %

\usepackage{booktabs} %

\usepackage{microtype}
\usepackage[ruled]{algorithm2e} %
\usepackage[shortlabels]{enumitem}

\SetAlFnt{\small}
\SetAlCapFnt{\small}
\SetAlCapNameFnt{\small}
\SetAlCapHSkip{0pt}
\SetKwInput{KwData}{Input}
\SetKwInput{KwResult}{Output}

\setcopyright{rightsretained}
\acmJournal{TOG}
\acmYear{2024} \acmVolume{43} \acmNumber{4} \acmArticle{127} \acmMonth{7}\acmDOI{10.1145/3658139}

\begin{document}
\input{title}

\input{authors}

\input{abstract}

\input{keywords}

\input{teaser}

\maketitle

\input{intro}

\input{relatedwork}
\input{overview}

\input{background_SH}

\input{background_polar}

\input{challengesinSH}

\input{theory}

\input{theory_convolution}

\input{pprt}

\input{discussion}

\input{conclusion}

\input{acknowledgments}

\bibliographystyle{ACM-Reference-Format}
\bibliography{bibliography}


\onecolumn
\appendix
\section*{\textbf{Supplemental Document:\\
Spin-Weighted Spherical Harmonics for Polarized Light Transport}}
\vspace{2cm}

\input{title-supple}

\input{intro-supp}


\input{background1-supp} %
\input{background2-supp} %
\input{stokesfield-supp}

\input{theory-supp}
\input{application1-supp}

\bibliographystyle{ACM-Reference-Format}
\bibliography{bibliography}


\end{document}

%% file: mypackages.tex
\usepackage{import} %
\graphicspath{{figs/}}

\usepackage{accents}

\usepackage[ruled,noend]{algorithm2e} %

\SetAlFnt{\small}
\SetAlCapFnt{\small}
\SetAlCapNameFnt{\small}
\SetAlCapHSkip{0pt}
\IncMargin{-\parindent}
\usepackage{ifpdf}
\usepackage{graphicx}
\ifpdf
  
\else
\fi
\definecolor{cyan}{cmyk}{1,0,0,0}
\definecolor{darkred}{rgb}{0.7,0,0}
\definecolor{darkgreen}{rgb}{0,0.7,0}
\definecolor{darkblue}{rgb}{0,0,0.7}
\definecolor{orange}{rgb}{1,0.5,0}
\definecolor{magenta}{cmyk}{0,1,0,0}
\definecolor{darkyellow}{cmyk}{0,0,0.75,0}
\definecolor{lightgray}{rgb}{0.8,0.8,0.8}

\usepackage{amsmath} %

\usepackage[toc,page]{appendix} %
\usepackage{wrapfig} %
\usepackage{comment}
\usepackage{cuted} %
\usepackage{lipsum} %
\usepackage{mathtools} %
\newcommand{\mparagraph}[1]{\vspace{1mm}\noindent\textsf{\textbf{#1.}\hspace{2mm}}}

\usepackage{algorithmicx}
\usepackage{algpseudocode}

\usepackage{gensymb}
\usepackage{float}
\usepackage{soul}
\usepackage{array}
\usepackage{multirow}
\usepackage{hhline}
\usepackage{setspace}
\usepackage{enumerate}
\usepackage{pdflscape} %

\algdef{SE}[DOWHILE]{Do}{doWhile}{\algorithmicdo}[1]{\algorithmicwhile\ #1}%

\usepackage{amsthm}
\usepackage[many]{tcolorbox}
\tcbuselibrary{theorems}

\newtheorem{theorem}{Theorem}[section]

\newtheorem{lemma}{Lemma}[theorem]

\newtcbtheorem[number within=section,use counter=theorem]{definition}{Definition}%
{colback=black!5,colframe=blue!35!black,fonttitle=\bfseries}{def}
\newtcbtheorem[number within=section,use counter=theorem]{proposition}{Proposition}%
{colback=black!5,colframe=blue!35!black,fonttitle=\bfseries}{prop}

\usepackage{footnote}
\usepackage{etoolbox}
\BeforeBeginEnvironment{definition}{\savenotes}
\AfterEndEnvironment{definition}{\spewnotes}
\BeforeBeginEnvironment{proposition}{\savenotes}
\AfterEndEnvironment{proposition}{\spewnotes}

\makeatletter
\renewenvironment{proof}[1][\proofname]{\par
	\pushQED{\qed}%
	\normalfont \topsep6\p@\@plus6\p@\relax
	\trivlist
	\item\relax
	{\bfseries\itshape#1\@addpunct{:}}
}{%
	\popQED\endtrivlist\@endpefalse
}

\tcolorboxenvironment{proof}{
	beforeafter skip balanced=25pt,
	colback={White!90!CornflowerBlue},
	boxrule=0pt, %
	breakable
}

\makeatother

\usepackage{arydshln}

\usepackage{array}

\newcolumntype{P}[1]{>{\centering\arraybackslash}p{#1}}
\newcolumntype{M}[1]{>{\centering\arraybackslash}m{#1}}

\newcolumntype{?}{!{\vrule width 1pt}}

\def\thickhline{\noalign{\hrule height 1pt}}

\makeatletter
\renewcommand{\ALG@beginalgorithmic}{\small}
\makeatother

\newcommand{\DELETE}[1]{} %
\newcommand{\IGNORE}[1]{}
\usepackage{datenumber}
\usepackage{calc}
\usepackage[mmddyyyy]{datetime}

\newcounter{datetoday}
\newcounter{diffyears}
\newcounter{diffmonths}
\newcounter{diffdays}

\newcommand{\difftoday}[3]{%
      \setmydatenumber{datetoday}{\the\year}{\the\month}{\the\day}%
      \setmydatenumber{diffdays}{#1}{#2}{#3}%
      \addtocounter{diffdays}{-\thedatetoday}%
      \ifnum\value{diffdays}>0
        \def\diffbefore{}%
        \def\diffafter{left}%
      \else
        \def\diffbefore{}%
        \def\diffafter{ago}%
        \setcounter{diffdays}{-\value{diffdays}}%
      \fi
      \setcounter{diffyears}{\value{diffdays}/365}%
      \setcounter{diffdays}{\value{diffdays}-365*\value{diffyears}}%
      \setcounter{diffmonths}{\value{diffdays}/30}%
      \setcounter{diffdays}{\value{diffdays}-30*\value{diffmonths}}%
      \diffbefore
      \ifnum\value{diffyears}=0
      \else
        \ifnum\value{diffyears}>1
            \thediffyears\space years,
        \else
            \thediffyears\space year,
        \fi
      \fi
      \ifnum\value{diffmonths}=0
      \else
        \ifnum\value{diffmonths}>1
            \thediffmonths\space months
        \else
            \thediffmonths\space month
        \fi
      \fi
      \ifnum\value{diffdays}=0
      \else
        \ifnum\value{diffdays}>1
            \thediffdays\space days
        \else
            \thediffdays\space day
        \fi
      \fi
      \diffafter
}

%% file: mymacros.tex
\newcommand{\NEW}[1]{{\color{black}{#1}}}%
\newcommand{\NEWJ}[1]{{\color{black}{#1}}}%

\def\C{\mathbb{C}}
\def\R{\mathbb{R}}
\def\Z{\mathbb{Z}}

\def\rmd{\mathrm{d}}
\def\rmf{\mathrm{f}}
\def\rmk{\mathrm{k}}

\def\rms{\mathrm{s}}
\def\rmK{\mathrm{K}}
\def\rmL{\mathrm{L}}
\def\rmM{\mathrm{M}}
\def\rmP{\mathrm{P}}
\def\rmT{\mathrm{T}}

\def\bfe{\mathbf e}
\def\bff{\mathbf f}
\def\bfg{\mathbf g}
\def\bfk{\mathbf k}

\def\bfp{\mathbf p}

\def\bfs{\mathbf{s}}
\def\bft{\mathbf{t}}

\def\bfx{\mathbf x}
\def\bfy{\mathbf y}

\def\bfA{\mathbf A}
\def\bfB{\mathbf B}
\def\bfC{\mathbf C}

\def\bfF{\mathbf F}

\def\bfI{\mathbf I}
\def\bfJ{\mathbf J}
\def\bfK{\mathbf K}
\def\bfM{\mathbf M}
\def\bfN{\mathbf N}
\def\bfR{\mathbf R}
\def\bfT{\mathbf T}
\def\bfU{\mathbf U}

\def\calB{\mathcal{B}}

\def\calF{\mathcal{F}}
\def\calH{\mathcal{H}}
\def\calL{\mathcal{L}}
\def\calM{\mathcal{M}}
\def\calS{\mathcal{S}}

\def\bbK{\mathbb{K}}
\def\bbS{\mathbb{S}}

\def\frF{\vec{\mathbf F}}
\def\frG{\vec{\mathbf G}}
\def\frtp{\vec{\mathbf F}_{\theta\phi}} %

\def\dvf{\accentset{\leftrightarrow}{f}}
\def\dvg{\accentset{\leftrightarrow}{g}}
\def\dvk{\accentset{\leftrightarrow}{k}}
\def\dvs{\accentset{\leftrightarrow}{s}}
\def\dvt{\accentset{\leftrightarrow}{t}}
\def\dvx{\accentset{\leftrightarrow}{x}}
\def\dvy{\accentset{\leftrightarrow}{y}}
\def\dvA{\accentset{\leftrightarrow}{A}}
\def\dvB{\accentset{\leftrightarrow}{B}}
\def\dvI{\accentset{\leftrightarrow}{I}}
\def\dvK{\accentset{\leftrightarrow}{K}}
\def\dvM{\accentset{\leftrightarrow}{M}}
\def\dvP{\accentset{\leftrightarrow}{P}}

\def\dvY{\accentset{\leftrightarrow}{Y}}

\def\abs#1{\left|#1\right|}
\def\norm#1{\left\|#1\right\|}

\def\lrangle#1{\left\langle{#1}\right\rangle}
\def\difrac#1#2{\frac{\rmd{#1}}{\rmd{#2}}}
\def\pfrac#1#2{\frac{\partial{#1}}{\partial{#2}}}

\def\pownu#1{\left(-1\right)^{#1}}

\def\hn{\hat{n}}
\def\homega{\hat{\omega}}

\def\brahomega{\left(\hat{\omega}\right)}
\def\brai{\left(\homega_i\right)}
\def\brao{\left(\homega_o\right)}
\def\braio{\left(\homega_i, \homega_o\right)}

\def\Ciso{\C_{\mathrm{iso}}}
\def\Cconj{\C_{\mathrm{conj}}}
\def\Rtt{\R^{2\times 2}}
\def\ppi{\bfp\bfp\mathrm{i}}
\def\ppc{\bfp\bfp\mathrm{c}}

\def\Rspv{\vec{\mathbb{R}}^3} %
\def\Sspv{\hat{\mathbb{S}}^2} %

\def\Rns#1{\mathbf{R}_{1:2}\left(#1\right)} %

\def\FRsp{\vec{\mathbb F}^3}
\def\FRspo#1{\vec{\mathbb F}_{{#1}}^3}
\def\SOgroup{\mathbf{SO}\left(3\right)}
\def\SOgroupv{\overrightarrow{SO}\left(3\right)}

\def\STKsp#1{\mathcal{S}_{#1}} %
\def\MUEsp#1#2{\mathcal{M}_{{#1}\to{#2}}} %

\def\extColx#1{{#1}\left[:,1\right]} %
\def\extColy#1{{#1}\left[:,2\right]} %
\def\extColz#1{{#1}\left[:,3\right]} %

\def\bfMat{\mathbf{Mat}}

\def\sY#1{{}_2 Y_{#1}}

\def\underarrowref#1#2{\underset{\substack{\color{darkgray}\uparrow \\ \color{darkgray}\text{#2}}}{#1}}

%% file: szpackage.tex
\definecolor{grayL}{rgb}{.90, .90, .90}
\definecolor{purpleL}{rgb}{.9735, .95, .9761}
\definecolor{purpleD}{rgb}{.8941, .8, .9043}
\definecolor{greenL}{rgb}{.92, .96, .93}
\definecolor{greenD}{rgb}{.7145, .9249, .7999}
\definecolor{orangeLL}{rgb}{0.9991, 0.9846, 0.9759} %
\definecolor{orangeL}{rgb}{.997, .94, .93}
\definecolor{orangeD}{rgb}{.9929, .8766, .8071}

\definecolor{redL}{rgb}{1.0, 0.95, 0.95}
\definecolor{redD}{rgb}{1.0, 0.8, 0.8}
\definecolor{yellowL}{rgb}{0.98, 0.98, 0.8}
\definecolor{yellowD}{rgb}{0.95, 0.95, 0.6}
\definecolor{blueLL}{rgb}{0.98, 0.98, 1.0}
\definecolor{blueL}{rgb}{0.95, 0.95, 1.0}
\definecolor{blueD}{rgb}{0.8, 0.8, 1.0}

\definecolor{col0}{RGB}{218, 240, 210}
\definecolor{col2}{RGB}{212, 238, 255}
\definecolor{col3}{RGB}{255, 224, 219}
\definecolor{col4}{RGB}{255, 243, 183}

\newtcolorbox{szBox}{%
	colback=blueLL,colframe=lightgray,top=1mm,bottom=1mm,left=1mm,right=1mm%
}
\newtcolorbox{szMathBox}{%
	colback=blueLL,colframe=lightgray,top=-1mm,bottom=1mm,left=1mm,right=1mm%
}
\newtcolorbox{szTitledBox}[2][]{%
	colback=white,colframe=lightgray,top=1mm,bottom=1mm,left=1mm,right=1mm,title={#2},fonttitle=\bfseries\color{black},#1%
}

\newcommand{\tcboxmathStS}[1]{
	\tcboxmath[size=small, opacityframe=0,colback=col0]{#1}
}

\newcommand{\tcboxmathStV}[1]{
	\tcboxmath[size=small, opacityframe=0,colback=col4]{#1}
}

\newcommand{\tcboxmathVtS}[1]{
	\tcboxmath[size=small, opacityframe=0,colback=col3]{#1}
}

\newcommand{\tcboxmathVtV}[1]{
	\tcboxmath[size=small, opacityframe=0,colback=col2]{#1}
}

\newcommand{\tcboxStS}[1]{
	\tcboxmath[size=tight, opacityframe=0,colback=col0]{#1}
}

\newcommand{\tcboxStV}[1]{
	\tcboxmath[size=tight, opacityframe=0,colback=col4]{#1}
}

\newcommand{\tcboxVtS}[1]{
	\tcboxmath[size=tight, opacityframe=0,colback=col3]{#1}
}

\newcommand{\tcboxVtV}[1]{
	\tcboxmath[size=tight, opacityframe=0,colback=col2]{#1}
}

%% file: title.tex
\title{Spin-Weighted Spherical Harmonics for Polarized Light Transport}

%% file: authors.tex
\author{Shinyoung Yi}
\orcid{0000-0003-1312-657X}
\affiliation{%
  \institution{KAIST}
  \country{South Korea}}
\email{syyi@vclab.kaist.ac.kr}

\author{Donggun Kim}
\orcid{0000-0002-6670-6263}
\affiliation{%
  \institution{KAIST}
  \country{South Korea}}
\email{dgkim@vclab.kaist.ac.kr}

\author{Jiwoong Na}
\orcid{0000-0002-9685-1146}
\affiliation{%
  \institution{KAIST}
  \country{South Korea}}
\email{jwna@vclab.kaist.ac.kr}  

\author{Xin Tong}
\orcid{0000-0001-8788-2453}
\affiliation{%
  \institution{Microsoft Research Asia}
  \country{China}}
\email{xtong@microsoft.com}

\author{Min H. Kim}
\orcid{0000-0002-5078-4005}
\affiliation{%
  \institution{KAIST}
  \country{South Korea}}
\email{minhkim@vclab.kaist.ac.kr}

%% file: abstract.tex
\begin{abstract}
The objective of polarization rendering is to simulate the interaction of light with materials exhibiting polarization-dependent behavior. However, integrating polarization into rendering is challenging and increases computational costs significantly. The primary difficulty lies in efficiently modeling and computing the complex reflection phenomena associated with polarized light. Specifically, frequency-domain analysis, essential for efficient environment lighting and storage of complex light interactions, is lacking. To efficiently simulate and reproduce polarized light interactions using frequency-domain techniques, we address the challenge of maintaining continuity in polarized light transport represented by Stokes vectors within angular domains. The conventional spherical harmonics method cannot effectively handle continuity and rotation invariance for Stokes vectors. To overcome this, we develop a new method called polarized spherical harmonics (PSH) based on the spin-weighted spherical harmonics theory. Our method provides a rotation-invariant representation of Stokes vector fields. Furthermore, we introduce frequency domain formulations of polarized rendering equations and spherical convolution based on PSH. We first define spherical convolution on Stokes vector fields in the angular domain, and it also provides efficient computation of polarized light transport, nearly on an entry-wise product in the frequency domain. Our frequency domain formulation, including spherical convolution, led to the development of the first real-time polarization rendering technique under polarized environmental illumination, named precomputed polarized radiance transfer, using our polarized spherical harmonics. Results demonstrate that our method can effectively and accurately simulate and reproduce polarized light interactions in complex reflection phenomena, including polarized environmental illumination and soft shadows.
\end{abstract}

%% file: keywords.tex
\ccsdesc[300]{Mathematics of computing~Functional analysis}
\ccsdesc[300]{Computing methodologies~Rendering}
\ccsdesc[300]{Computing methodologies~Appearance and texture representations}

\keywords{spherical harmonics, polarized rendering, polarimetric imaging, polarimetric appearance, theory of light transport, basis function}

%% file: teaser.tex
\begin{teaserfigure}
	\centering
	\vspace{-2mm}%
	\includegraphics[width=\linewidth]{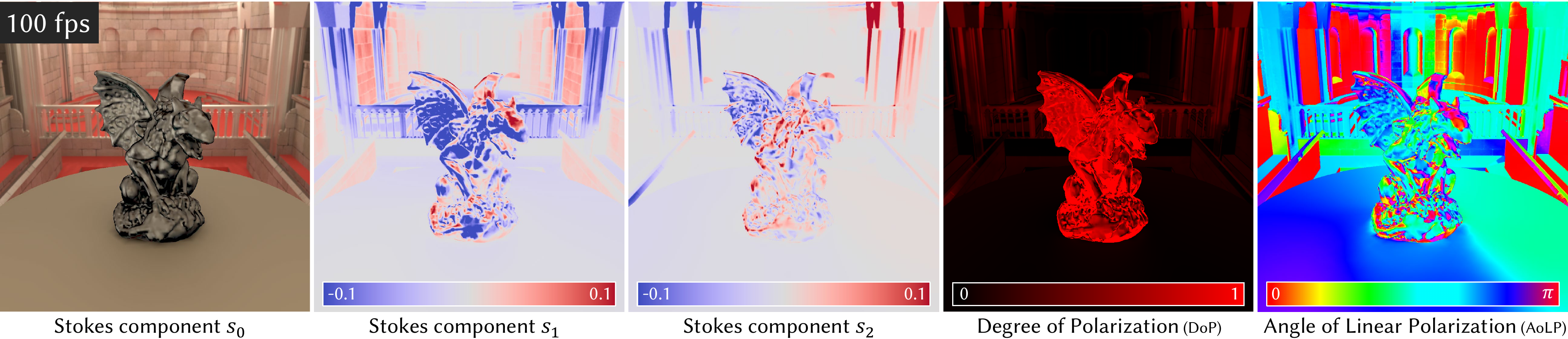}%
	\vspace{-2mm}%
	\caption[]{\label{fig:teaser}%
Real-time rendering of polarized light has been unavailable due to its higher dimensional space of polarized light. We introduce a new frequency-domain analysis of polarized light transport and 
propose a new method, called \emph{polarized spherical harmonics}, based on the spin-weighted spherical harmonics theory in physics.
Our method provides a rotation-invariant representation and spherical convolution of Stokes vector fields, enabling efficient simulation and reproduction of polarized light interactions
We demonstrate the first real-time polarization rendering under polarized environmental illumination through polarized spherical harmonics. 
Refer to the supplemental video for real-time video results.}
\vspace{2mm}%
\end{teaserfigure}

%% file: intro.tex
\section{Introduction}
\label{sec:intro}

Polarization, imperceptible to the human eye, offers a wealth of auxiliary information about an object's shape and physically meaningful material characteristics. 
Consequently, polarization has been increasingly exploited in both the fields of computer graphics and vision for tasks of geometry modeling and appearance acquisition in recent years. 
It has shown extensive applications, including multispectral ellipsometry to obtain a polarimetric bidirectional reflectance distribution function (pBRDF) dataset~\cite{baek2020image}, inverse rendering for acquiring polarimetric appearance and geometry~\cite{baek2018simultaneous, hwang2022sparse}, and physically-based polarization rendering through the synthesis and analysis of polarized light transport \cite{Mitsuba3}.

Polarization rendering, as opposed to traditional rendering that calculates light intensity, necessitates additional computation of further information. 
First, polarization rendering simulates a four-dimensional vector that consists of light intensity, two linear polarizations in horizontal/vertical and diagonal/antidiagonal directions, and circular polarization, commonly represented as a Stokes vector. 
This fully characterizes the polarization state of light as it travels along a ray. 
Second, different from the conventional vector, the Stokes vector physically quantifies the sinusoidal oscillation of light waves.
Consequently, when the coordinate system rotates, the elements of the Stokes vector change at a rate twice as fast as the components of a conventional vector change. 
It is critical to account for this fact when transforming a Stokes vector for polarized light simulation. 
Third, the unique features of polarization rendering not only escalate the computational expenses significantly but also necessitate special transformation.
The Stokes vector is determined within a local coordinate system of the progressing ray. 
Therefore, the polarimetric reflectance function must be configured in accordance with the coordinate systems of the incident and exitant Stokes vectors.
This implies that additional coordinate conversions are required for polarization rendering. 
Lastly, the formulation of the polarimetric reflectance must consider both the incident and exitant Stokes vectors. This process is often presented as a Mueller matrix, a structure of sixteen components arranged in a four-by-four matrix format, of which components change depending on the incident/exitant light angles. The complexity of this matrix makes it sixteen times larger than a scalar value used in conventional reflectance functions, significantly increasing the computational cost of ray samples in polarization rendering \cite{wilkie2012polarised}.

Due to the unique attributes of Stokes vectors in the representation of light polarization, 
conventional frequency-domain analysis of environment lighting~\cite{ramamoorthi2001signal}, does not guarantee smooth transformations and invariance under rotations in polarization rendering.
Moreover, the process of rendering specular reflections using spherical harmonics (SH), which necessitates spherical convolution~\cite{sloan2002precomputed}, functions efficiently for scalar fields that do not change under rotation. 
However, these conventional spherical harmonics cannot be applied for the transportation of polarized light. 
This is because, in polarization rendering, light is represented not as a scalar intensity but as a Stokes vector. 
In this work, we focus on addressing the following two main challenges with the goal of facilitating real-time polarization rendering via frequency-domain analysis.

\paragraph{Rotation invariance}
Different from the conventional SH-based rendering that computes light intensity as a scalar value on a sphere, 
polarization rendering needs to simulate a Stokes vector field on a sphere. %
However, dealing with Stokes vector fields using conventional basis functions, including SH, which are designed for scalar fields, results in a \emph{singularity problem}.
This is commonly known as the Hairy Ball Theorem \cite{nash1983topology}. 
Representing a Stokes vector field using conventional SH requires separating it into four scalar fields and one frame field, which assigns local frames for tangent planes for each point on a sphere. The Hairy Ball Theorem implies we cannot assign a smooth and consistent direction of unit tangent vectors at every point on the sphere without at least one singular point,
so that the resulting Stokes vector field combined with continuous scalar basis functions should have a singular point.
This is a critical problem for rotation transformation in light transport.
For instance, suppose we want to rotate Stokes vector fields by transforming basis functions.
The transformation cannot guarantee the rotation invariance because this operator to the basis functions will create another singularity point in another location on the sphere.

To address the issue of rotation invariance in polarization rendering, we introduce a new method utilizing a spin-weighted spherical harmonics (SWSH) theory~\cite{scanio1977spin} in physics. 
These SWSH serve as orthonormal basis functions, and they can be classified based on how different spin fields behave on a sphere.
The spin-0 spherical harmonics, which are equivalent to traditional spherical harmonics, represent scalar fields that remain unchanged under the rotation of local frames. 
On the other hand, spin-1 spherical harmonics represent vector fields on a sphere. These can transform under rotations in the same way as a typical vector, which indicates they possess a certain spin orientation.
Finally, spin-2 spherical harmonics (S2SH) represent fields of such quantities as neither scalar nor ordinary vectors, which are characterized as multiple directions associated with each point, mirroring the properties of Stokes vector fields.
To resolve the fundamental rotation invariance problem in polarization rendering, we employ S2SH, which handles Stokes vector fields by extending the domain of the basis function space from the sphere to the frame space while maintaining a spin-2 constraint, rather than improperly separating them into scalars and frame fields with singularity. This approach paves the way for real-time polarization rendering.

\paragraph{Spherical convolution}
To accomplish efficient real-time rendering of specular reflection in conventional rendering, SH-based rendering has utilized the scalar spherical convolution of light intensity~\cite{sloan2002precomputed}. 
For real-time polarization rendering to be feasible, it is also crucial to establish a spherical convolution of Stokes vector fields.
In polarization rendering, the input and output for spherical convolution are represented as Stokes vectors. 
Therefore, the convolution kernel needs to be defined as a Mueller matrix. 
However, we observe that the Mueller matrix domain should have only one degree of freedom of the zenith angle of the kernel in relation to the zonal axis of spherical harmonics when generalizing conventional spherical convolution as rotation equivariant linear operators.
The spherical convolution of a Mueller matrix to Stokes vector fields, which we formulate in this paper, has not yet been addressed in the field of computer graphics research. 

We, therefore, introduce a new frequency-domain method for spherical convolution for Stokes vector fields.
This method allows for the efficient yet precise convolution of approximated Stokes vectors, thereby enabling real-time rendering of polarized light.
Our approach, which is based on polarized spherical harmonics (PSH), facilitates efficient computation, operating nearly on an element-by-element product basis.
To this end, we jointly combine the spin-0 and spin-2 cases of spin-weighted spherical harmonics, incorporating a new theory concerning the frequency-domain analysis of pBRDF and spherical convolution of Stokes vector fields.

Further, we demonstrate a real-time technique for polarization rendering, the so-called precomputed polarized radiance transfer (PPRT), using our polarized spherical harmonics.
\NEW{See Figure~\ref{fig:teaser} for an example.}
Our proposed method can efficiently and accurately simulate and replicate the approximated interactions of polarized light in complex reflection phenomena, including polarized environmental illumination and soft shadows.

%% file: relatedwork.tex
\section{Related Work}
\label{sec:relatedwork}

\subsection{Spherical Harmonics}

A frequency-domain framework
using spherical harmonics is introduced by Ramamoorthi and Hanrahan~\shortcite{ramamoorthi2001efficient, ramamoorthi2001signal} to computer graphics community. They represent environment maps~\cite{ramamoorthi2001efficient} into SH coefficients and render environment map lighting by the product of coefficient vectors.
Extending SH coefficients of diffuse albedo to store radiance self-transfer, including self-shadow and interreflection, their framework has been extended to precomputed radiance transfer (PRT)~\cite{sloan2002precomputed}. 
The PRT method has various extensions which deal with dynamic shadow~\cite{zhou2005precomputed}, deformable objects~\cite{sloan2005local}, and polygonal lights~\cite{wang2018analytic, wu2020analytic}. 
Benefits of some of these methods come from not only algorithmic enhancement but also analytic integrals related to SH, such as triple product~\cite{zhou2005precomputed} and integrals on spherical polygons~\cite{wang2018analytic}. 
\NEW{We refer to \citet{kautz2005precomputed} and \citet{ramamoorthi2009precomputation} for more history and overview of the field of precomputation-based rendering.} 
Note that not only real-time rendering methods, application of SH to rendering also include physically based ray tracing~\cite{belcour2018integrating}, which uses SH products as control variates and inverse rendering of reflectance~\cite{ramamoorthi2001signal} which projects BRDF and normal vectors into SH coefficients.

Other bases for spherical functions, including the Haar wavelet \NEW{\cite{ng2003all, ng2004triple, lessig2008soho}}, spherical Gaussians \cite{ritschel2012state}, and neural bases~\cite{xu2022lightweight} have been discussed. Still, only spherical harmonics simultaneously hold orthonormality, rotation invariance, and a coefficient-wise product of spherical convolution. There is another recent approach to learning basis functions on the sphere rather than defining analytically by \citet{xu2022lightweight}, but their work produces no genuine basis that should satisfy linearity. 

While SH provides a wide range of applications in computer graphics, as discussed above, there has been no extension of any of these methods to polarized light transport due to the difficulty of the continuity structure of Stokes vector fields.

\subsection{Polarization}
Polarization has played an important role in computer graphics.
For example, polarized illumination enhances the reconstruction quality of 3D geometry and reflectance
\NEW{\cite{ghosh2011multiview, kadambi2015polarized, ba2020deep}}. 
In addition, rendering~\cite{mojzik2016bi, jarabo2018bidirectional} and reconstructing \NEW{in both explicit geometry}~\cite{baek2018simultaneous, hwang2022sparse} \NEW{and radiance fields~\cite{kim2023nespof}} polarized quantities themselves have also \NEW{been} investigated recently. 
These problems handle polarized appearance, which captures what traditional scalar intensity-based appearance has not done and has been addressed as challenging problems due to more parameters and unconventional coordinate conversion problems. 
However, no frequency-domain methods have been developed. \citet{mojzik2016bi, jarabo2018bidirectional} introduce polarized ray tracing methods that consider the light source and material appearance as Stokes vectors and Mueller matrices, respectively, but there are no precomputed methods through basis functions that achieve real-time performance. 
\citet{baek2020image} captured image-based pBRDF datasets, but there is still a lack of methods of how to render their materials in runtime efficiently. 
In this context, we propose a new frequency-domain framework of polarized light transport, which implies polarized precomputed rendering, so that our novel rendering method achieves real-time performance and provides a novel way to render \citet{baek2020image}'s data-based pBRDF.

\NEW{Certain studies have utilized polarized gradient illumination to capture the appearance of objects \cite{ma2007rapid, ghosh2009estimating, ghosh2011multiview}, which is related to spherical harmonics up to order 2. The utilization of polarized light in these studies is specific to scenarios where it is necessary to separate two scalar fields of diffuse and specular reflection. However, bases of Stokes vector fields have not been addressed in these studies.}

\NEW{The works mentioned above, including this one, use Mueller calculus formulations to deal with polarized light. However, physical light transport methods, such as those presented in recent works~\cite{steinberg2021generic, steinberg2021physical, steinberg2022towards}, have introduced a generalized Stokes parameters formulation based on optical coherence theory. This formulation combines the strengths of both Mueller and Jones calculus. However, it does not address the challenges in the angular domain, and its contributions are not relevant to our current scope.} 

For more concepts, history, and applications in computer graphics of polarization, we refer to \citet{collett2005field}, \citet{wilkie2012polarised}, and \citet{baek2023polarization}.

\subsection{\NEW{Spin-Weighted Spherical Harmonics}}
\label{sec:relatedwork-SWSH}

\NEW{Spin-weighted spherical harmonics theory is originally introduced by \citet{newman1966note, goldberg1967spin} to handle the symmetry of gravitational radiation in physics.  \citet{zaldarriaga1997all} point out that spin-2 SH can encode the all-sky information of polarized light to the frequency domain in the context of the cosmic microwave background. Rotation invariance and coefficient rotations of SWSH are shown by \citet{boyle2013angular}.

Note that SWSH has also been referred to as \emph{generalized spherical harmonics} in some literature~\cite{kuvsvcer1959matrix, phinney1973representation, garcia1986generalized, keegstra1997generalized}, and the relation between these names is pointed out by \citet{rossetto2009general}.

While SWSH formulation of Stokes vector fields already exists, to the best of our knowledge, we first formulate linear operators on Stokes vectors, including pBRDF, into SWSH coefficients.

\citet{zaldarriaga1997all} and \citet{ng1999correlation} establish the SWSH formulation of the \emph{correlation} operation between two Stokes vector fields in the perspective to analyze statistics of given data. While the correlations have some similarities to \emph{convolutions}, these are inherently different operations in terms of types of inputs and outputs. We focus on the convolution operation from the perspective of image processing and computer graphics, especially PRT.

Spherical convolution of Stokes vector fields has been discussed in \citet{garcia1986generalized}, \citet{ng1999correlation}, and \citet{tapimo2018discrete}. However, their formulations are subsets of our formulation of polarized spherical convolution. Specifically, their polarized convolution kernels have one degree of freedom (DoF)~\cite{ng1999correlation} or six DoF~\cite{garcia1986generalized, tapimo2018discrete}  for each frequency band, while ours has 16 DoF. Based on this generalization, we first discover that polarized spherical convolution is equivalent to rotation equivariance linear operators on Stokes vector fields, with a proper sense of such linearity.

We refer to Section~\ref{sec:discussion-theory-novelty} for a more technical description of our novelty against existing work on SWSH.}

%% file: overview.tex
\section{Overview}
\label{sec:overview}
The following is a brief outline of our paper's organization. In Section~\ref{sec:background}, we provide the theoretical foundations of traditional spherical harmonics, spherical convolution, and polarization of light %
in Mueller calculus. This section is included for the sake of readability, but expert readers may skip it, 
while  Section~\ref{sec:mueller_num_geo} gives a brief introduction to the mathematical notations used in this paper. It will help the readers to understand the mathematical concepts presented in the paper.
In Section~\ref{sec:stokes-on-sphere}, we discuss the challenges of applying existing spherical harmonics to Stokes vector fields.
Our main method is presented in Section~\ref{sec:our_theory}, which consists of the polarized spherical harmonics theory (Section~\ref{sec:theory_rotinv}) and polarized spherical convolution (Section~\ref{sec:theory_conv}).
In Section~\ref{sec:pprt}, we demonstrate the first real-time polarized rendering method, followed by a discussion in Section~\ref{sec:discussion} and a conclusion in Section~\ref{sec:conclusion}.
\NEW{Tables~\ref{tb:quantities_gen} and~\ref{tb:rot_inner} provide notations, symbols, and operators used in this paper.} \NEW{We also make our code available on our project website (\url{https://vclab.kaist.ac.kr/siggraph2024/}), which includes a step-by-step tutorial to help understand various quantities and equations. }

%% file: background_SH.tex
\input{notations_all_v2}

\section{Background}
\label{sec:background}

\subsection{Spherical Harmonics}
\label{sec:background_sh}
This subsection briefly reviews the definition and core properties of spherical harmonics.
In Supplemental Sections~\ref{sec:bkgnd_func_anal} and~\ref{sec:bkgnd_SH}, we additionally provide a general theory of function spaces and bottom-up mathematical description of SH, including how some properties of SH are inherited from the general theory.

Spherical harmonics are spherical functions $Y_{lm}\in \calF\left(\Sspv,\C\right)$, where $\calF\left(\Sspv,\C\right)\coloneqq \left\{f\colon \Sspv\to\C\right\}$, which can be evaluated in spherical coordinates $\left(\theta,\phi\right)$ as follows:
\begin{szMathBox}%
\begin{equation}\label{eq:bkgnd_sh_def}
	Y_{lm}\left(\theta,\phi\right) = A_{lm}P_l^m\left(\cos\theta\right)e^{im \phi},
\end{equation}
\vspace{-4mm}
\end{szMathBox}
\noindent where $A_{lm}=\sqrt{\frac{2l+1}{4\pi}\frac{\left(l-m\right)!}{\left(l+m\right)!}}$ and $P_l^m$ denotes the associated Legendre function of order $l$ and degree $m$ (Supplemental Equation~\eqref{eq:bkgnd_sh_def_final}).
$\left\{ Y_{lm} \mid \left(l,m\right)\in \Z^2,\, \abs{m}\le l\right\}$
is an orthonormal basis of $\calF\left(\Sspv,\C\right)$.
In other words, any spherical function $f\in \calF\left(\Sspv,\C\right)$ is equal to an infinite number of the linear combination of SH as
\begin{align}\label{eq:bkgnd_sh_lincomb}
	f&=\sum_{l=0}^\infty\sum_{m=-l}^l \rmf_{lm}Y_{lm},
\end{align}
and the \emph{coefficient} $\rmf_{lm}$ is computed as
\begin{align}\label{eq:bkgnd_sh_coeff}
	\rmf_{lm}&=\lrangle{Y_{lm},f}_{\calF\left(\Sspv,\C\right)} \coloneqq \int_{\Sspv}{Y_{lm}^*\left(\homega\right)f\left(\homega\right)\rmd\homega},
\end{align}
where the integration over the sphere~$\Sspv$ is defined with the solid angle measure $\rmd\homega=\sin\theta\rmd \theta\rmd \phi$,
and $z^*$ indicates the complex conjugate of an arbitrary $z\in\C$.
Note that when the domain of an inner product is clear in context, we just write the inner product as $\lrangle{Y_{lm},f}_{\calF}$ for the sake of simplicity.

From Equation~\eqref{eq:bkgnd_sh_coeff}, a numeric vector which consists of such $\rmf_{lm}$ called \emph{coefficient vector}, which encodes frequency-domain information of the spherical function $f$. 
While an infinite dimensional coefficient vector $\left[\rmf_{00}, \rmf_{1,-1}, \rmf_{10}, \rmf_{11}, \cdots\right]^T$ represents continuously defined $f$ without loss of information, 
we can take the {projection of} $f$ {on SH up to} order $l_{\mathrm{max}}$,
 and store it into a finite coefficient vector%
 $\left[\rmf_{00},\cdots, \rmf_{l_{\mathrm{max}},l_{\mathrm{max}}}\right]^T$ of $O\left(l_{\mathrm{max}}^2\right)$ entries.

\subsubsection{Coefficient matrix and radiance transfer}
\NEW{In rendering pipelines or other frequency-domain analysis, many methods can be represented as functions of spherical functions (linear operator).}
SH also represents linear operators on spherical functions into discrete coefficients, called \emph{coefficient matrix}.
Suppose that $T\colon \calF\left(\Sspv,\C\right) \to \calF\left(\Sspv,\C\right)$ be a linear operator on spherical functions.
Similar to Equation~\eqref{eq:bkgnd_sh_coeff}, the linear operator $T$ can be represented by discrete SH coefficients $\rmT_{l_om_o,l_im_i}$ as
\begin{equation} \label{eq:bkgnd_SHcoeff_operator}
	\rmT_{l_om_o,l_im_i} = \lrangle{Y_{l_om_o}, T\left[Y_{l_im_i}\right]}_\calF,
\end{equation}
where the subscript $i$ and $o$ in $l$ and $m$ stands for input and output.
The evaluation of $T$ at a function $f\in\calF\left(\Sspv,\C\right)$ can be considered as a matrix-vector multiplication in the SH coefficient space as
\begin{equation}\label{eq:bkgnd_SHcoeff_matmul}
	\lrangle{Y_{l_om_o}, T\left[f\right]}_{\calF} =  \sum_{l_i,m_i} \rmT_{l_om_o,l_im_i} \rmf_{l_im_i},
\end{equation}
where $\lrangle{Y_{l_om_o}, T\left[f\right]}_\calF$ is the coefficient of the output function $T\left[f\right]$, obtained by Equation~\eqref{eq:bkgnd_sh_coeff}.

In computer graphics, a BRDF\footnote{{We consider a cosine-weighted BRDF which already contains the term $\abs{\hn \cdot \homega_i}$.}} $\rho\colon \Sspv\times\Sspv\to\R$ can be characterized by a linear operator $\rho_{\calF}\colon \calF\left(\Sspv,\C\right) \to \calF\left(\Sspv,\C\right)$ which acts as the rendering equation:
\begin{equation} \label{eq:rendering_eq}
	\rho_{\calF}\left[L^{\mathrm{in}}\right]\left(\homega_o\right) =\int_{\Sspv}{ \rho\left(\homega_i,\homega_o\right) L^{\mathrm{in}}\left(\homega_i\right) \rmd\homega_i},
\end{equation}

\noindent for any incident radiance function of a direction $L^{\mathrm{in}}$.
\NEW{Taking the matrix product of the SH coefficient matrix of $\rho_\calF$, also called the radiance transfer matrix, and the coefficient vector of $L^{\mathrm{in}}$  is the core operation in the efficient environment lighting \cite{ramamoorthi2001signal} and PRT \cite{sloan2002precomputed} methods.}

Moreover, the isotropy constraint of the BRDF (in general, an azimuthal symmetric operator) yields increasing the sparsity of SH coefficients, which can be written with fewer indices as~\cite{ramamoorthi2001signal,ramamoorthi2002frequency}
\begin{equation} \label{eq:bkgnd_SHcoeff_isobrdf}
	\rho_{l_om_o,l_im_i}=\delta_{m_om_i}\rho_{l_ol_im_i},
\end{equation}
where $\delta_{m_om_i}$ indicates the Kronecker delta.
Note that while a general linear operator requires 
$O\left(l_\mathrm{max}^4\right)$ 
SH coefficients in Equation~\eqref{eq:bkgnd_SHcoeff_operator}, azimuthal symmetry described in Equation~\eqref{eq:bkgnd_SHcoeff_isobrdf} reduces the number of coefficients to 
$O\left(l_\mathrm{max}^3\right)$.

\subsubsection{Rotation invariance}
One of the most important properties of SH is rotation invariance, 
\NEW{which} allows us to efficiently convert SH coefficients with respect to another frame without loss of information.

A rotation can be considered as a linear operator. Given rotation transform $\vec R\in \SOgroupv$, the rotation on spherical functions rather than vectors is denoted by $\vec R_{\calF}\colon \calF\left(\Sspv,\C\right)\to \calF\left(\Sspv,\C\right)$ and acts as
\begin{equation}\label{eq:bkgnd_rot_sph_func}
	\vec R_{\calF}\left[f\right]\left(\homega\right)=f\left(\vec R^{-1}\homega\right).
\end{equation}

The coefficient matrix of the rotation $\vec R_\calF$ is obtained from Equation~\eqref{eq:bkgnd_SHcoeff_operator}. It can be written with the Kronecker delta and a special function $D_{mm'}^l$, which is called a Wigner D-function as
\begin{szMathBox}%
	\vspace{1mm}
	\begin{equation} \label{eq:wignerD_def}
		\lrangle{Y_{lm}, \vec R_{\calF}\left[Y_{l'm'}\right]}_\calF = \delta_{ll'}D_{mm'}^l\left(\vec R\right).
	\end{equation}
	\vspace{-3mm}
\end{szMathBox}
The rotation invariance of SH is stated as the block diagonal constraint of the coefficient matrices of rotations due to the term $\delta_{ll'}$ in Equation~\eqref{eq:wignerD_def}, which is also visualized in %
Figure~\ref{fig:theory_rotmat_compare}(a).
This property also implies that we can commute the SH projection of a function and a rotation without loss of information. 
We refer to Supplemental Figures~\ref{fig:bkgnd_sphfunc_rotation}(a),~\ref{fig:bkgnd_SH_rotmat}, and~\ref{fig:bkgnd_SH_rotation_property} in Supplemental Section~\ref{sec:bkgnd_SH_rotation} for further description and visualization.

\subsubsection{Spherical convolution}
\NEW{Spherical convolution is defined for a kernel $k\colon \left[0,\pi\right] \to \C$, a spherical function with azimuthal symmetry $k\left(\theta,\phi\right)=k\left(\theta\right)$, and any spherical function $f$ as follows.}
\begin{equation}\label{eq:bkgnd_sph_conv}
	k*f\left(\homega\right) = \int_{\Sspv}{ k\left(\cos^{-1}\left(\homega\cdot\homega'\right)\right)f\left(\homega'\right) \rmd\homega'}.
\end{equation}
The definition of spherical convolution in 
Equation~\NEW{\eqref{eq:bkgnd_sph_conv}} is determined from its important properties, linearity, and rotation equivariance \NEW{for $f$}. 
Conversely, it is known that a rotation equivariant linear operator on spherical functions is equivalent to a convolution with some kernel $k$.

SH provide an efficient computation of this convolution. The SH coefficients of the convolution result, $\rmf_{lm}'\coloneqq \lrangle{Y_{lm}, k\ast f}_\calF$ is evaluated by
\begin{szMathBox}%
\vspace{1mm}
\begin{equation}\label{eq:bkgnd_SH_sph_conv}
	\rmf_{lm}' = \sqrt\frac{4\pi}{2l+1} \rmk_{l0}\rmf_{lm},
\end{equation}
\vspace{-2mm}
\end{szMathBox}
\noindent which is just an \NEW{element}-wise product of the kernel and the input function in SH coefficients. Note that it is analogous to the convolution theorem of the Fourier transform in Euclidean domains.

In \NEW{a} rendering context, a BRDF is encoded to a coefficient matrix with $O\left(l_{\mathrm{max}}^4\right)$ space complexity. 
However, assuming \NEW{P}hong-like BRDFs with rotation equivariance whose reflected lobe just rotates as the incident ray rotates, a BRDF can be represented as a spherical convolution kernel~\cite{sloan2002precomputed}, which can lead to more efficient computation from its $O\left(l_{\mathrm{max}}\right)$ sparsity.

\subsubsection{Real and complex SH}

While SH defined in Equation~\eqref{eq:bkgnd_sh_def} are complex-valued functions, 
real-SH $Y_{lm}^R$ are \NEW{also} defined as follows:%
\begin{equation}\label{eq:bkgnd_real_SH}
	Y_{lm}^R  = \begin{cases}
		\sqrt{2}\Re Y_{lm} = \frac1{\sqrt2}\left( Y_{lm} + \left(-1\right)^mY_{l,-m} \right)& m>0\\
		Y_{lm}& m=0 \\
		\sqrt{2}\Im Y_{l\abs{m}} = \frac i{\sqrt2}\left( \left(-1\right)^mY_{lm} - Y_{l,-m} \right) & m < 0
	\end{cases}.
\end{equation}
We will sometimes call $Y_{lm}$ defined in Equation~\eqref{eq:bkgnd_sh_def} \emph{complex} SH to distinguish \NEW{from} real ones. 
Note that the real SH also satisfy orthonormality and rotation invariance, but \NEW{they} always convert real-valued functions into real-valued coefficients.

For the rotation transform of real SH coefficients, it can be written similarly to complex SH as
\begin{equation}\label{eq:bkgnd_real_wignerD}
	\lrangle{Y_{lm}^R, \vec R_{\calF}\left[Y_{l'm'}^R\right]}_\calF = \delta_{ll'}D_{mm'}^{l,R}\left(\vec R\right),
\end{equation}
where  $D_{mm'}^{l,R}$ is named {real Wigner D-functions}, and it
can be evaluated simply as a linear combination of complex-valued $D_{\pm m, \pm m'}^l$ (Supplemental~Equation~\eqref{eq:wignerD_RC}). %
See Supplemental Section~\ref{sec:bkgnd_SH_RC} for more details.

For computational efficiency, most existing computer graphics works use real SH. %
However, both real and complex SH should be considered for our polarized SH, which will be introduced in Section~\ref{sec:our_theory}.

%% file: notations_all_v2.tex
\begin{table}[tbp]\small
	\caption{\label{tb:quantities_gen} 
		Lists of notations and symbols used in this paper.}
	\vspace{-4mm}
	\rowcolors{2}{lightgray!0}{lightgray!50}
	\setlength{\tabcolsep}{2pt}
	\renewcommand{\arraystretch}{1.15}
	\begin{tabular}{m{0.12\linewidth} m{0.17\linewidth} m{0.67\linewidth}} %
		\thickhline
		\textbf{\sffamily Notation} & & \\
		\hline
		
		$\bfx, \bfy, \cdots$
		& $\in\R^N$
		& \emph{Numeric} $N$-dimensional vectors, lowercase Latin letters with boldface (including Stokes component vector) \\
		
		$\bfA, \bfB, \cdots$
		& $\in\R^{M\times N}$
		& \emph{Numeric} $M\times N$ matrices, uppercase Latin letters with boldface (including Mueller matrices) \\
		
		$\vec x, \vec y, \cdots$
		& $\in\vec \R^N$
		& \emph{Geometric} $N$-dimensional vectors, lowercase Latin letters accented single side arrow \\
		
		$\vec A, \vec B, \cdots$
		& $\in\vec \R^{M\times N}$
		& \emph{Geometric} $M\times N$ matrices, uppercase Latin letters accented single side arrow \\
		
		$\dvx, \dvy,\cdots$
		& $\in\STKsp{\homega}$
		& \emph{Stokes vectors} (geometric), lowercase Latin letters accented both side arrow \\
		
		$\dvA, \dvB,\cdots$
		& $\in\MUEsp{\homega_i}{\homega_o}$
		& \emph{Mueller transforms} (geometric), uppercase Latin letters accented both side arrow  \\

		\thickhline
	\end{tabular}
	\rowcolors{2}{lightgray!0}{lightgray!50}
	\begin{tabular}{m{0.12\linewidth} m{0.17\linewidth} m{0.67\linewidth}} %
		\thickhline
		\textbf{\sffamily Symbol} & & \\
		\hline
		$\hat \omega$
		& $\in\Sspv$ %
		& Directions (unit vector), where $\Sspv$ is unit sphere \\
		
		$\frF$
		& $\in\FRsp$ 
		& Orthonormal \emph{frames} in 3D, uppercase Latin letter F with boldface accented single side arrow\\
		
		$\bfR$
		& $\in\SOgroup $ 
		& \emph{Numeric} 3D rotation matrices \\
		
		$\vec R$
		&$\in\SOgroupv $ 
		& \emph{Geometric} 3D rotation transforms \\
		
		$\calF\left(X, Y\right)$
		&
		& \emph{Function space} from $X$ into $Y$, for any sets $X$ and $Y$ \\
		
		$\STKsp{\homega}$
		&
		& \emph{Stokes space}: set of all Stokes vectors of a ray along direction $\homega$ \\
		
		$\MUEsp{\homega_i}{\homega_o}$
		&
		& \emph{Mueller space} from $\STKsp{\homega_i}$ to $\STKsp{\homega_o}$ \\
		
		\thickhline
	\end{tabular}

	\rowcolors{2}{lightgray!0}{lightgray!50}
	\begin{tabular}{m{0.15\linewidth} m{0.14\linewidth} m{0.67\linewidth}}
		\thickhline
		\textbf{\sffamily Operator} & & \\
		\hline
		
		$\left[\bfs\right]_{\frF}$
		& $=\dvs$ %
		& Stokes component vector $\bfs$ to Stokes vector $\dvs$ w.r.t. frame $\frF$ \\
		
		$\left[\dvs\right]^{\frF}$
		& $=\bfs$ %
		& Stokes vector $\dvs$ to Stokes component vector $\bfs$  w.r.t. a frame $\frF$ \\
		
		$\left[\bfM\right]_{\frF_1\to\frF_2}$
		& $=\dvM$ %
		& Mueller matrix $\bfM$ to geometric Mueller transform $\dvM$ w.r.t. frames $\frF_1,\frF_2$\\
		
		$\left[\dvM\right]^{\frF_1\to\frF_2}$
		& $=\bfM$ %
		& Mueller transform $\dvM$ to numeric Mueller matrix $\bfM$ w.r.t. frames $\frF_1,\frF_2$\\
		
		$z^*$
		& $=x-yi$
		& Complex conjugation of $z=x+yi\in\C$ \\
		
		\NEW{$\Re z$, $\Im z$}
		& \NEW{$=x, y$ }
		& \NEW{Real and imaginary parts of $z=x+yi\in \C$} \\
		
		\multicolumn{2}{l}{$\R^2\left(z\right)$, $\C\left(\left[x,y\right]^T\right)$}
		& Conversion between complex number $z=x+yi\in\C$ and $\left[x, y\right]^T\in\R^2$ (Eq.~\eqref{eq:theory_complex_to_real}) \\
		
		$\R^{2\times2}\left(z\right)$
		& %
		& Eq.~\eqref{eq:theory_comp2mat}, Conversion from complex number to 2D real numeric matrix \\
		
		\multicolumn{2}{l}{$\Ciso\left(\bfM\right)$, $\Cconj\left(\bfM\right)$}
		& Eq.~\eqref{eq:theory_mat2comp}, Conversion from $2\times 2$ real matrix $\bfM$ to two complex numbers respectively \\
		
		\thickhline
	\end{tabular}
	\vspace{+6mm}
	\caption{\label{tb:rot_inner} 
		List of rotations and inner products in various quantities.
 	}
	\vspace{-4mm}
	\rowcolors{2}{lightgray!0}{lightgray!50}
	\setlength{\tabcolsep}{2pt}
	\renewcommand{\arraystretch}{1.2}
	\begin{tabular}{m{0.2\linewidth} m{0.65\linewidth} m{0.11\linewidth}} %
		\thickhline
		Symbol   & Operand & Eq.\,num. \\
		
		\thickhline
		
		$\vec R \vec x$,\quad $\vec R \frF$
		& Geometric vectors $\vec x\in \Rspv$, and frames $\frF\in\FRsp$
		& \\
		
		$\vec R_\calS\dvs$
		& Stokes vectors $\dvs \in \calS$
		& Eq.~\eqref{eq:stokes_rotation} \\

		$\vec R_\calM \left[\dvM\right]$
		& Mueller transforms $\dvM \in \calM$
		& Eq.~\eqref{eq:theory_rot_mueller}\\

		$\vec R_\calF \left[f\right]\left(\homega\right)$
		& Scalar fields $f:\Sspv\to \C$
		& Eq.~\eqref{eq:bkgnd_rot_sph_func} \\

		$\vec R_\calF \left[\dvf\right]\left(\homega\right)$
		& Stokes vector fields $\dvf:\Sspv\to\STKsp{\homega}$
		& Eq.~\eqref{eq:theory_rot_STKfield} \\
		
		\hline

		$\lrangle{\dvs, \dvt}_\calS$
		& Stokes vectors $\dvs,\dvt\in\STKsp{\homega}$ (identical direction)
		& Eq.~\eqref{eq:bkgnd-stokes-inner} \\

		$\lrangle{f, g}_{\calF}$
		& Scalar fields $f,g:\Sspv\to \C$
		& Eq.~\eqref{eq:bkgnd_sh_coeff} \\

		$\lrangle{\dvf, \dvg}_{\calF} $
		& Stokes vector fields $\dvf,\dvg:\Sspv\to\STKsp{\homega}$
		& Eq.~\eqref{eq:inner_stokes_field}\\
		
		\thickhline
	\end{tabular}
\end{table}

%% file: background_polar.tex
\subsection{Polarization and Mueller Calculus}
\label{sec:background_polar}

\NEW{Given a local frame $\frF=\left[\hat x,\hat y,\hat z\right]$, the intensity of a polarized ray along the propagation direction $\hat z$ is characterized by the four Stokes parameters $\bfs=\left[s_0,s_1,s_2,s_3\right]^T$. Here, each component $s_0$ to $s_3$ indicates total intensity, linear polarization in horizontal/vertical direction, linear polarization in diagonal/anti-diagonal direction, and circular polarization, respectively. We refer novice readers to Supplemental Section~\ref{sec:background_polar_intro} for more introduction.}

When taking another local frame $\frF' = \vec R_{\hat z}\left(\vartheta\right)\frF$, obtained by rotating $\frF$ by $\vartheta$ along its $z$ axis, the Stokes parameters with respect to the new frame $\frF'$ is evaluated as
\begin{equation} \label{eq:stk_coord_convert}
	\bfs' = \bfC_{\frF\to \frF'}\bfs =  \begin{bmatrix}
		1&0&0&0 \\
		0&\cos2\vartheta&\sin2\vartheta&0 \\
		0&-\sin2\vartheta&\cos2\vartheta&0 \\
		0&0&0&1
	\end{bmatrix} \bfs.
\end{equation}
We can observe here that $s_0$ and $s_3$ behave as \emph{scalar}s, which are measured independent of local frames.
On the other hand, $s_1$ and $s_2$ are neither scalars nor coordinates of an ordinary vector, which must have $\vartheta$ rather than $2\vartheta$ in Equation~\eqref{eq:stk_coord_convert}.
This twice rotation property of $s_1$ and $s_2$ under coordinate conversion will be dealt as \emph{spin-2 functions} in %
Section~\ref{sec:stokes-on-sphere}.

\begin{figure}[tbp]
	\centering
	\vspace{-2mm}
	\includegraphics[width=\columnwidth]{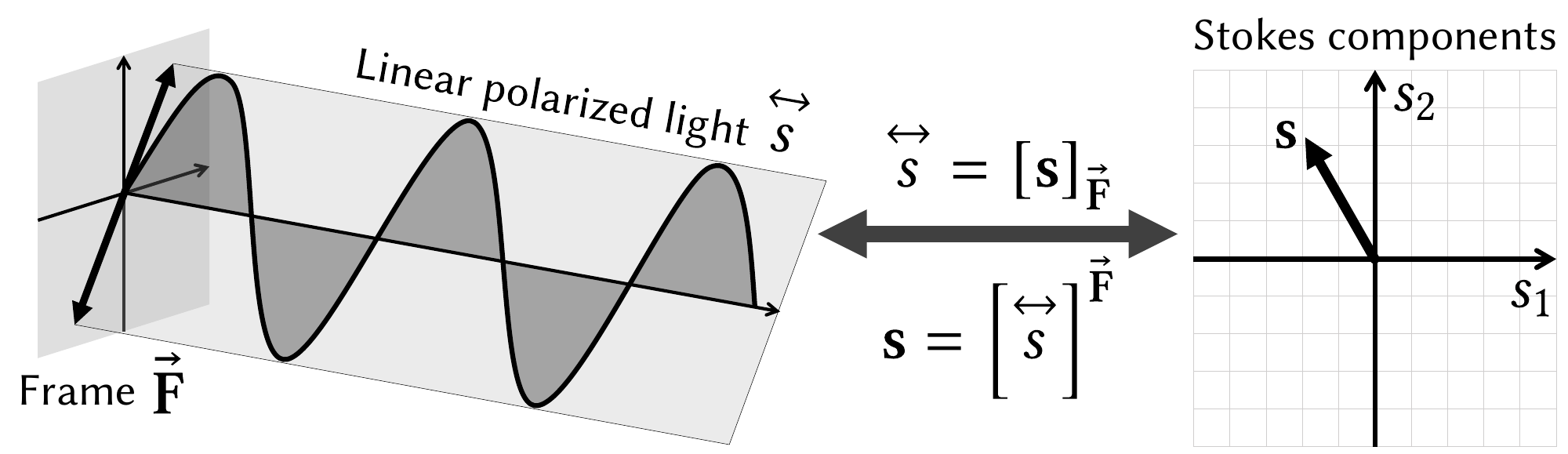}%
	\vspace{-2mm}
	\caption[]{\label{fig:bkgnd_stokes_rotation1}
		\NEW{Intensity of a polarized ray visualized in the left is characterized by a Stokes vector $\dvs$. While $\dvs$ is defined without any measurement frame, it can be measured into a Stokes component vector $\bfs$ under such a frame.}}
	\vspace{-2mm}		
\end{figure}

\subsubsection{Stokes vectors in numeric vs. geometric quantities}
\label{sec:mueller_num_geo}
As discussed before, dealing with polarized radiance needs careful attention for whether focusing on a ray itself as a physical object or Stokes parameter values $s_0,\cdots,s_3$, only 
\NEW{defined relative to a measurement frame associated with the ray.}
Note that we distinguish \emph{numeric} and \emph{geometric} quantities in this paper.
Due to the twice rotation property described in Equation~\eqref{eq:stk_coord_convert}, the polarized intensity of a ray should be considered as a novel type of geometric quantity, named \emph{Stokes vector} and denoted by $\dvs$.
Note that $\bfs$ and $\bfs'$ in Equation~\eqref{eq:stk_coord_convert} are numeric quantities and not geometric ones themselves since they depend on observing local frames.
Combining data of $\bfs$ and $\frF$ yields the geometric quantity $\dvs$, but it is not a matrix-vector product as ordinary vectors.
Thus, we write it in a novel notation as
\begin{szMathBox}
\begin{equation} \label{eq:stk_num2geo}
	\dvs=\left[\bfs\right]_{\frF} = \left[\bfs'\right]_{\frF'}.
\end{equation}
\end{szMathBox}
\noindent In addition, we call such numeric vector $\bfs$, the Stokes parameters observed under a certain frame, as \emph{Stokes component vector}\footnote{Note that we try to distinguish terminologies \emph{Stokes vectors} and \emph{Stokes components} as geometric and numeric quantities, respectively, so this distinction is not common in other literature. See also Supplemental Figure~\ref{fig:bkgnd_vector_coord}.}.
We also define the notation that evaluates the Stokes component vector of a given Stokes vector and the frame as
\begin{szMathBox}
\begin{equation} \label{eq:stk_geo2num}
	\bfs = \left[\dvs\right]^{\frF}.
\end{equation}
\end{szMathBox}
\noindent \NEW{Figure~\ref{fig:bkgnd_stokes_rotation1} visualizes it where the two-sided arrow in the left indicates the actual oscillation direction of a polarized ray characterized by a Stokes vector and the right plot shows \NEW{the} Stokes component vector under a local frame.}

We also denote $\STKsp{\homega}=\left\{ \left[\bfs\right]_{\frF} \mid \frF\in\FRsp, \extColz{\frF}=\homega \right\}$ as the \emph{Stokes space}, the set of all Stokes vector of rays along direction $\homega$,
where $\extColx{\frF}$, $\extColy{\frF}$, and $\extColz{\frF}$ indicate the local $x$, $y$, and $z$ axes of given frame $\frF$, respectively.

\begin{figure}[tbp]
	\centering
	\vspace{-2mm}
	\includegraphics[width=\columnwidth]{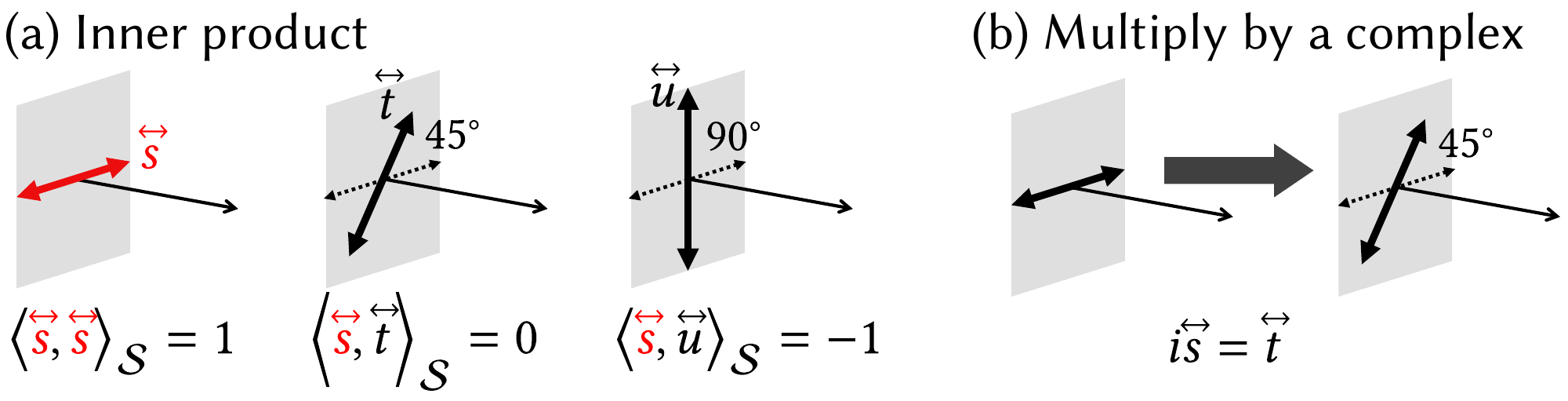}%
	\vspace{-3mm}
	\caption[]{\label{fig:bkgnd_stokes_operations}
		Additional basic operations on Stokes vectors are defined in (a) Equation~\eqref{eq:bkgnd-stokes-inner} and (b) Equation~\eqref{eq:bkgnd-stokes-compmult}.
	}
	\vspace{-3mm}		
\end{figure}
\paragraph{Stokes vector operations}
Binary operations on two Stokes vectors $\dvs=\left[\bfs\right]_{\frF_1}$ and $\dvt=\left[\bft\right]_{\frF_2}$ are defined only if they belong to the identical Stokes space. i.e., the ray directions are same ($\extColz{\frF_1}=\extColz{\frF_2}$).
If so, the addition \NEW{and the inner product are} defined by converting the Stokes vectors to the same frame as

\noindent
\NEW{\begin{minipage}{.5\linewidth}
	\begin{equation} \label{eq:bkgnd-stokes-addition}
		\dvs+\dvt \coloneqq \left[\bfs + \left[\dvt\right]^{\frF_1}\right]_{\frF_1},
	\end{equation}
\end{minipage}%
\begin{minipage}{.5\linewidth}
	\begin{equation} \label{eq:bkgnd-stokes-inner}
		\langle \dvs,\dvt \rangle_{\STKsp{\homega}} \coloneqq \bfs \cdot\left[\dvt\right]^{\frF_1},
	\end{equation}
\end{minipage}}
\NEW{respectively.}
\begin{figure}[tbp]
	\centering
	\vspace{-3mm}	
	\includegraphics[width=\columnwidth]{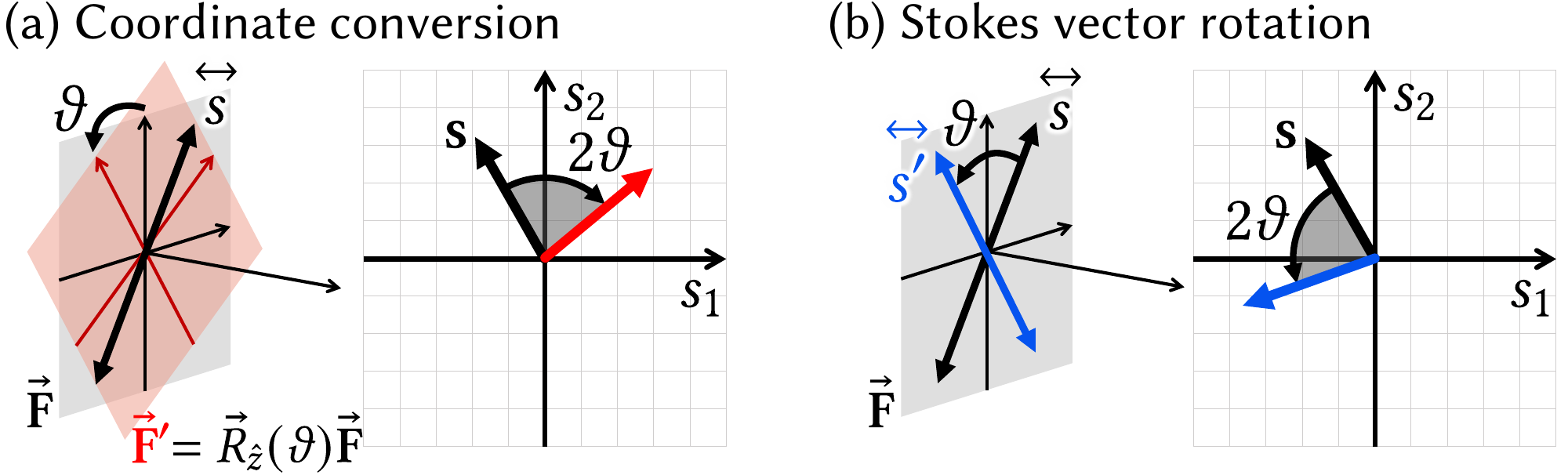}%
	\vspace{-4mm}
	\caption[]{\label{fig:bkgnd_stokes_rotation2}
		(a) When we fix the Stokes vector $\dvs$ and rotate the frame by $\vartheta$, the (numeric) Stokes components of $\dvs$ rotate by $-2\vartheta$.
		(b) Rotating the (geometric) Stokes vector itself by $\vartheta$ is equivalent to rotating its Stokes components by $2\vartheta$ with a fixed frame.}
	\vspace{-2mm}		
\end{figure}
\NEW{We also define the rotation of the underlying polarized ray of $\dvs$ itself. For a rotation $\vec R\in \SOgroupv$, to avoid confusion, we denote $\vec R_\calS$ as Stokes vector version of $\vec R$. Then for any Stokes vector $\dvs \in \STKsp{\homega}$, $\vec R_\calS$ acts as
\begin{equation}\label{eq:stokes_rotation}
	\vec R_\calS \dvs = \left[\left[\dvs\right]^{\frF}\right]_{\vec R\frF}\in\STKsp{\vec R\homega},
\end{equation}
where $\frF\left[{:},3\right]=\homega$. Figure~\ref{fig:bkgnd_stokes_rotation2} visualizes the difference between a coordinate conversion and a rotation around $\homega$.}
\NEW{Note} that these are frame-independently well-defined.

\paragraph{Spin-2 vs. full Stokes vectors}
Suppose that we have a Stokes vector $\dvs=\left[\bfs\right]_{\frF}$ with $\bfs=\left[s_0, s_1, s_2, s_3\right]^T$. 
To handle the special behaviors of linear polarization components $s_1$ and $s_2$, we sometimes need to process only these two separately from the four components. To do so, we define a \emph{spin-2} Stokes vector (in \emph{spin-2} Stokes space) %
as $\left[\left[s_1, s_2\right]^T\right]_{\frF}\in \calS_{\homega}^2$ in a similar way to Equation~\eqref{eq:stk_num2geo}. Then the original Stokes vector, also called a \emph{full Stokes vector} to be clear and written as $\dvs = s_0 \oplus \left[\left[s_1, s_2\right]^T\right]_{\frF} \oplus s_3$, where $\oplus$ symbol indicates the direct sum in linear algebra, which also can be considered as vector concatenation in numerical programming tools.

Now a spin-2 Stokes vector can also be written with a complex component as $\left[\left[s_1, s_2\right]^T\right]_{\frF} = \left[s_1+is_2\right]_{\frF}$. With this representation, multiplication by a complex number:
\begin{equation} \label{eq:bkgnd-stokes-compmult}
	z\left[s_1+is_2\right]_{\frF} = \left[z\left(s_1+is_2\right)\right]_{\frF},
\end{equation}
is well defined independent of choice of the frame $\frF$, which indicates scaling by $\abs{z}$ followed by rotating $\arg{z}/2$ around its ray direction, as illustrated in Figure~\ref{fig:bkgnd_stokes_operations}(b). Note that \NEW{while} other operations such as addition (Equation~\eqref{eq:bkgnd-stokes-addition}) and inner product (Equation~\eqref{eq:bkgnd-stokes-inner}) are defined both for spin-2 and full Stokes vectors, the complex multiple is only defined for spin-2 Stokes vectors.

\subsubsection{Mueller transform in numeric vs. geometric quantities}
Linear maps from Stokes vectors along $\homega_i$ to Stokes vectors along $\homega_o$, such as polarimetric BRDF and other polarized light interactions, are called \emph{Mueller transforms}. The set of these Mueller transforms is called a \emph{Mueller space} and written as
\begin{align}\nonumber
	\MUEsp{\homega_i}{\homega_o}\coloneqq& \left\{\dvM :\STKsp{\homega_i} \to \STKsp{\homega_o}  \mid \dvM\left(a \dvs + b\dvt\right) = a\dvM\dvs + b\dvM\dvt \right. \\
	& \left. \text{for any }a,b\in\R \text{ and } \dvs,\dvt\in\STKsp{\homega_i} \right\}.
\end{align}
Similar to Stokes vector, a Mueller transform $\dvM\in \MUEsp{\homega_i}{\homega_o}$ is a geometric quantity, and it can be measured into a numeric matrix $\bfM\in\R^{4\times 4}$, named \emph{Mueller matrix}\footnote{Similar to Stokes vector and Stokes component vectors, we distinguish terminologies \emph{Mueller transforms} and \emph{Mueller matrices}.}
with respect to observing local frames. Here, we need two frame\NEW{s} $\frF_i$ and $\frF_o$ with $\extColz{\frF_i}=\homega_i$ and $\extColz{\frF_o}=\homega_o$ and relations between $\dvM$ and $\bfM$ is notated as follows:
\begin{szMathBox}
\begin{equation} \label{eq:mue_geo_num}
	\dvM = \left[\bfM\right]_{\frF_1\to\frF_2}, \quad \bfM =\left[\dvM\right]^{\frF_1\to\frF_2},
\end{equation}
\end{szMathBox}
\noindent similar to Equations~\eqref{eq:stk_num2geo} and \eqref{eq:stk_geo2num}.

%% file: challengesinSH.tex
\section{Challenges of Stokes vector fields in angular domain}
\label{sec:stokes-on-sphere}

\NEW{Stokes vector radiance as a function on an angular domain, called a \emph{Stokes vector field}, is a fundamental quantity to describe polarized transport. It has been the subject of previous work such as polarized environment illumination, including the sky dome~\cite{wilkie2004analytical, riviere2017polarization, wilkie2021fitted} and polarized perspective images in all the existing polarization renderers. However, the challenges of dealing with Stokes vector fields have rarely been discussed. In this section, we introduce such challenges in terms of different continuity conditions from scalar fields in Section~\ref{sec:stokes-on-sphere-conti}.
It raises the necessity of novel basis functions rather than scalar SH for frequency domain methods of polarized light.
In Section~\ref{sec:stokes-on-sphere-operation}, we additionally define basic operations on Stokes vector fields, which are required for frequency domain analysis. %
}

\subsection{Continuity of Scalar vs. Stokes Vector Fields}
\label{sec:stokes-on-sphere-conti}
A Stokes vector field on the unit sphere\footnote{Rigorously, it should be written as $\left\{ \dvf\colon \Sspv\to \cup_{\homega\in\Sspv}\STKsp{\homega} \mid \forall\homega\in\Sspv, \dvf\left(\homega\right)\in \STKsp{\homega} \right\}$, but we write as the main text for the sake of simplicity and better intuition.} \NEW{is formulated} as $\dvf\colon \Sspv\to \calS_{\homega}$.
Here we can observe that, unlike scalar radiance, the value of the Stokes vector field at each direction $\homega$ lies on the different Stokes space, i.e., $\dvf\left(\homega\right)\in\STKsp{\homega}$, depending on the direction $\homega$.

The simple way to measure a Stokes vector field is to assign local frames for each direction $\homega$.
We call this type of function the function from directions $\homega\in\Sspv$ to local frames $\frF\left(\homega\right)\in\FRsp$ with $\extColz{\frF\left(\homega\right)}=\homega$ as a \emph{frame field}.

\begin{figure}[t]
	\centering
	\includegraphics[width=\columnwidth]{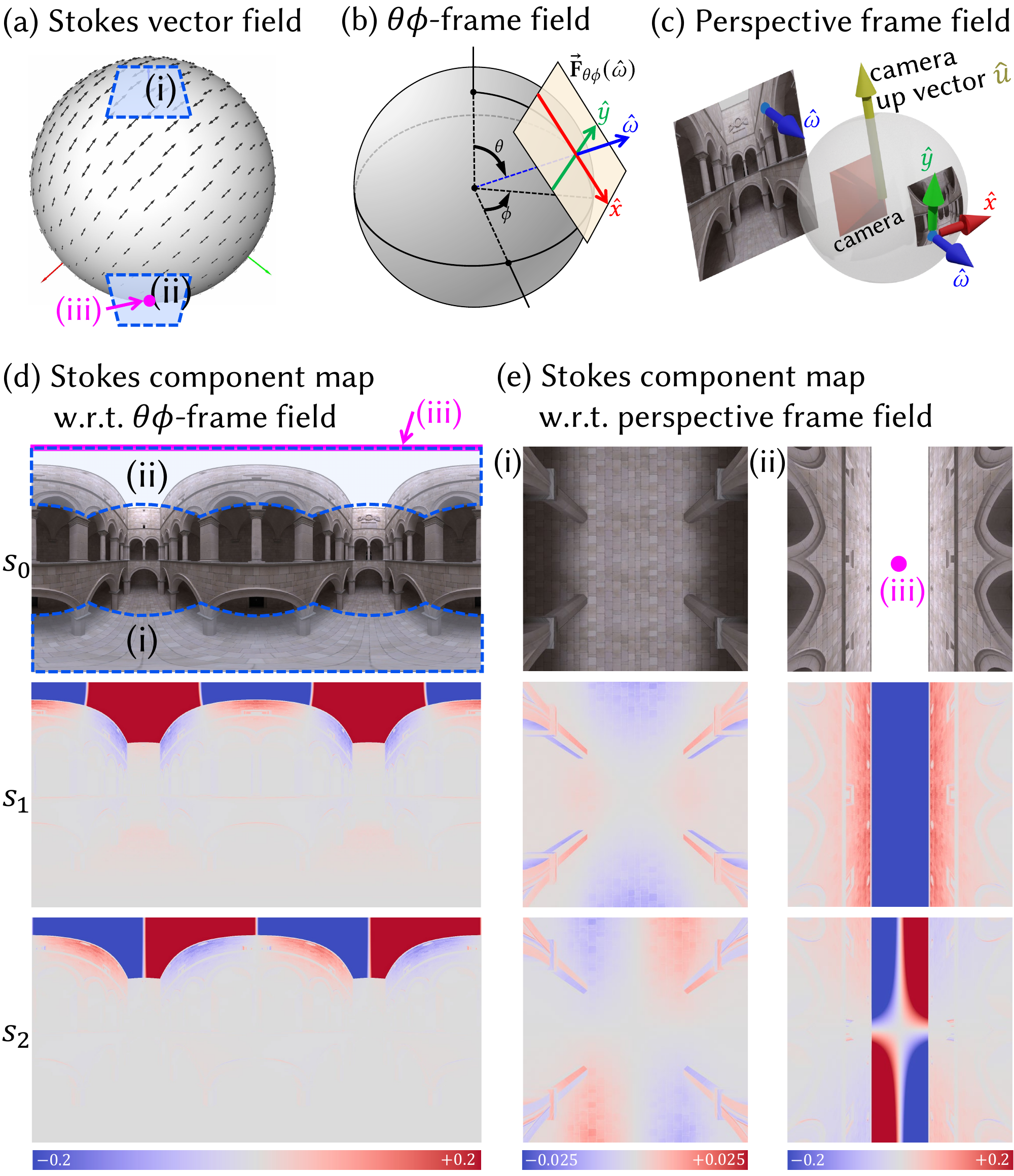}%
	\vspace{-3mm}
	\caption[]{\label{fig:bkgnd_penvmap_visualize}
		Visualizing a Stokes vector field (polarized environment map) depends on the choice of frame fields. Taking Stokes components of Stokes vector field (a) with respect to a typical $\theta\phi$-frame field (b)
		yields equirectangular images shown in (d). Using a perspective frame field used in Mitsuba~3 renderer,
		several perspective images are visualized as (e). \NEW{Note that while the $s_1$ component (ii) %
		in (e) at the sky, especially (iii), has consistent signs of values, and the component in (d) under a different frame field has a different trend of values.}}
	\vspace{-2mm}		
\end{figure}

\begin{figure}[t]
	\centering
	\includegraphics[width=\columnwidth]{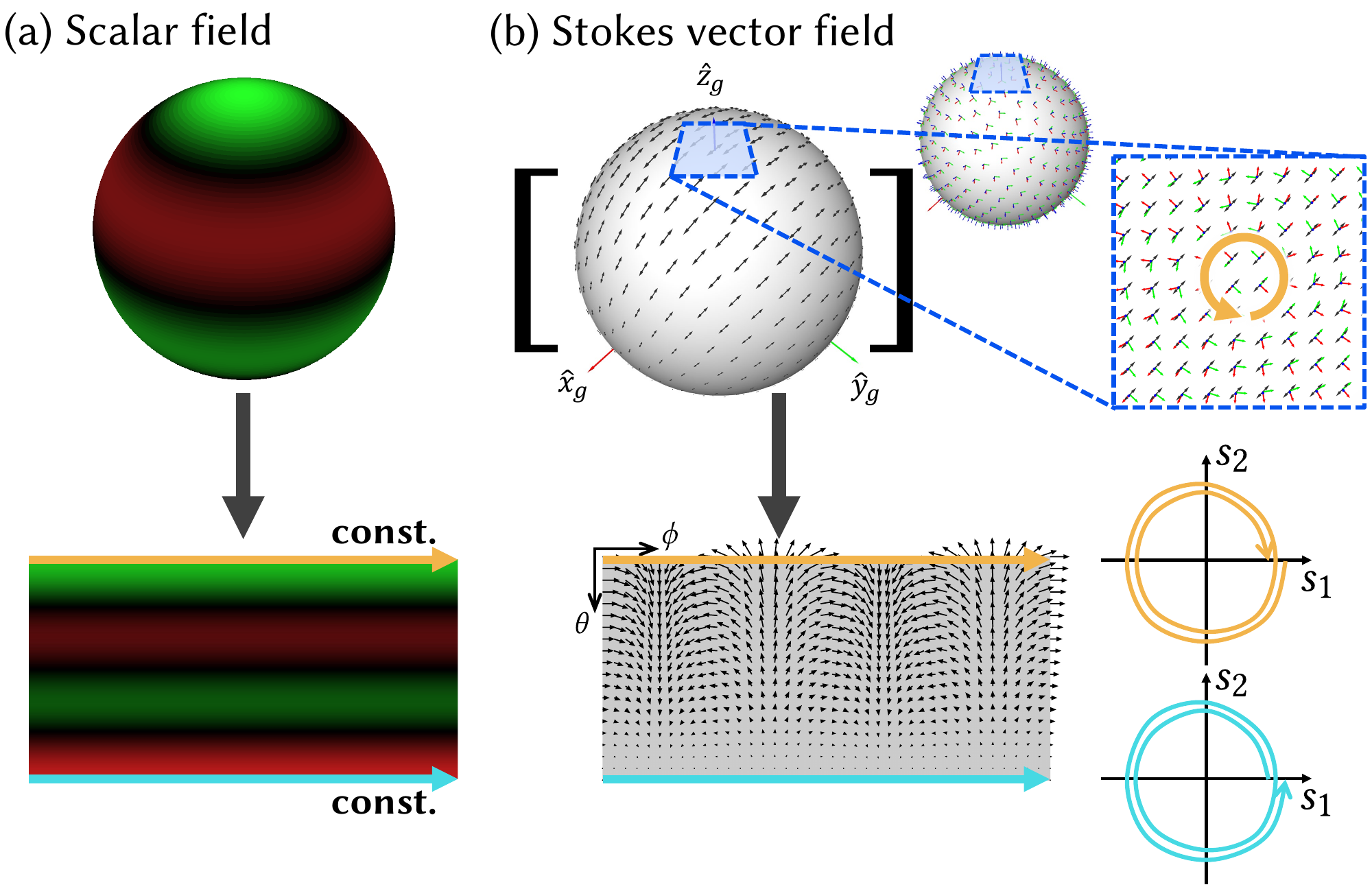}%
	\vspace{-3mm}
	\caption[]{\label{fig:bkgnd_continuity_stokes_field}
		Analyzing continuity and smoothness for Stokes vector fields in $\theta\phi$ domain.
		(a) The visualization of a Stokes vector field.
		As a \emph{geometric} quantity, Stokes vector fields are continuous and smooth on the entire sphere, including the zenith.
		(b) To make the \emph{geometric} Stokes vector fields to \emph{numeric} Stokes components, we can assign the specific frame field, named \NEW{$\theta\phi$-frame field.}%
}
	\vspace{-2mm}		
\end{figure}

Among the various choices of frame field, one choice is a $\theta\phi$ frame field $\frF_{\theta\phi}$, 
defined by aligning local $x$ and $y$ axes along longitudinal and latitudinal directions as \NEW{shown in Figure~\ref{fig:bkgnd_penvmap_visualize}(b) and Supplemental Equation~\eqref{eq:field-tp-frame}.}

\NEW{Note that $\dvf$ can be visualized as double-sided arrows (following Figure~\ref{fig:bkgnd_stokes_rotation1}) on tangent planes of the sphere, as shown in Figure~\ref{fig:bkgnd_penvmap_visualize}(a).}
After choosing the frame field, 
$\dvf$ can be converted into four scalar fields on the sphere based on the numeric--geometric conversion notation we defined in Equation~\eqref{eq:stk_num2geo} as
\begin{equation}\label{eq:stkfield-component}
	\left[\dvf\left(\homega\right)\right]^{\frF_{\theta\phi}\left(\homega\right)} = \begin{bmatrix}
		f_0\left(\theta,\phi\right) & f_1\left(\theta,\phi\right) & f_2 \left(\theta,\phi\right) & f_3 \left(\theta,\phi\right)
	\end{bmatrix}^T.
\end{equation}
\NEW{We can visualize scalar fields of each component $f_i$ as equirectangular images by unwrapping the spherical domain into the rectangle of spherical coordinates $\theta$ and $\phi$ as Figure~\ref{fig:bkgnd_penvmap_visualize}(d).}

However, there is an issue that \emph{any} frame field always has a \emph{singularity}, which means a local frame cannot be \NEW{continuously} defined due to the Hairy Ball Theorem~\cite{nash1983topology}.
For example, $\frF_{\theta\phi}\left(\homega\right)$ has two singularities\footnote{We let axis symbols without subscripts such as $\hat x$ and $\hat y$ denote values of a frame field, which are used to measure Stokes vectors along each direction, while those with subscript $g$ such as $\hat z_g$ denote a fixed global frame which is used to assign spherical coordinates on a sphere.} at 
\NEW{$\hat z_g$ ($\theta=0$)} and 
\NEW{$-\hat z_g$ ($\theta=\pi$)}.
\NEW{In the rectangle domain}, the top (and bottom, respectively) 
edge indicates just a single point $\hat z_g$ ($-\hat z_g$, respectively) 
but has different local frames that rotate one turn in counterclockwise (clockwise, respectively) as $\phi$ increases. 
\NEW{It yields different continuity conditions for scalar and Stokes vector fields.}
\NEW{While scalar fields (e.g., scalar radiance, $f_0$ or $f_3$) have constant values at those top and bottom edges, a two-dimensional numeric vector $\left[f_1\left(0,\phi\right), f_2\left(0,\phi\right)\right]$ rotates twice in clockwise as $\phi$ increases from $0$ to $2\pi$ due to rotation of the frame field $\frF_{\theta\phi}\left(0,\phi\right)$, and similarly for $\theta=\pi$. These difference are illustrated in Figure~\ref{fig:bkgnd_continuity_stokes_field}.}

Note that \NEW{such pair of} spherical functions with the continuity condition of twice rotation 
such as $s_1$ and $s_2$ Stokes components are called \emph{spin-2} functions, 

and scalar functions with the \NEW{constant} continuity condition 
are called \emph{spin-0} functions.

\begin{figure}[tpb]
	\centering
	\vspace{-3mm}
	\includegraphics[width=\linewidth]{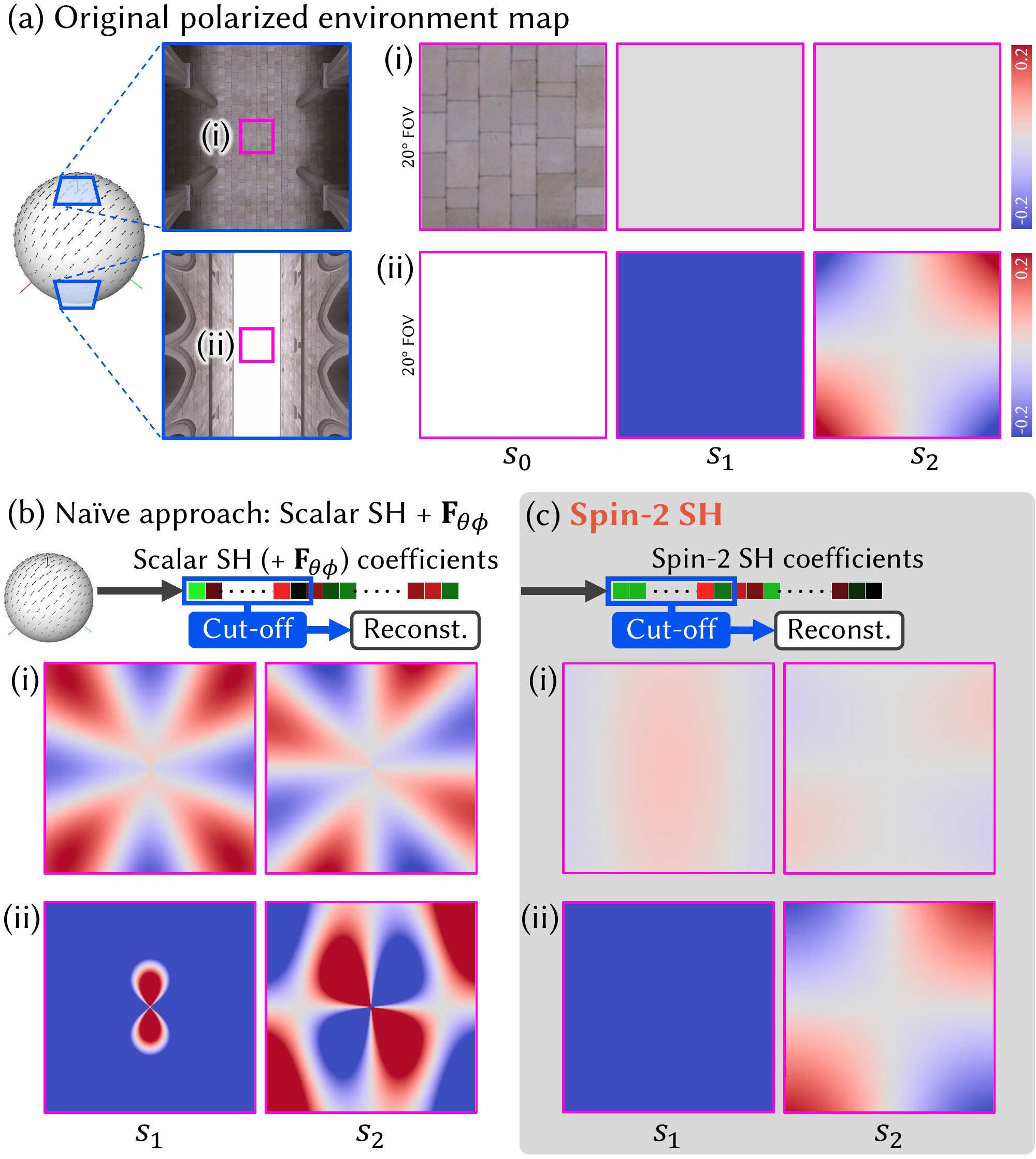}%
	\vspace{-3mm}
	\caption[]{\label{fig:theory_penvmap_singularity}
		We propose a frequency-domain analysis framework of Stokes vector fields, which is represented by a polarized environment map here. Then, we need spin-2 spherical harmonics rather than conventional ones to avoid the singularity problem.
		See Figure~\ref{fig:theory_penvmap_rotation} for rotation invariance of spin-2 SH, and see Figure~\ref{fig:theory_rotmat_compare} and Equation~\eqref{eq:theory_PSH_rotmat} for how the coefficient matrix of rotation under conventional SH (Wigner D-functions) can be utilized to spin-2 SH.
	}
	\vspace{-4mm}		
\end{figure}

\NEW{To construct a frequency domain method similar to ones based on scalar SH in scalar rendering, one may consider a na\"ive approach to apply scalar SH combined with the $\theta\phi$-frame field as a basis of Stokes vector fields. However, this approach raises the singularity problem due to the different continuity conditions between scalar and Stokes vector fields.
In Figure~\ref{fig:theory_penvmap_singularity}, the $s_1$ and $s_2$ components of the original Stokes vector field are nearly flat around $\pm \hat z_g$ (views (i) and (ii)), but its %
\NEW{projection onto the basis obtained by the na\"ive approach} yields (i) too high-frequency change
at (b) or (ii) singularity at (b). This is a fundamentally different feature from how the conventional SH behaved on scalar fields, which always converts finite coefficients to continuous functions and has a smoothing role.
We also point out that this singularity problem also implies a violation of rotation invariance. We refer to Figure~\ref{fig:theory_penvmap_rotation}, which is described in Section~\ref{sec:theory-rotinv-valid}, and Supplemental Section~\ref{sec:stokes-vector-fields} for more discussion.}

In summary, the different continuity conditions %
are an essential difference in the 
\NEW{nature} of Stokes vector fields.
Although we only show the case of the $\theta\phi$ frame field here, Stokes vector fields \emph{always} have different properties in terms of continuity regardless of which frame field is used.

\subsection{\NEW{Stokes Vector Fields Operations}}
\label{sec:stokes-on-sphere-operation}
\NEW{To discuss bases for Stokes vector fields, we should define several operations on Stokes vector fields. It can be done by generalizing scalar field operations in Section~\ref{sec:background_sh}, based on Stokes vectors operations in Section~\ref{sec:background_polar}.
The inner product of two Stokes vector fields $\dvf$ and $\dvg:\Sspv\to\STKsp{\homega}$ is defined as follows:
\begin{equation}\label{eq:inner_stokes_field}
	\lrangle{\dvf, \dvg}_\calF \coloneqq \int_{\Sspv}{\lrangle{\dvf\left(\homega\right), \dvg\left(\homega\right)}_\calS \rmd\homega}.
\end{equation}
In addition, the rotation acting on Stokes vector fields by $\vec R\in\SOgroupv$ is defined by
\begin{equation}\label{eq:theory_rot_STKfield}
	\vec R_\calF \left[\dvf\right]\brahomega =\vec R_\calS \left(\dvf\left(\vec R^{-1}\homega\right)\right),
\end{equation}
for any $\dvf$.
We summarize inner products and rotations on different types of quantities in Table~\ref{tb:rot_inner}.}
	
\begin{figure}[tp]
	\centering
	\includegraphics[width=\linewidth]{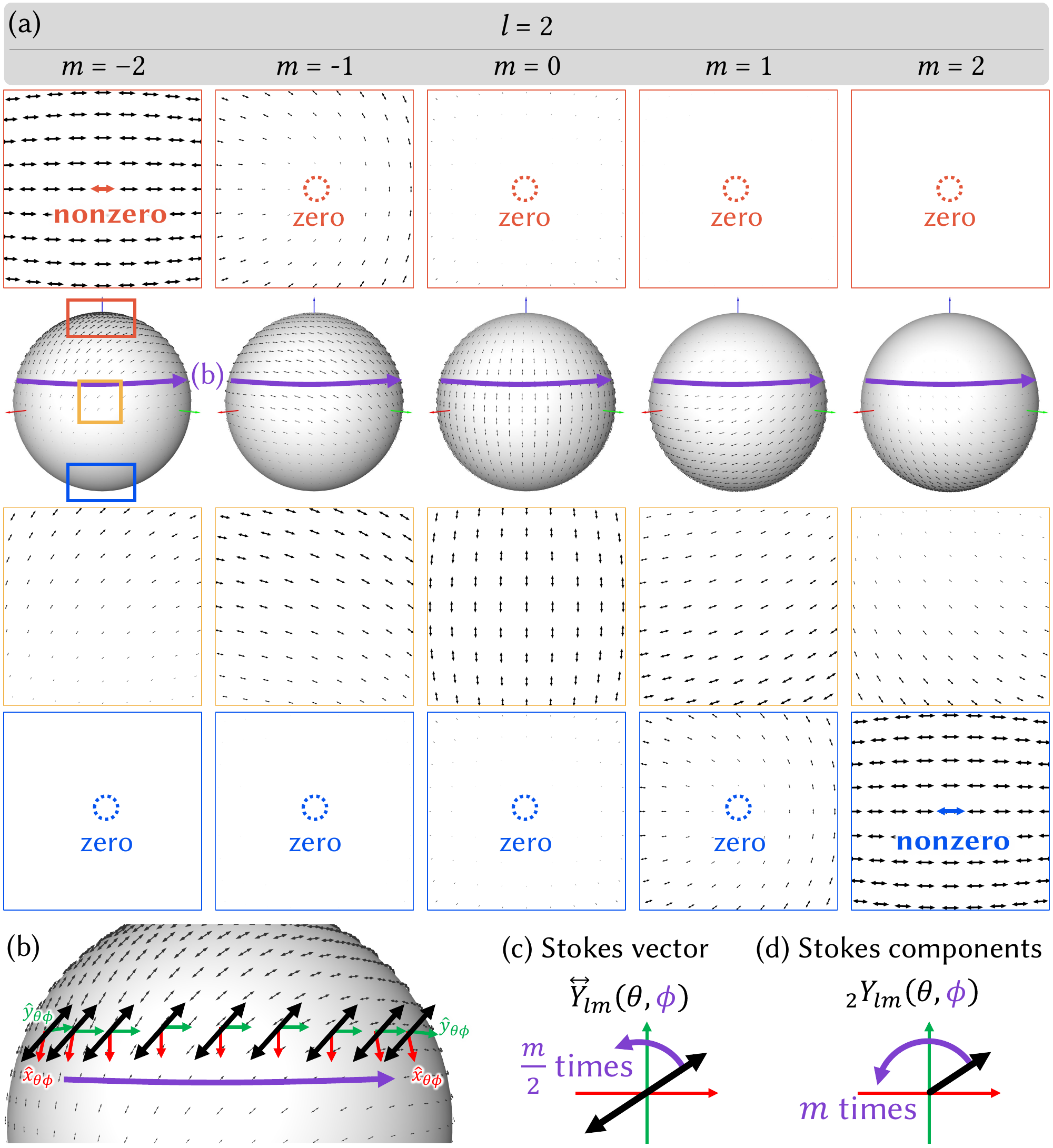}%
	\vspace{-3mm}
	\caption[]{\label{fig:theory_s2SH_nvSH_b}
		Visualization of the first order ($l=2$) of spin-2 spherical harmonics, defined in Eqs.~\eqref{eq:theory-spin2-SH} and~\eqref{eq:theory_s2SH_geo}, which are the basis functions of the space of linear polarization as functions of directions (spin-2 Stokes vector fields).
		(a)~The first, third, and fourth rows show the closeup of the region indicated in the second row.
		Note that spin-2 SH only have nonzero values at 
		\NEW{the north pole for $m=-2$ and the south pole for $m=2$}, ($\homega=\pm\hat z_g$, i.e., $\theta=0$ and $\pi$) respectively. In addition, tracing a line with fixed $\theta$ by increasing $\phi$, as the blue curves in (a) and (b), can be seen as (c) a Stokes vector rotating $\frac m2$ times and (d) Stokes components $\sY{lm}$ rotating $m$ times.
	}
	\vspace{-4mm}
\end{figure}

%% file: theory.tex
\section{Polarized spherical harmonics}
\label{sec:our_theory}

To overcome the challenges described in Section~\ref{sec:stokes-on-sphere} and bring benefits of frequency-domain framework to polarized radiance functions, we need a novel set of basis functions, %
\emph{polarized spherical harmonics}.
In Section~\ref{sec:spin_sph_harm}, we introduce spin-weighted SH and show how it plays a role in the basis functions for polarized light transport in computer graphics.
Although spin-weighted SH are an existing theory in physics~\cite{newman1966note,goldberg1967spin}, it has never been used in rendering pipelines to describe full Stokes vectors and Mueller transforms. 
In Section~\ref{sec:theory_rotinv}, we introduce our polarized spherical harmonics, combining spin-0 (scalar) SH and spin-2 SH, which can fully describe Stokes vector fields for polarized frequency domain analysis. 

Moreover, we will also show how to perform rotation (Section~\ref{sec:theory_rotinv}), linear operators (e.g., general pBRDFs and radiance transfer, Section~\ref{sec:theory_linop}), and convolution (Section~\ref{sec:theory_conv}) in the PSH domain, which are inevitable operations in frequency-domain analysis.
These three main operations are not only the theoretical foundation but also the main building blocks of our PSH rendering pipeline. 
See Section~\ref{sec:pprt} for our real-time polarized rendering results based on our theory described in overall Section~\ref{sec:our_theory}.

\subsection{Spin-Weighted Spherical Harmonics}
\label{sec:spin_sph_harm}

\NEW{The spin-weighted spherical harmonics ${}_s Y_{lm}$ are the basis for spin-$s$ functions on the sphere, and they have continuity conditions
depicted in Figure~\ref{fig:bkgnd_continuity_stokes_field} by replacing the double rotation by $s$ times rotation~\cite{newman1966note,goldberg1967spin}.
As a brief introduction, SWSH can be derived from the basis for functions on higher dimensional space, rotation transforms $\SOgroupv$, and introducing appropriate constraints that make these higher dimensional functions equivalent to spin-$s$ functions on the sphere.}

For more motivation and derivation of SWSH, refer to Supplemental Section~\ref{sec:theory-SWSH}, and here we focus on the usage of SWSH.

To handle Stokes vectors, we focus on spin $s=0$ and $s=2$.
With $s=0$, SWSH are exactly the same as conventional SH $({}_0 Y_{lm}=Y_{lm})$, so SWSH can be considered as a generalization of SH.
When $s=2$, spin-2 SH $(\sY{lm})$ become an orthonormal basis for spin-2 functions such as Stokes vector fields.
While there are several types of formulae to evaluate spin-2 SH, we introduce a way by utilizing scalar (spin-0) SH as follows:

\begin{szTitledBox}{Spin-2 SH (w.r.t. $\theta\phi$-frame field)}
\vspace{-1mm}
\begin{subequations}\label{aaaa}
	\begin{align}\label{eq:theory-spin2-SH}
	\sY{lm}\!\left(\theta,\phi\right) &=\! \sqrt{\frac{
			\left(l-2\right)!
		}{
			\left(l+2\right)!	
	}} \!\left[ \alpha_{lm}\!\left(\theta\right)Y_{lm}\!\left(\theta,\phi\right) + \beta_{lm}\!\left(\theta\right) Y_{l-1,m}\!\left(\theta,\phi\right) \right]\!, \\
	\alpha_{lm}\left(\theta\right) &= \frac{2m^2 - l\left(l+1\right)}{\sin^2\theta} - 2m\left(l-1\right) \frac{\cot\theta}{\sin\theta} + l\left(l-1\right)\cot^2\theta, \\
	\beta_{lm}\left(\theta\right) &= 2\sqrt{
		\frac{2l+1}{2l-1} \left(l^2-m^2\right)
	} \left( \frac{m}{\sin^2\theta} + \frac{\cot\theta}{\sin\theta} \right).
\end{align}
\end{subequations}
\vspace{-2mm}
\end{szTitledBox}
\noindent Note that $\sY{lm}$ here is complex-valued Stokes \emph{components} of a basis for spin-2 Stokes \emph{vector} fields with respect to the $\theta\phi$ frame field $\frF_{\theta\phi}$.
Thus, $\sY{lm}$
satisfies 
\NEW{the following condition, which indicates the double rotation at the north and south poles as visualized in Figure~\ref{fig:bkgnd_continuity_stokes_field}(b) as follows:}
\begin{equation} \label{eq:theory_s2sh_conti}
	\begin{alignedat}{2}
		\sY{lm}\left(0,\phi\right) &=0, & \text{ if }m\ne - 2 \\
		\sY{l,-2}\left(0,\phi\right) &= e^{-2i\phi}\cdot \text{const.} \ne 0, & \\
		\sY{lm}\left(\pi,\phi\right) &=0, &\text{ if }m\ne 2 \\
		\sY{l,2}\left(\pi,\phi\right) &= e^{2i\phi}\cdot \text{const.} \ne 0. &
	\end{alignedat}
\end{equation}
Now, using the numeric--geometric conversion (Equation~\eqref{eq:stk_num2geo}), we can define a (geometric) Stokes vector version of spin-2 SH as
\begin{equation}\label{eq:theory_s2SH_geo}
	\dvY_{lm}\left(\homega\right)\coloneqq \left[\sY{lm}\left(\theta,\phi\right)\right]_{\frF_{\theta\phi}\left(\theta,\phi\right)}.
\end{equation}
The first order $l=2$ of spin-2 SH $\dvY_{lm}$ are visualized in Figure~\ref{fig:theory_s2SH_nvSH_b}.
Note that due to the nature of spin-2 functions, there are no orders $l=0$ and $l=1$.

\NEW{Both terms $Y_{lm}\left(\theta,\phi\right)$ and $Y_{l-1,m}\left(\theta,\phi\right)$ in Equation~\eqref{eq:theory-spin2-SH} have $e^{im\phi}$ terms originated from Equation~\eqref{eq:bkgnd_sh_def}.
From the $e^{im\phi}$ term,} we can observe that, following the circle formulated by some fixed $\theta$ on the sphere, the geometric Stokes vector (double-sided arrow) rotates $\frac m2$ times. In contrast, numeric Stokes components rotate $m$ times, as shown in Figure~\ref{fig:theory_s2SH_nvSH_b}(b).
	
Additionally, when comparing spin-2 SH with scalar SH, spin-2 SH are similar in that they have azimuthal symmetry at $m=0$.
However, one difference is the condition of non-zero value at $\homega=\pm \hat z_g$; for spin-2 SH, it occurs at $m=-2$ or $m=2$ (Figure~\ref{fig:theory_s2SH_nvSH_b}), while for scalar SH, it occurs at $m=0$.

\begin{figure}[tpb]
	\centering
	\includegraphics[width=\linewidth]{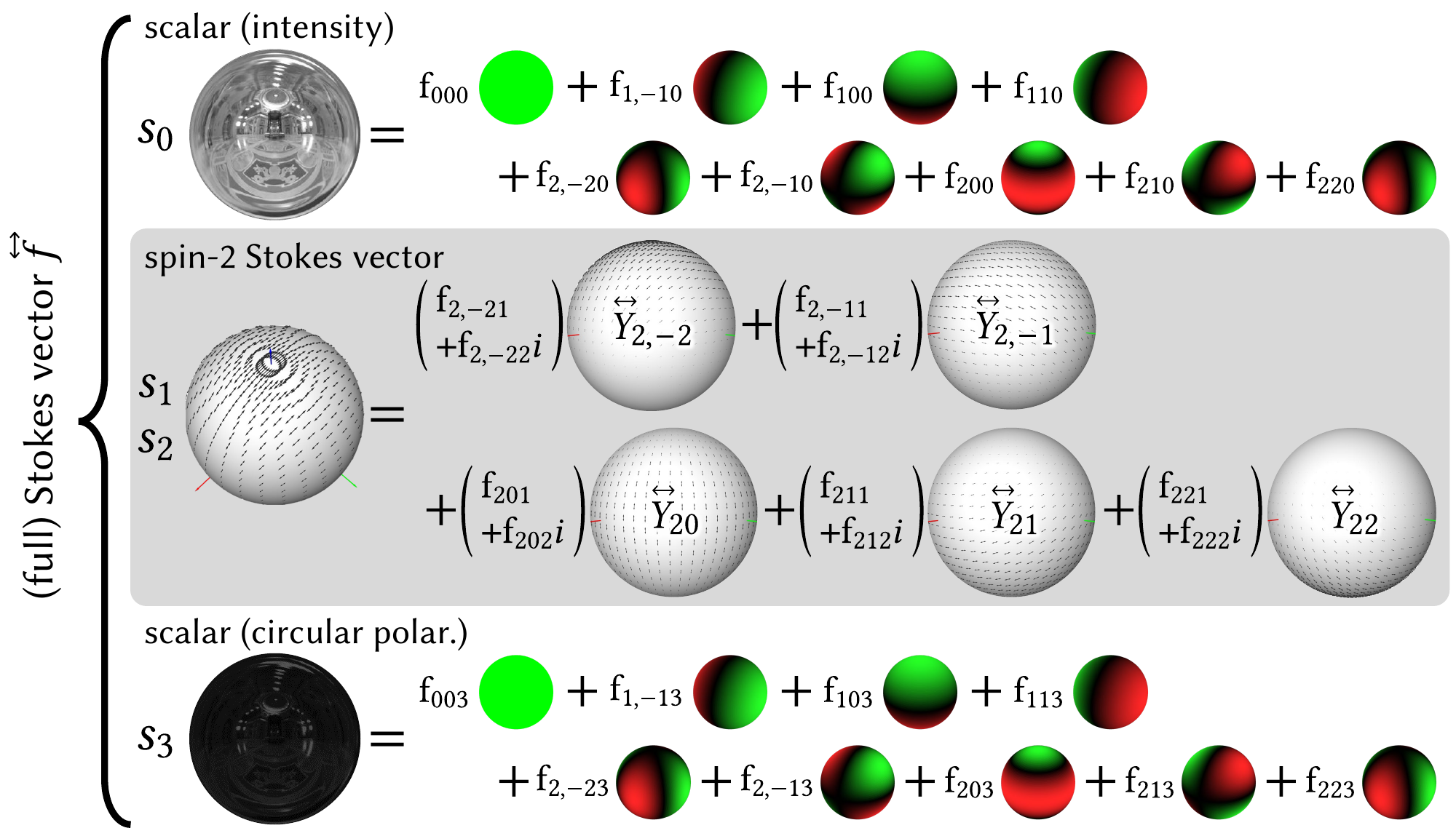}%
	\vspace{-3mm}
	\caption[]{\label{fig:theory_s2sh_complexmultiple}
		Any full Stokes vector field $\dvf$ can be linearly decomposed using scalar (spin-0) SH $Y_{lm}^R$ for $s_0$ (intensity) and $s_3$ (circular polarization) and spin-2 SH $\dvY_{lm}$ (or $\dvY_{lm1}$ and $\dvY_{lm2}$, equivalently) for $s_1$ and $s_2$ (linear polarization) as described in Equations~\eqref{eq:theory_PSH_basis} and~\eqref{eq:theory_PSH_lincomb}.
		The coefficient vector, which consists of such $\rmf_{lmp}$, becomes our frequency domain representation of given $\dvf$.
		Note that while $s_0$ and $s_3$ components in the original angular domain completely correspond to coefficients with $p=0$ and $p=3$, respectively, $s_1$ and $s_2$ components do not exactly correspond to $p=1$ and $p=2$, respectively, due to their values depend on the choice of frame fields.
	}
	\vspace{0mm}
\end{figure}

\subsection{Polarized Spherical Harmonics}
\label{sec:theory_rotinv}
Now we combine Stokes components spin-0 functions $s_0$ (total intensity) and $s_3$ (circular polarization) with spin-2 functions $s_1$ and $s_2$ (linear polarization).
Then we define the orthonormal basis \emph{polarized spherical harmonics}, which span the full Stokes vectors fields $\calF\left(\Sspv,\STKsp{\homega}\right)$ over real coefficients.
By using the additional index $p={0,1,2,3}$ that indicates the index of polarization components $s_0$, $s_1$, $s_2$, $s_3$ respectively, the PSH $\dvY_{lmp}$ 
\NEW{are defined by}

\begin{szTitledBox}{Polarized spherical harmonics}
\vspace{-1mm}
\begin{equation}\label{eq:theory_PSH_basis}
	\begin{split}
		\dvY_{lm0}\!\left(\homega\right) = \begin{bmatrix} Y_{lm}^R\brahomega \\0\\0\\0 \end{bmatrix}_{\frtp\!\left(\homega\right)} \!\!\!\!\!\!\!\!\!\!\!\!\!\!,& \dvY_{lm1}\!\left(\homega\right) = \begin{bmatrix} 0 \\ \Re \left[_2Y_{lm}\brahomega\right] \\ \Im \left[_2Y_{lm}\brahomega\right] \\0 \end{bmatrix}_{\frtp\!\left(\homega\right)} \!\!\!\!\!\!\!, \\
		\dvY_{lm2}\!\left(\homega\right) =  \begin{bmatrix}
		0\\ -\Im \left[_2Y_{lm}\brahomega\right] \\ \Re \left[_2Y_{lm}\brahomega\right] \\0 \end{bmatrix}_{\frtp\!\left(\homega\right)} \!\!\!\!\!\!\!\!\!\!\!\!\!\!,& \dvY_{lm3}\!\left(\homega\right) = \begin{bmatrix}
		0 \\ 0 \\ 0 \\ Y_{lm}^R\brahomega \end{bmatrix}_{\frtp\!\left(\homega\right)}\!\!\!\!\!\!\!.
	\end{split}
\end{equation}
\vspace{-2mm}
\end{szTitledBox}

\noindent Here, $\Re[\cdot]$ and $\Im[\cdot]$ %
indicate real and imaginary part of some scalar complex number $z$ where $z = \Re\left[z\right]+i\Im\left[z\right]$.

Using these bases, any Stokes vector field $\dvf$ can be written as
\begin{equation}\label{eq:theory_PSH_lincomb}
	\dvf\left(\homega\right) = \sum_{\left(l,m,p\right)\in I_{PSH}} \rmf_{lmp} \dvY_{lmp},
\end{equation}
where $I_{PSH}$ denotes the set of the indices $l$, $m$, and $p$:
\begin{equation}
\begin{aligned}
		I_{PSH} =& \left\{\left(l,m,p\right)\in \Z^3 \mid l\ge 0, \left|m\right|\le l, p=\{0,3\}\right\} \\
	&\cup \left\{\left(l,m,p\right)\in \Z^3 \mid l\ge 2, \left|m\right|\le l, p=\{1,2\}\right\},
\end{aligned}
\end{equation}
and the coefficient $\rmf_{lmp}$ can be computed as
\begin{equation} \label{eq:theory-PSH-coeff}
	\rmf_{lmp} = \lrangle{\dvY_{lmp}, \dvf}_{\calF}.
\end{equation}
By using PSH, the decomposition example is illustrated in Figure~\ref{fig:theory_s2sh_complexmultiple}, when $\dvf$ is a polarized environment map.

\paragraph{Real coefficient formulation}
One important adaption from spin-2 SH to our PSH is the separation of the complex part to make the coefficient a real number.
Suppose we have a spin-2 Stokes vector field $\dvf$, which only considers the linear polarization part.
Generally, using spin-2 SH, we can write frequency domain representation with complex number coefficient as
\begin{equation} \label{eq:theory-s2SH-Ccoeff}
	\dvf=\sum_{l,m}\underbrace{\left(\rmf_{lm1}+\rmf_{lm2}i\right)}_{\text{complex coeff.}} \underbrace{\dvY_{lm}}_\text{basis}.
\end{equation}
In contrast, using PSH, we can write the real number coefficient using the form as
\begin{equation} \label{eq:theory-s2SH-Rcoeff}
	\dvf=\sum_{l,m} \underbrace{\rmf_{lm1}}_\text{real coeff.} \underbrace{\dvY_{lm1}}_\text{basis} + \underbrace{\rmf_{lm2}}_\text{real coeff.} \underbrace{\dvY_{lm2}}_\text{basis}.
\end{equation}
Although it looks trivially identical, there are some reasons why this real-valued adaptation is important.
First, since the real-world quantities (Stokes vectors and Mueller transforms) have a real value, using real-valued representation allows us to easily manage the consistency when computing such quantities in the frequency domain.
Second, the formulation in Equation~\eqref{eq:theory-s2SH-Ccoeff} actually loses the information for representing Mueller transforms, while Equation~\eqref{eq:theory-s2SH-Rcoeff} does not. This will be introduced in later Section~\ref{sec:theory_linop}.
For spin-0 components, we use real SH $Y_{lm}^R$ rather than complex SH $Y_{lm}$ not only for the consistency to the angular domain but also to take algebraic closedness of induced coefficients matrices into account, which is discussed in Supplemental Section~\ref{sec:theory-PSH_RC}. %
Hence, we choose the real-valued formulation to build a solid theory for our PSH, except for some intermediate representations for efficient derivations that do not violate the reasons for choosing the real-valued formulation.

\subsubsection{Rotation invariance of PSH}
Since PSH are an orthonormal basis, the PSH coefficient rotation can also be done with the coefficient matrix similar to scalar SH (Equation~\eqref{eq:wignerD_def}).
For given Stokes vector field $\dvf$ and rotation $\vec R_\calF$, the rotated coefficient $\rmf_{lmp}'$ can be computed as
\begin{equation}\label{eq:theory_rotate_PSH_coeff}
	\rmf_{l_om_o p_o}' = \sum_{l_i,m_i,p_i} \underbrace{\lrangle{\dvY_{l_om_o p_o}, \vec R_\calF\left[\dvY_{l_im_i p_i}\right]}_\calF}_\text{coefficient matrix} \rmf_{l_im_i p_i},
\end{equation}
where subscript $o$ at the indices notes output (rotated) and subscript~$i$ at the indices notes input.
By using the definitions, the coefficient matrix at $p_o$-th row and $p_i$-th column can be calculated as
\begin{szTitledBox}{\NEW{Coefficient matrix of the rotation in PSH}}
	\vspace{-1mm}
	\begin{equation}\label{eq:theory_PSH_rotmat}
		\small
		\begin{aligned}
		&\left[\lrangle{\dvY_{l_om_op_o}, \vec R_\calF\left[\dvY_{l_im_ip_i}\right]}\right]_{p_o,p_i}= \quad\\
		&\delta_{l_il_o}\begin{bmatrix}
			\tcboxmathStS{\underbracket{D_{m_om_i}^{l,R}\left(\vec R\right)}_{p_o=0,\ p_i=0}} & \mathbf{0}_{1\times 2} & 0  \\
			\mathbf{0}_{2\times 1} & \tcboxmathVtV{\underbracket{\R^{2\times2}\left(D_{m_om_i}^{l,C}\left(\vec R\right)\right)}_{p_o=\{1,2\},\ p_i=\{1,2\}}} & \mathbf{0}_{2\times 1} \\
			0 & \mathbf{0}_{1\times 2} & \tcboxmathStS{\underbracket{D_{m_om_i}^{l,R}\left(\vec R\right)}_{p_o=3,\ p_i=3}}
		\end{bmatrix}.
		\end{aligned}
	\end{equation}
	\vspace{-2mm}
\end{szTitledBox}
\noindent Note that $D_{m_om_i}^{l,C}$ and $D_{m_om_i}^{l,R}$ are complex and real Wigner-D functions defined in Equations~\eqref{eq:wignerD_def} and~\eqref{eq:bkgnd_real_wignerD} respectively, and $\R^{2\times 2}$ indicates an operator that convert complex numbers to $2\times 2$ real matrices as
\begin{equation}\label{eq:theory_comp2mat}
	\R^{2\times 2}\left(x+yi\right)\coloneqq \begin{bmatrix}
		x & -y \\
		y & x
	\end{bmatrix}.
\end{equation}

As a result, we can observe that the resulting coefficient matrix of rotation on Stokes vector fields in Equation~\eqref{eq:theory_PSH_rotmat} only computes the same order $l$ for $\rmf$ and $\rmf'$.
This means the resulting matrix is block diagonal, and PSH satisfy the rotation invariance property.
Moreover, %
since Equation~\eqref{eq:theory_PSH_rotmat} consists of existing Wigner D-functions,
another advantage is that we can utilize existing formulas and computation methods from scalar SH \NEW{rotation}.
For more details and derivations of proving rotation invariance, refer to Supplemental Section~\ref{sec:theory_rotinv}.

\subsubsection{Rotation invariance validation}
\label{sec:theory-rotinv-valid}

\begin{figure}[tpb]
	\centering
	\includegraphics[width=\columnwidth]{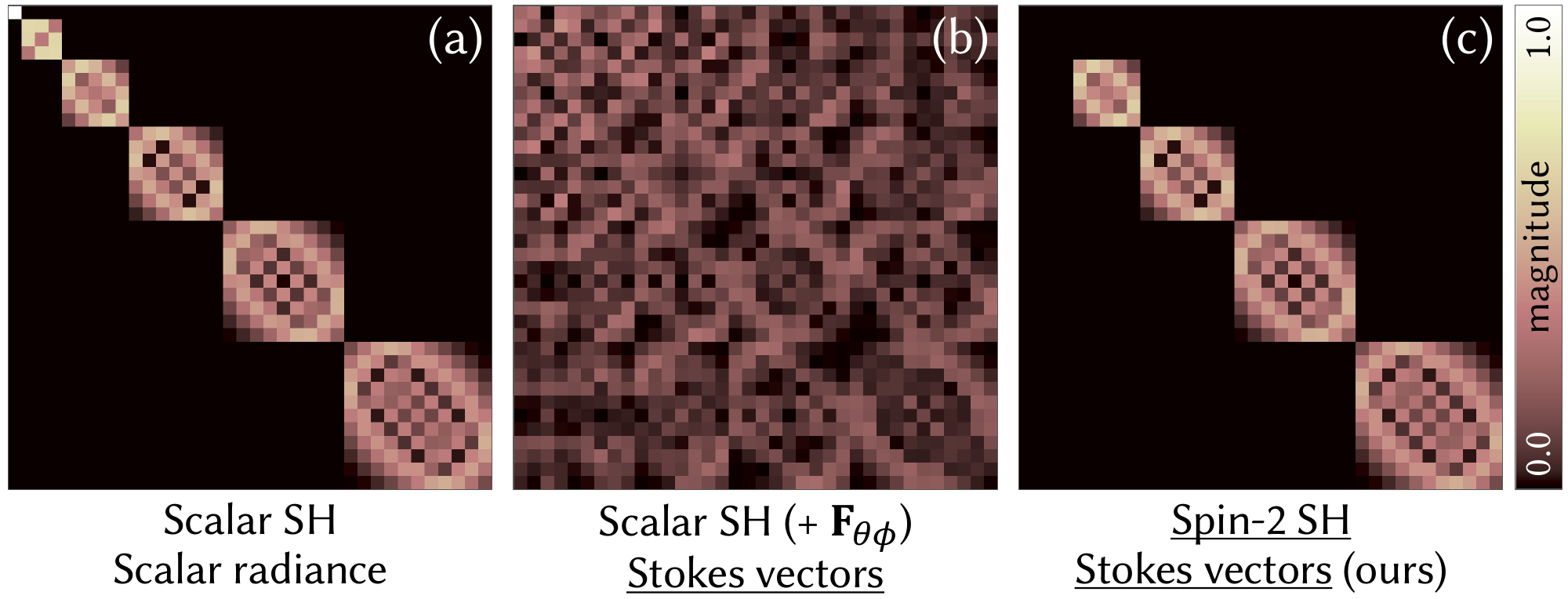}%
	\vspace{-3mm}
	\caption[]{\label{fig:theory_rotmat_compare}
		Comparison of coefficient matrices of a particular rotation with respect to each basis.
		Each matrix indicates the coefficient matrix of the rotation with respect to \NEW{scalar} SH (Equation~\eqref{eq:bkgnd_rot_sph_func}, also visualized in Supplemental Figure~\ref{fig:bkgnd_SH_rotmat}), \NEW{scalar SH with $\theta\phi$-frame field},
		and spin-2 SH (Equation~\eqref{eq:theory_PSH_rotmat}) for (a), (b), and (c), respectively. Each row and column indicates enumerated pairs of $\left(l,m\right)$ indices.
		(a) \NEW{In the case of scalar SH and scalar radiance, the coefficient matrix (Wigner D-function) shows block-diagonal behavior, and it yields the rotation invariance.}
		(b) However, if we na\"ively apply rotation using \NEW{scalar} SH \NEW{with $\theta\phi$-frame field} to Stokes vectors, the rotation invariance does not hold anymore.
		(c) By changing the basis to spin-2 SH, the rotation invariance holds on Stokes vectors.
		Note that spin-2 SH starts from $l=2$, so the first two block diagonals are empty in (c).}
	\vspace{-4mm}		
\end{figure}

\paragraph{Numerical validation}
\NEW{
So far, we have shown the theoretical guarantee of rotation invariance of PSH; here, we will show it numerically.
For the given rotation transform, we can compute the corresponding coefficient matrix that rotates some physical quantity with respect to some basis function.
Then, we can validate rotation invariance by checking the block diagonal behavior of the computed coefficient matrix.
Figure~\ref{fig:theory_rotmat_compare} shows such false-color magnitude visualization of the complex-numbered coefficient matrix, with rotation transform $\vec R=\vec R_{\hat u}\left(\theta\right)$ with $\theta\hat u = \frF_g\left[10,0.1,0.2\right]^T$.
Figure~\ref{fig:theory_rotmat_compare}(a) shows the coefficient matrix of rotating scalar radiance projected on the scalar SH.
It can be clearly shown that the coefficient matrix is block diagonal.
Figures~\ref{fig:theory_rotmat_compare}(b) and \ref{fig:theory_rotmat_compare}(c) shows the case of rotating spin-2 part of Stokes vector projected on the scalar SH and spin-2 SH, respectively.
Note that here we use complex-valued representation (Equation~\eqref{eq:theory-s2SH-Ccoeff}) to compare with the scalar radiance case in Figure~\ref{fig:theory_rotmat_compare}(a).
Also, we use the $\theta\phi$-frame field for scalar SH projection of Stokes vectors since we need to specify the frame field as described in Equation~\eqref{eq:stkfield-component}.
As shown in Figure~\ref{fig:theory_rotmat_compare}(b), using scalar SH on Stokes vectors never becomes block diagonal so that implies no rotation invariance.
In contrast, as shown in Figure~\ref{fig:theory_rotmat_compare}(c), the coefficient matrix of spin-2 SH on Stokes vector is block diagonal, which implies the rotation invariance and even computed value are the same as scalar radiance case in Figure~\ref{fig:theory_rotmat_compare}(a).
}

\begin{figure*}[tpb]
	\centering
	\includegraphics[width=\linewidth]{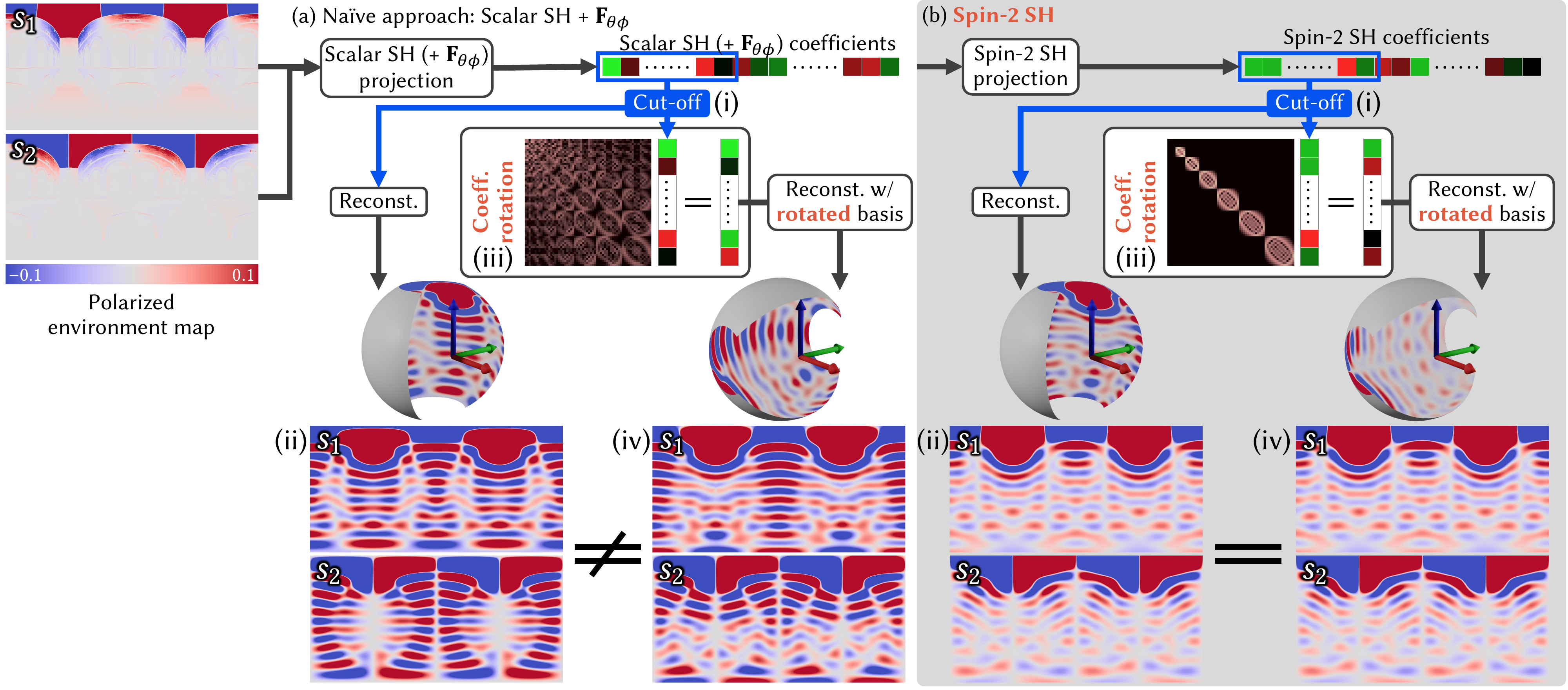}%
	\vspace{-3mm}
	\caption[]{\label{fig:theory_penvmap_rotation}
		Numerical validation of rotation invariance. First, get band-limited Stokes vector field (polarized environment map) (ii) from finite coefficients (i) under each basis: (a) \NEW{a na\"ive approach that combines scalar SH with the $\theta\phi$-frame field} %
		and (b) spin-2 SH from Equations~\eqref{eq:theory-spin2-SH} and~\eqref{eq:theory_PSH_basis}. Then (iii) applying rotation transform in the frequency domain, i.e., multiplying a coefficient matrix for a rotation (See Figure~\ref{fig:theory_rotmat_compare} for more details). Finally, the rotated coefficient vectors are reconstructed, and the inverse rotation is applied in the final angular domain. Then, while (a) the na\"ive approach gives inconsistent results, (b) our spin-2 SH give rotation-invariant results.
	}
	\vspace{-2mm}		
\end{figure*}

\paragraph{Polarized environment map reconstruction}
\NEW{
We also validate the rotation invariance with the polarized environment map, as shown in Figure~\ref{fig:theory_penvmap_rotation}.
Similar to the numerical validation above, we only show the spin-2 part of the Stokes vector.
For the given polarized environment map, we initially project it to the basis function such as scalar SH (with $\theta\phi$-frame field) or spin-2 SH, and \emph{cut-off} the coefficient vector to take the finite coefficient vector.
First, we reconstruct that finite coefficient vector into an angular domain, which yields a band-limited environment map.
On the other hand, we rotate that finite coefficient vector with the given rotation transform $\vec R$ and perform reconstruction with the \emph{rotated} basis with the same rotation transform $\vec R$.
Since we rotate both coefficients and basis with the same rotation transform, the reconstruction result should be the same as the vanilla cut-off reconstruction case.
As a result, the na\"ive approach using scalar SH with $\theta\phi$-frame field shows inconsistent behavior (Figure~\ref{fig:theory_penvmap_rotation}(a)), while using spin-2 SH results between two reconstructions is identical (Figure~\ref{fig:theory_penvmap_rotation}(b)).
}

\subsection{Coefficient Matrices for pBRDF and Radiance Transfer}
\label{sec:theory_linop}

Beyond coefficient vector representation of polarized environment map, our PSH \NEW{also} provide frequency domain representation for polarized light interaction such as pBRDF or radiance transfer operator 
into \emph{coefficient matrices}.
\NEW{Here we derive a general formulation of PSH coefficient matrices that extends scalar quantities described in} Section~\ref{sec:background_sh}, Equations~\eqref{eq:bkgnd_SHcoeff_operator} to~\eqref{eq:bkgnd_SHcoeff_isobrdf}.
\NEW{Recall that the coefficient matrix generally represents linear operators on Stokes vector fields in the angular domain.
Hence, they can be characterized as a \emph{Mueller transform field} $\dvP$, which is a function from given two directions $\homega_i$ and $\homega_o$ to a Mueller transform as}
\begin{equation} \label{eq:theory_mueller_field}
	\dvP \colon \Sspv\times \Sspv \to \calM_{\homega_i\to\homega_o}.
\end{equation}
Note that a Mueller transform field can be considered as a (cosine-weighted) pBRDF, \NEW{that} can act on a Stokes vector field $\dvf$ as a linear operator as
\begin{equation} \label{eq:theory_mueller_linop}
	\dvP_{\calF}\left[\dvf\right]\left(\homega_o\right) = \int_{\bbS^2}{\dvP\left(\homega_i,\homega_o\right)\dvf\left(\homega_i\right)\rmd \homega_i}.
\end{equation}
\NEW{As a result, by using the appropriate type of inner product described in Equations~\eqref{eq:bkgnd-stokes-inner} and~\eqref{eq:inner_stokes_field}, the PSH coefficients of the Mueller transform can be directly extended from scalar SH coefficients (Equation~\eqref{eq:bkgnd_SHcoeff_operator}) as}
\begin{szMathBox}
	\begin{multline}\label{eq:theory_psh_coeff_linop}
			\rmP_{l_om_op_o,l_im_ip_i} = \lrangle{\dvY_{l_om_op_o}, \dvP_\calF\left[\dvY_{l_im_ip_i}\right]}_{\calF} \\
			= \!\int_{\Sspv\times \Sspv}{\!\!\lrangle{\dvY_{l_om_op_o}\left(\homega_o\right), \dvP\left(\homega_i,\homega_o\right) \! \dvY_{l_im_ip_i}\left(\homega_i\right)}_{\calS} \!\! \rmd \homega_i \rmd \homega_o}.
	\end{multline}
\end{szMathBox}
\NEW{\noindent Similar to scalar SH, regarding the indices $\left(l_o,m_o,p_o\right)$ as rows and $\left(l_i,m_i,p_i\right)$ as columns, we can obtain the \emph{coefficient matrix} of $\dvP$.}
\NEW{Now suppose that} we have a coefficient vector $\rmf_{l_im_ip_i}$ from a polarized incident radiance $\dvf$,
obtained by Equation~\eqref{eq:theory_PSH_lincomb} and a coefficient matrix $\rmP_{l_om_op_o,l_im_ip_i}$ from a pBRDF $\dvP$.
\NEW{Then, similar to the conventional scalar SH-based rendering pipeline (Equation~\eqref{eq:bkgnd_SHcoeff_matmul}), the coefficient vector of reflected radiance $\lrangle{\dvY_{l_om_op_o}, \dvP_{\calF}\left[\dvf\right]}$ is evaluated \NEW{by} a matrix-vector product as}
\begin{equation}\label{eq:theory-PSH-matmul}
	\lrangle{\dvY_{l_om_op_o}, \dvP_{\calF}\left[\dvf\right]} = \sum_{\left(l_i,m_i,p_i\right)\in I_{PSH}} \rmP_{l_om_op_o,l_im_ip_i}\rmf_{l_im_ip_i} .
\end{equation}

\subsubsection{Submatrices of Mueller transforms and coefficient matrices}
\NEW{Due to the nature of Mueller transform, there are additional indices $p_o$ and $p_i$ in Equation~\eqref{eq:theory_psh_coeff_linop}.
Consequently, we have $16 (= 4 \times 4)$ times more coefficients than the coefficient matrices in scalar SH.
For further analysis and efficient computation in a constant factor, we can split a Mueller transform and the corresponding coefficient matrix.
From the given Mueller transform $\dvP$ in the angular domain and a single pair of directions $\left(\homega_i,\homega_o\right)$, we can denote a single Mueller transform $\dvM=\dvP\left(\homega_i,\homega_o\right)$.
By using the numeric--geometric conversion, the numeric Mueller matrix $\bfM$ can be computed as $\bfM = \left[\dvM\right]^{\frF_i\to \frF_o}$.
Now recall that ${s_1, s_2}$ components are dependent to frame (spin-2), and ${s_0, s_3}$ are independent to frame (spin-0).
In this context, the Mueller matrix can be split into 9 submatrices according to dependency on the $\frF_i$ and $\frF_o$ as
\begin{equation}\label{eq:theory-submat-mueller}
	\arraycolsep=2pt
	\bfM = \left[\begin{array}{c;{2pt/2pt}c;{2pt/2pt}c}
		\tcboxmathStS{\begin{matrix}\rmM_{00}\end{matrix}}
		& \tcboxmathVtS{\begin{matrix} \rmM_{01} & \rmM_{02}\end{matrix}}
		& \tcboxmathStS{\begin{matrix}\rmM_{03}\end{matrix}} \\ \hdashline[2pt/2pt]
		\tcboxmathStV{\begin{matrix}\rmM_{10} \\ \rmM_{20}\end{matrix}} 
		& \tcboxmathVtV{\begin{matrix} \rmM_{11} & \rmM_{12} \\ \rmM_{21} & \rmM_{22}\end{matrix}}
		& \tcboxmathStV{\begin{matrix} \rmM_{13} \\ \rmM_{23}\end{matrix}} \\ \hdashline[2pt/2pt]
		\tcboxmathStS{\begin{matrix}\rmM_{30}\end{matrix}}
		& \tcboxmathVtS{\begin{matrix}\rmM_{31} & \rmM_{32}	\end{matrix}}
		& \tcboxmathStS{\begin{matrix}\rmM_{33}\end{matrix}}
	\end{array}\right].
\end{equation}
By following its spin-weights, we call the submatrices \tcboxStS{\emph{spin 0-to-0}}, \tcboxVtS{\emph{spin 2-to-0}}, \tcboxStV{\emph{spin 0-to-2}}, and \tcboxVtV{\emph{spin 2-to-2}} blocks.
This decomposition is also valid to evaluate the \emph{coefficient matrix} of $\dvP$.
By fixing indices $l_o$, $m_o$, $l_i$, and $m_i$ for the coefficients defined in Equation~\eqref{eq:theory_psh_coeff_linop}, we can split the coefficient matrix in the same way as
\begin{equation}\label{eq:theory-submat-coefficient}
	\arraycolsep=1pt
	\left[\begin{array}{c;{2pt/2pt}c;{2pt/2pt}c}
		\tcboxmathStS{\begin{matrix}\rmP_{l_om_o0,l_im_i0}\end{matrix}}
		& \tcboxmathVtS{\begin{matrix} \rmP_{l_om_o0,l_im_i1} & \rmP_{l_om_o0,l_im_i0} \end{matrix}}
		& \tcboxmathStS{\begin{matrix}\rmP_{l_om_o0,l_im_i3}\end{matrix}} \\ \hdashline[2pt/2pt]
		\tcboxmathStV{\begin{matrix} \rmP_{l_om_o1,l_im_i0} \\ \rmP_{l_om_o2,l_im_i0} \end{matrix}}
		& \tcboxmathVtV{\begin{matrix} \rmP_{l_om_o1,l_im_i1} & \rmP_{l_om_o1,l_im_i2} \\ \rmP_{l_om_o2,l_im_i1} & \rmP_{l_om_o2,l_im_i2} \end{matrix}}
		& \tcboxmathStV{\begin{matrix}	\rmP_{l_om_o1,l_im_i3} \\ \rmP_{l_om_o2,l_im_i3}\end{matrix}} \\ \hdashline[2pt/2pt]
		\tcboxmathStS{\begin{matrix}\rmP_{l_om_o3,l_im_i0}\end{matrix}}
		& \tcboxmathVtS{\begin{matrix} \rmP_{l_om_o3,l_im_i1} & \rmP_{l_om_o3,l_im_i2} \end{matrix}}
		& \tcboxmathStS{\begin{matrix}\rmP_{l_om_o3,l_im_i3}\end{matrix}}
	\end{array}\right].
\end{equation}
What we can observe here is each of the nine submatrices in the Mueller matrix in \NEW{the} angular domain (Equation~\eqref{eq:theory-submat-mueller}) only affects the corresponding submatrix in the coefficient matrix in the frequency domain (Equation~\eqref{eq:theory-submat-coefficient}). This fact allows us to compute the coefficient matrix of each block separately, with less memory requirement for simulating numerical integration for Equation~\eqref{eq:theory_psh_coeff_linop}. \NEW{In other words, the matrix product with sizes $1\times 4$, $4\times 4$, and $4\times 1$ in the integrand of Equation~\eqref{eq:theory_psh_coeff_linop} can be reduced to $1\times 2$, $2\times 2$, and $2\times 1$, respectively.}
}

\begin{figure}[tpb]
	\centering
	\includegraphics[width=\columnwidth]{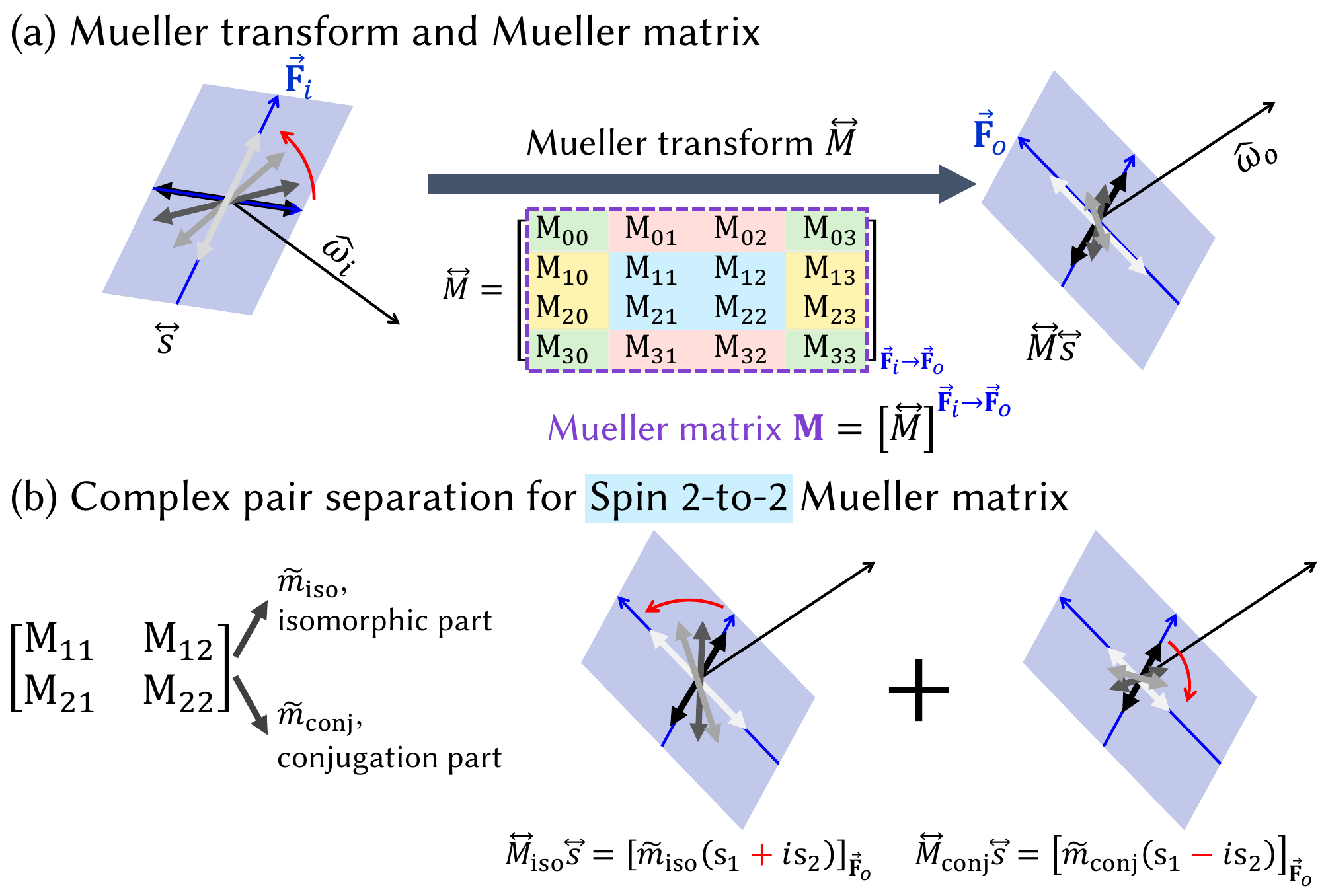}%
	\vspace{-3mm}
	\caption[]{\label{fig:theory_muel2isoconj}
		(a) When polarized light is reflected, the output Stokes vector $\dvM \dvs$ changes its magnitude and direction, and even direction change is not constant for the general Mueller transform.
		(b) The spin 2-to-2 block of a Mueller transform from $\homega_i$ to $\homega_o$ can be represented into two complex numbers: the \emph{isomorphic} part, denoted by $\rmM_{\mathrm {iso}}$, and the \emph{conjugation} part, denoted by $\rmM_{\mathrm {conj}}$. The isomorphic part indicates a Mueller transform, which preserves the self-rotation of the input Stokes vector $\dvs$. In contrast, the conjugation part indicates one which rotates the output $\dvM_{\mathrm {conj}}\dvs$ CW around $\homega_o$ as $\dvs$ rotates CCW around~$\homega_i$.}
	\vspace{-2mm}		
\end{figure}

\subsubsection{Complex pair separation of spin 2-to-2 Mueller transform}

\NEW{In addition to separating the full Mueller transform into nine blocks, we find that \tcboxVtV{\text{spin 2-to-2}} part $\dvM$ can once more separated into two frame-independent parts $\dvM_{\mathrm{iso}}$ and $\dvM_{\mathrm{conj}}$.
However, such separation is not as simple as the full Mueller transform, which splits the matrix into submatrices.
For example, taking only $\rmM_{11}$ and replacing $\rmM_{12}$, $\rmM_{21}$, and $\rmM_{22}$ to zero results in different Mueller transforms depending on the choice of frames.}

\NEW{While we find such a separation between theoretical and computational convenience, we also introduce a way to understand it intuitively.
Suppose that there is a \tcboxVtV{\text{spin 2-to-2}} Mueller transform $\dvM\in\MUEsp{\homega_i}{\homega_o}$, that transforms a spin-2 Stokes vector $\dvs_i$ to $\dvs_o$.}
Imagine the rotation of $\dvs_i$ around $\homega_i$, as depicted in Figure~\ref{fig:theory_muel2isoconj}.
As shown in Figure~\ref{fig:theory_muel2isoconj}(a), both magnitude and direction are changed in the output $\dvs_o$. %
\NEW{Decomposing it into $\dvM_{\mathrm{iso}}$ and $\dvM_{\mathrm{conj}}$,} the output Stokes vectors $\dvM_{\mathrm{iso}}\dvs_i$ and $\dvM_{\mathrm{conj}}\dvs_i$ rotate around $\homega_o$ in opposite directions without changing their magnitude as shown in Figure~\ref{fig:theory_muel2isoconj}(b).

\NEW{To obtain such two parts of the Mueller transform, we define the following conversion functions that convert $2\times 2$ real matrices to complex numbers as}
\begin{equation}\label{eq:theory_mat2comp}
	\begin{split}
		\C_{\mathrm{iso}}\left(\bfM\right) &\coloneqq \frac{\rmM_{11} + \rmM_{22}}{2} + \frac{\rmM_{21} - \rmM_{12}}{2}i, \\
		\C_{\mathrm{conj}}\left(\bfM\right) &\coloneqq \frac{\rmM_{11} - \rmM_{22}}{2} + \frac{\rmM_{21} + \rmM_{12}}{2}i.
	\end{split}
\end{equation}
The output pair of complex numbers from this conversion is denoted by $\tilde m_\mathrm{iso}\coloneqq \C_{\mathrm{iso}}\left(\bfM\right)$ and $\tilde m_\mathrm{conj}\coloneqq \C_{\mathrm{conj}}\left(\bfM\right)$.
Conversely, we can reconstruct to the original $2\times 2$ real matrix as
\begin{equation}\label{eq:theory_comppair2mat}
		\bfM = \R^{2\times 2} \left(\tilde m_\mathrm{iso}\right) + \R^{2\times 2} \left(\tilde m_\mathrm{conj}\right) \bfJ \mathrm{,\ \ where\ }
		\bfJ \coloneqq \begin{bmatrix}
			1&0\\0&-1
		\end{bmatrix}.
\end{equation}
\NEW{Then we can separate the Mueller transform as
\begin{equation}\label{eq:theory_compt2muellertform}
	\begin{split}
		\dvM_\mathrm{iso}&\coloneqq\left[\R^{2\times 2}\left(\tilde m_\mathrm{iso}\right)\right]_{\frF_i\to\frF_o}, \\
		\dvM_\mathrm{conj}&\coloneqq\left[\R^{2\times 2}\left(\tilde m_\mathrm{conj}\right)\bfJ\right]_{\frF_i\to\frF_o}.
	\end{split}
\end{equation}
}The key property of this representation is that it converts the \NEW{product between the matrix $\bfM$ and the vector $\R^2\left(z\right)$}
\NEW{ or the matrix $\R^{2 \times 2}\left(z\right)$}
into complex products as
\begin{subequations} 
	\begin{align}\label{eq:theory_mat2comp_property}
		\bfM \R^2\left(z\right) &= \R^2\left(\tilde m_\mathrm{iso} z + \tilde m_\mathrm{conj} z^*\right), \\
		\label{eq:theory_mat2comp_property2}
		\bfM \R^{2\times 2}\left(z\right)&= \R^{2\times 2}\left(\tilde m_\mathrm{iso}z\right) + \R^{2\times 2}\left(\tilde m_\mathrm{conj}z^*\right)\bfJ, \quad \forall z\in\C
	\end{align}
\end{subequations}
where $\R^2$ here denotes the nature conversion from a complex number to a 2-dimensional real vector:
\begin{equation}\label{eq:theory_complex_to_real}
	\R^2\left(x+yi\right) = \begin{bmatrix}
		x \\ y
	\end{bmatrix}.
\end{equation}

\NEW{Now, we will show that these are well-defined frame-independent quantities. If we rotate the frames $\frF_i$ and $\frF_o$ around their $z$ axes by $\alpha$ and $\beta$, respectively, the new Mueller matrix under the rotated frames can be evaluated as follows:
\begin{equation}
	\begin{split}
		\bfM' =& \R^{2\times 2}\left(e^{-2i\alpha}\right) \bfM \R^{2\times 2}\left(e^{2i \beta}\right) \\
		\underarrowref{=}{\kern-3em Eq.~\eqref{eq:theory_comppair2mat}\kern-3em}& \R^{2\times 2}\left(e^{-2i\alpha}\right)\left[ \R^{2\times 2} \left(\tilde m_\mathrm{iso}\right) + \R^{2\times 2} \left(\tilde m_\mathrm{conj}\right) \bfJ\right]\R^{2\times 2}\left(e^{2i \beta}\right) \\
		\underarrowref{=}{\kern-3emEq.~\eqref{eq:theory_mat2comp_property2}\kern-3em}& \R^{2\times 2}\left(e^{-2i\alpha}\right)\left[ \R^{2\times 2} \left(\tilde m_\mathrm{iso}e^{2i \beta}\right) + \R^{2\times 2} \left(\tilde m_\mathrm{conj}e^{-2i \beta}\right) \bfJ\right] \\
		=&  \R^{2\times 2} \left(\tilde m_\mathrm{iso}e^{2i \left(-\alpha+\beta\right)}\right) + \R^{2\times 2} \left(\tilde m_\mathrm{conj}e^{-2i \left(\alpha+\beta\right)}\right) \bfJ.
	\end{split}
\end{equation}
We note here that it is identical to Equation~\eqref{eq:theory_comppair2mat} by replacing $\tilde m_{\mathrm{iso}}$ with $\tilde m_\mathrm{iso}e^{2i \left(-\alpha+\beta\right)}$ and $\tilde m_\mathrm{conj}$ with $\tilde m_\mathrm{conj}e^{-2i \left(\alpha+\beta\right)}$. Since $\tilde m_{\mathrm{iso}}$ and $\tilde m_\mathrm{conj}$ do not affect each other, this separation is well-defined independent of the choice of frames.

}
Finally, we obtain the following property:
\begin{equation}
	\begin{split}
		\bfM \R^2\left(e^{i\vartheta}\left(\rms_{i1}+i\rms_{i2}\right)\right) &= \R^2\left(e^{i\vartheta} \tilde m_\mathrm{iso} \left(\rms_{i1}+i\rms_{i2}\right)\right) \\
		&+ \R^2\left( e^{-i\vartheta} \tilde m_\mathrm{conj} \left(\rms_{i1}-i\rms_{i2}\right)\right).
	\end{split}
\end{equation}
\NEW{By using this property and Equation~\eqref{eq:theory_compt2muellertform}, it implies that $\dvM_\mathrm{iso}$ preserves the rotation direction of the input, and $\dvM_\mathrm{conj}$ reverses the rotation direction.}
\NEW{We call this \tcboxVtV{\text{spin 2-to-2}} Mueller matrix $\bfM$}\NEW{ (spin 2-to-2 Mueller transform $\dvM$, respectively)} \NEW{to two complex numbers $\tilde m_\mathrm{iso}$ and $\tilde m_\mathrm{conj}$ \NEW{($\dvM_{\mathrm{iso}}$ and $\dvM_{\mathrm{conj}}$, respectively)} conversion as \emph{complex pair separation}.
And we call each resulting complex number as the \emph{isomorphic part} and \emph{conjugation part}, respectively.

The important property of this separation is that the isomorphic and conjugation parts of the spin 2-to-2 Mueller transform in the angular domain only affect the corresponding spin 2-to-2 submatrix of the coefficient matrix in the frequency domain.
Consequently, we can reduce direct 4 integrals in Equation~\eqref{eq:theory_psh_coeff_linop} for $p_o,p_1=1,2$ into only 2 integrals as separating coefficient matrix as}
\begin{szMathBox}
	\small
	\begin{subequations}\label{eq:theory_mueller_coeff}
	\begin{multline} \label{eq:theory_mueller_coeff_C2R_LP2LP}
		\tcboxmathVtV{\begin{bmatrix}
				\rmP_{l_om_o1,l_im_i1} & \rmP_{l_om_o1,l_im_i2} \\ \rmP_{l_om_o2,l_im_i1} & \rmP_{l_om_o2,l_im_i2}
		\end{bmatrix}} = \\
		\Rtt\left(\tcboxmathVtV{\tilde\rmP_{l_om_o,l_im_i,\mathrm{iso}}}\right) + \Rtt\left(\tcboxmathVtV{\tilde\rmP_{l_om_o,l_im_i,\mathrm{conj}}}\right)\bfJ,
	\end{multline}
	\begin{multline}
		\label{eq:theory_mueller_coeff_V2Vi}
		\tcboxmathVtV{\tilde\rmP_{l_om_o,l_im_i,\mathrm{iso}}} \coloneqq \\
		\int_{\Sspv\times \Sspv}{ \tilde P_{\mathrm{iso}}\braio{}_2Y_{l_om_o}^*\brao {}_2Y_{l_im_i}\brai \rmd\homega_i\rmd\homega_o},
	\end{multline}
	\begin{multline}
		\label{eq:theory_mueller_coeff_V2Vc}
		\tcboxmathVtV{\tilde\rmP_{l_om_o,l_im_i,\mathrm{conj}}} \coloneqq \\
		\int_{\Sspv\times \Sspv}{ \tilde P_{\mathrm{conj}}\braio{}_2Y_{l_om_o}^*\brao {}_2Y_{l_im_i}^*\brai \rmd\homega_i\rmd\homega_o},
	\end{multline}
	\end{subequations}
\end{szMathBox}
\noindent where $\tilde P_{\mathrm{iso}}$ and $\tilde P_{\mathrm{conj}}$ denote isomorphic and conjugation parts of $\left[\dvP\left(\homega_i,\homega_o\right)\right]^{\frF_{\theta\phi}\left(\homega_i\right)\to \frF_{\theta\phi}\left(\homega_o\right)}$, respectively.
\NEW{Note that both $\sY{l_om_o}$ is complex conjugated in Equations~\eqref{eq:theory_mueller_coeff_V2Vi} and Equation~\eqref{eq:theory_mueller_coeff_V2Vc}, while $\sY{l_im_i}$ is complex conjugated only in Equation~\eqref{eq:theory_mueller_coeff_V2Vc}.
This difference comes from the property of the complex pair separation in Equation~\eqref{eq:theory_mat2comp_property}.
}

\NEW{Based on these formulations, we can now explain the information loss problem in Section~\ref{sec:theory_rotinv}, the reason for using the real coefficient formulation (Equation~\eqref{eq:theory-s2SH-Rcoeff}) rather than complex coefficient formulation (Equation~\eqref{eq:theory-s2SH-Ccoeff}).
For fixed order $l$ and degree $m$, Equation~\eqref{eq:theory-s2SH-Ccoeff} represents a spin-2 Stokes vector field into a single complex coefficient.
This implies the spin 2-to-2 block of a Mueller transform field is also represented as a single complex coefficient.
Since a single complex number is equivalent to the isomorphic part $\tilde\rmP_{l_om_o,l_im_i,\mathrm{iso}}$ in Equations~\eqref{eq:theory_mueller_coeff_C2R_LP2LP} and~\eqref{eq:theory_mueller_coeff_V2Vi}, it only has half the information.}

Another further interesting property of the complex pair separation is that we can utilize the commutativity of the complex product, while the original matrix product is non-commutative.
It is the main key to proving our polarized spherical convolution theorem, which will be introduced in Section~\ref{sec:theory_conv}.

\subsubsection{Isotropic pBRDF}
\label{sec:theory-pbrdf-iso}
\NEW{
Similar to the sparsity condition of isotropic BRDF (Equation~\eqref{eq:bkgnd_SHcoeff_isobrdf}), the PSH coefficients of isotropic pBRDF have a sparsity condition.
Such constraints can be easily obtained from Equations~\eqref{eq:theory_mueller_coeff_V2Vi} and~\eqref{eq:theory_mueller_coeff_V2Vc} using $e^{im\phi}$ term in Equation~\eqref{eq:theory-spin2-SH}.
As a result, the sparsity constraint of the PSH coefficients of isotropic pBRDF can be written as
\begin{equation}\label{eq:theory_iso2coeff}
	\rmP_{l_om_op_o,l_im_ip_i} = 0 \mathrm{,\ if\ } \abs{m_i} \ne \abs{m_o}.
\end{equation}
Not only the above constraints but there are also additional linear constraints for the \tcboxVtV{\text{spin 2-to-2}} submatrix:
\begin{equation}\label{eq:theory-pBRDF-isoV2V}
	\begin{split}
		\tcboxmathVtV{\tilde\rmP_{l_om_o,l_im_i,\mathrm{iso}}} &= 0 \mathrm{,\ if\ } m_i \ne m_o, \\
		\tcboxmathVtV{\tilde\rmP_{l_om_o,l_im_i,\mathrm{conj}}} &= 0 \mathrm{,\ if\ } m_i\ne -m_o.
	\end{split}
\end{equation}
By using those constraints from isotropy, the complexity of pBRDF coefficient matrix $O\left(4\times 4l_\mathrm{max}^4\right)$ reduces to $O\left(4\times 4 l_\mathrm{max}^3\right)$, which is also similar to the scalar SH coefficients of BRDF.}

\input{tripleproduct}

%% file: tripleproduct.tex
\subsubsection{Shadowed radiance transfer via triple products}
\label{sec:theory_shadow_tp}

\begin{figure}[tpb]
	\centering
	\vspace{-3mm}	
	\includegraphics[width=\linewidth]{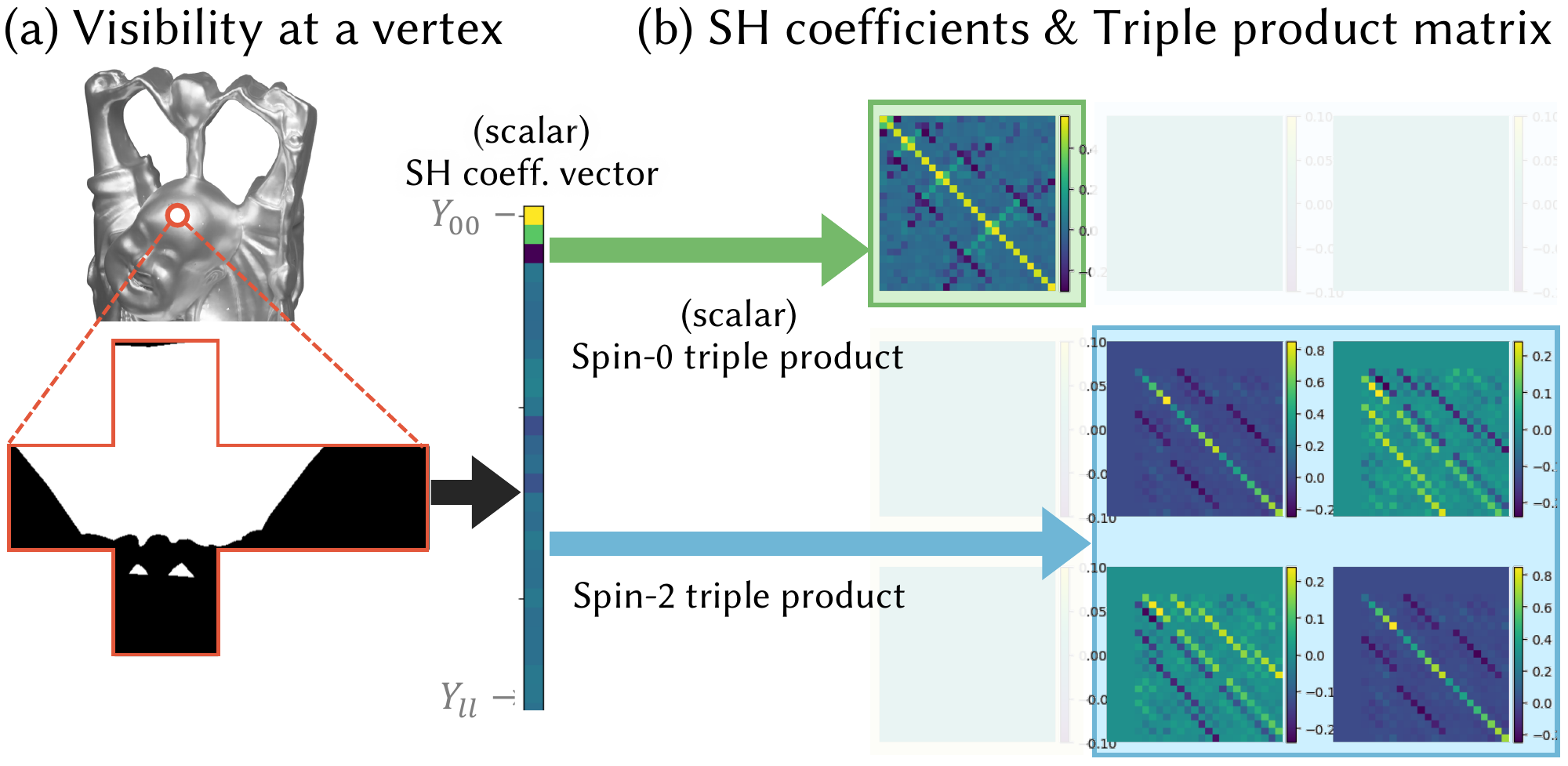}%
	\vspace{-3mm}
	\caption[]{\label{fig:theory_triple_product}
		For precomputation of self-shadow in radiance transfer matrices, (a) a visibility function $V\left(\homega\right)$ at a vertex can be converted into (b) SH coefficients $\mathrm{v}_{lm}$ first and then these are expanded to the radiance transfer matrix using the triple product equations.
		The \tcboxStS{\text{spin 0-to-0 submatrix}} can be obtained by the conventional triple product described in Equation~\eqref{eq:theory_tp_s2s}, and the \tcboxVtV{\text{spin 2-to-2 submatrix}} can be obtained by the triple product of spin-0, spin-2, and spin-2 functions described in Equation~\eqref{eq:theory_tp_v2v}.
	}
	\vspace{-2mm}
\end{figure}

\NEW{For more realistic rendering, the self-shadow at a vertex of an object can be considered.
To compute the self-shadowing radiance transfer coefficients $\mathrm{V}_{l_om_op_o,l_im_ip_i}$ on the vertex, we first consider the binary visibility mask $V\colon\Sspv\to\R$ at the vertex.
Then we consider such binary visibility mask as a linear operator $V_\calF\colon \calF\left(\Sspv,\STKsp{\homega}\right) \to \calF\left(\Sspv, \STKsp{\homega}\right)$, which acts on a polarized illumination (Stokes vector field) $\dvf$ as $V_\calF\left[\dvf\right]\left(\homega\right)\coloneqq V\left(\homega\right)\dvf\left(\homega\right)$.
Consequently, the coefficients can be computed as
\begin{equation}\label{eq:theory_shadow_direct}
	\mathrm{V}_{l_om_op_o,l_im_ip_i} = \int_{\Sspv}{\lrangle{\dvY_{l_om_op_o}\left(\homega\right), V\left(\homega\right)\dvY_{l_im_ip_i}\left(\homega\right) }_\calS \rmd \homega}.
\end{equation}
Note that this equation can be also considered as Equation~\eqref{eq:theory_psh_coeff_linop} with a Dirac delta Mueller transform.

\NEW{While Equation~\eqref{eq:theory_shadow_direct} can be evaluated in $O\left(n_\mathrm{ray}l_\mathrm{max}^4\right)$ times, it has a useful relationship with the scalar SH coefficients $\mathrm v_{lm}$ of $V$, which has $O\left(n_\mathrm{ray}l_\mathrm{max}^2\right)$ complexity. Here note that $n_\mathrm{ray}$ indicates the number of ray castings in numerical computation (i.e., number of discrete samples for the integrals in Equations~\eqref{eq:bkgnd_sh_coeff} and~\eqref{eq:theory_shadow_direct}).}
We can compute the submatrices of $\mathrm{V}_{l_om_op_o,l_im_ip_i}$ separately \NEW{using the identities of triple products of spin-weighted spherical harmonics} as depicted in Figure~\ref{fig:theory_triple_product}.

The \tcboxStS{\text{spin 0-to-0}} submatrix of $\mathrm{V}_{l_om_op_o,l_im_ip_i}$ can be computed using the triple product of three spin-0 (scalar) SH functions.
In other words, it can be computed by the scalar SH coefficient of the point-wise product of two scalar SH bases as
\begin{equation}\label{eq:theory_tp_s2s}
	\int_{\Sspv}{Y_{l_om_o}^*Y_{l'm'}Y_{l_im_i}\rmd\homega},
\end{equation}
which has a known analytic formula. Here, $l'$ and $m'$ corresponds to indices of $\mathrm v_{lm}$.
For the \tcboxVtV{\text{spin 2-to-2}} part, it can be computed using the triple product of one spin-0 and two spin-2 functions.
In other words, it can be computed by the spin-2 SH coefficient of the point-wise product of the scalar SH basis and the spin-2 SH basis functions as
\begin{equation}\label{eq:theory_tp_v2v}
	\int_{\Sspv}{\sY{l_om_o}^*Y_{l'm'}\sY{l_im_i}\rmd\homega}.
\end{equation}
Finally, the \tcboxVtS{\text{spin 2-to-0}}, \tcboxStV{\text{spin 0-to-2}} parts are zero since the point-wise product between spin-0 and spin-0 functions, and spin-0 and spin-2 functions are spin-0 function and spin-2 function, respectively.

Precomputing the shadowed radiance transfer using Equations~\eqref{eq:theory_tp_s2s} and~\eqref{eq:theory_tp_v2v} rather than direct computation using Equation~\eqref{eq:theory_shadow_direct} requires less computation as the number of ray castings for visibility test increases. This is because expanding the coefficient vector $\mathrm{v}_{lm}$ to the coefficient matrix $\mathrm{V}_{l_om_op_o,l_im_ip_i}$ does not depend on the number of rays.
Moreover, triple product relations in Equation~\eqref{eq:theory_tp_v2v} will be used \NEW{to extend our} polarized PRT, which enables the dynamic self-shadowing based on previous techniques~\cite{zhou2005precomputed,xin2021fast}.

Note that we do not describe the exact computation of the above triple product integrals here, but we only point out that the spin-0 triple product described in Equation~\eqref{eq:theory_tp_s2s} has already been used in existing SH-based methods, including \citet{zhou2005precomputed}.
The spin-0 and spin-2 triple product described in Equation~\eqref{eq:theory_tp_v2v} can be easily implemented once the implementation of Equation~\eqref{eq:theory_tp_s2s} is given.
For detailed explanation and computation, refer to Supplemental Section~\ref{sec:theory-tp}.
}

%% file: theory_convolution.tex
\subsection{Polarized Spherical Convolution}
\label{sec:theory_conv}

\NEW{A strength of the frequency domain analysis (e.g., Fourier transform, spherical harmonics) is that it converts the convolution between two functions into an element-wise product, allowing efficient computation.
However, even though spin-weighted SH themselves have been already invented in physics, the spherical convolution on polarized light has not been defined, analyzed, or discussed.
Hence, we will start by defining a polarized spherical convolution operation in Section~\ref{sec:theory_conv_def}.
After that, we will show how to represent polarized convolution kernels as coefficients in Section~\ref{sec:theory_conv_kernel}, by investigating the subspace of PSH.
Finally, we propose the polarized spherical convolution theorem in Section~\ref{sec:theory_conv_polarsh}, which is the frequency domain analysis of polarized spherical convolution in PSH.
Note that we only introduce the theorem statement and its experimental validation in Section~\ref{sec:theory_conv_polarsh}, but the derivation of such a theorem is a core contribution of this paper.
The detailed derivation and step-by-step proof can be found in Supplemental Section~\ref{sec:theory-conv}.
}

\subsubsection{Definition of spherical convolution on Stokes vector fields}
\label{sec:theory_conv_def}
\begin{figure}[tp]
	\centering
	\includegraphics[width=\columnwidth]{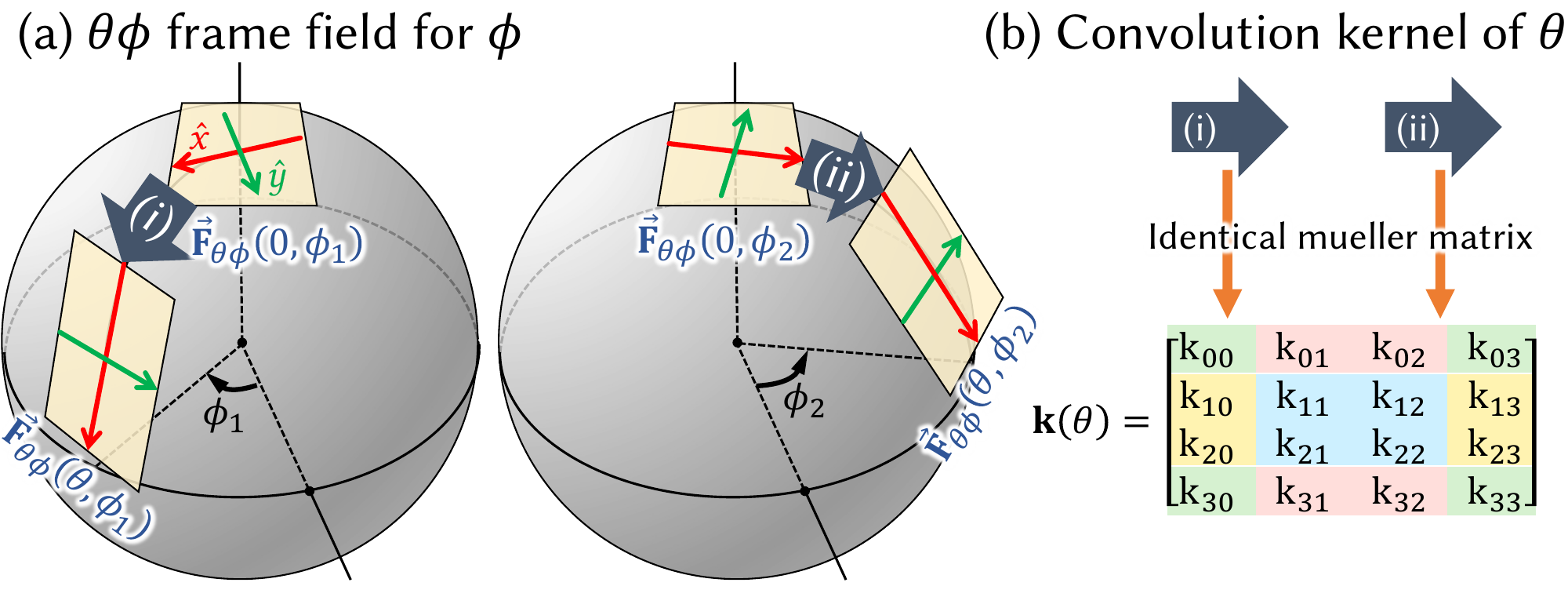}%
	\vspace{-4mm}
	\caption[]{\label{fig:theory_convolution_kernel}
		We propose the concept of spherical convolution of Stokes fields. A convolution kernel is defined as Mueller transform as a function of a single direction $\dvk\left(\homega\right)\in\MUEsp{\hat z_g}{\homega}$. Due to rotation equivariance, its numeric Mueller matrix under the $\theta\phi$-frame field has azimuthal symmetry: $\bfk\left(\theta\right)\in\R^{4\times 4}$. Concretely, the Mueller matrix under $\frF_{\theta\phi}\left(0,\phi_1\right)\to \frF_{\theta\phi}\left(\theta,\phi_1\right)$, illustrated as (a) (i), and the Mueller matrix under $\frF_{\theta\phi}\left(0,\phi_2\right)\to \frF_{\theta\phi}\left(\theta,\phi_2\right)$, illustrated as (a) (ii) become an identical matrix, as illustrated in (b). Similar to the general Mueller transform field described in Equation~\eqref{eq:theory-submat-mueller}, the convolution kernel also can be separated in spin 0-to-0, 0-to-2, 2-to-0, and 2-to-2 submatrices (b). }
	\vspace{-4mm}		
\end{figure}

\NEW{While scalar spherical convolution (Equation~\eqref{eq:bkgnd_sph_conv}) can be naturally defined without considering its rotation equivariance, extending such definition to Stokes vector fields are not trivial.
When we try to extend Equation~\eqref{eq:bkgnd_sph_conv} to Stokes vector fields, a question may be asked: What will be the kernel $k$?
Will it still be a scalar? Otherwise, will it be a Stokes vector field or a Mueller transform field?
Although we can answer the question with heuristic choice, we will build a general and standard definition here.
To do so, we will start with the \emph{linearity} and \emph{rotation equivariance}, which also defines scalar spherical convolution as described in Supplemental Section~\ref{sec:spherical-convolution}.

Suppose there is a linear and rotation equivariant operator on Stokes vector fields.
Since it is a linear operator, it can be characterized as a Mueller transform field $\dvK\colon \Sspv\times \Sspv\to \MUEsp{\homega_i}{\homega_o}$, as discussed in Equations~\eqref{eq:theory_mueller_field} and~\eqref{eq:theory_mueller_linop}.
In the beginning, %
we simply write the result of the convolution as
\begin{equation}\label{eq:theory-conv-first}
	\int_{\Sspv}{\dvK\left(\homega_i, \homega_o\right)\dvf\left(\homega_i\right)\rmd \homega_i},
\end{equation}
where $\dvf$ is the input Stokes vector field.
On the other hand, the rotation equivariance yields:
\begin{equation} \label{eq:theory_mueller_rotequi}
	\dvK\left(\vec R \homega_i, \vec R\homega_o, \right) = \vec R_\calM \left[\dvK\left(\homega_i,\homega_o\right)\right]\quad \forall \vec R\in\SOgroupv,
\end{equation}
where $\vec R_\calM \left[\cdot\right]$ is the rotation on Mueller transforms.
Here, $\vec R_\calM \left[\cdot\right]$ is defined as the composition of three Mueller transforms $\vec R_{\calS}$, $\dvK$, and $\vec R_{\calS}^{-1}$ via matrix multiplication as
\begin{equation}\label{eq:theory_rot_mueller}
	\vec R_\calM \left[\dvK\right]\coloneqq \vec R_\calS \dvK \vec R_\calS^{-1},
\end{equation}
where $\vec R_{\calS}$ is defined in Equation~\eqref{eq:stokes_rotation}.

Now we will define the corresponding \emph{kernel} from the above linear and rotation equivariant operator.
Moving back to the scalar spherical convolution, the kernel can be obtained by using Supplemental Equation~\eqref{eq:sph_conv_roteqv_delta}.
In particular, the scalar spherical convolution kernel can be obtained from the output of the convolution operation when the input source $f\left(\homega_i\right)$ is a Dirac delta at the north pole $\delta\left(\homega_i,\hat z_g\right)$.
Similarly, we can naturally extend this to the polarization by using a Dirac delta Stokes vector $\dvf\left(\homega_i\right)=\dvs_i\delta\left(\homega_i,\hat z_g\right)$ with any Stokes vector $\dvs_i\in \STKsp{\hat z_g}$.
Substituting this Dirac delta Stokes vector into Equation~\eqref{eq:theory-conv-first}, we can get the resulting Stokes vector field as
\begin{equation}\label{eq:theory_kernelconv_result}
	\dvK\left(\hat z_g, \homega_o\right)\dvs_i.
\end{equation}
Here we have two choices to define the kernel, the Stokes vector fields in Equation~\eqref{eq:theory_kernelconv_result} itself with a fixed $\dvs_i$ or a Mueller transform $\dvK\left(\hat z_g, \homega_o\right)$ as a function of $\homega_o$.
Among the choices, it is natural to choose the latter one since the former choice cannot store all the information of the spin \tcboxStS{\text{0-to-0}}, \tcboxStV{\text{0-to-2}}, \tcboxVtS{\text{2-to-0}}, and \tcboxVtV{\text{2-to-2}} parts of $\dvK$.
Finally, we define the kernel using rotation equivariance in Equation~\eqref{eq:theory_mueller_rotequi} to $\dvK\left(\hat z_g, \homega_o\right)$, which reduces $\dvK\left(\hat z_g, \homega_o\right)$ into a Mueller matrix as a function of single zenith angle~$\theta$.
As a result, the kernel is defined as
\begin{equation} \label{eq:theory-conv-kernel}
	\bfk\left(\theta\right)\coloneqq\left[\dvK\left(\hat z_g, \homega_{\mathrm{sph}}\left(\theta,\phi\right)\right)\right]^{\frF_{\theta\phi}\left(0,\phi\right)\to \frF_{\theta\phi}\left(\theta,\phi\right)} \in \R^{4\times 4},
\end{equation}
which is independent of the choice of $\phi$.
Note that when evaluating the above equation to obtain the numeric Mueller matrix from the Mueller transform $\dvK$ under the $(\theta,\phi)$, we have to consider the alignment between incident and outgoing frames.
In particular, we have to rotate the incident frame by $\phi$ to get $\frF_{\theta\phi}\left(0,\phi\right)$, so that the incident and outgoing frames are always aligned along the great circle of the constant $\phi$, as illustrated in Figure~\ref{fig:theory_convolution_kernel}(a).

Finally, by using the defined kernel above and reformulating Equation~\eqref{eq:theory-conv-first}, we can define the \emph{polarized spherical convolution}, which is the spherical convolution on Stokes vector fields as follows.
}
\begin{szTitledBox}{Polarized spherical convolution}
	\vspace{-1mm}
	A \emph{spherical convolution kernel for Stokes vector fields} is defined as a function $\dvk\colon \Sspv\to \MUEsp{\hat z_g}{\homega}$ which maps a single direction to a Mueller transform and has an azimuthal symmetry, i.e., its numeric kernel $\bfk:\left[0,\pi\right]\to \R^{4\times 4}$ is defined independent of $\phi$ as:
	\begin{equation} \label{eq:theory_convkernel_geo2num}
		\bfk\left(\theta\right) \coloneqq  \left[\dvk\left(\theta,\phi\right)\right]^{\frtp\left(0,\phi\right)\to\frtp\left(\theta,\phi\right)}.
	\end{equation}
	Then the convolution of the kernel $\dvk$ and a Stokes vector field $\dvf\colon \Sspv\to \STKsp{\homega}$ is defined as:
	\begin{equation}\label{eq:theory_conv_def}
		\begin{split}
			\left(\dvk\ast \dvf\right)\left(\homega\right) &= \int_{\Sspv}{\left(\vec R_{\hat z_g\to\homega'}\right)_\calM\left[\dvk\left(\vec R_{\hat z_g\to\homega'}^{-1}\homega\right)\right]\dvf\left(\homega'\right) \rmd \homega'} \\
			&= \int_{\Sspv}{\left[\bfk\left(\cos^{-1}\homega\cdot\homega'\right)\right]_{\frF_i\to \frF_o}\dvf\left(\homega'\right)\rmd \homega'}.
		\end{split}
	\end{equation}
	Here, $\vec R_{\hat z_g \to\homega'}$ is a rotation transform satisfying $\vec R_{\hat z_g \to\homega'}\hat z_g = \homega'$, and choices of $\vec R_{\hat z_g \to\homega'}$ does not affect on the definition of the convolution. $\frF_i$ and $\frF_o$ are local frames at $\homega'$ and $\homega$, respectively, such that their $x$ axes are aligned along the common great circle of $\homega'$ and $\homega$.
	\vspace{-1mm}
\end{szTitledBox}

\subsubsection{Polarized SH coefficients for convolution kernels}
\label{sec:theory_conv_kernel}
\begin{figure}[tp]
	\centering
	\includegraphics[width=\columnwidth]{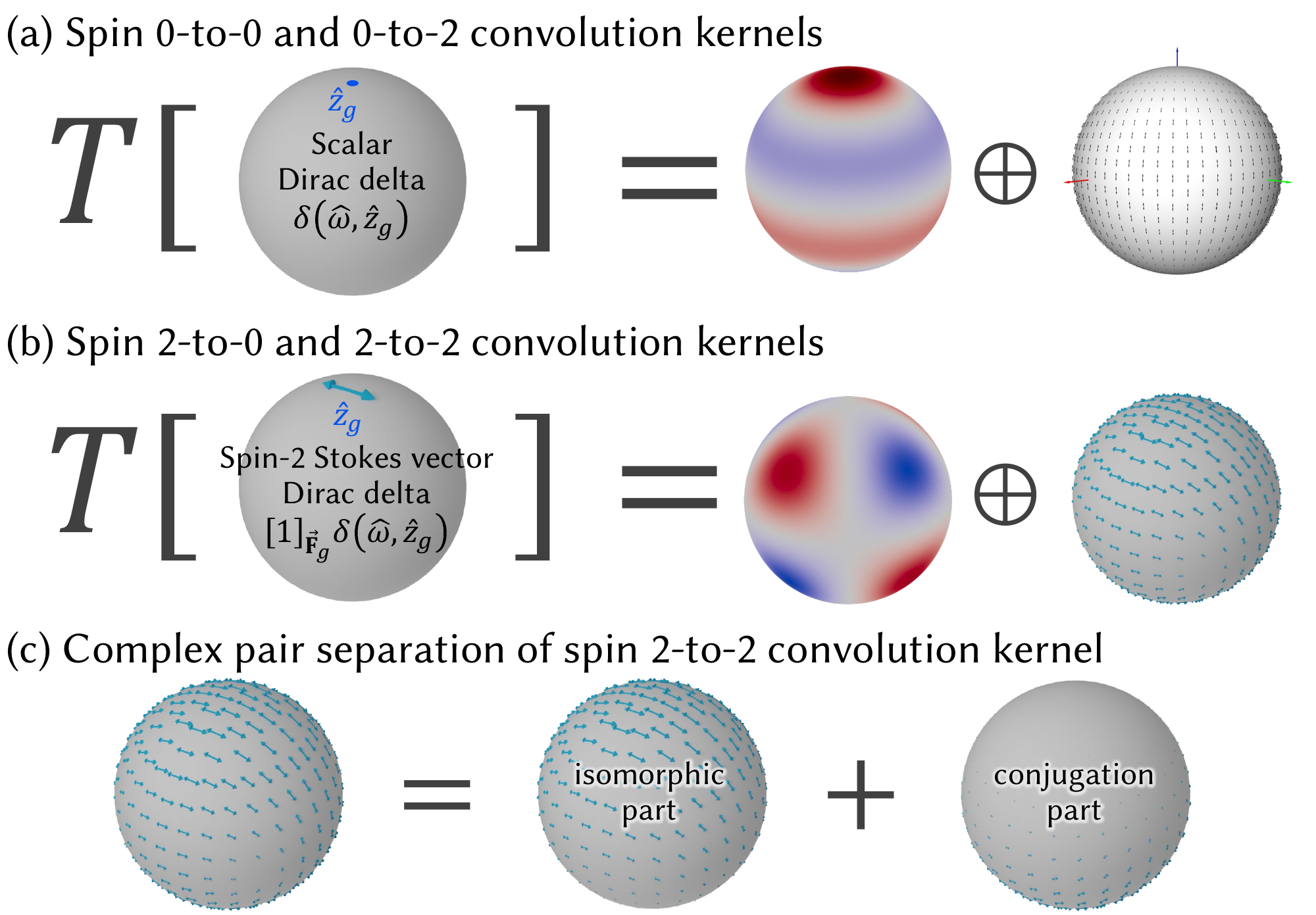}%
	\vspace{-3mm}
	\caption[]{\label{fig:theory_convolution_kernel_parts}
		\NEW{As conventional convolution kernel operates,} %
		(a) spin 0-to-0 and 0-to-2 convolution kernels for polarized spherical convolution can be characterized as the output of a rotation equivariant linear operator on a Dirac delta scalar field. Due to rotation equivariance, output Stokes vector fields are azimuthally symmetric for spin-0 and spin-2 components, i.e., expand with $Y_{l0}$ and $\sY{l0}$ bases. On the other hand, (b) gives Stokes vector fields with $e^{\pm 2i\phi}$ dependency, i.e., the spin 2-to-0 kernel expands with $Y_{l,-2}^C$ basis. 
		(c)~The spin 2-to-2 kernel should even be split into two parts: an isomorphic part and a conjugation part using complex pair separation, which is discussed in Equations~\eqref{eq:theory_mat2comp} and~\eqref{eq:theory_comppair2mat}.}
	\vspace{-2mm}		
\end{figure}

\NEW{Recall that the scalar spherical convolution kernel $k\colon \left[0,\pi\right] \to \R$, which is an operand for convolution, can be converted to scalar SH coefficients.
Similarly, we will show how to convert the polarized spherical convolution kernel $\dvk$ to the PSH coefficients.
While it requires rigorous derivation steps, we only provide a comprehensive observation-based description.
Refer to Supplemental Section~\ref{sec:theory-conv} for the full derivation.

We start by considering the polarized spherical convolution kernel $\dvk$ as a function.
Then the domain of the $\dvk$ is simply $\left[0,\pi\right]\ni \theta$, and its PSH coefficient has a single $l$-index and no $m$-index similar to conventional kernels as Equation~\eqref{eq:bkgnd_SH_sph_conv}.
On the other hand, the codomain of the $\dvk$ is Mueller transforms, and its PSH coefficient has both $p_o$ and $p_i$ similar to polarized coefficient matrices as described in Section~\ref{sec:theory_linop}.
Hence, we can write the convolution coefficients in PSH as $\rmk_{lp_op_i}$, and the remaining question is where these coefficients come from (i.e., from which basis function and which part of the kernel on the angular domain).

We first have a look at the resulting Stokes vector field in Equation~\eqref{eq:theory_kernelconv_result}, which can be written as $\dvk\left( \homega\right)\dvs_i$.
Taking $\theta\phi$-frame field, we have: 
\begin{equation} \label{eq:theory-conv-kernel-stk}
	\bfk\left(\theta\right) \bfC\left(\phi\right)\bfs_i,
\end{equation}
where $\bfC$ indicates the frame conversion matrix defined in Equation~\eqref{eq:stk_coord_convert} and $\bfs_i\coloneqq \left[\dvs_i\right]^{\frF_g}$.
Note that the frame conversion matrix should be inserted to convert the Stokes component representation~$\bfs_i$ from $\frF_g$ to $\frF_{\theta\phi}\left(0,\phi\right)$.
Now, similar to the previous derivations, we will consider the spin-0 ($s_0, s_3$) and the spin-2 ($s_1, s_2$) parts of the incident Stokes vector $\bfs_i$ separately.

For the spin-0 Stokes components of $\bfs_i$, we can consider $\bfs_i = \left[1, 0, 0, 0\right]^T$ or $\left[0, 0, 0, 1\right]^T$.
Then the conversion matrix $\bfC\left(\phi\right)$ vanishes so that Equation~\eqref{eq:theory-conv-kernel-stk} turns into an azimuthally symmetric full Stokes vector field, which can be expanded by zonal harmonics $Y_{l0}$ (scalar SH kernel coefficients),
and a subset of spin-2 SH $\sY{l0}$.
Recall that the $\phi$ dependency of spin-2 SH $\sY{lm}$ is characterized as $e^{im\phi}$ in Equation~\eqref{eq:theory-spin2-SH}. Thus, similar to scalar SH, spin-2 SH $\sY{l0}$ with $m=0$ also have azimuthal symmetry, shown in Figure~\ref{fig:theory_s2SH_nvSH_b}.
In summary, Figure~\ref{fig:theory_convolution_kernel_parts}(a) illustrates the symmetry of the kernel Stokes vector field.}

\NEW{%
For the spin-2 Stokes components of $\bfs_i$, we can consider $\bfs_i=\left[0, 1, 0, 0\right]^T$.
Then the $s_0$ component of Equation~\eqref{eq:theory-conv-kernel-stk} turns into:
\begin{equation} \label{eq:theory-conv-kstk-v2s}
	\begin{bmatrix}
		\rmk_{01}\left(\theta\right) & \rmk_{02}\left(\theta\right)
	\end{bmatrix} \R^2\left(e^{-2i\phi}\right).
\end{equation}
From the $\phi$ dependency which comes from $\bfC\left(\phi\right)$, this scalar field can be expanded by $Y_{l,\pm2}^R$.
However, through some derivation details, we find that the best way to describe it is using $Y_{l,-2}^C$.
Note that the $s_3$ component for Equation~\eqref{eq:theory-conv-kernel-stk} is also expanded with the same bases.
Now considering the spin-2 ($s_1, s_2$) component of Equation~\eqref{eq:theory-conv-kernel-stk}, it turns into similarly as follows:
\begin{equation}
	\begin{bmatrix} \label{eq:theory-conv-kstk-v2v}
		\rmk_{11}\left(\theta\right) & \rmk_{12}\left(\theta\right) \\
		\rmk_{21}\left(\theta\right) & \rmk_{22}\left(\theta\right)
	\end{bmatrix} \R^2\left(e^{-2i\phi}\right).
\end{equation}
These Stokes vectors in Equations~\eqref{eq:theory-conv-kstk-v2s} and~\eqref{eq:theory-conv-kstk-v2v} are also illustrated in Figure~\ref{fig:theory_convolution_kernel_parts}(b).
In addition, we can apply the complex pair separation described in Equations~\eqref{eq:theory_mat2comp} and~\eqref{eq:theory_comppair2mat}.
As a result, we can split Equation~\eqref{eq:theory-conv-kstk-v2v} into two spin-2 Stokes vector fields:
\begin{equation}\label{eq:theory-conv-kstk-comppair}
	\R^2\left(\tilde\rmk_\mathrm{iso}\left(\theta\right) e^{-2i\phi}\right) + \R^2\left(\tilde\rmk_\mathrm{conj}\left(\theta\right) e^{2i\phi}\right),
\end{equation}
which is also described in Figure~\ref{fig:theory_convolution_kernel_parts}(c).
From the $\phi$ dependency here, we observe that the isomorphic part expands using $\sY{l,-2}$ bases, and the conjugation part does using $\sY{l2}$.

Eventually, the coefficients of the convolution kernel $\bfk\left(\theta\right)$ with respect to PSH are defined as follows.

\begin{szTitledBox}{Convolution coefficients in polarized spherical harmonics}
	\vspace{-1mm}
	\begin{subequations}\label{eq:theory_conv_coeff}
	\begin{align}
		\label{eq:theory_conv_coeff_s2s}
		\tcboxmathStS{\rmk_{l,\left\{0,3\right\},\left\{0,3\right\}}} &\coloneqq \tcboxmathStS{\int_{\Sspv}{ Y_{l0}^*\left(\homega\right)\rmk_{\left\{0,3\right\},\left\{0,3\right\}}\left(\homega\right) \rmd \homega}} \\
		\label{eq:theory_conv_coeff_s2v}
		\tcboxmathVtS{\tilde \rmk_{l,\left\{0,3\right\},\bfp}} &\coloneqq \tcboxVtS{\int_{\Sspv}{ Y_{l,-2}^{C,*}\left(\homega\right)\tilde \rmk_{\left\{0,3\right\},\bfp}\left(\homega\right) \rmd \homega}} \\
		\label{eq:theory_conv_coeff_v2s}
		\tcboxmathStV{\tilde \rmk_{l,\bfp,\left\{0,3\right\}}} &\coloneqq \tcboxStV{\int_{\Sspv}{ \sY{l0}^{*}\left(\homega\right)\tilde \rmk_{\bfp,\left\{0,3\right\}}\left(\homega\right) \rmd \homega}} \\
		\label{eq:theory_conv_coeff_v2vi}
		\tcboxmathVtV{\tilde \rmk_{l,\mathrm{iso}}} &\coloneqq \tcboxVtV{\int_{\Sspv}{ \sY{l,-2}^{*}\left(\homega\right)\tilde \rmk_{\mathrm{iso}}\left(\homega\right) \rmd \homega}} \\
		\label{eq:theory_conv_coeff_v2vc}
		\tcboxmathVtV{\tilde \rmk_{l,\mathrm{conj}}} &\coloneqq \tcboxVtV{\int_{\Sspv}{ \sY{l,2}^{*}\left(\homega\right)\tilde \rmk_{\mathrm{conj}}\left(\homega\right) \rmd \homega}}
	\end{align}
	\end{subequations}
	\vspace{-2mm}
\end{szTitledBox}
\noindent While $\bfk\left(\theta\right)$ is defined on $\left[0,\pi\right]$, Equation~\eqref{eq:theory_conv_coeff} considers each part of $\bfk$ as a function defined on $\Sspv$. Here, the $\phi$ dependency of each part can be assumed to vanish the $\phi$ dependency of the entire integrand. For instance, the $\phi$ dependency of $\tilde \rmk_\mathrm{iso}$ and $\tilde \rmk_\mathrm{conj}$ is considered as $e^{-2i\phi}$ and $e^{2i\phi}$, respectively, as described in Equation~\eqref{eq:theory-conv-kstk-comppair}.

On both hand sides in Equations~\eqref{eq:theory_conv_coeff_s2v} and~\eqref{eq:theory_conv_coeff_v2s}, the subscript $\bfp$ indicates the collection of the indices 1 and 2 in the Mueller matrix $\bfk$ and the tilde symbol converts it into a single complex number.
In Equation~\eqref{eq:theory_conv_coeff_v2s}, we explicitly write the superscript $C$ to avoid confusion with the real SH $Y_{lm}^R$.
Recall that we mentioned the convolution coefficient can be written as $\rmk_{lp_op_i}$.
It can be constructed directly from the above five types of complex coefficients by converting them into $\R^2$ or $\Rtt$.
However, we found that the complex-valued forms in Equations~\eqref{eq:theory_conv_coeff_s2s} to~\eqref{eq:theory_conv_coeff_v2vc} are more convenient for evaluating the convolution operation in the polarized SH domain.
}

\subsubsection{Polarized spherical convolution in polarized spherical harmonics}
\label{sec:theory_conv_polarsh}
\NEW{Using the convolution coefficients, we can now perform spherical convolution on a Stokes vector field $\dvf\colon\Sspv\to\STKsp{\homega}$ with PSH coefficients.
Recall that scalar spherical convolution is evaluated as an element-wise product between the kernel coefficient and the coefficient of the input, as described in Equation~\eqref{eq:bkgnd_SH_sph_conv}.
Similarly, polarized spherical convolution is evaluated by an element-wise product with the coefficients $\rmf_{lmp}$ of $\dvf$ and other coefficients obtained by flipping the sign of the $m$ index from $\rmf_{lmp}$ as follows.
\begin{szTitledBox}{Polarized spherical convolution theorem}
	\vspace{-1mm}
	The PSH coefficients of the convolution of a kernel $\dvk$ and a Stokes vector field $\dvf$, denoted by $\rmf_{lmp}' \coloneqq \lrangle{\dvY_{lmp}, \dvk \ast \dvf}_\calF$, is evaluated as 
	\begin{subequations}\label{eq:theory_conv_to}
		\begin{equation}\label{eq:theory_conv_to0}
		\begin{split}
			\rmf_{lm,\left\{0,3\right\}}' = \sqrt\frac{4\pi}{2l+1} \bigg[&\sum_{p_i=0,3}\tcboxmathStS{\tilde \rmk_{l,\left\{0,3\right\},p_i}} \rmf_{lmp_i} \\
			+&\!\!\sum_{m'\in\left\{\pm m\right\}}\Re\left(\tcboxVtS{W_{mm'}^{2\to 0, *}\tilde \rmk_{l,\left\{0,3\right\},\bfp}^*} \tilde \rmf_{lm'\bfp}\right)\bigg],
		\end{split}
	\end{equation}
	\begin{equation}\label{eq:theory_conv_to2}
		\begin{split}
			\tilde \rmf_{lm\bfp}'= \sqrt\frac{4\pi}{2l+1} \bigg[&\sum_{p_i=0,3}\sum_{m'\in\left\{\pm m\right\}}\tcboxmathStV{W_{mm'}^{0\to 2}\tilde \rmk_{l,\bfp,p_i}} \rmf_{lm'p_i} \\
			+&\tcboxVtV{\tilde \rmk_{l,\mathrm{iso}}\tilde\rmf_{lm\bfp}+\left(-1\right)^{m}\tilde k_{l,\mathrm{conj}}\tilde\rmf_{l,-m\bfp}^*}\bigg],
		\end{split}
	\end{equation}
	where
	\begin{equation}\label{eq:theory_conv_minor}
		\tilde \rmf_{lm\bfp} \coloneqq \rmf_{lm1}+\rmf_{lm2}i,
	\end{equation}
	\end{subequations}
	and $W_{mm'}^{2\to0}$ and $W_{mm'}^{0\to2}$ are simple constants taking values of $0$, $\pm \frac1{\sqrt2}$, or $\pm\frac i{\sqrt2}$ following Supplemental Equations~\eqref{eq:conv0p_phase} and~\eqref{eq:convp0_phase} in Supplemental Section~\ref{sec:theory-conv-SH}.
	\vspace{-2mm}
\end{szTitledBox}
\noindent
Note that the constant weights $W_{mm'}^{2\to 0}$ and $W_{mm'}^{0\to 2}$ become zero when $\abs{m}\ne\abs{m'}$, one when $m=m'=0$, and are evaluated as Supplemental Equations~\eqref{eq:conv0p_phase} and~\eqref{eq:convp0_phase} otherwise.
This polarized spherical convolution is nearly an element-wise product for the indices $l$ and~$m$, but similar to $4\times 4$ matrix-vector product for the index $p$.

}

\begin{figure}[tp]
	\centering
	\includegraphics[width=\columnwidth]{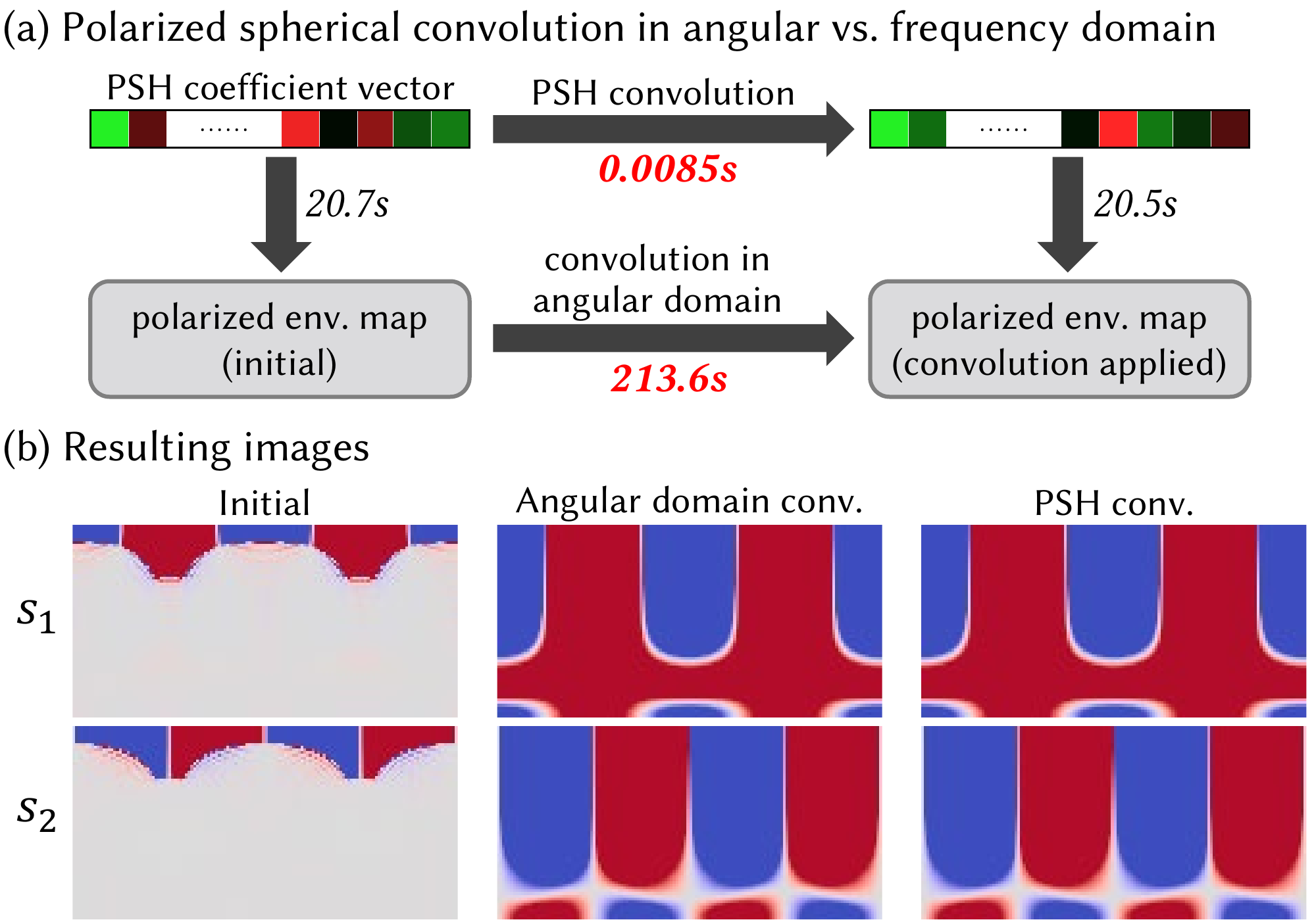}%
	\vspace{-3mm}
	\caption[]{\label{fig:theory_conv_validation}
		(a) We compare polarized spherical convolution performed in angular and frequency (polarized SH) domains to validate our polarized spherical convolution theorem in Equation~\eqref{eq:theory_conv_to}.
		For fair validation, methods on both domains start from the finite (band limited) coefficient vectors of a polarized environment map, which are performed once reconstructing into the angular domain and once the convolution operation is performed. We test frequency levels $l<100$ and reconstruct pixel numbers $= \left(l+1\right)^2$. We observe that convolution in the frequency domain is significantly faster while two operations give identical results  (b).}
	\vspace{-2mm}		
\end{figure}

\paragraph{Validation between the angular and frequency domains}
\NEW{We here provide a numerical experiment that compares polarized spherical convolution in the angular and frequency domains, and also Supplemental Section~\ref{sec:theory-conv-SH} provides a complete step-by-step derivation to validate our polarized spherical convolution theorem.
For the computation in the angular domain, we use an analytic kernel $\bfk\left(\theta\right)=\mathrm{diag}\left(\pi - \theta\right)$ for convolution.
First, we project the polarized environment map onto the PSH coefficient vector and take the finite (band-limited) coefficient vector \NEW{for a fair comparison with the computation in the frequency domain}.
Then, we reconstruct the polarized environment map to the angular domain and perform convolutions on it.
For the frequency domain,
we first perform convolution on the frequency domain and then reconstruct the polarized environment map.
Figure~\ref{fig:theory_conv_validation}(a) depicts the validation pipeline and computation time.
The convolution in PSH is significantly faster than the angular domain operation.
\NEW{In addition,} the two results are identical, as shown in Figure~\ref{fig:theory_conv_validation}(b).
}

\begin{figure}[tp]
	\centering
	\includegraphics[width=\columnwidth]{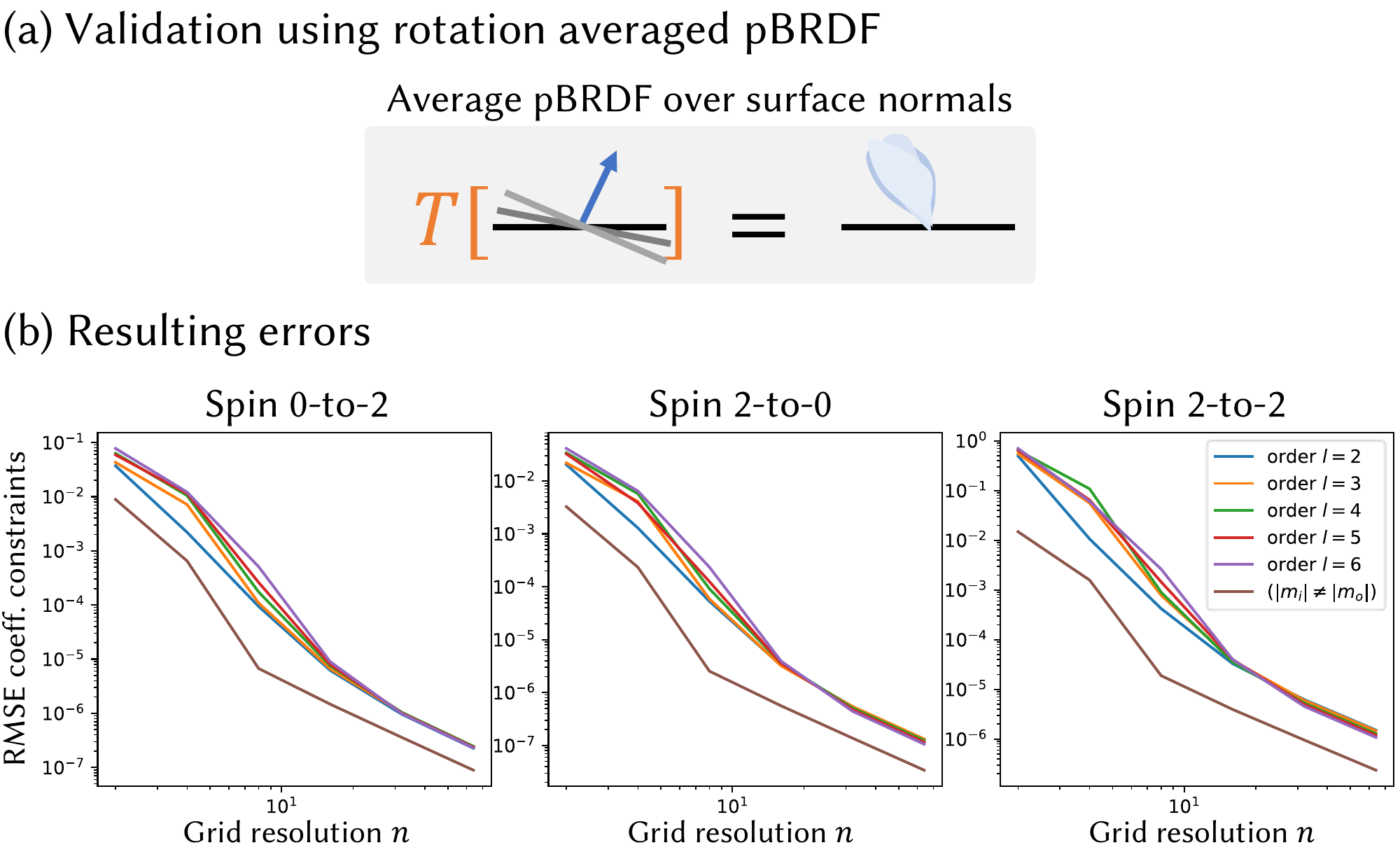}%
	\vspace{-3mm}
	\caption[]{\label{fig:theory_pBRDF_approx_conv}
	As scalar sphere convolution theorem in Equation~\eqref{eq:bkgnd_SH_sph_conv} can be expanded as a diagonal coefficient matrix, our polarized spherical convolution theorem in Equations~\eqref{eq:theory_conv_to0} to~\eqref{eq:theory_conv_minor} can also be expanded to a coefficient matrix with some linear constraints. Averaging a pBRDF for each normal vector of the material as depicted in (a), we can enforce rotation equivariance to the pBRDF. We validate our convolution theorem by measuring the projection errors of the coefficient matrix of the rotation averaged pBRDF to the linear constraints of convolution operators. (b) provides the error virtually converges to zero as the grid resolution $n$, which indicates the number of samples of normal vectors, increases. Note that we separate the projection error into each of spin 0-to-2, 2-to-0, and 2-to-2 submatrices (each of three plots), coefficients at $\abs{m_i}=\abs{m_o}$ for each order $l$ (first five curves in the legend), and coefficients at $\abs{m_i}\ne \abs{m_o}$, for better analysis.}
	\vspace{-2mm}		
\end{figure}

\paragraph{Validation using rotation averaged pBRDF}
Note that the scalar sphere convolution theorem in Equation~\eqref{eq:bkgnd_SH_sph_conv} can be expanded as a coefficient matrix with linear constraints since the entry-wise product of two vectors is equivalent to the product of a diagonal matrix and a vector.
\NEW{Similarly}, our polarized spherical convolution theorem from Equations~\eqref{eq:theory_conv_to0} to~\eqref{eq:theory_conv_minor} can also be expanded to a coefficient matrix with some linear constraints, \NEW{described} in Supplemental Section~\ref{sec:theory-conv-SH}.
We can approximate a pBRDF coefficient matrix into convolution coefficients by averaging each normal vector of the material, as described in Figure~\ref{fig:theory_pBRDF_approx_conv}(a), to ensure rotation equivariance.
\NEW{Figure~\ref{fig:theory_pBRDF_approx_conv}(b) shows} the projection error of the rotation averaged pBRDF to the linear constraints of convolution.
\NEW{We can observe that} RMS errors virtually converge to zero, which supports our polarized spherical convolution theorem in the frequency domain.

%% file: pprt.tex
\begin{figure*}[tpb]
	\centering
	\includegraphics[width=\linewidth]{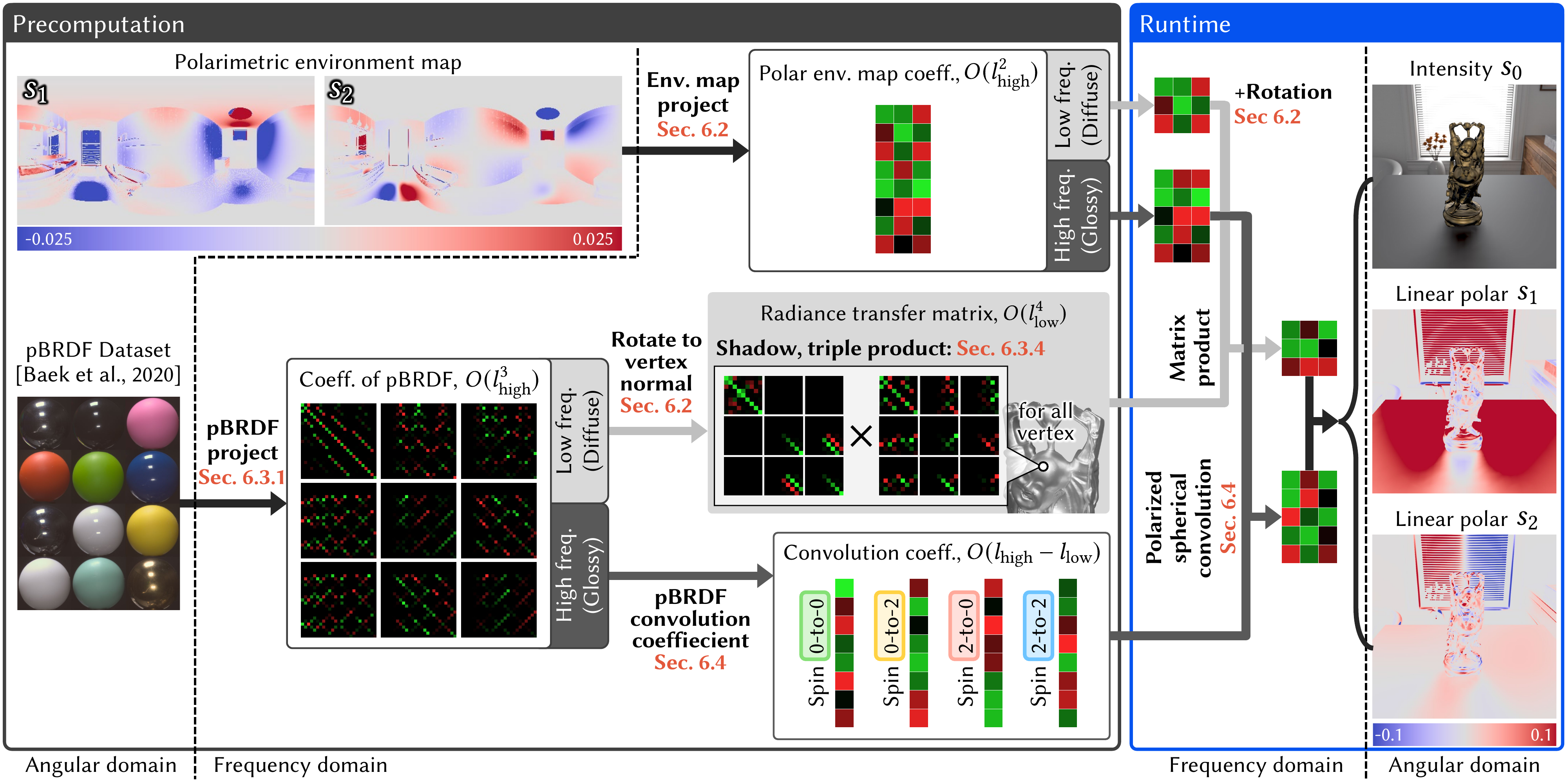}%
	\vspace{-3.5mm}
	\caption[]{\label{fig:theory_rendering_pipeline}
		Our real-time polarized rendering pipeline consists of precomputation and run time parts. Each process described as arrows corresponds to our spin-weighted spherical harmonics operations discussed in Sections~\ref{sec:theory_rotinv} to \ref{sec:theory_conv}. %
		We use $l_\mathrm{low}=4$ and $l_\mathrm{high}=9$ for our rendering results.
	}
	\vspace{-2mm}		
\end{figure*}

\section{Precomputed Polarized Radiance Transfer}
\label{sec:pprt}

This section presents a \NEW{real-time} rendering pipeline and results of our precomputed polarized radiance transfer, which utilizes PSH and operations in Section~\ref{sec:our_theory}. \NEW{How each theoretical component in Section~\ref{sec:our_theory} contributes to our rendering pipeline is summarized in Figure~\ref{fig:theory_rendering_pipeline}.}
Note that our main challenge is related to the linear polarization components $s_1$ and $s_2$; we omit the circular polarization component $s_3$ in rendering results since it can be simply processed like total intensity $s_0$. We also refer to our supplemental video for real-time rendering results.

\paragraph{Processing polarized environment map}
We generate a synthetic polarized environment map using the polarized variant and polarization-aware materials in Mitsuba 3~\cite{Mitsuba3}.
In the precomputation stage, we store the PSH coefficient vector of the environment map up to orders (frequency bands) $l\le l_\mathrm{high}=9$ using Equation~\eqref{eq:theory-PSH-coeff}.
Then, in the runtime, %
these coefficients are rotated to each object frame using Equations~\eqref{eq:theory_rotate_PSH_coeff} and \eqref{eq:theory_PSH_rotmat} (Section~\ref{sec:theory_rotinv}).

\paragraph{pBRDF projection to PSH coefficients} 
In the precomputation stage, we also convert data-based isotropic pBRDFs from~\citet{baek2020image} into PSH coefficient matrices using Equation~\eqref{eq:theory_psh_coeff_linop}.
When converting and storing the pBRDF coefficient matrix, we utilize the sparsity from the isotropy of pBRDF described in Equations~\eqref{eq:theory_iso2coeff} and~\eqref{eq:theory-pBRDF-isoV2V}.
For the cut-off order, we select $l_\mathrm{high}=9$, same as the environment map (Section~\ref{sec:theory_linop}).

\paragraph{Low--high frequency separation}
If we increase the order $l_\mathrm{high}$, the radiance transfer result will converge the the ground truth.
However, the BRDF coefficient matrix requires the complexity of $O\left(l^4\right)$, and simply increasing the order by utilizing a full radiance transfer matrix might \NEW{significantly reduce} the computational efficiency.
Therefore, we divide the coefficients into low-frequency and high-frequency parts.
Then, we apply the O$\left(l^4\right)$ radiance transfer matrix only to the low-frequency part rather than utilizing full coefficients.
The remaining high-frequency part will be handled in a distinct convolution pipeline.
In our implementation, such separation is done in $l_\mathrm{low}=4$, so the low-frequency part contains $0\le l \le 4$ and $5 \le l\le 9$ for the high-frequency part.

\paragraph{Radiance transfer using PSH coefficients} 
Now we rotate the low-frequency part of projected pBRDF coefficients to each vertex normal, yielding the simple unshadowed version of polarized radiance transfer operators.
In the runtime, similar to the low--high-frequency separation in pBRDF, the coefficient from the environment map can also be separated by simply splitting the coefficient vector.
After that, radiance transfer can be done by a simple matrix-vector product between the radiance transfer operator and the low-frequency part of the environment map, as described in Equation~\eqref{eq:theory-PSH-matmul} (Section~\ref{sec:theory_linop}).

\paragraph{Shadowed transfer using triple product} 
In the previous paragraph, we propose the unshadowed version of the radiance transfer operator.
However, the shadow can also be considered using the triple product as described in Section~\ref{sec:theory_shadow_tp}.
To do so, we evaluate visibility for each vertex by casting 2,000 rays from the vertex in the precomputation stage.
Then we convert it into the SH coefficient vector and convert it again to a coefficient matrix using SH and PSH triple product in Equations~\eqref{eq:theory_tp_s2s} and~\eqref{eq:theory_tp_v2v}.
Finally, applying the matrix product of the projected shadow map coefficient and the unshadowed transfer matrix yields the shadowed transfer matrix that can replace the unshadowed transfer matrix (Section~\ref{sec:theory_shadow_tp}).

\begin{figure}[tp]
	\centering
	\includegraphics[width=\columnwidth]{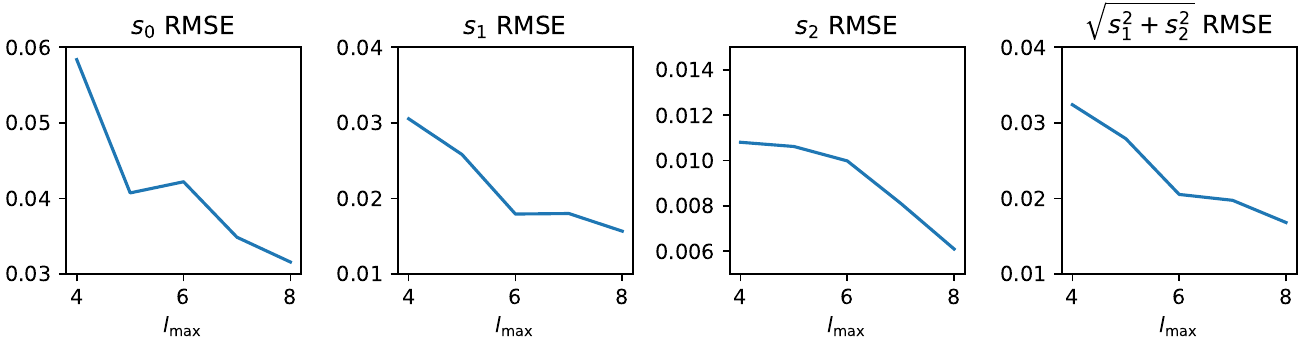}%
	\vspace{-3mm}
	\caption[]{\label{fig:result_GT_comparison_plot}
		\NEW{We validate our real-time polarized rendering with shadowed radiance transfer compared with Mitsuba~3 ray tracer. We report RMSE error, which decreases close to zero as the cut-off frequency $l_{\mathrm{max}}$ increases. We refer to Supplemental Figure~\ref{fig:result_GT_comparison} for resulting rendered images and difference maps.}}
	\vspace{-2mm}		
\end{figure}

\paragraph{\NEW{Validation with shadowed transfer}}
\NEW{We also provide a validation experiment by comparing ours with a physically-based polarization ray tracer, Mitsuba~3. Since our spherical convolution method in the PPRT assumes additional symmetry for pBRDFs and it is already validated in different experiments in Figures~\ref{fig:theory_conv_validation} and~\ref{fig:theory_pBRDF_approx_conv}, we experiment our shadowed transfer without high-frequency convolution approximation. Figure~\ref{fig:result_GT_comparison_plot} compares RMSE values between each Stokes component of the rendered images of Mitsuba~3 and our method. We observe that the error for each component decreases close to zero as the cut-off frequency $l_\mathrm{max}$ increases. Note that the errors will ideally converge to zero when the vertex resolution of the scene additionally increases. We refer to Supplemental Figure~\ref{fig:result_GT_comparison} for rendered images and difference maps.}

\begin{figure*}[tp]
	\centering
	\vspace{-4mm}	
	\includegraphics[width=\linewidth]{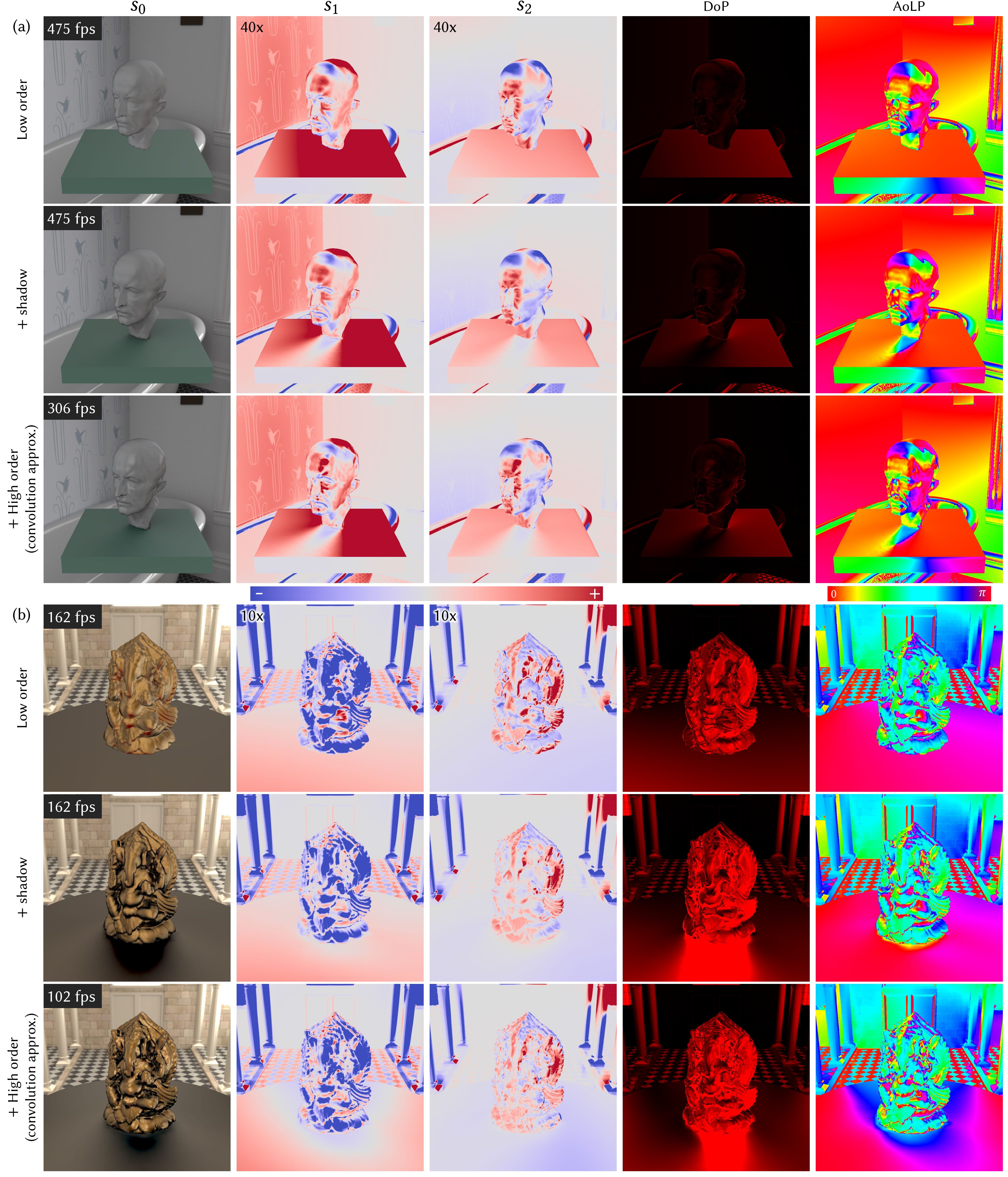}%
	\vspace{-4mm}
	\caption[]{\label{fig:result_ablations}
		Ablation study. We provide a real-time frequency-domain environment map lighting method with linear polarization. We use the PSH coefficient vector of a polarized environment map computed in Section~\ref{sec:theory_rotinv}. The coefficient matrix (radiance transfer matrix) of \citet{baek2020image}'s data-based pBRDF computed in Section~\ref{sec:theory_linop}, which yields unshadowed transfer shown in the first rows in (a) and (b). We also provide shadowed transfer using Supplemental Section~\ref{sec:theory-tp} and efficient pBRDF approximation for specular appearance using polarized spherical convolution in Section~\ref{sec:theory_conv}.
		}
	\vspace{-2mm}		
\end{figure*}

\begin{figure*}[tp]
	\centering
	\vspace{-3.5mm}	
	\includegraphics[width=\linewidth]{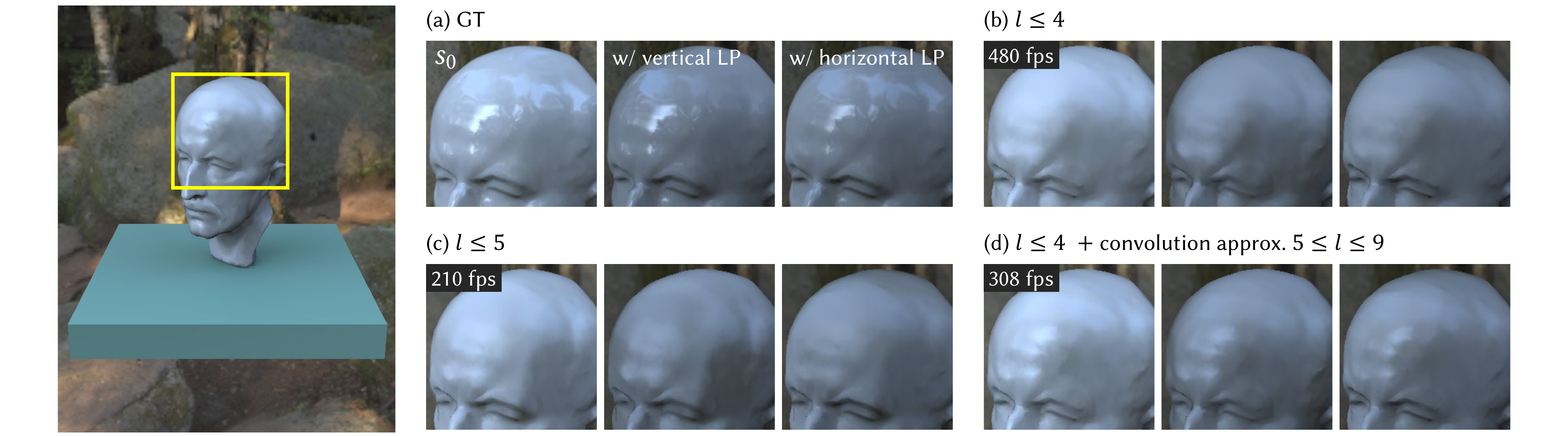}%
	\vspace{-3.5mm}
	\caption[]{\label{fig:app_prt_result}
		Although increasing the number of orders $l$ makes the result converge to the actual appearance (a), it suffers from quartic computational complexity as (c) reports less than half fps than (b) even though it uses one more frequency band. Rather than using the full radiance transfer matrix for high orders, (c) we can project the coefficient matrix into convolution coefficients to achieve efficient high order $l<10$ appearance, which provides much higher performance than the full matrix of $l<6$ in (b).}
	\vspace{-2mm}		
\end{figure*}

\paragraph{Efficient specular appearance using polarized spherical convolution} 
Now, for the remaining high-frequency part, we project the matrix into convolution coefficients by linear constraints of the convolution following the \citet{sloan2002precomputed}.
Note that following \citet{sloan2002precomputed}, convolution approximation of a reflected BRDF, which flips the reflected radiance with respect to the surface normal, is preferable to the original BRDF. Thus, we project the product of a reflection operator's coefficient matrix, introduced in Supplemental Section~\ref{sec:theory-refl}, and the radiance transfer matrix into a convolution coefficient.
This convolution approximation is based on the fact that a specular lobe of a BRDF usually has the \NEW{peak} at the mirror reflection direction so that we can approximate the flipped lobe along the normal as a rotation equivariant one.
Then, in runtime, we evaluate PSH values at the reflected direction of the view vector by normals rather than the view vector itself (Section~\ref{sec:theory_conv}).

To evaluate the impact of each rendering component, we conduct an ablation study as shown in Figure~\ref{fig:result_ablations}.
All experiments are done in the machine with an Intel i9-12900K CPU and an NVIDIA GeForce RTX 4090 GPU. All scenes are rendered in $1024\times 1024$ resolution.
\NEW{We refer to Supplemental Table~\ref{tb:result_info} for detailed specification the scene setups throughout the paper.}
The \emph{low-order} results only use low-frequency parts with unshadowed radiance transfer.
The \emph{+shadow} results use the same order as low-order results, but the shadow is considered.
The \emph{+high-order} results use our full pipeline, including the convolution approximation of the high-frequency part.
The result shows real-time performance in 102-475\,fps for polarized rendering, considering polarized environment lighting. From our shadowed light transport, we can see soft shadows due to environment maps not only in unpolarized $s_0$ images but also in linear polarization $s_1$ and $s_2$ images. When convolution approximation of pBRDF at high order is applied, specular behaviors are enhanced.

We also conduct another ablation experiment for convolution approximation, and the result is shown in Figure~\ref{fig:app_prt_result}.
We provide intensity images through two directions of linear polarizer for better intuition to see specular behavior.
If we use only low-order radiance transfer (Figure~\ref{fig:app_prt_result}(b)), it is computationally efficient that achieves 480\,fps, but it loses some high-frequency appearance.
Increasing the order to $0\le l\le 5$ makes the result close to the ground truth, but its performance is degraded to 210\,fps.
Finally, applying convolution approximation for $5 \le l \le 9$ and using full radiative transfer matrix for $0 \le l \le 4$ in Figure~\ref{fig:app_prt_result}(d) shows a much higher 308\,fps but a rich specular appearance than (c), which utilizes orders up to $\le 5$ for the transfer matrix.

%% file: discussion.tex
\section{Discussion}
\label{sec:discussion}

\subsection{Choice of PRT Framework}
There have been plenty of PRT methods and design choices for the PRT pipeline. For instance, \citet{sloan2005local} store BRDF into SH coefficient (frequency domain) along incident ray direction but tabulates several outgoing ray directions (angular domain). 
\citet{sloan2002precomputed} precompute coefficient matrix of self-shadow by directly simulating Equation~\eqref{eq:theory_shadow_direct} rather than converting coefficient vector of visibility mask followed by applying SH triple product. 
However, these choices are totally orthogonal to our main contribution.
For a better application of our method to polarization rendering,
our PPRT pipeline described in Figure~\ref{fig:theory_rendering_pipeline} is designed to be aimed to maximize usage of frequency domain operations (theoretical properties for polarized SH). %
For instance, to build a PPRT method with pBRDF tabulated for each outgoing radiance sample, any method can be plugged in, but to represent pBRDF into a full coefficient matrix, our method is required as described in Section~\ref{sec:theory_linop}. %

\subsection{\NEW{Physical Constraints}}
\label{sec:discuss-valid-range}

\begin{figure*}[tp]
	\centering
	\vspace{-1mm}
	\includegraphics[width=\linewidth]{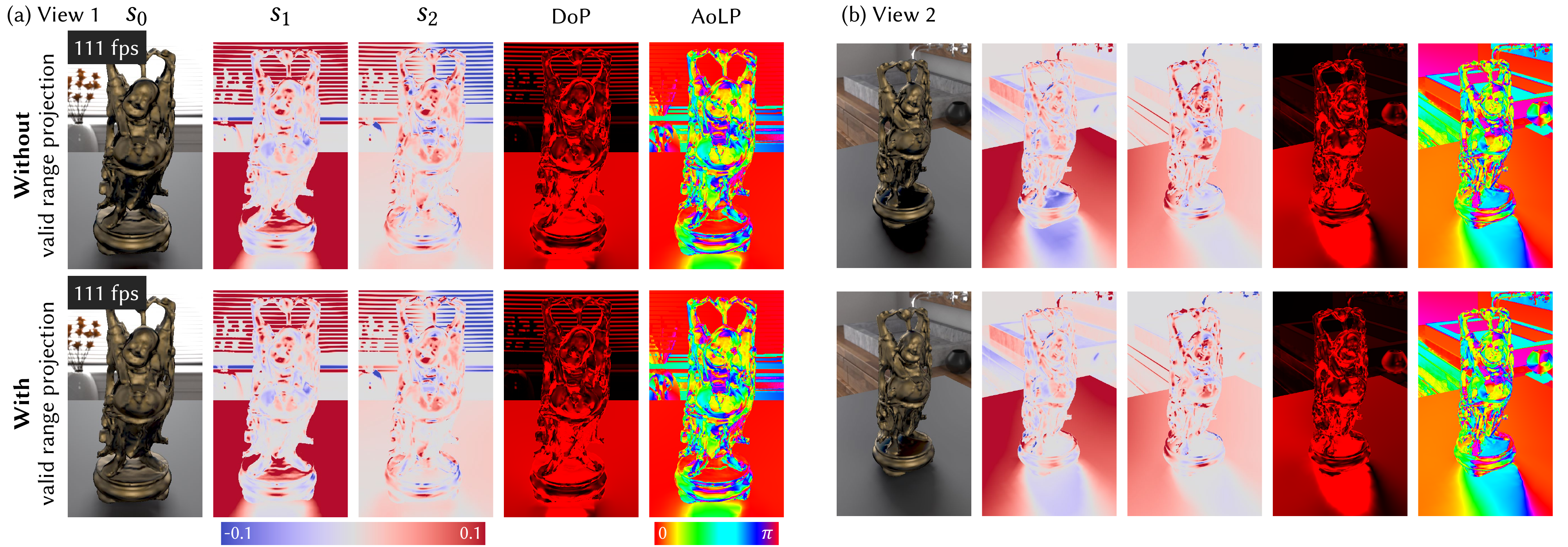}%
	\vspace{-4mm}
	\caption[]{\label{fig:result_validprojection}
		We can project the rendered image to the physically valid range of Stokes vectors as discussed in Section~\ref{sec:discuss-valid-range}, without loss of rendering time.
		\NEW{(a) and (b) show two views of the same scene, respectively.}
	}
	\vspace{-2mm}
\end{figure*}
\paragraph{\NEW{Valid range of Stokes vectors}}
It is known that the physically valid Stokes vectors should satisfy
\begin{equation}\label{eq:disc-physval}
	s_0 \ge \sqrt{s_1^2 +s_2^2+s_3^2}.
\end{equation}
There can be many sources of invalidity, such as invalid values in the pBRDF dataset we use~\cite{baek2020image}, and the characteristic of the frequency domain method itself. 
However, since frequency domain analysis decomposes Stokes components into linear factors, such nonlinear inequality is hard to represent in the frequency domain \NEW{so that the latter source invalidity cannot ideally vanish}.
\NEW{We regard Equation~\eqref{eq:disc-physval} as an extended constraint of positivity of radiance in unpolarized radiance transport. Note that SH produce negative values which are related to \emph{ringing} artifacts~\cite{ramamoorthi2001signal, sloan2002precomputed}, even if the original radiance is positive in any direction. There have been a variety of works to overcome negativity and ringing artifacts from conventional SH~\cite{boyd2001chebyshev, mcclarren2008filtered, sloan2008stupid, berger2011approximate, sloan2017deringing}. Extending them for PSH will be an interesting future research direction.}
\NEW{For the simplest example, }
Figure~\ref{fig:result_validprojection} shows the result of our PPRT followed by \NEW{simply} projecting Stokes components to the physically valid range.
\NEW{By enforcing the inequality in Equation~\eqref{eq:disc-physval}, $s_0$ components become slightly brighter, and $s_1$ and $s_2$ components do slightly darker, while DoP and AoLP are preserved.}
Except for this figure, we report our rendering result without this valid range projection to show the direct output of our method.

\paragraph{\NEW{Constraints for pBRDF}}
\NEW{We leave PSH formulation of physical constraints of pBRDF as future work while providing brief discussions. pBRDF should also satisfy energy conservation, but we consider reformulating it into the PSH domain will be a challenging problem since SH and PSH are related to $L^2$-norm, but energy conservation is related to $L^1$-norm. To the best of our knowledge, energy conservation of SH-projected BRDF is not guaranteed even in unpolarized light transport. Finding the PSH formulation of reciprocity of pBRDF is an interesting problem. Note that flipping the order of direction variables of a pBRDF makes it belong to a different Mueller space, i.e., $\dvP\left(\homega_i,\homega_o\right)\in\MUEsp{\homega_i}{\homega_o}\ne\MUEsp{\homega_o}{\homega_i}\ni \dvP\left(\homega_o,\homega_i\right)$, so that investigating reciprocity of pBRDF requires a solid theoretical foundation. We observe that only a few works address this obscure challenge~\cite{sekera1966scattering, ding2021polarimetric}.}

\subsection{Difference against the Traditional SWSH}
\label{sec:discussion-theory-novelty}

\NEW{The main difference of this work against traditional SWSH theory consists of the SWSH coefficient formulation for linear operators on Stokes vector fields, including pBRDF and polarized spherical convolution, which is generally equivalent to rotation equivariant linear operators both in the angular and frequency domains.}

\NEW{Before discussing the convolution in more detail}, we distinguish two senses to extend conventional convolution on Euclidean domains to others. First, let us denote an operation between two quantities as $k*f = g$. One defines the operation $*$ as an extension of convolution by assuming $k$ and $f$ as the same type of quantities, which we call \emph{correlation}. On the other hand, one can define the operation $*$ to have the same kind of input $f$ and the output $g$, which we call \emph{convolution} here. 
\NEW{In the spherical domain, the output of such \emph{correlation} between two Stokes vector fields should be} a function of single angle\NEW{~\cite{zaldarriaga1997all,ng1999correlation}} or a function of rotations, which is not compatible with our PPRT pipeline. In the perspective of image processing and computer graphics,
We must extend the \emph{convolution} rather than correlation for Stokes vector fields. 
\NEW{As discussed in Section~\ref{sec:relatedwork-SWSH}, existing convolution theories for Stokes vector fields are limited to one~\cite{ng1999correlation} (spin \tcboxStS{\text{0-to-0}} and only real part of the isomorphic part of spin \tcboxVtV{\text{2-to-2}}) or six~\cite{garcia1986generalized, tapimo2018discrete} degrees of freedoms of kernels at each frequency band, which correspond to subsets} of our full kernel formulation described in Equation~\eqref{eq:theory_convkernel_geo2num} and Equations~\eqref{eq:theory_conv_coeff_s2s} to~\eqref{eq:theory_conv_coeff_v2vc}. To the best of our knowledge, we define new spherical convolution on Stokes vector fields so that it is equivalent to rotation equivariant linear operators. We also establish its PSH formulation, which is applicable to pBRDF approximation.

\NEW{Our main contributions come from two novel technical details that may be hard to recognize at a high level. First, our \emph{real coefficient formulation} discussed in Section~\ref{sec:theory_rotinv} and Supplemental Section~\ref{sec:theory-PSH_RC} is a key part of constructing our PSH formulation of linear operators. It includes our discussion about which sense of linearity of Stokes vectors should be chosen to represent general Mueller matrices. The second technical novelty is the \emph{complex pair separation}, introduced in Section~\ref{sec:theory_linop}. It is critical to derive our polarized convolution theorem in Equation~\eqref{eq:theory_conv_to} through Supplemental Equations~\eqref{eq:convpp_org} to~\eqref{eq:convpp_convcoeff2}.
We refer to Supplemental Section~\ref{sec:discussion-SWSH} for more detailed discussion.
}

\subsection{Future Work}
\citet{wang2018analytic}'s analytic SH coefficient for polygonal lights can be directly applied to the PPRT method for unpolarized polygonal lights and polarized material.
However, finding analytic formulae for polarized polygonal lights is expected to be a further challenging problem.
\citet{xin2021fast} found a fast triple product method for conventional SH utilizing FFT.
While this method cannot be directly applicable to polarized SH, a similar method is expected to be found using a similar idea.

Applying another PRT pipeline to our PSH theory will be an interesting work. For instance, one can tabulate outgoing directions of pBRDFs rather than using full coefficient matrices as~\citet{sloan2005local}, or compute shadows in runtime following \citet{zhou2005precomputed} utilizing spin-2 SH triple product introduced in Equation~\eqref{eq:theory_tp_v2v}. 
Another possible application is combining physically-based ray tracing for polarized environment map lighting.
We can use low-order PSH coefficients for polarized environment maps as Monte Carlo control variates. %

In subsurface scattering, an analytic solution of the radiative transfer equation (volume rendering equation) for participating media utilizes the SH up to $l=1$~\cite{jensen2001practical}\footnote{\emph{Scalar irradiance} and \emph{vector irradiance} in that paper corresponds to $l=0$ and $l=1$ SH expansion.}\NEW{, and even at higher orders~\cite{zhao2014high}}. Similarly, finding an analytic solution to the polarized radiative transfer equation would be interesting for future work.
Without limiting forward rendering, our polarized SH can be used to \NEW{extend various SH-based methods to polarized states such as acquiring pBRDF based on \citet{ghosh2007brdf} and \citet{tunwattanapong2013acquiring}'s methods for scalar BRDF,} construct\NEW{ing} novel polarized spherical CNN, or enhancing polarized radiance field methods \NEW{based on existing SH-based methods~\cite{yu_and_fridovichkeil2021plenoxels, verbin2022ref} and polarized methods without utilizing basis functions~\cite{dave2022pandora, kim2023nespof}}.

Extending non-harmonic bases such as wavelets and spherical Gaussians to Stokes vector fields would be a completely different approach from this work, but it will be an interesting future work. 
Even though they are different types of basis functions, properties of Stokes vector fields, including continuity, discussed in Section~\ref{sec:stokes-on-sphere}, must be handled properly.

%% file: conclusion.tex
\section{Conclusion}
\label{sec:conclusion}

While spherical harmonics have been a powerful tool in conventional unpolarized light transport, such basis functions that provide frequency domain analysis for polarized light transport have been absent.
We have addressed Stokes vector fields' challenges regarding frame fields' choices and their singularities.
Also, we have presented spin-weighted spherical harmonics, 
which provide a rotation invariant orthonormal basis for Stokes vector fields.
Combining conventional spin-0 SH for $s_0$ and $s_3$ Stokes components and spin-2 SH for $s_1$ and $s_2$ components, we have provided our polarized spherical harmonics theory, including linear operator formulation for pBRDF and polarized spherical convolution.
Also, we have presented the precomputed polarized radiance transfer, which achieves the first real-time polarized rendering, considering environment lighting and shadows.
We expect SWSH and our PSH theory to become helpful in understanding the special nature of polarization and to be used in various applications in future work.

%% file: acknowledgments.tex
\begin{acks}
\NEW{
\noindent Min H.~Kim acknowledges the MSIT/IITP of Korea (RS-2022-00155620, RS-2024-00398830, 2022-0-00058, and 2017-0-00072), Microsoft Research Asia, LIG, and Samsung Electronics.
}
\end{acks}

%% file: title-supple.tex
\title[Supplemental Document: Spin-Weighted Spherical Harmonics for Polarized Light Transport]{Supplemental Document:\\Spin-Weighted Spherical Harmonics for Polarized Light Transport}

%% file: intro-supp.tex
\noindent
\section*{Preface}
\NEW{This supplemental document serves several purposes for different readers, with the exception of Sections~\ref{sec:bkgnd_geo_num} and~\ref{sec:bkgnd_sphere_rotation}, which are recommended for all readers. First, Sections~\ref{sec:bkgnd_SH_SH} to~\ref{sec:bkgnd_SH_rotation},~\ref{sec:spherical-convolution},~\ref{sec:background_polar_intro},~\ref{sec:continuity-stokes-vector-field},~\ref{sec:scalar_vector_fields} to~\ref{sec:naive-SH} provide some additional motivation and detail to the background described in Sections~\ref{sec:background} and~\ref{sec:stokes-on-sphere} of the main paper for readers who are not familiar with either spherical harmonics or polarization. %
Second, the remainder of this document contains formal definitions and detailed steps for proofs in a more axiomatic and rigorous manner. This remainder is intended for more dedicated readers who want to verify the mathematical properties of polarized spherical harmonics presented in the main paper. Since each of Sections~\ref{sec:bkgnd_SH},~\ref{sec:background_polar}, and~\ref{sec:stokes-vector-fields} contains subsections intended for different readers, we also clarify the purpose of each subsection at the beginning of each of these sections.

Since our work deals with extensions of quantities and equations that have been previously treated in spherical harmonics and polarization, it is helpful to see Table~\ref{tb:elem_set_desc}, which compares the formulae proposed in this work with the existing formulae to which each corresponds.
}

%% file: background1-supp.tex
\input{equation_table}

\section{Preliminaries}
\label{sec:preliminary}

\subsection{Geometric and Numeric Quantities}
\label{sec:bkgnd_geo_num}

In this paper, we investigate various categories of quantities such as vectors, Stokes vectors, transforms, and functions on the unit sphere to these quantities. Before discussing individual concepts of them, we first distinguish them into two categories, \emph{geometric} quantities and \emph{numeric} quantities, inspired by a computer graphics textbook~\cite{gortler2012foundations}.

Geometric quantities can be easily understood as physical quantities, which we can see in the real world, and numeric quantities can be considered as just arrays of numbers. For example, we call \emph{vectors} (or \textit{geometric vectors} to clearly avoid confusion of terminology), denoted by $\vec a,\vec v,\homega,\cdots\in\Rspv$, as geometric quantities, and  \emph{numeric vectors}, denoted by $\mathbf{a}, \mathbf{c},\mathbf{x}\in\R^n$, as numeric quantities. While vectors discussed in this paper are always three-dimensional quantities, numeric vectors include four-dimensional Stokes component vectors, which is discussed in Section~\ref{sec:background_polar} in the main paper, and spherical harmonics coefficient vectors with arbitrary dimension, which is discussed in Section~\ref{sec:background_sh} in the main paper. We will call the numerical representation of vectors as \emph{coordinate vectors}, which are special cases of numeric vectors.

Regardless of whether geometric or numeric, we call a set with well-defined addition and scalar multiplication a \emph{linear space}\footnote{This is more frequently called \emph{vector space} in other literature, but we do not use it since the word 'vector' might be misunderstood as a geometric quantity.}, in the sense of linear algebra. Both the set of (geometric) vectors and the set of numeric vectors are linear spaces.

\begin{figure}[tbp]
	\centering
	\includegraphics[width=0.6\columnwidth]{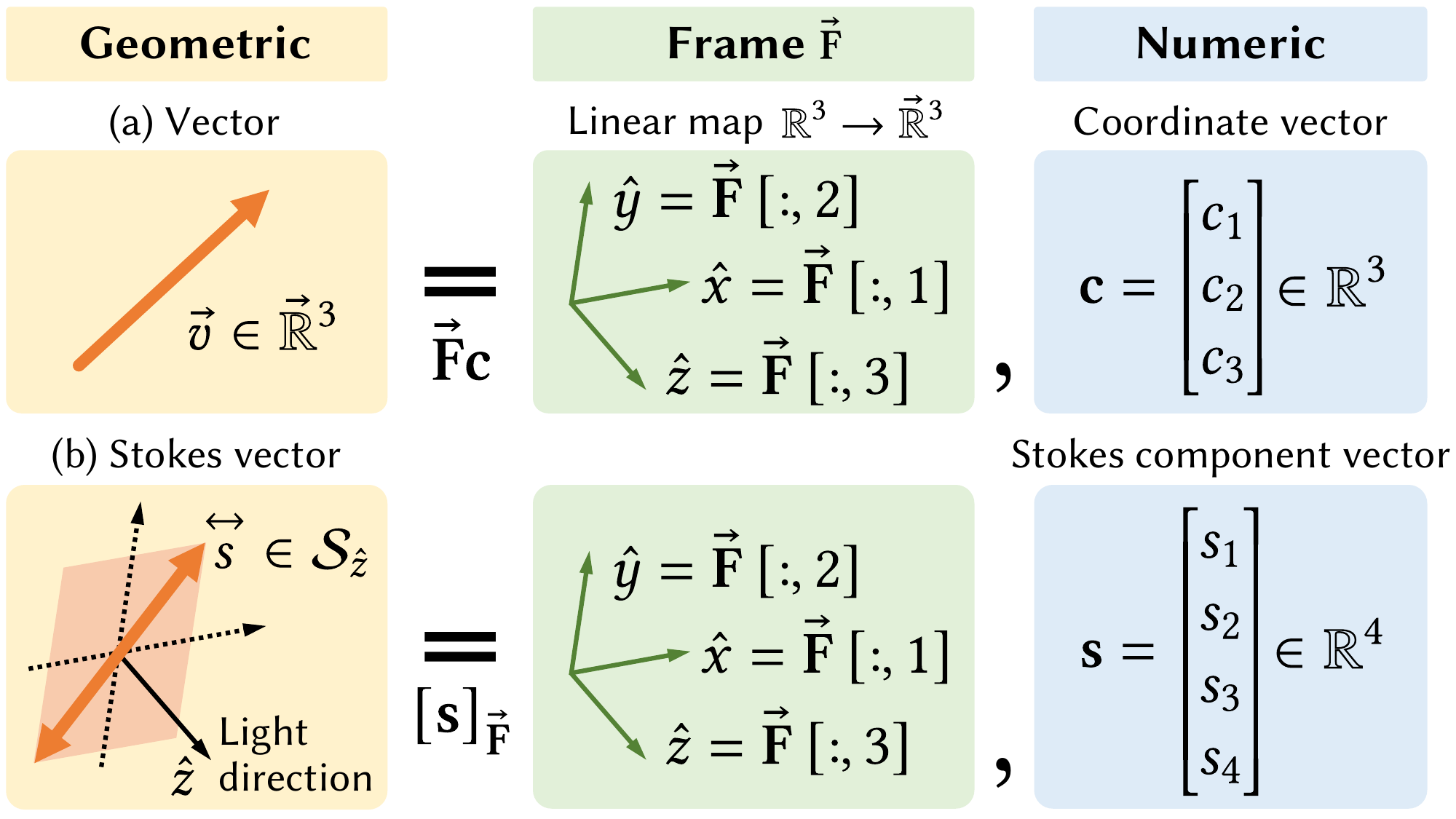}%
	\vspace{-3mm}
	\caption[]{\label{fig:bkgnd_vector_coord}
		We distinguish \emph{geometric} and \emph{numeric} quantities. (a) A (geometric) vector is equal to the product of an orthonormal \emph{frame}, which is a linear map from numeric vectors to geometric vectors, and a coordinate vector, which is a kind of numeric vector. (b) Combining a frame $\frF$ and a numeric vector $\bfs$, named a \emph{Stokes component vector}, we get a geometric quantity \emph{Stokes vector}, which indicates a polarized intensity of a ray. Here, it is essentially different from the product of a frame and a numeric vector, we write the relationship of these quantities with our novel notation $\dvs = \left[\bfs\right]_{\frF}$.
	}
	\vspace{-2mm}		
\end{figure}

For sets $X$ and $Y$, $\calF\left(X,Y\right)=\left\{f:X\to Y\right\}$
\footnote{To consider it as an inner product space in later sections, $\calF$ should contain additional conditions such as L2 integrability for mathematical rigor. For the sake of simplicity, however, we have omitted such conditions as they are always satisfied in practical cases. \NEW{We refer to \citet{groemer1996geometric} for complete mathematical rigor of the theory of spherical harmonics}} 
denotes the set of all functions from $X$ into $Y$. If $X$ and $Y$ are linear spaces, $\calL\left(X,Y\right)=\left\{f\in \calF\left(X,Y\right) \mid f\left(ax+by\right) = af\left(x\right)+bf\left(y\right) \right\}$ indicates the set of \emph{linear maps} from $X$ into $Y$, regardless whether geometric or numeric. We call linear maps between numeric vectors \emph{matrices} and those between geometric vectors \emph{transforms}. Moreover, a \emph{frame} indicates an orthonormal\footnote{It may not be orthonormal in general, but we only consider orthonormal frames in this work for simplicity.} linear map from coordinate vectors to geometric vectors, and the set of frames is denoted by $\FRsp\coloneqq \left\{\frF\in\calL\left(\Rspv,\R^3\right) \mid \frF\text{ is orthonormal}\right\}$. Then, we observe that a vector is equal to the matrix product of a frame and a coordinate vector as described in Figure~\ref{fig:bkgnd_vector_coord}(a). Note that a coordinate vector itself does not have any physical meaning in the real world, but it can be converted into a geometric vector by combining it with a frame.

Similar to the multiplication of frames and coordinate vectors, we have several kinds of multiplications as follows:
\begin{align}\nonumber
	\mathrm{matrix}\ (\in \calL\left(\R^3, \R^3\right)) &\times \mathrm{coordinate\ vector} \ (\in \R^3) &= &\mathrm{coordinate\ vector}\ (\in \R^3) \\ \nonumber
	\mathrm{frame}\ (\in \calL\left(\R^3, \Rspv\right))&\times \mathrm{coordinate\ vector}\ (\in \R^3) &= &\mathrm{vector}\ (\in \Rspv) \\ \nonumber
	\mathrm{transform}\ (\in \calL\left(\Rspv, \Rspv\right))&\times \mathrm{vector}\ (\in \Rspv)) &= &\mathrm{vector}\ (\in \Rspv) \\ 
	\mathrm{matrix}\ (\in \calL\left(\R^3, \R^3\right))&\times \mathrm{matrix}\ (\in \calL\left(\R^3, \R^3\right)) &= &\mathrm{matrix}\ (\in \calL\left(\R^3, \R^3\right)) \\ \nonumber
	\mathrm{frame}\ (\in \calL\left(\R^3, \Rspv\right))&\times \mathrm{matrix}\ (\in \calL\left(\R^3, \R^3\right)) &= &\mathrm{frame}\ (\in \calL\left(\R^3, \Rspv\right)) \\ \nonumber
	\mathrm{transform}\ (\in \calL\left(\Rspv, \Rspv\right))&\times \mathrm{frame}\ (\in \calL\left(\R^3, \Rspv\right)) &= &\mathrm{frame}\ (\in \calL\left(\R^3, \Rspv\right)) \\ \nonumber
	\mathrm{transform}\ (\in \calL\left(\Rspv, \Rspv\right))&\times \mathrm{transform}\ (\in \calL\left(\Rspv, \Rspv\right)) &= &\mathrm{transform}\ (\in \calL\left(\Rspv, \Rspv\right)).
\end{align}
Note that we also denote $\calL\left(\R^m,\R^n\right)\eqqcolon \R^{m\times n}$.
These multiplications are well defined in the sense of the action of linear maps on linear spaces and the composition of linear maps. Note that the multiplication of some pairs of quantities, which is not included above, is usually not allowed.
For example to distinguish numeric matrices and geometric transforms, we can imagine a rotation.
We denote $\SOgroup\subset \R^{3 \times 3}$ and $\SOgroupv\subset \calL\left(\Rspv,\Rspv\right)$ as the sets of (numeric) rotation matrices and (geometric) rotation transforms, respectively. When a frame $\frF=\begin{bmatrix}
	\hat x & \hat y & \hat z
\end{bmatrix}$ is given, the rotation transforms around the axis $\hat x$, $\hat y$, and $\hat z$ by angle $\theta$ can be written as follows:
\begin{equation}
	\label{eq:bkgnd_rot_geo_num}
	\vec R_{\left\{\hat x,\hat y,\hat z\right\}} = \frF \bfR_{\left\{x,y,z\right\}} \frF^{-1}, \quad \text{where }
\end{equation}
\begin{equation}
	\bfR_x = \begin{bmatrix}
		1&0&0\\0&\cos\theta & -\sin\theta \\ 0&\sin\theta & \cos\theta
	\end{bmatrix}, \quad
	\bfR_y = \begin{bmatrix}
		\cos\theta & 0 &\sin\theta \\ 0 & 1& 0 \\ -\sin\theta & 0 & \cos \theta
	\end{bmatrix}, \quad
	\bfR_z = \begin{bmatrix}
		\cos\theta & -\sin\theta &0 \\ \sin\theta & \cos\theta & 0\\0&0&1
	\end{bmatrix}.
\end{equation}
Note that while subscripts $\hat x$, $\hat y$, and $\hat z$ in the left-hand side of Equation~\ref{eq:bkgnd_rot_geo_num} indicate the axis vectors of $\frF$ which has been defined in this context, subscripts $x$, $y$, and $z$ in the right-hand side just symbols which means the first, second, and third of a frame which do not have to be given in advanced. %
Also note that conversion between a rotation transform $\vec R\in\SOgroupv$ and $\bfR\in\SOgroup$ with respect to a frame $\frF\in\FRsp$ can be done by $\vec R = \frF \bfR \frF^{-1}$ and $\bfR = \frF^{-1}\vec R \frF$.
For compactness, we often write consecutive rotation transforms about some axes $\hat u_1$, $\hat u_2$, $\cdots \in \Sspv$ and rotation matrices about $u_1$, $u_2$, $\cdots\in\left\{x,y,z\right\}$ as following ways, respectively:
\begin{equation}
	\vec R_{\hat u_1}\left(\theta_1\right) \vec R_{\hat u_2}\left(\theta_2\right) \cdots \eqqcolon \vec R_{\hat u_1 \hat u_2 \cdots}\left(\theta_1,\theta_2,\cdots\right), \quad
	\bfR_{u_1}\left(\theta_1\right)\bfR_{u_2}\left(\theta_2\right)\cdots \eqqcolon \bfR_{u_1u_2 \cdots}\left(\theta_1, \theta_2 \cdots\right)
\end{equation}
which also can represent Euler angles. 

For numeric quantities, we will write NumPy style indexing notation such as:
\begin{equation}
	\bfx=\begin{bmatrix}
		\bfx\left[1\right] \\ \cdots \\ \bfx\left[n\right]
	\end{bmatrix}, \text{ and }
	\bfA = \begin{bmatrix}
		\bfA\left[1,1\right] && \\
		& \ddots & \\
		&& \bfA\left[m,n\right]
	\end{bmatrix}.
\end{equation}
$\bfA\left[i,:\right]$ and $\bfA\left[:,i\right]$ denote $i$-th row and column vectors of a matrix $\bfA$. Referring to $i$-th (or $i,j$-th) entries of geometric quantities are illegal. Since a frame is both related to numeric and geometric vectors, referring its $i$-th row is illegal while its $i$-th column is well defined. For example, we have $\extColz{\frF}=\hat z$ for a frame $\frF=\begin{bmatrix}\hat x & \hat y & \hat z\end{bmatrix}\in\FRsp$.

Notations of sets of each type of quantity and notation convention for them are summarized in %
Table~\ref{tb:elem_set_desc}.

\begin{equation}
	\bfMat\left[A_{ijkl} \mid j=1,\cdots,n, l=1,\cdots, n \right] = \begin{bmatrix}
		A_{i1k1} & \cdots & A_{i1kn} \\
		\vdots & \ddots & \vdots \\
		A_{ink1} & \cdots & A_{inkn} \\
	\end{bmatrix}.
\end{equation}
If the range of two indices is the same, then we sometimes write it as $j,l=1,\cdots,n$ simply, and we sometimes even omit the range if it is clear in context. Moreover, we can also take the range of indices that is not an interval, such as:
\begin{equation}
	\bfMat\left[A_{ij} \mid i,j=+m, -m \right]  = \begin{bmatrix}
		A_{+m,+m} & A_{+m,-m} \\
		A_{-m,+m} & A_{-m,-m}
	\end{bmatrix}.
\end{equation}

\subsection{Unit Sphere, Frames, and Rotations}
\label{sec:bkgnd_sphere_rotation}
As a subset of the space of 3D geometric vectors $\Rspv$, the \emph{unit sphere} (or just \emph{sphere}) $\Sspv=\left\{\homega \in \Rspv\mid \norm{\homega}=1\right\}$\footnote{While denoting by $\mathbb{S}^2$ or $S^2$ is more common in other text, but we write with the $\hat{ }$ symbol to clarify it is a set of geometric vectors, not numeric ones.} indicates the set of all vectors with unit norms. It also can be considered as the set of all directions in $\Rspv$ in the context of computer graphics. It usually parameterized by spherical coordinates of a zenith angle $\theta$ and a azimuth angle $\phi$ as follows:
\begin{equation} \label{eq:sph_coord}
	\homega_{\mathrm{sph}}\left(\theta,\phi\right) = \frF_g \begin{bmatrix}
		\sin\theta\cos\phi \\ \sin\theta\sin\phi \\ \cos\theta
	\end{bmatrix},
\end{equation}
where a global frame $\frF_g = \begin{bmatrix}\hat x_g & \hat y_g & \hat z_g\end{bmatrix}$ is given.

\begin{figure}[tbp]
	\centering
	\includegraphics[width=\columnwidth]{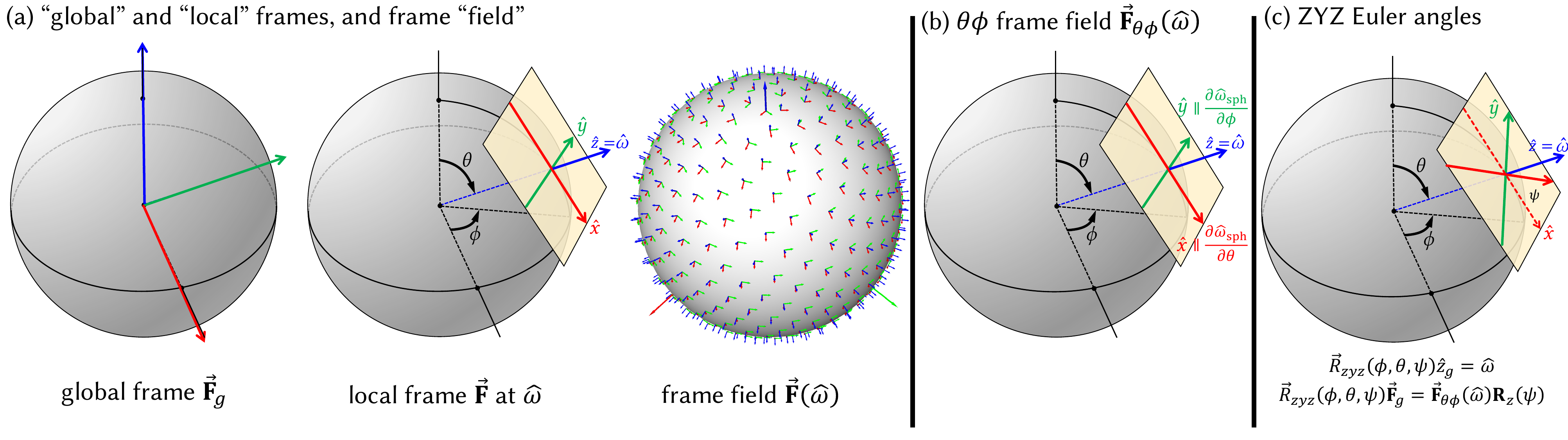}%
	\vspace{-3mm}
	\caption[]{\label{fig:bkgnd_frame_terminology}
		Some definitions and useful identities about \emph{frames}. (a) In this paper, we distinguish a \emph{global frame}, a \emph{local frame at $\homega$}, and a \emph{frame field}. (b) We usually use the \emph{$\theta\phi$ frame field}, which is defined in Equation~\eqref{eq:bkgnd_tp_moving_frame}. (c) Spherical coordinates and a local frame can be evaluated with ZYZ Euler angles as Equation~\eqref{eq:bkgnd_euler_fr}.
	}
	\vspace{-2mm}		
\end{figure}

In this paper, we will distinguish the terms global, local frames, and frame field as Figure~\ref{fig:bkgnd_frame_terminology}(a).
We call a \emph{global frame} as a frame independent of a particular direction $\homega\in\Sspv$, and the global frame, which is often used to assign spherical coordinates on $\Sspv$. A \emph{local frame at $\homega$} indicates a frame with local $\hat z$ axis as $\homega$, i.e., $\frF\left[{:},3\right]=\homega$, which is used to measure Stokes vectors along $\homega$, and \emph{frame field} (or moving frame) as a function from $\Sspv$ to $\FRsp$.
We also define $\FRspo{\homega} \coloneqq\left\{\frF\in \FRsp\mid \extColz{\frF}=\homega\right\}\subset \FRsp$.

There are infinitely many choices to assign a frame field on the sphere $\Sspv$, we use a typical example which we call the \emph{$\theta\phi$ frame field} and denote by $\frF_{\theta\phi}\left(\homega\right)$. Using spherical coordinates, it can be defined as follows:
\begin{equation}\label{eq:bkgnd_tp_moving_frame}
	\frF_{\theta\phi}\left(\theta,\phi\right) = \begin{bmatrix}
		\pfrac{\homega_{\mathrm{sph}}}{\theta}\left(\theta, \phi\right) &
		\mathrm{normalized}\pfrac{\homega_{\mathrm{sph}}}{\phi}\left(\theta, \phi\right) &
		\homega_{\mathrm{sph}}\left(\theta,\phi\right)
	\end{bmatrix},
\end{equation}
which is visualized in Figure~\ref{fig:bkgnd_frame_terminology} (b). We observe that $\theta\phi$ frame field $\frF_{\theta\phi}$ has two singularities at $\homega_{\mathrm{sph}}\left(0, \phi\right) = \hat z_g$ and $\homega_{\mathrm{sph}}\left(\pi, \phi\right) = -\hat z_g$, where the function $\frF_{\theta\phi}$ cannot be continuously defined. Not only the $\theta\phi$ frame field, but any frame field on the sphere always has a singularity due to the hairy ball theorem, which is common in differential geometry.

\subsection{\NEW{Useful Identities}}
\label{sec:bkgnd_useful}
\mparagraph{\NEW{Identities using rotations}}
\NEW{Note that inner products on $\R^3$ and $\Rspv$ are preserved under rotation. In other words,
\begin{equation}\label{eq:prem_inner_rot}
	\forall \bfx,\bfy\in\R^3,\, \forall \bfR\in\SOgroup, \quad \bfx^T\bfy = \left(\bfR \bfx\right)^T \left(\bfR \bfy\right), \qquad
	\forall \vec x,\vec y\in\Rspv,\, \forall \vec R\in\SOgroupv, \quad \vec x\cdot \vec y = \left(\vec R\vec x\right)\cdot \left(\vec R\vec y\right).
\end{equation}
It can be directly proven by the fact that rotations are orthogonal matrix so that $\bfR\bfR^T = \bfR^T \bfR = \bfI$.
}

It is often useful that a global frame $\frF_g$ can also be considered as a local frame at the zenith ($\hat z_g$), and using ZYZ Euler angle rotation spherical coordinates and the $\theta\phi$ frame field can be rewritten as:

\vspace{2mm}
\begin{minipage}{0.48\linewidth}
	\begin{equation}\label{eq:bkgnd_euler_sph}
	\homega_{\mathrm{sph}} \left(\theta,\phi\right) = \vec R_{\hat z_g\hat y_g\hat z_g}\left(\phi,\theta,\psi\right) \hat z_g,
\end{equation}
\end{minipage}
\begin{minipage}{0.48\linewidth}
	\begin{equation}\label{eq:bkgnd_euler_fr}
		\frF_{\theta\phi}\left(\theta,\phi\right) = \vec R_{\hat z_g\hat y_g\hat z_g}\left(\phi,\theta,0\right) \frF_g,
	\end{equation}
\end{minipage}
\vspace{2mm}

\noindent
while $R_{\hat z_g\hat y_g\hat z_g}\left(\phi,\theta,\psi\right) \frF_g$ represents an arbitrary local frame at $\homega_{\mathrm{sph}}\left(\theta,\phi\right)$.

Another choice of a frame field is the perspective frame field $\frF_\mathrm{pers}$ shown in Figure~\ref{fig:bkgnd_penvmap_visualize}(c) in the main paper, characterized by the virtual perspective camera.
Note that there are several choices of such camera-based frame field conventions.
We follow the convention of Mitsuba~3 renderer~\cite{Mitsuba3}, which utilizes the up-axis of camera $\hat{u}$ to define $\frF_\mathrm{pers}$ as

\begin{equation}\label{eq:bkgnd_mitsuba_moving_frame}
	\frF_{\mathrm{pers}}\left(\homega; \hat u\right) = \begin{bmatrix}
		\hat x & \hat y & \homega
	\end{bmatrix} =\begin{bmatrix}
		\mathrm{normalize}\left(\hat u\times \homega\right) & \homega \times \hat x & \homega
	\end{bmatrix}.
\end{equation}

While the $\theta\phi$ and the perspective frame fields are highly related, as $\frF_{\mathrm{pers}}\left(\homega; \hat z_g\right) = \frF_{\theta\phi}\left(\homega\right)\bfR_z\left(\frac\pi 2\right)$, we use the both since they have their own convenience. Formulas of special functions, including SWSH and Wigner D-functions, are usually written related to $\theta\phi$ frame field, while it is natural to use perspective frame fields, whose local $y$ axes are close to the camera up vector, for perspective view.

\mparagraph{Integral formulae}
To derive some identities for spherical harmonics and our polarized spherical harmonics, we sometimes need to integrate some functions over the space of rotation transforms $\SOgroupv$. The differential measure $\rmd \vec R$ for $\vec R\in\SOgroupv$ is evaluated as follows using ZYZ Euler angles with respect to a frame $\frF = \begin{bmatrix}\hat x & \hat y & \hat z\end{bmatrix}\in\FRsp$ is given:
\begin{equation}\label{eq:def_rot_measure}
	\int_{\SOgroupv}{f\left(\vec R\right)\rmd \vec R} = \int_0^{2\pi}\int_0^\pi\int_0^{2\pi}{ f\left(\vec R_{\hat z \hat y \hat z}\left(\alpha,\beta,\gamma\right)\right) \sin \beta \rmd \alpha\rmd\beta\rmd\gamma}.
\end{equation}
Note that this measure is equivalent to a constant multiple of the subspace measure by identifying $\SOgroupv$ to a subset of $\R^9$, and it is invariant under choice of the frame $\frF$.

Several integration techniques for the sphere $\Sspv$ and rotation transforms $\SOgroupv$ are used to prove the important properties of conventional and our polarized SH as: 
\begin{gather}\label{eq:inttech_sph2rot}
	\int_{\Sspv}{f\left(\homega\right) \rmd \homega} = \frac1{2\pi}\int_{\SOgroupv}{f\left(\vec R\hat z_g\right)\rmd \vec R}, \\
	\label{eq:inttech_rotinv}
	\int_{\Sspv}{f\left(\homega\right) \rmd \homega} = \int_{\Sspv}{f\left(\vec R\homega\right)\rmd \homega} \text{ for any }\vec R\in\SOgroupv .
\end{gather}
Readers who are not about to verify the proof of this paper and just want to use the results can skip this part.

\subsection{Linear Algebra on Function Spaces}
\label{sec:bkgnd_func_anal}

We call an algebraic object equipped with addition and scalar multiplication as \emph{linear space} while other literature more frequently calls it \emph{vector space}. To avoid confusion, the term \emph{vector} is usually used to discuss numeric vectors and geometric vectors in this paper.

This paper investigates several function spaces such as spherical harmonics, spin-weighted spherical harmonics, and naively applied spherical harmonics to Stokes vectors fields. To distinguish properties inherited from general properties of orthonormal basis and properties of a certain individual basis, we recall the general theory of linear algebra on function spaces in this section. Then, we will describe the properties of spherical harmonics as examples of general theory. Later, we introduce spin-weighted spherical harmonics in Section~\ref{sec:our_theory} in the main paper, also based on the language defined in this section.

First of all, We will discuss function spaces, including the set of spherical functions (or scalar fields on the sphere) $\calF\left(\Sspv,\C\right)$ and the set of Stokes vector fields on the sphere in this paper.
They are important in computer graphics since a spherical function can represent radiance as a function of directions, such as an environment map and a 2D slice of a BSDF with a fixed incoming or outgoing direction, and the set of Stokes vector fields can represent polarized versions of those quantities.

These function spaces are inner product spaces so they can be described by the general theory of linear algebra. Even though those function spaces have infinite dimensionality, fundamental properties of linear spaces are well extended to function spaces, as described in this section.

Spherical harmonics are known as \emph{bases} of function spaces, so we first define bases and coefficient representation with respect to them.
\begin{definition}{Bases and coefficient vectors}{basis_coeff}%
	For a countable index set $I$ and an inner product space $\calH$ (usually a function space) over scalar $\bbK$ (=$\R$ or $\C$), an indexed collection $\calB\coloneqq\left\{b_i\mid i\in I\right\}\subset \calH$ is called a \textit{basis}
	\footnote{Rigorously, it should be called a Hilbert basis since Equation~\eqref{eq:basis_axiom} includes not only finite summations but also countably infinite ones, but we simply call basis for simplicity.}
	of $\calH$ if and only if for any $f \in \calH$ there uniquely exists an indexed collection of scalars $\left\{a_{i}\mid i\in I\right\}$ such that:
	\begin{equation} \label{eq:basis_axiom}
		f = \sum_{i\in I} \rmf_{i}b_i.
	\end{equation}
	Here, $a_i$ is called the \textit{coefficient of} $f$ \textit{with respect to} $b_i$ or the $i$\textit{-th coefficient of} $f$ \textit{with respect to} $\calB$. When order on $I$ is given in context, $\bff \coloneqq \bfMat\left[\rmf_{i}\mid i\in I\right]$ is called the \textit{coefficient vector} of $f$ with respect to the basis $\calB$.
\end{definition}
While bases are usually defined without admitting such index sets as above, having them in the definition of bases makes writing statements about spherical harmonics and further bases, including our polarized spherical harmonics, much more convenient. Note that converting a vector $f$ into its coefficient vector is linear, so the coefficient vector can be considered to be equivalent to the original vector $f$. For the sake of simplicity, we consistently denote $I$, $\calH$, $\bbK$, and $\calB$ with the conditions stated in Definition~\ref{def:basis_coeff}.

In Section~\ref{sec:bkgnd_func_anal} italic characters such as $f$ and $b_i$ usually denote elements of a vector space, which also can be functions, Roman characters such as $\rmf_i$ denote coefficients with respect to some basis, and bold roman characters such as $\bff$ do coefficient vectors. While $f$ may be a geometric or numeric quantity depending on its space $\calH$, $\rmf_i$, and $\bff$ can be considered as numeric quantities since they are indexed collections of scalars.

\begin{proposition}{Coefficient for a basis}{basis_coeff_eval} %
	In Definition~\ref{def:basis_coeff}, suppose that $\lrangle{\cdot,\cdot}_\calH$ denotes the inner product on $\calH$ and the basis $\left\{b_i\mid i\in I\right\}$ is orthonormal, i.e., $\lrangle{b_i,b_j}_{\calH}=\delta_{ij}$. Then the coefficient in Equation~\eqref{eq:basis_axiom} is evaluated as $\rmf_i = \lrangle{b_i, f}_\calH$, i.e.,
	\begin{equation}\label{eq:basis_coeff_eval}
		\forall f\in \calH,\quad f=\sum_{i\in I} \lrangle{b_i, f}_\calH b_i.
	\end{equation}
\end{proposition}

\begin{definition}{projection on subsets of bases}{basis_projection} %
	Suppose that $\calB' =\left\{b_i\mid i \in J\right\}$ be a subset of a basis $\calB$ of a linear space $\calH$, which is characterized by $J \subset I$. A \textit{projection of} $f \in \calH$ \textit{on} $\calB'$ is defined as:
	$$
	\tilde f=\sum_{i\in J} \rmf_i b_i,
	$$
	where $\rmf_i$ is the coefficient of $f$ with respect to $b_i$. This is also called the projection of $f$ on the basis of $\calB$ \textit{up to} $J$.
\end{definition}

Note that the projection of $f$ up to $J$ sometimes indicates the coefficients $\left\{\rmf_i \mid i\in J\right\}$ rather than $\sum_{i\in J}\rmf_i b_i$, and it will be distinguished according to the context.

Note that the space of linear map $\calL\left(X,Y\right)$ is still well defined even if $X$ is an infinite-dimensional function space. However, in this case, we usually call such linear maps as \emph{linear operators} conventionally to emphasize that $X$ may be a function space. %

\begin{proposition}{Coefficient matrices of linear operators}{linalg-coeff-linop}%
	Suppose that $\left\{b_i\mid i\in I\right\}\subset \calH$ is an orthonormal basis, there is a linear operator $T:\calH\to\calH$. When denoting the coefficients of $f\in\calH$ and $T\left[f\right]$ by $\left\{\rmf_i\mid i\in I\right\}$ and $\left\{\rmf_i'\mid i\in I\right\}$, respectively, $\rmf_i'$ is evaluated as:
	\begin{equation}\label{eq:coeff_transform}
		\rmf_i' = \sum_{j\in I} \lrangle{b_i, T\left[b_j\right]}_\calH \rmf_j.
	\end{equation}
	Here, $\lrangle{b_i, T\left[B_j\right]}$ is called as the \emph{coefficient of} $T$ \emph{with respect to} $\left(b_i,b_j\right)$ or ($i$,$j$)-\emph{th coefficient of the linear operator $T$} with respect to the basis $\calB$. When order on $I$ is given in context, $\bfT \coloneqq\bfMat\left[\lrangle{b_i,T\left[B_j\right]}_\calH \mid i,j\in I\right]$ is called the \emph{coefficient matrix} of $T$ \emph{with respect to} $\calB$.
	
\end{proposition}

\vspace{-10mm}
\begin{proof}
	By Proposition~\ref{prop:basis_coeff_eval},
	$$
	\rmf_i = \lrangle{b_i,f}_\calH, \text{ and } \rmf_i' = \lrangle{b_i,T\left[f\right]}_\calH.
	$$
	From the later equation, substituting the formal equation and the definition of basis (Equation~\eqref{def:basis_coeff}) yields:
	$$
	\rmf_i' = \lrangle{b_i, T\left[f\right]}_\calH = \lrangle{b_i, T\left[\sum_{j\in I}\rmf_j b_j\right]}_\calH = \sum_{j\in I}\lrangle{b_i,T\left[b_j\right]}_\calH \rmf_j.
	$$
	Here, the rightmost implication comes from the linearity of $T$ and the inner product.
\end{proof}

Note that Equation~\eqref{eq:coeff_transform} can be rewritten as the matrix-vector product of the coefficient matrix of $T$ and the coefficient vector of $f$. For coefficients of linear operators, the following properties are useful.

\begin{proposition}{Identities for linear operator coefficients}{linop_identity}
	Suppose that $\calB = \left\{b_i\mid i\in I\right\}\subset \calH$ is an orthonormal basis on $\calH$ and there are linear operators $T,T_1,T_2:\calH\to\calH$. Denote their coefficient matrices with respect to $\calB$ by $\bfT$, $\bfT_1$, and $\bfT_2$. The following properties hold.
	
	\begin{enumerate}
		\item For the identity operator $I:\calH \to \calH$ with $I\left[f\right]=f$, the coefficient matrix w.r.t. $\calB$ is the identity matrix, i.e., $\lrangle{b_i, I\left[b_j\right]}_\calH = \delta_{ij}$.
		\item The coefficient matrix of $T_1 \circ T_2$ w.r.t. $\calB$ is the matrix product of $\bfT_1$ and $\bfT_2$, i.e.:
		\begin{equation} \nonumber
			\sum_{k\in I}\lrangle{b_i, T_1\left[b_k\right]}_\calH \lrangle{b_k,  T_2\left[b_j\right]}_\calH = \lrangle{b_i, T_1\circ T_2\left[b_j\right]}_\calH.
		\end{equation}
		\item If $T^{-1}$ exists, then the coefficient matrix of $T^{-1}$ w.r.t. $\calB$ is the inverse matrix of $\bfT$, i.e.:
		\begin{equation} \nonumber
			\sum_{k\in I}\lrangle{b_i, T\left[b_k\right]}_\calH \lrangle{b_k,  T^{-1}\left[b_j\right]}_\calH = \sum_{k\in I}\lrangle{b_i, T^{-1}\left[b_k\right]}_\calH \lrangle{b_k,  T\left[b_j\right]}_\calH = \delta_{ij}
		\end{equation}
		\item If $T$ is a symmetric operator, i.e., $\lrangle{f, T\left[g\right]}_\calH = \lrangle{T\left[f\right], g}_\calH$ for any $f,g\in\calH$, then its coefficient matrix $\bfT$ is a Hermitian matrix ($\bfT^T = \bfT^*$), i.e., $\lrangle{b_i,T\left[b_j\right]}_\calH = \lrangle{b_j,T\left[b_i\right]}_\calH^*$.
		\item If $T$ is a unitary operator, i.e., $\lrangle{f, T\left[g\right]}_\calH = \lrangle{T^{-1}\left[f\right], g}_\calH$ for any $f,g\in\calH$, then its coefficient matrix $\bfT$ is a unitary matrix ($\bfT^{-1} = \left(\bfT^{T}\right)^*$), i.e., $\lrangle{b_i,T^{-1}\left[b_j\right]}_\calH = \lrangle{b_j,T\left[b_i\right]}_\calH^*$.
	\end{enumerate}
\end{proposition}

What we deal with as an important desirable property of spherical harmonics is rotation invariance. For a generalized description, we first formulate \emph{transform invariance} for given transforms and discuss the rotation invariance of spherical harmonics in the later section. First, the invariance of a subset of a space is naturally defined.

\begin{definition}{Transform invariance of a subset}{}\label{prop:transform_invariance_subset}
	A set $A \subset \calH$ is called to be invariant under a linear operator $T:\calH\to\calH$ if $T\left(A\right)=A$.
\end{definition}

Here, we also call the linear operator $T$ as a \emph{transform} conventionally when we are interested in invariance.

Now, a basis can be called to be invariant if it can be separated into a partition of finite sets so that these finite subsets of the basis span invariant subspaces.

\begin{definition}{Transform invariance}{transform_invariance}
	A basis $\left\{b_i\mid i\in I\right\}\subset \calH$ is called to be invariant under a linear operator (transform) $T:\calH\to\calH$ if there exists a partition of the index set $I$ into finite subsets, i.e., $I = \bigcup_{k=0}^\infty J_k $ with $J_i\cap J_j = \emptyset$, such that $\mathrm{span}\left\{b_i \mid i \in J_k\right\}$ is invariant under $T$ for any $k$.
\end{definition}

\begin{proposition}{Equivalent conditions for transform invariance}{transform_invariance}
	Suppose that there is an orthonormal basis $\calB = \left\{b_i\mid i\in I\right\}$ of an inner product space $\calH$ and a linear operator (transform) $T:\calH\to\calH$. Then, the following statements are equivalent to each other.
	\begin{enumerate}[label=(\roman*)]
		\item The basis is invariant under $T$, by Definition~\ref{def:transform_invariance}.
		\item Let $\calB_k \coloneqq \left\{b_i \mid i\in J_k\right\}$. For any $f\in \calH$ and $k\ge 0$ the projection of $T\left[f\right]$ on $\calB_k$ is equal to $T\left[f'\right]$ where $f'$ is the projection of $f$ on $\calB_k$.
		\item Let $\calB_{\le k} \coloneqq \left\{b_i \mid i\in J_j \text{ for some }j\le k\right\}$. For any $f\in \calH$ and $k\le 0$ the projection of $T\left[f\right]$ on $\calB_{\le k}$ is equal to $T\left[f'\right]$ where $f'$ is the projection of $f$ on $\calB_{\le k}$.
		\item
		\begin{equation} \label{eq:tf_invariance}
			\forall k\ne k'\ge 0,\ \forall \left(i,j\right)\in J_k \times J_{k'},\ \lrangle{b_i,T\left[B_j\right]}_\calH = 0.
		\end{equation}
	\end{enumerate}
\end{proposition}
\vspace{-10mm}
\begin{proof}
	For simplicity, we will briefly show a few implications among (i) and (iv) rather than full proof.
	
	(i) $\Longrightarrow$ (iv): For $i \in J_k$, there exist some $a_{ij}$ for $j\in J_k$ such that $T\left[b_i\right]=\sum_{j\in J_k}a_{ij} b_j$ since $T\left[b_i\right]\in \mathrm{span}\left\{b_i \mid i \in J_k\right\}$ by invariance in Definition~\ref{def:transform_invariance}. Since $\calB$ is an orthonormal basis of $\calH$, we can rewrite: $T\left[b_i\right]=\sum_{i\in I}\lrangle{b_j,T\left[b_i\right]}_\calH b_j$. Note that basis yields the unique linear coefficients so that we finally get $a_{ij}=\lrangle{b_j,T\left[b_i\right]}_\calH$ for $j\in J_k$ and $\lrangle{b_j,T\left[b_i\right]}_\calH=0$ for $j\notin J_k$. The latter one implies $\mathrm P_4$.
	
	(iv) $\Longrightarrow$ (iii): Note that $f=\sum_{i\in I} \lrangle{b_i,f}_\calH b_i$ by Equation~\eqref{eq:basis_coeff_eval}. By linearity of $T$, we get $T\left[f\right]= \sum_{i\in I}\lrangle{b_i,f}_{\calH}T\left[b_i\right]$. Expanding $T\left[b_i\right]$ using Equation~\eqref{eq:basis_coeff_eval} yields:
	\begin{equation}
		T\left[f\right] = \sum_{i\in I}\sum_{j \in I} \lrangle{b_i,f}_{\calH} \lrangle{b_j,T\left[b_i\right]}_{\calH}b_j.
	\end{equation}
	Since it is a linear combination of basis $b_j$, $\sum_{i\in I}\lrangle{b_i,f}_{\calH} \lrangle{b_j,T\left[b_i\right]}_{\calH}$ is the coefficient of $T\left[f\right]$ w.r.t. $b_j$. Changing letters for summation indices and using (iv), we finally get the projection of $T\left[f\right]$ on $\calB_{\le k}$ is:
	\begin{equation} \label{eq:prop_transform_invariance_1}
		\sum_{j\le k, i' \in J_j} \sum_{i\in I} \lrangle{b_{i},f}_{\calH} \lrangle{b_{i'},T\left[b_{i}\right]}_{\calH} b_{i'} = \sum_{j\le k, i' \in J_j} \sum_{i\in J_j} \lrangle{b_{i},f}_{\calH} \lrangle{b_{i'},T\left[b_{i}\right]}_{\calH} b_{i'}.
	\end{equation}
	On the other hand:
	\begin{align}
		f' &= \sum_{j \le k, i\in J_j} \lrangle{b_i, f}_\calH b_i, \\
		\label{eq:prop_transform_invariance_2}
		T\left[f'\right] &= \sum_{j \le k, i\in J_j} \lrangle{b_i, f}_\calH T\left[b_i\right] = \sum_{j \le k, i\in J_j} \sum_{i'\in I}\lrangle{b_i, f}_\calH \lrangle{b_{i'},T\left[b_i\right] }_\calH b_{i'} \\
		&= \sum_{j \le k, i\in J_j} \sum_{i'\in J_j}\lrangle{b_i, f}_\calH \lrangle{b_{i'},T\left[b_i\right] }_\calH b_{i'}. \nonumber
	\end{align}
	Here, the last implication comes from (iv). Now we observe that Equations~\eqref{eq:prop_transform_invariance_1} and~\eqref{eq:prop_transform_invariance_2} are equal.
	
	(iii) $\Longrightarrow$ (ii): It is straightforward since a projection is a linear operation and a projection on $\calB_k$ is identical to the subtraction of the projection on $\calB_{\le k-1}$ from that on $\calB_{\le k}$
\end{proof}

We observe here the matrix representation of $\lrangle{b_i,T\left[b_j\right]}$ satisfying Equation~\eqref{eq:tf_invariance} becomes a block diagonal matrix. Conditions (i) and (iv) can be determined only with the basis and the transform themselves, while (ii) and (iii) show why this invariance is important when one applies the transform for a given vector. In other words, for an invariant basis and a transform, projecting on a subset of the basis and applying the transform commute without any loss of information. Moreover, this commutativity allows us to reduce the transform applied on a projected vector to a finite computation even if the vector space $\calH$ has an infinite dimensionality.

\begin{proposition}{Finite matrix for invariant transform}{transform_invariance_finite}
	There is an orthonormal basis $\calB=\left\{b_i \mid i\in I\right\}$ of an inner product space $\calH$ and a linear operator $T:\calH\to\calH$. Suppose that $\calB$ is invariant under $T$ with a partition of finite indices $I=\bigcup_{k=0}^\infty J_k$. For any $f\in \calH$, let $\bff_{\le k}\coloneqq\left\{\lrangle{b_i, f} \mid b_i \in \calB_{\le k}\right\}$ denote the (finite) coefficient vector of $f$ projected onto $\calB_{\le k}\coloneqq \left\{b_i \mid j\le k,\ i\in J_j\right\}$. Then the $T\left[f\right]$ projection on $\calB_{\le k}$ is evaluated as the following finite matrix-vector product.
	\begin{equation}
		\bfT_{\le k}\bff_{\le k} \text{, where } \bfT_{\le k}\coloneqq \bfMat\left[\lrangle{b_i, T\left[b_j\right]}\mid b_i,b_j \in \calB_{\le k}\right].
	\end{equation}
\end{proposition}

Proposition~\ref{prop:transform_invariance_finite} is a necessary but not sufficient condition of invariance described in Definition~\ref{def:transform_invariance} and Proposition~\ref{prop:transform_invariance}, but it is related to what actually a rendering pipeline computes. Thus, Figures~\ref{fig:theory_rotmat_compare} and~\ref{fig:theory_penvmap_rotation} in our main paper shows experimental validation of Proposition~\ref{prop:transform_invariance_finite}.

\subsubsection{Linear spaces over $\R$ vs. $\C$}
\label{sec:bkgnd_linsp_R_vs_C}
In this paper, we sometimes consider a linear space with the scalar as $\R$ and sometimes do so with the scalar as $\C$.
Then some relations discussed here will be useful.

\begin{proposition}{Linear spaces over $\R$ vs. $\C$}{linalg-linsp-rc}
	If $\calB$ is a basis for a linear space $V$ over $\C$ then the set $\calB' \coloneqq\left\{e, ie\mid e\in \calB\right\}$ is a basis for $V$ as a linear space over $\R$. Concretely, if an arbitrary vector $v\in V$ is represented as a linear combination over complex coefficients by Equation~\eqref{eq:basis_axiom} as:
	\begin{equation}
		v = \sum_{i} c_i e_i,
	\end{equation}
	then it can be rewritten using the new basis $\calB'$ and real coefficients as follows:
	\begin{equation}
		v = \sum_{i} a_i e_i + b_i \left(ie_i\right), \quad \text{where}\quad a_i \coloneqq \Re c_i \quad \text{and}\quad b_i \coloneqq \Im c_i.
	\end{equation}
	Moreover, $V$ over $\C$ is equipped with an inner product $\lrangle{\cdot,\cdot}_{V|\C}$, an inner product on $V$ over $\R$ is canonically induced as $\lrangle{\cdot,\cdot}_{V|\R}\coloneqq \Re \lrangle{\cdot,\cdot}_{V|\C}$. $\calB$ is orthonormal (w.r.t. $\lrangle{\cdot,\cdot}_{V|\C}$) then the new basis $\calB'$ is orthonormal with respect to $\lrangle{\cdot,\cdot}_{V|\R}$.
\end{proposition}
Here, $\Re$ and $\Im$ denote taking real and imaginary parts of given complex numbers, respectively. We often write each inner product as $\lrangle{\cdot,\cdot}_{\C}$ and $\lrangle{\cdot,\cdot}_{\R}$, respectively, for simplicity when it is clear in context. The following relations between coefficients and bases are useful.
\begin{align}
	c_i &= \lrangle{B_i, v}_\C, \\
	a_i &= \Re\lrangle{B_i, v}_\C = \lrangle{B_i, v}_\R, \\
	b_i &= \Re\lrangle{iB_i, v}_\C = \lrangle{iB_i, v}_\R = \Im \lrangle{B_i, v}_\C.
\end{align}
Please be careful that while $B_i$ and $iB_i$ are not orthogonal in $V$ over $\C$ (i.e., $\lrangle{B_i,iB_i}_{\C} = i \ne 0$), these are orthogonal in $V$ over $\R$ (i.e., $\lrangle{B_i, iB_i}_{\R}=0$).

\clearpage

\section{Background: Spherical Harmonics}
\label{sec:bkgnd_SH}

\subsection{Spherical Harmonics}
\label{sec:bkgnd_SH_SH}

Spherical harmonics is described as a special case of Definition~\ref{def:basis_coeff}, as:
\begin{proposition}{Spherical harmonics}{bkgnd_def_SH}
	\emph{Spherical harmonics} are spherical functions $Y_{lm}\in \calF\left(\Sspv,\C\right)$ which can be evaluated in spherical coordinates $\left(\theta,\phi\right)$ as follows:
	\begin{subequations} \label{eq:bkgnd_sh_def0}
	\begin{equation} \label{eq:bkgnd_sh_def}
		Y_{lm}\left(\theta,\phi\right) = A_{lm}P_l^m\left(\cos\theta\right)e^{im \phi},
	\end{equation}
	\begin{align} \label{eq:bkgnd_sh_def_final}
		A_{lm} =& \sqrt{\frac{2l+1}{4\pi}\frac{\left(l-m\right)!}{\left(l+m\right)!}}, & P_l\left(x\right) =& \frac1{2^l l!}\difrac{^l}{x^l}\left(x^2-1\right)^l, \\ \nonumber
		P_l^m\left(x\right)=& \left(-1\right)^m \left(1-x^2\right)^{m/2}\difrac{^m}{x^m}P_l\left(x\right), & P_l^{-m}\left(x\right) =& \left(-1\right)^m\frac{
			\left(l-m\right)!
		}{
			\left(l+m\right)!
		} P_l^m\left(x\right)
		\text{, for }m\ge0. 
	\end{align}
	\end{subequations}
	With an index set $I_\mathrm{SH}=\left\{\left(l,m\right)\in \Z^2 \mid \left|m\right|\le l\right\}$, $\left\{ Y_{lm} \mid \left(l,m\right)\in I_\mathrm{SH}\right\}$ is an orthonormal basis of $\calF\left(\Sspv,\C\right)$.
\end{proposition}

Here, $P_l$ is called the \emph{Legendre function of order} $l$ and $P_l^m$ is called the \emph{associated Legendre function of order} $l$ \emph{and degree} $m$. \footnote{Unfortunately, there is a difference in terminologies \emph{order} and \emph{degree} depending on each research field. Mathematics and physics such as~\cite{canzani2013analysis, hall2013quantum} usually call $l$ and $m$ by \emph{degree} and \emph{order} respectively. We follow computer graphics convention as~\cite{ramamoorthi2001efficient, sloan2002precomputed, xin2021fast}.} The first few spherical harmonics functions can easily be evaluated using the recurrence relations above as follows. 
\begin{szMathBox}
\begin{align}\nonumber
	Y_{00}\left(\theta,\phi\right) &= \sqrt{\frac{1}{4\pi}}, &  Y_{1,-1}\left(\theta,\phi\right) &= \sqrt{\frac3{8\pi}}\sin\theta e^{-i\phi}, &Y_{10}\left(\theta,\phi\right)&=\sqrt{\frac3{4\pi}}\cos\theta, \\
	Y_{11}\left(\theta,\phi\right) &= -\sqrt{\frac3{8\pi}}\sin\theta e^{i\phi},
	& Y_{2,-2}\left(\theta,\phi\right)&=\sqrt\frac{15}{32\pi}\sin^2\theta e^{-2i\phi} ,& Y_{2,-1}\left(\theta,\phi\right)&= \sqrt\frac{15}{8\pi}\sin\theta\cos\theta e^{-i\phi}, \\
	Y_{20}\left(\theta,\phi\right)&= \sqrt\frac{5}{16\pi}\left(3\cos^2\theta-1\right),& Y_{21}\left(\theta,\phi\right)&= -\sqrt\frac{15}{8\pi}\sin\theta\cos\theta e^{i\phi},& Y_{22}\left(\theta,\phi\right)&= \sqrt\frac{15}{32\pi}\sin^2\theta e^{2i\phi} \nonumber
\end{align}
\end{szMathBox}
Be careful that other literature and programming libraries sometimes use different conventions in Equation~\eqref{eq:bkgnd_sh_def0}, so that they might have slightly different formulae such as multiplying $\pownu{m}$ or $\sqrt{4\pi}$.

\NEW{Orthonormality defined in Proposition~\ref{prop:basis_coeff_eval} assumes the set of spherical functions $\calF\left(\Sspv,\C\right)$ as an inner product space. An inner product of two spherical functions $f$ and $g\in\calF\left(\Sspv,\C\right)$ is an integral of the product of the values of the given two functions in each direction:}
\begin{equation}
	\lrangle{Y_{lm},f}_{\calF\left(\Sspv,\C\right)} = \int_{\Sspv}{f^*\left(\homega\right)g\left(\homega\right)\rmd\omega},
\end{equation}
where $\rmd\homega=\sin\theta\rmd \theta\rmd \phi$ is the solid angle measure on the sphere $\Sspv$. \NEW{Note the presence of the complex conjugation, whereas it can be ignored for real-valued functions. Note that inner products on other function spaces can be defined in a similar way in Section~\ref{sec:our_theory}.}

Applying Equation~\eqref{eq:basis_axiom} in Definition~\ref{def:basis_coeff} and Proposition~\ref{prop:basis_coeff_eval} implies that any spherical function $f\in \calF\left(\Sspv,\C\right)$ is equal to an infinite number of linear combination of spherical harmonics as:

\begin{equation}\label{eq:bkgnd_SH_expand}
	f=\sum_{l=0}^\infty\sum_{m=-l}^l \rmf_{lm}Y_{lm},
\end{equation}
\NEW{and the \emph{coefficient} $\rmf_{lm}$ is computed as}
\begin{equation}\label{eq:bkgnd_SH_coeff}
	\rmf_{lm}=\lrangle{Y_{lm},f}_{\calF\left(\Sspv,\C\right)}. %
\end{equation}
 An infinite dimensional numeric vector $\left[\rmf_{00}, \rmf_{1,-1}, \rmf_{10}, \rmf_{11}, \cdots\right]^T$\NEW{, which is called the \emph{coefficient vector} of $f$,} represents continuously defined $f$ without loss of information. However, we can take the \emph{projection of} $f$ \emph{on spherical harmonics up to} $l=l_{\mathrm{max}}$ by Definition~\ref{def:basis_projection} so that store it into a finite numeric vector $\begin{bmatrix}\rmf_{00} & \cdots & \rmf_{l_{\mathrm{max}},l_{\mathrm{max}}} \end{bmatrix}^T$ of \NEW{$\left(l_{\mathrm{max}}+1\right)^2=O\left(l_{\mathrm{max}}^2\right)$} entries. It can also be understood as a \emph{smoothed} data of $f$ up to the $l_{\mathrm{max}}$-th frequency band.

We observe that spherical harmonics satisfy the following identities, \NEW{which will be used later.}

\begin{proposition}{Spherical harmonics identities}{SH_identity}
	\begin{align} \label{eq:bkgnd_SH_conj}
		Y_{lm}^* &= \pownu{m} Y_{l,-m} \\
		\label{eq:bkgnd_SH_antipodal}
		Y_{lm}\left(\homega\right) &= \pownu{l+m}Y_{lm}\left(-\homega\right)
	\end{align}
\end{proposition}

\subsubsection{Zonal harmonics} \label{sec:bkgnd_ZH}
There is an important subset of spherical harmonics, which is useful for spherical functions with some symmetry. When a global frame $\frF_g$ is fixed, a spherical function $f\in\calF\left(\Sspv, \bbK\right)$ ($\bbK=\R$ or $\C$) is called to be \emph{azimuthally \NEW{(axially)} symmetric} if $f\left(\vec R_{\hat z_g}\left(\alpha\right)\homega\right)=f\left(\homega\right)$ for any $\alpha\in \R$ and $\homega\in\Sspv$. Note that such a function can be simply written as $f\left(\theta\right)$, a function of the single zenith angle $\theta$.
\NEW{Note that the two formulations of an azimuthally symmetric function about $\theta\in\left[0,\pi\right]$ and $\homega\in\Sspv$, respectively, are interchangeable using the following relation.

\begin{subequations}\label{eq:bkgnd_azimforms}
	\begin{minipage}{.35\linewidth}
	\begin{equation} \label{eq:bkgnd_azimforms_s2z}
		\underbrace{f\left(\theta\right)}_{\text{domain }\left[0,\pi\right]} = f\left(\cos^{-1}\hat z_g \cdot \homega\right),
	\end{equation}
\end{minipage}%
\begin{minipage}{.63\linewidth}
	\begin{equation} \label{eq:bkgnd_azimforms_z2s}
		\underbrace{f\left(\homega\right)}_{\text{domain }\Sspv} = f\left(\homega_{\mathrm{sph}}\left(\theta,\phi\right)\right)\text{, with any choice of }\phi\in\R.
	\end{equation}
\end{minipage}
\end{subequations}
}

Spherical harmonics $Y_{l0}$ with zero degrees ($m=0$) is called \emph{Zonal harmonics}, and the set of Zonal harmonics is a basis of the space of azimuthally symmetric spherical functions. \NEW{In other words, $Y_{l0}$ has azimuthal symmetry, and conversely any function $f\in\calF\left(\Sspv,\C\right)$ can be represented as $f=\sum_{l=0}^\infty {\rmf_{l0}Y_{l0}}$. Note that in contrast to SH $Y_{lm}$ for $m\ne 0$, Zonal harmonics basis $Y_{l0}$ always has real values so that $\rmf_{l0}$ is also real for any real-valued function $f\in\calF\left(\Sspv,\R\right)$.}

\subsection{Linear Operators in Spherical Harmonics}
\label{sec:SH_linop}
First, let's investigate the desirable properties of linear operators on spherical functions.

\begin{definition}{Linear operators and kernels}{sph_linop_kernel}
	Suppose there is a function $K\in\calF \left(\Sspv\times \Sspv,\bbK\right)$, where $\bbK=\R$ or $\C$. The \emph{linear operator with the kernel $K$},  denoted by $K_\calF\in \calL\left(\calF\left(\Sspv,\bbK\right), \calF\left(\Sspv,\bbK\right)\right)$, is defined as follows:
	\begin{equation}
		\forall f\in \calF\left(\Sspv,\bbK\right),\quad K_\calF\left[f\right]\left(\homega_i\right) = \int_{\Sspv}{ K\left(\homega_i,\homega_o\right)f\left(\homega_i\right) \rmd \homega_i}.
	\end{equation}
	If a linear operator $K_\calF$ was given first, a function $K$ satisfying the above equation is called the \emph{operator kernel} (or simply \emph{kernel}) of the operator $K_\calF$.
\end{definition}
\NEW{Here, we slightly abuse the notation of the symbol $\calF$. While on the first page $\calF\left(X, Y\right)$ is defined as the set of functions from $X$ to $Y$ for given sets $X$ and $Y$, in Definition~\ref{def:sph_linop_kernel} $K_\calF$ denotes a \emph{functional version} of the given $K$. Note that we will define such functional versions of a given quantity in different ways depending on the type of the given quantity. While such different ways will share the notation of the subscript $\calF$ in this paper, they will be clearly distinguished in context.}

In Section~\ref{sec:SH_linop}, we usually call the operator kernels simply \emph{kernels}, but in later sections, we often refer to them as \emph{operator kernel} to distinguish them from convolution kernels which will be introduced in Section~\ref{sec:spherical-convolution}.

As a special case of Proposition~\ref{eq:coeff_transform}, spherical harmonics provide frequency-domain formulations of spherical functions and linear operators on these spherical functions.

In the context of computer graphics, while a spherical function can be radiance as a function of directions, including an environment map, a linear operator on spherical functions can be a light interaction effect.

One of the simplest cases of it is surface reflection determined by a bidirectional reflectance distribution function (BRDF). Assuming we have a BRDF $\rho\colon \Sspv\times\Sspv\to\R$, its surface reflection can be considered as a linear operator $\rho_{\calF}^\perp\in \calL\left(\calF\left(\Sspv,\C\right), \calF\left(\Sspv,\C\right)\right)$ which maps incident radiance to outgoing radiance through the rendering equation as follows:
\begin{equation} \label{eq:rendering_eq}
	\rho_{\calF}^\perp\left[L^{\mathrm{in}}\right]\left(\homega_o\right) = L^{\mathrm{out}}=\int_{\Sspv}{ \rho\left(\homega_i,\homega_o\right)\abs{\hn\cdot\homega_i} L^{\mathrm{in}}\left(\homega_i\right) \rmd\homega_i},
\end{equation}
where the superscript $\perp$ denotes cosine-weighted.\footnote{\NEW{Note that in our main paper, we assume that the notation $\rho$ denotes a cosine-weighted BRDF for the sake of simplicity}}
Not only reflection due to a BRDF, other light interaction effects, including self-shadowing and self-transfer, can also be described as linear operators in similar ways by replacing $\rho\left(\homega_i,\homega_o\right)\abs{\hn\cdot\homega_i}$ to other functions.

Once we have a linear operator $\rho_\calF^\perp$, we can convert both the operator itself and the evaluation of the operator on a spherical function into frequency domain formulation using spherical harmonics. First, the \emph{coefficient of} $\rho_{\calF}^\perp$ \emph{with respect to} $\left(Y_{l_om_o},Y_{l_im_i}\right)$ or the $\left(l_o,m_o\right)-\left(l_i,m_i\right)$\emph{-th coefficient of} $\rho_\calF^\perp$ with respect to SH is defined as:
\begin{equation}\label{eq:bkgnd_SH_coeffmat}
	\rho_{l_om_o,l_im_i}\coloneqq\lrangle{Y_{l_om_o},\rho_\calF^\perp\left[Y_{l_im_i}\right]}_\calF.
\end{equation}
\NEW{Considering each pair of indices $\left(l,m\right)\in I_{\mathrm{SH}}$ to be linearly enumerated, Equation~\eqref{eq:bkgnd_SH_coeffmat} converts the linear operator $\rho_\calF^\perp$ into a (either finite or infinite) numeric matrix with the elements $\rho_{l_om_ol_im_i}$ in the $\left(l_o,m_o\right)$-th row and the $\left(l_i,m_i\right)$-th column, called the \emph{coefficient matrix} of $\rho_\calF^\perp$.}

In the case of the operator $\rho_\calF^\perp$, it has a kernel. Then, the coefficient can also evaluated from the kernel as follows.
\begin{equation}
	\rho_{l_om_o,l_im_i} = \int_{\Sspv\times \Sspv}{Y_{l_om_o}^*\left(\homega_o\right) \rho^{\perp}\left(\homega_i,\homega_o\right)Y_{l_im_i}\left(\homega_i\right)\rmd \homega_i \rmd\homega_o}.
\end{equation}
Then the rendering equation in Equation~\eqref{eq:rendering_eq} is reformulated as the following by Equation~\eqref{eq:coeff_transform}:
\begin{equation} \label{eq:bkgnd_SH_matmul}
	\rmL_{l_om_o}^{\mathrm{out}} =\sum_{\left(l_i,m_i\right)\in I_{\mathrm{SH}}}\rho_{l_om_o,l_im_i}\rmL_{l_im_i}^{\mathrm{in}},
\end{equation}
where $\rmL_{lm}^{\left\{\mathrm{in},\mathrm{out}\right\}} \coloneqq \lrangle{Y_{lm},L^{\left\{\mathrm{in},\mathrm{out}\right\}}}$ denotes the $\left(l,m\right)$-th SH coefficient of incident and outgoing radiance, respectively.

Note that the above equation can be considered as a matrix multiplication with the integer pairs $\left(l_o,m_o\right)$ as rows and the integer pairs $\left(l_i,m_i\right)$ as columns. %
\NEW{Figure~\ref{fig:bkgnd_SH_rotmat} illustrates a coefficient matrix of a linear operator (Equation~\eqref{eq:bkgnd_SH_coeffmat}) and how its action on a spherical function (Equation~\eqref{eq:bkgnd_SH_matmul}) can be converted into a matrix-vector product in the SH coefficient domain. Note that the special case of the given linear operator in Figure~\ref{fig:bkgnd_SH_rotmat}, including its sparsity, will be explained in the next subsection.

Taking finite coefficients up to orders $l\le l_{\mathrm{max}}$, the SH coefficient matrix of a linear operator consists of $\left(l_\mathrm{max}+1\right)^4=O\left(l_\mathrm{max}\right)$ in general, since it consists of $\left(l_\mathrm{max}+1\right)^2$ rows and columns.

Encoding linear operators into coefficient matrices as described in this subsection follows directly from the general theory described in Section~\ref{sec:bkgnd_func_anal}, so it can be applied in a similar way to other types of basis in a similar way. However, the strengths of SH appear when investigating sparsity and analytic formulations for coefficient matrices of special kinds of linear operators. In the next subsections, except for Section~\ref{sec:bkgnd_SH_RC}, we will investigate coefficient matrices of the functional version of rotation transforms (Section~\ref{sec:bkgnd_SH_rotation}), operators with azimuthal symmetry (isotropic BRDF, Section~\ref{sec:bkgnd_SH_azim}) and rotation equivariance (Section~\ref{sec:spherical-convolution}), and the functional version of the reflection operation which flips a direction vector to its antipodal direction (Section~\ref{sec:sh-reflection}). Note that the main theoretical purpose of this paper is to extend the desirable properties found in these subsections to the domain of a novel basis introduced in Section~\ref{sec:our_theory} taking Mueller calculus (Section~\ref{sec:background_polar}) into account.}

\mparagraph{\NEW{Application in precomputation-based rendering}}
\NEW{When the SH coefficient vector of $L^{\mathrm{in}}$ and the SH coefficient matrix of $\rho_\calF^\perp$ have been precomputed, environment map lighting can be computed efficiently as a matrix-vector product in runtime~\cite{ramamoorthi2001efficient}. In the precomputed radiance transfer (PRT) methods, the coefficient matrix, which is also called the \emph{radiance transfer matrix}, can contain further light transport effects, such as self-shadowing and inter-reflection, by replacing $\rho\left(\homega_i, \homega_o\right)$ in precomputation time~\cite{sloan2002precomputed}. In particular, self-shadowing can be achieved by replacing $\rho\left(\homega_i, \homega_o\right)$ with $\rho\left(\homega_i, \homega_o\right)V\left(\homega_i,\homega_o\right)$, where $V$ is the binary visibility function.}

\subsection{Rotation of Spherical Harmonics}
\label{sec:bkgnd_SH_rotation}

\begin{figure}[tbp]
	\centering
	\includegraphics[width=\columnwidth]{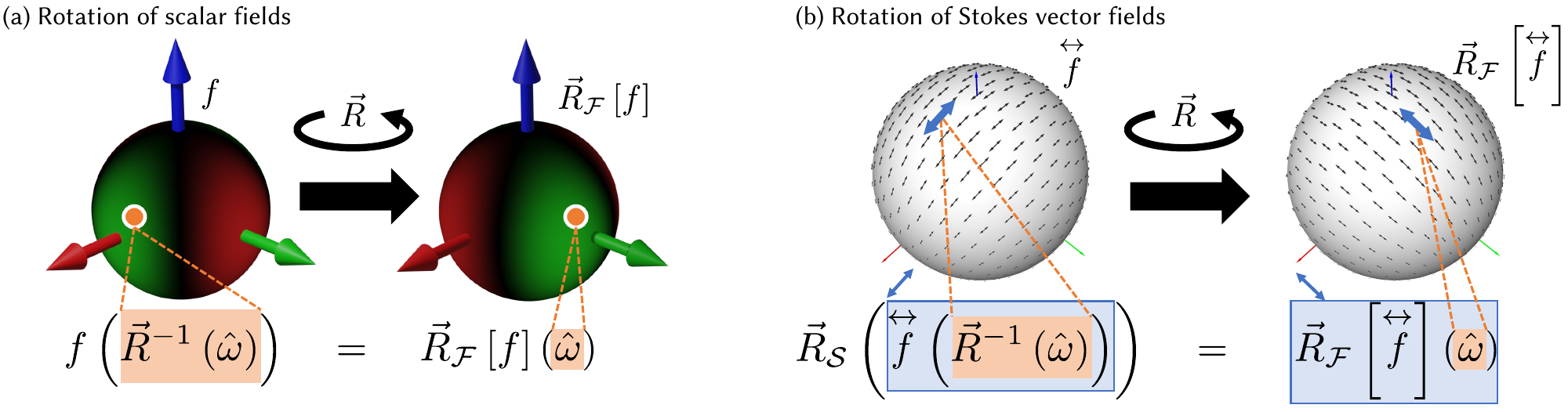}%
	\vspace{-3mm}
	\caption[]{\label{fig:bkgnd_sphfunc_rotation}
		Given a rotation transform $\vec R\in\SOgroupv$, (a) a rotation of a spherical function $f:\Sspv\to \R$ by $\vec R$ can be naturally defined by considering functions as textured spherical objects, which yields Equation~\eqref{eq:rotation_on_scalar_field}. (b) In later Section~\ref{sec:stokes-vector-field-rot}, 
		(b) We can similarly define a rotation of a Stokes vector field $\dvf:\Sspv\to\STKsp{\homega}$ by considering it as a spherical object attached with two-sided arrows on their surface points, which is represented by Equation~\eqref{eq:field-stk-rotation}. 
		To distinguish from the original rotation transform $\vec R$, which is defined as a function from single vectors to single vectors, we denote the induced rotation from functions to functions by $\vec R_\calF$.
	}
	\vspace{-2mm}		
\end{figure}

\begin{figure}
	\centering
	\begin{minipage}{0.60\textwidth}
		\centering
		\includegraphics[width=1.0\textwidth]{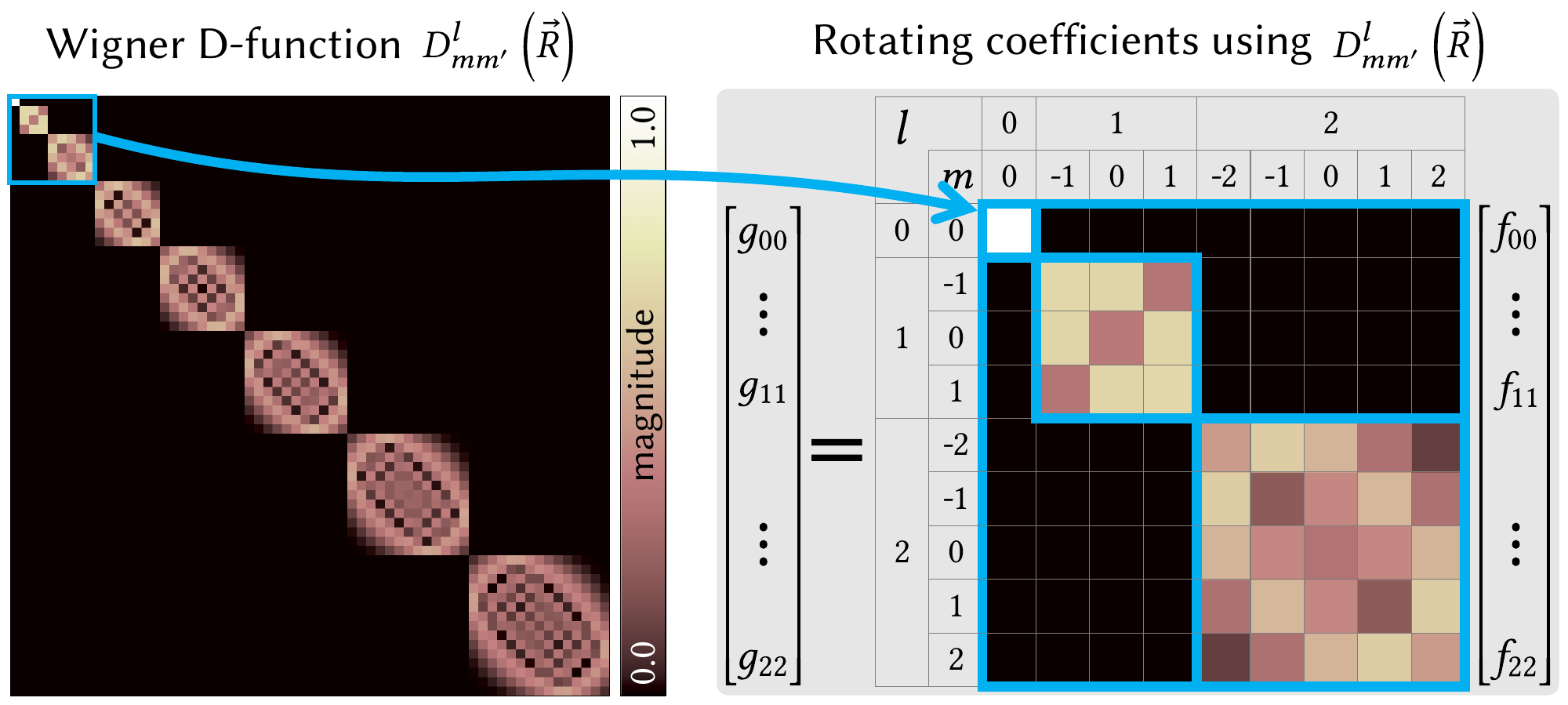} %
		\caption[]{ \label{fig:bkgnd_SH_rotmat}
			Visualization of the Wigner D-function of given such $\vec{R}$ and rotating the SH coefficients.
			Note that the elements of the Wigner D-function are complex numbers. Thus, we visualize the matrix element by its magnitude.
			The matrix values are $0$ when $l\neq l'$ (block-diagonal behavior) due to the Kronecker delta term, which yields the rotation invariance.
			The SH coefficients are rotated by simply multiplying the corresponding Wigner D-function as a coefficient matrix to the original SH coefficients without loss of information.
		}
	\end{minipage} \hfill
	\begin{minipage}{0.36\textwidth}
		\centering
		\vspace{0mm}
		\includegraphics[width=1.0\textwidth]{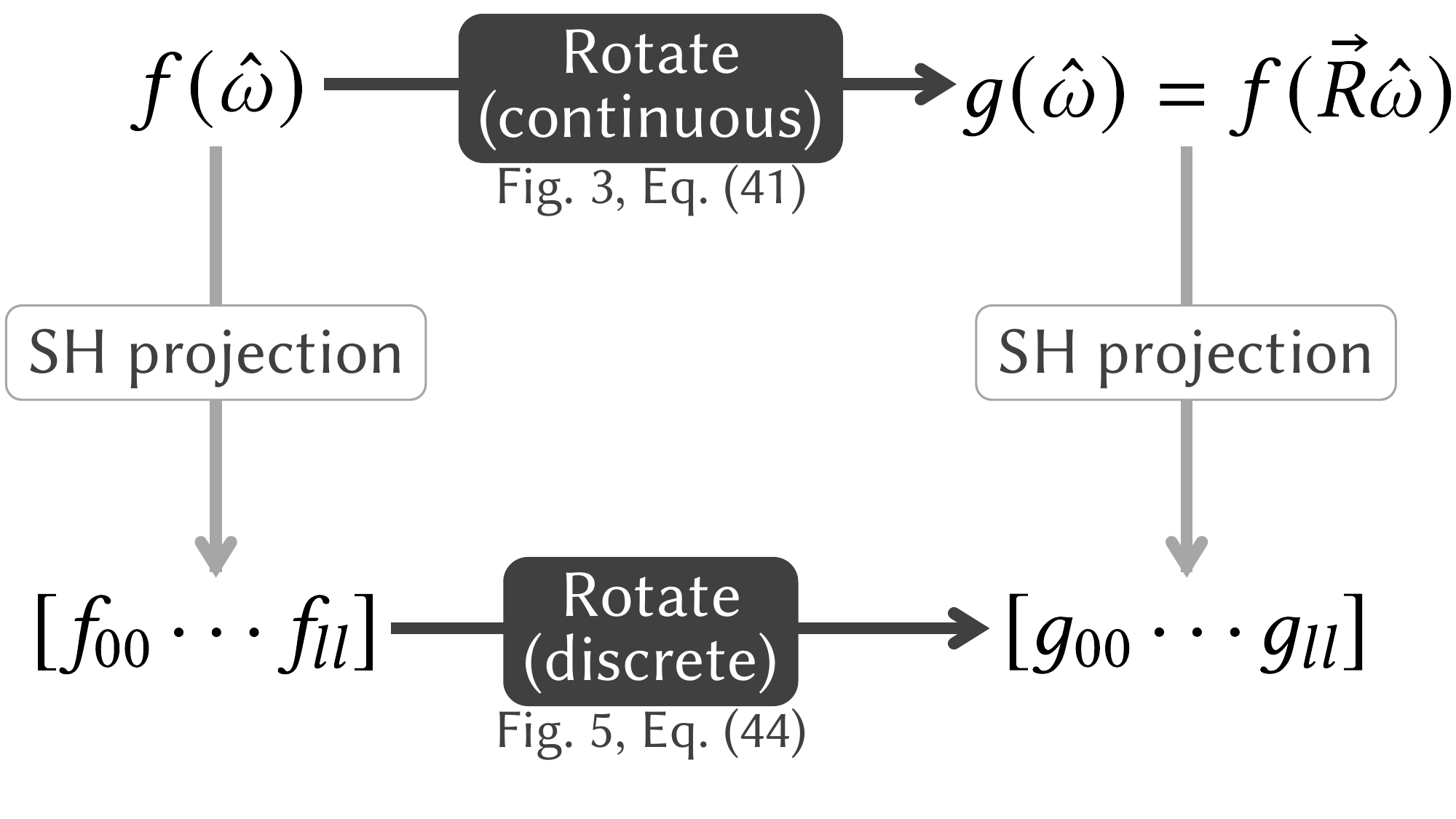} %
		\vspace{0mm}
		\caption[]{\label{fig:bkgnd_SH_rotation_property}
			The illustrative description of rotation invariance in spherical harmonics
			The upper path in the figure (rotate $\to$ SH projection) should be identical to the bottom path in this figure (SH projection $\to$ rotate).
			For rotating the discrete SH coefficients. %
			
		}
	\end{minipage}
\end{figure}

One of the most important properties of spherical harmonics, which is not satisfied by another basis, such as spherical wavelets \NEW{and spherical Gaussian}, is rotation invariance. We first formulate how a rotation transform can act on \emph{functions}, not only individual vectors, and then investigate the rotation of spherical harmonics.

First of all, given a rotation transform $\vec R\in \SOgroupv$, which is a function from $\Rspv$ onto $\Rspv$ \NEW{(restricted to a function from $\Sspv$ to $\Sspv$)}, we naturally define a rotation of \emph{functions}, denoted by $\vec R_\calF$, as follows:

\begin{equation} \label{eq:rotation_on_scalar_field}
	\vec R_\calF : \calF\left(\Sspv,\C\right) \to \calF\left(\Sspv,\C\right), \quad \vec R_\calF \left[f\right]\left(\homega\right) = f\left(\vec R^{-1} \homega\right),
\end{equation}
where this rotation on functions is also described in Figure~\ref{fig:bkgnd_sphfunc_rotation} (a), and $\vec R_\calF$ can also be considered as functions on real-valued functions, i.e., $\vec R_\calF:\calF\left(\Sspv,\R\right) \to \calF\left(\Sspv,\R\right)$.

We observe that Equation~\eqref{eq:rotation_on_scalar_field} is linear about $f$, so $\vec R_\calF$ is a linear operator on the space of spherical functions $\calF\left(\Sspv,\C\right)$. Then, we can formulate the rotation invariance of spherical harmonics in the same manner as Definition~\ref{def:transform_invariance}. Using (iv) in Proposition~\ref{prop:transform_invariance}, we can formulate the invariance property as follows:

\begin{proposition}{Rotation invariance of spherical harmonics}{rotinv_sh}
	Spherical harmonics $\left\{Y_{lm} \mid \left(l,m\right)\in I_\mathrm{SH}\right\}$, a basis of $\calF\left(\Sspv, \C\right)$, is invariant under a linear operator $\vec R_\calF$ for any rotation $\vec R\in\SOgroupv$ with a partition of index set $\left\{I_{\mathrm{SH}, l} = \left\{\left(l',m\right)\in I_\mathrm{SH} \mid l' = l\right\}\right\}$. In other words, the coefficient of the linear operator $\vec R_\calF$ with respect to spherical harmonics can be written as:
	\begin{equation} \label{eq:rotinv_sh}
		\lrangle{Y_{lm}, \vec R_\calF\left[Y_{l'm'}\right]}_{\calF\left(\Sspv,\C\right)} = 0, \quad \text{whenever } l\ne l'.
	\end{equation}
\end{proposition}
\vspace{-10mm}
\begin{proof}
	We will not cover a symbolic integration-based proof here, but there is a simple way to understand this invariance in a few steps. First, each SH function is an eigenfunction of the Laplace-Beltrami operator on $\Sspv$ corresponding to an eigenvalue $-l\left(l+1\right)$, which does not depend on $m$. Then the subspace of spherical functions spanned by $\left\{Y_{lm}\mid m\in\Z \text{ with } \abs{m}\le l \right\}$ for fixed $l$ is a degenerated eigenspace. Since the Laplace-Beltrami operator commutes with any rotation, the eigenspace is invariant under rotation.
\end{proof}

For the actual computation of rotation in the SH coefficient space, we need to know the non-zeros inner product value in the left-hand side of Equation~\eqref{eq:rotinv_sh} in the case of $l=l'$. This is an important special function called a \emph{Wigner D-function}\footnote{\NEW{Alternatively, it is known as \emph{Wigner D-matrix} in other literature.} Terminology \emph{matrix} comes from viewing $m$ and $m'$ in $D_{mm'}^l$ as row and column indices of a matrix.}, which is also common in mathematics and physics. It is defined as follows:
\begin{definition}{Wigner D-function}{wignerD}
For indices $l,m,m'\in \mathbb{Z}$ with $\abs{m}\le l$ and $\abs{m'}\le l$, \emph{Wigner D-function} $D_{mm'}^l: \SOgroupv\to\C$ is defined as follows:
\begin{equation}\label{eq:def_wignerD}
	D_{mm'}^l\left(\vec R\right) = \lrangle{Y_{lm}, \vec R_\calF\left[Y_{lm'}\right]}_{\calF\left(\Sspv,\C\right)}.
\end{equation}
\end{definition}
Combining Equations~\eqref{eq:rotinv_sh} and~\eqref{eq:def_wignerD} with the Kronecker delta notation, the coefficient of a rotation transform with respect to SH can be generally rewritten as follows:
\begin{equation} \label{eq:def_delta_wignerD}
	\lrangle{Y_{lm}, \vec R_\calF\left[Y_{l'm'}\right]}_{\calF\left(\Sspv,\C\right)} = \delta_{ll'} D_{mm'}^l\left(\vec R\right).
\end{equation}

\NEW{
The coefficient matrix of Equation~\eqref{eq:def_delta_wignerD} for a particular rotation transform is shown in Figure~\ref{fig:bkgnd_SH_rotmat}. The rotation invariance of SH also appears as the block diagonal constraint on the coefficient matrix, as shown in the figure. This property also implies that we can commute the SH projection of a function and a rotation without loss of information. If one wants to obtain the SH coefficients of a function $g=\vec R_\calF\left[f\right]$, the discrete computation between Wigner D-functions and the SH coefficients of $f$ gives the exact same result. This process is also illustrated in Figure~\ref{fig:bkgnd_SH_rotation_property}. Note that for finite coefficients up to $l\le l_\mathrm{max}$, the block diagonal sparsity produces at most $\left(l_\mathrm{max}+1\right)\left(2l_\mathrm{max}+1\right)\left(2l_\mathrm{max}+3\right)/3=O\left(l_\mathrm{max}^3\right)$ nonzero elements.
}
\subsubsection{Properties of Wigner D-functions} \label{sec:bkgnd_wigner_detail}

Following the definition, exact formulae for the first few Wigner D-functions are obtained as following equations using ZYZ Euler angle parameterization $D_{mm'}^l\left(\alpha,\beta,\gamma\right)\coloneqq D_{mm'}^l\left(\vec R_{\hat z_g \hat y_g \hat z_g}\left(\alpha,\beta,\gamma\right)\right)$.
\begin{szMathBox}
\begin{equation} \label{eq:first_few_wignerD}
	D_{00}^0\left(\alpha,\beta,\gamma\right) = 1, \qquad\qquad\qquad \quad
	\begin{array}{@{}cccc@{}}
		\multicolumn{4}{c@{}}{D_{mm'}^1\left(\alpha,\beta,\gamma\right)=}
		\\
		\midrule
		m & m'=1 & m'=0 & m'=-1 \\
		\cmidrule(l){2-4}
		1 & \frac{1+\cos\beta}{2}e^{-i\left(\alpha+\gamma\right)}
		&  -\frac1{\sqrt2}\sin\beta e^{i\alpha} & \frac{1-\cos\beta}{2}e^{-i\left(\alpha-\gamma\right)}\\
		0 & \frac1{\sqrt2}\sin\beta e^{-i\gamma}
		& \cos\beta & -\frac1{\sqrt2}\sin\beta e^{i\gamma}\\
		-1 & \frac{1-\cos\beta}{2}e^{i\left(\alpha-\gamma\right)}
		& \frac1{\sqrt2}\sin\beta e^{i\alpha} & \frac{1+\cos\beta}{2}e^{i\left(\alpha+\gamma\right)}\\
	\end{array}
\end{equation}
\end{szMathBox}
As seen in examples of Wigner D-functions in the above, $\alpha$ and $\gamma$ dependencies of them can be separated as the following equation.
\begin{equation}
	D_{mm'}^l \left(\alpha,\beta,\gamma\right) = e^{-im\alpha} d_{mm'}^l\left(\beta\right)e^{-im'\gamma}.
\end{equation}
It can be directly derived from the integral (inner product) in Equation~\eqref{eq:def_wignerD} by separating $\theta$ and $\phi$ dependencies of SH using Equation~\eqref{eq:bkgnd_sh_def}. Note that the remaining $\beta$ dependency, denoted by $d_{mm'}^l\left(\beta\right)$, is called a Wigner (small) d-function, but we do not need such complicated recurrence relations for it.

Additionally, note the following identities for Wigner D-functions.
\begin{proposition}{Wigner D-function indentities}{wignerD_identity}
	\begin{enumerate}
		\item $D_{mm'}^l \left(\vec I\right) = \delta_{mm'}$, where $\vec I\in \SOgroupv$ denotes the identity rotation.
		\item $\sum_{m_2=-l}^l D_{m_1 m_2}^l\left(\vec R_1\right)D_{m_2m_3}\left(\vec R_2\right) = D_{m_1 m_3}^l\left(\vec R_1 \vec R_2\right)$
		\item $\sum_{m_2=-l}^l D_{m_1 m_2}^l\left(\vec R\right)D_{m_2m_3}\left(\vec R^{-1}\right) = \delta_{m_1m_3}$
		\item $D_{mm'}^l\left(\vec R^{-1}\right) = D_{m'm}^{l}\left(\vec R\right)^*$
		\item $D_{-m,-m'}^l\left(\vec R\right) = \pownu{m+m'}D_{mm'}^l\left(\vec R\right)^*$
		\item $D_{m0}^l\left(\vec R_{z_gy_gz_g}\left(\phi,\theta,\psi\right)\right) = \sqrt\frac{4\pi}{2l+1}Y_{lm}^*\left(\theta,\phi\right) = \pownu{m}\sqrt\frac{4\pi}{2l+1}Y_{l,-m}\left(\theta,\phi\right)$
		\item $\left\{D_{mm'}^l \mid \abs{m},\abs{m'}\le l \right\}$ is an orthogonal basis on $\calF\left(\SOgroupv,\C\right)$, especially:
		\begin{equation}
			\lrangle{D_{m_1m_1'}^{l_1}, D_{m_2m_2'}^{l_2}}_{\calF\left(\SOgroupv,\C\right)} = \int_{\SOgroupv}{D_{m_1m_1'}^{l_1} \left(\vec R\right)^* D_{m_2m_2'}^{l_2}\left(\vec R\right)\rmd \vec R} = \frac{8\pi^2}{2l_1+1}\delta_{l_1l_2}\delta_{m_1m_2}\delta_{m_1'm_2'}
		\end{equation}
	\end{enumerate}
\end{proposition}
\vspace{-10mm}
\begin{proof}
	(1)---(4): Straightforward from Proposition~\ref{prop:linop_identity} (1)---(3) and (5), respectively.
	
	(5):
	\begin{equation}\nonumber
			D_{-m,-m'}^l\left(\vec R\right) \underarrowref{=}{Def.~\ref{def:wignerD}} \lrangle{Y_{l,-m},\vec R_\calF\left[Y_{l,-m'}\right]} \underarrowref{=}{Eq.~\eqref{eq:bkgnd_SH_conj}} \lrangle{\pownu{m}Y_{lm}^*,\vec R_\calF\left[\pownu{m'}Y_{lm'}^*\right]}
			= \pownu{m+m'} \lrangle{Y_{lm}^*,\vec R_\calF\left[Y_{lm'}^*\right]}.
	\end{equation}
	Here, we observe $\lrangle{Y_{lm}^*,\vec R_\calF\left[Y_{lm'}^*\right]} = \lrangle{Y_{lm}^*,\left(\vec R_\calF\left[Y_{lm'}\right]\right)^*} = \lrangle{Y_{lm},\vec R_\calF\left[Y_{lm'}\right]}^*$. Now, we finally get the given equation.
	
	(6) and (7): We refer to a book \cite{edmonds1996angular}. Note that Equations~(2.5.17) on p.23 and (2.5.29) on p.24 in the textbook provide an equivalent definition of SH to ours in Proposition~\ref{prop:bkgnd_def_SH}. Equation~(4.1.10) on p.55 in the book also provides the equivalent definition of Wigner D-functions to ours in Definition~\ref{def:wignerD}. Then, we can find that our propositions (6) and (7) are shown in Equations (4.1.25) \NEW{on} p.59 and (4.6.1) \NEW{on} p.62 in the book, respectively.
\end{proof}

\subsection{Complex and Real Spherical Harmonics}
\label{sec:bkgnd_SH_RC}

Spherical harmonics defined in Equation~\eqref{eq:bkgnd_sh_def} are complex functions spaning complex-valued functions $\calF\left(\Sspv,\C\right)$ with complex coefficients. However, radiometric intensity in the real world only consists of real numbers, so \emph{real spherical harmonics}, defined as follows, sometimes makes computational efficiency.
\begin{definition}{Real spherical harmonics}{}
	\begin{equation}\label{eq:bkgnd_real_SH}
		Y_{lm}^R  = \begin{cases}
			\sqrt{2}\Re Y_{lm}^C = \frac1{\sqrt2}\left( Y_{lm}^{C} + \left(-1\right)^mY_{l,-m}^{C} \right)& m>0\\
			Y_{lm}^C& m=0 \\
			\sqrt{2}\Im Y_{l\abs{m}}^C = \frac i{\sqrt2}\left( \left(-1\right)^mY_{lm}^C - Y_{l,-m}^C \right) & m < 0
		\end{cases}.
	\end{equation}
\end{definition}
Here, $Y_{lm}^C$ is just equal to $Y_{lm}$ defined in Equation~\eqref{eq:bkgnd_sh_def}, and we will sometimes call it \emph{complex} spherical harmonics when we need to distinguish them from \emph{real} ones. Note that the real spherical harmonics are also an orthonormal basis for spherical functions and have rotation invariance, but they always produce real-valued functions whenever real coefficients are given. Due to the efficiency of representing real-valued functions, most of the existing computer graphics works have used real spherical harmonics, and we also use it for some parts of polarization. However, we should know both real and complex spherical harmonics since spin-2 spherical harmonics, which will be introduced in a later section, are related to the complex ones.

The relation between complex and real spherical harmonics can be rewritten shortly by introducing a symbol $M_{m_1m_2}^{C\to R}$ defined as:
\begin{equation}\label{eq:bkgnd_SH_C2R_22}
	\mathbf{Mat}\left[M_{m_1m_2}^{C\to R} \mid m_1,m_2=+\abs{m}, -\abs{m}\right] = \frac1{\sqrt2}\begin{bmatrix}
		1 & \left(-1\right)^{m} \\
		-i &  \left(-1\right)^{m}i
	\end{bmatrix} \text{, for }\abs{m} \ne 0,
\end{equation}
, $M_{00}^{C\to R}=1$, and $M_{m_1m_2}^{C\to R}=0$ if $\abs{m_1}\ne\abs{m_2}$. Similarly, a symbol $M_{m_1m_2}^{R\to C}$ is defined as follows:
\begin{equation}\label{eq:bkgnd_SH_R2C_22}
	\mathbf{Mat}\left[M_{m_1m_2}^{R\to C} \mid m_1,m_2=+\abs{m}, -\abs{m}\right] = \frac1{\sqrt2}\begin{bmatrix}
		1 & i \\
		\left(-1\right)^{m} & - \left(-1\right)^{m}i
	\end{bmatrix}, \text{ for }\abs{m}\ne 0.
\end{equation}
Note that Equations~\eqref{eq:bkgnd_SH_C2R_22} and~\eqref{eq:bkgnd_SH_R2C_22} are unitary matrices which are the inverse of each other, and it can be written as:
\begin{equation}
	M_{mm'}^{R\to C} = \left(M_{m'm}^{C\to R}\right)^*,\quad \sum_{m'\in\left\{\pm m\right\}} M_{mm'}^{R\to C} M_{m'm''}^{C\to R} = \delta_{m m''}.
\end{equation}
Here, we are using the summation symbol with $\sum_{m'\in\left\{\pm m\right\}}$ rather than much common $\sum_{m'=\pm m}$ to clarify $\sum_{m'\in\left\{\pm 0\right\}}f\left(m'\right)=f\left(0\right)$ rather than $f\left(0\right)+f\left(0\right)$.
Now Equation~\eqref{eq:bkgnd_real_SH} can be rewritten as follows:
\begin{equation}\label{eq:SH_basis_RC_convert}
	Y_{lm}^R = \sum_{m'\in\left\{\pm m\right\}} M_{mm'}^{C\to R} Y_{lm'}^C, \quad\quad Y_{lm}^C = \sum_{m'\in\left\{\pm m\right\}} M_{mm'}^{R\to C} Y_{lm'}^R.
\end{equation}
On the other hand, converting coefficients of a spherical function with respect to complex real SH requires an extra complex conjugation. Suppose that $f\in\calF\left(\Sspv,\C\right)$ is a spherical function, and $\rmf_{lm}^C$ and $\rmf_{lm}^R$ are coefficients of $f$ with respect to $Y_{lm}^C$ and $Y_{lm}^R$, respectively. The following relation is obtained by the definition of SH coefficients and Equation~\eqref{eq:SH_basis_RC_convert}:
\begin{equation}
	\rmf_{lm}^R = \sum_{m'\in\left\{\pm m\right\}} \left(M_{mm'}^{C\to R}\right)^* \rmf_{lm'}^C, \quad \rmf_{lm}^C = \sum_{m'\in\left\{\pm m\right\}} \left(M_{mm'}^{R\to C}\right)^* \rmf_{lm'}^R.
\end{equation}

\mparagraph{Complex and real SH coefficients for linear operators}
Similarly, we can also obtain the relation between the coefficients of a linear operator with respect to complex and real SH. Denoting a linear operator on spherical functions by $T:\calL\left(\calF\left(\Sspv,\C\right), \calF\left(\Sspv,\C\right)\right)$, its $\left(l_o,m_o\right)-\left(l_i,m_i\right)$-th complex and real SH coefficients by $\rmT_{l_om_o,l_im_i}^C$ and $\rmT_{l_om_o,l_im_i}^R$, respectively, the following holds.
\begin{equation}\label{eq:SH_op_coeff_C2R}
	\begin{split}
	\rmT_{l_om_o,l_im_i}^R &=  \lrangle{Y_{l_om_o}^R, T\left[Y_{l_im_i}^R\right]} =\lrangle{\sum_{m\in\left\{\pm m_o\right\}}M_{m_o m}^{C\to R}Y_{l_om}^C, T\left[ \sum_{m'\in\left\{\pm m_i\right\}} M_{m_i m'}^{C\to R}Y_{l_i m'}^C \right]} \\
	&= \sum_{m\in\left\{\pm m_o\right\}}\sum_{m'\in\left\{\pm m_i\right\}} \left(M_{m_o m}^{C\to R}\right)^* M_{m_i m'}^{C\to R}\rmT_{l_om,l_im'}^C.
	\end{split}
\end{equation}
Conversely, the following also holds.
\begin{equation}\label{eq:SH_op_coeff_R2C}
	\rmT_{l_om_o,l_im_i}^C = \sum_{m\in\left\{\pm m_o\right\}}\sum_{m'\in\left\{\pm m_i\right\}} \left(M_{m_o m}^{R\to C}\right)^* M_{m_i m'}^{R\to C}\rmT_{l_om,l_im'}^R.
\end{equation}

\mparagraph{Real Wigner-D functions}
Similar to Equation~\eqref{eq:wignerD_def} in the main paper and Definition~\ref{def:wignerD} in this document, we can also define rotation transform for real spherical harmonics, which we call \emph{real Wigner-D functions}, as follows:
\begin{equation}\label{eq:real_wignerD}
	D_{mm'}^{l,R}\left(\vec R\right) = \lrangle{Y_{lm}^R, \vec R_\calF\left[Y_{lm'}^R\right]} .
\end{equation}
Relation between real and complex Wigner-D functions is just a special case of Equations~\eqref{eq:SH_op_coeff_C2R} and \eqref{eq:SH_op_coeff_R2C} is found by the relation between real and complex SH.
\begin{equation}\label{eq:wignerD_RC}
	\begin{split}
		D_{mm'}^{l,R}\left(\vec R\right) &= \sum_{m_c\in\left\{\pm m\right\}}\sum_{m_c'\in\left\{\pm m'\right\}}\left(M_{mm_c}^{C\to R}\right)^* M_{m'm_c'}^{C\to R} D_{m_cm_c'}^{l,C}\left(\vec R\right), \\
		D_{mm'}^{l,C}\left(\vec R\right) &= \sum_{m_r\in\left\{\pm m\right\}}\sum_{m_r' \in\left\{\pm m'\right\}} \left(M_{mm_r}^{R\to C}\right)^* M_{m'm_r'}^{R\to C} D_{m_r m_r'}^{l,R}\left(\vec R\right) .
	\end{split}
\end{equation}

Using this result, the relation between real Wigner-D functions and real SH (real SH version of Proposition~\ref{prop:wignerD_identity} (6)) comes from the relation between complex ones:
\begin{equation}
	\begin{split}
		D_{m0}^{l,R}\left(\vec R_{z_g y_g z_z}\left(\alpha,\beta,\gamma\right)\right) &= \sum_{m_c\in\left\{\pm m\right\}}\left(M_{mm_c}^{C\to R}\right)^* D_{m_c0}^{l,C}\left(\vec R_{z_g y_g z_z}\left(\alpha,\beta,\gamma\right)\right) = \sqrt\frac{4\pi}{2l+1} \sum_{m_c\in\left\{\pm m\right\}}\left(M_{mm_c}^{C\to R}\right)^* Y_{lm_c}^{C,*}\left(\beta,\alpha\right) \\
		&=  \sqrt\frac{4\pi}{2l+1} Y_{lm}^{R,*}\left(\beta,\alpha\right) = \sqrt\frac{4\pi}{2l+1} Y_{lm}^{R}\left(\beta,\alpha\right) .
	\end{split}
\end{equation}

\subsection{Azimuthally Symmetric Operators (Isotropic BRDFs)} \label{sec:bkgnd_SH_azim}
While a general linear operator can be represented by its SH coefficients, it requires too many numbers, $\left(l_{\mathrm{max}}+1\right)^4$ for the maximum order $l_{\mathrm{max}}$, of coefficients. Several symmetry conditions for such an operator yield linear constraints on its SH coefficients, so we obtain much fewer degrees of freedom for the coefficients.

One of the common constraints of linear operators on spherical functions is azimuthal symmetry. It is defined as follows.
\begin{definition}{Azimuthally symmetric operators}{azim_sym_op}
	Suppose that a global frame $\frF_g = \begin{bmatrix} \hat x_g & \hat y_g & \hat z_g \end{bmatrix}$ is fixed. Then a linear operator $K\colon \calL\left(\calF\left(\Sspv,\C\right), \calF\left(\Sspv,\C\right)\right)$ on scalar fields is called to be \emph{azimuthally symmetric} if it commutes with any rotation along $\hat z_g$, i.e.:
	\begin{equation}\label{eq:def_azim_sym_op}
		\vec R_{\hat z_g}\left(\alpha\right)_\calF \left[K\left[f\right]\right] = K\left[\vec R_{\hat z_g}\left(\alpha\right)_\calF \left[f\right]\right], \quad \forall \alpha\in\R,\ \forall f\in \calF\left(\Sspv,\C\right) .
	\end{equation}
\end{definition}

When the given linear operator indicates surface interaction due to a BRDF in the rendering context, then this constraint is equivalent to the isotropy of BRDF. Suppose that the operator $K$ has a kernel $k:\Sspv\times \Sspv\to \C$, (again, cosine-weighted BRDF in a rendering context), then the azimuthal symmetry defined in Definition~\ref{def:azim_sym_op} is equivalent to the following condition:
\begin{equation}
	k\left(\homega_i,\homega_o\right) = k\left(\vec R_{\hat z_g}\left(\alpha\right)\homega_i, \vec R_{\hat z_g}\left(\alpha\right)\homega_o\right), \quad \forall\alpha\in\R, \ \quad \forall \homega_i,\homega_o\in\Sspv.
\end{equation}
\NEW{In the spherical coordinates, using the relation $\vec R_{\hat z_g}\homega_{\mathrm{sph}}\left(\theta,\phi\right)=\homega_{\mathrm{sph}}\left(\theta,\phi+\alpha\right)$ and substituting $\alpha=-\phi_i$ the above equation can be rewritten in more familiar form in computer graphics as
\begin{equation}
	k\left(\theta_i,\phi_i,\theta_o,\phi_o\right) = k\left(\theta_i,0,\theta_o,\phi_o-\phi_i\right).
\end{equation}
}
Now, we investigate how the symmetry condition makes a linear constraint on SH coefficients.

\begin{proposition}{Coefficients of azimuthally symmetric operators (isotropic BRDFs)}{bkgnd_isoBRDF}
	Suppose that $K\colon \calL\left(\calF\left(\Sspv,\C\right), \calF\left(\Sspv,\C\right)\right)$ is an azimuthally symmetric operator and $\rmK_{l_om_o,l_im_i}\coloneqq \lrangle{Y_{l_om_o}, K\left[Y_{l_im_i}\right]}$ denotes the $\left(l_o,m_o\right)-\left(l_i,m_i\right)$-th coefficient of $K$ with respect to complex SH. Then the coefficient vanishes whenever $m_i\ne m_o$, so that it can be denoted by a coefficient $\rmK_{l_ol_im}$ with three indices such that:
	\begin{equation}\label{eq:bkgnd_SH_iso}
		\rmK_{l_om_o,l_im_i} = \delta_{m_om_i} \rmK_{l_o l_i m}.
	\end{equation}
\end{proposition}
\vspace{-10mm}
\begin{proof}
	Start from Equation~\eqref{eq:def_azim_sym_op}. First, the equation holds for any function $f$ so that it can be rewritten as an equality of two operators. Then, taking $\left(l_o,m_o\right)-\left(l_i,m_i\right)$-th SH coefficients for both hand sides of them followed by applying Proposition~\ref{prop:linop_identity} (2) yields:
	\begin{equation}
		\begin{split}
		\sum_{\left(l,m\right)\in I_{\mathrm{SH}}} \lrangle{Y_{l_om_o}, \vec R_{\hat z_g}\left(\theta\right)_\calF\left[Y_{lm}\right]} \lrangle{Y_{lm}, K\left[Y_{l_im_i}\right]} &= \sum_{\left(l,m\right)\in I_{\mathrm{SH}}} \lrangle{Y_{l_om_o}, K\left[Y_{lm}\right]} \lrangle{Y_{lm}, R_{\hat z_g}\left(\theta\right)_\calF\left[Y_{l_im_i}\right]}, \\
		\Rightarrow\quad \sum_{\left(l,m\right)\in I_{\mathrm{SH}}} \delta_{l_ol}D_{m_om}^{l_o}\left(\vec R_{\hat z_g}\left(\theta\right)\right) \rmK_{lml_im_i} &= \sum_{\left(l,m\right)\in I_{\mathrm{SH}}} \rmK_{l_om_olm} \delta_{ll_i}D_{mm_i}^{l_i}\left(\vec R_{\hat z_g}\left(\theta\right)\right).
		\end{split}
	\end{equation}
	From definition of Wigner D-functions in Equation~\eqref{eq:def_wignerD} we easily get $D_{mm'}^l\left(\vec R_{\hat g}\left(\theta\right)\right)=\delta_{mm'}e^{-im\theta}$. Using it makes the above equation as follows:
	\begin{equation}
		\begin{split}
			\sum_{\left(l,m\right)\in I_{\mathrm{SH}}} \delta_{l_ol}\delta_{m_om}e^{-im_o \theta} \rmK_{lml_im_i} &= \sum_{\left(l,m\right)\in I_{\mathrm{SH}}}  \delta_{ll_i}\delta_{mm_i}e^{-im_i \theta} \rmK_{l_om_olm}, \\
			\Rightarrow e^{-im_o \theta} \rmK_{l_om_ol_im_i} &= e^{-im_i \theta} \rmK_{l_om_ol_im_i}, \\
			\Rightarrow \left(e^{-im_i \theta}-e^{-im_o \theta}\right) \rmK_{l_om_ol_im_i} &= 0.
		\end{split}
	\end{equation}
	Here, we observe that $\rmK_{l_om_ol_im_i}$ should be zero for $m_i \ne m_o$ to make the above equation hold for all $\theta$.
\end{proof}
Note that this property is used in \citet{ramamoorthi2001signal}. \NEW{From the sparsity in Equation~\eqref{eq:bkgnd_SH_iso}, the finite SH coefficient matrix of an azimuthally symmetric operator up to $l_i,l_o\le l_\mathrm{max}$ has $\left(l_\mathrm{max}+1\right)\left(2l_\mathrm{max}^2+4l_\mathrm{max}+3\right)/3$ $=O\left(l_\mathrm{max}^3\right)$ nonzero elements.}

Real-SH coefficients satisfy slightly different constraints, but their constraints also have the same degree of freedom as complex ones.

\begin{proposition}{Real-SH coefficients of azimuthally symmetric operators}{bkgnd_isoBRDF_real}
	Suppose that $K\colon \calL\left(\calF\left(\Sspv,\C\right), \calF\left(\Sspv,\C\right)\right)$ is an azimuthally symmetric operator and $\rmK_{l_om_ol_im_i}^R\coloneqq \lrangle{Y_{l_om_o}^R, K\left[Y_{l_im_i}^R\right]}$ denotes the $\left(l_o,m_o\right)-\left(l_i,m_i\right)$-th coefficient of $K$ with respect to real SH. Then, the coefficient satisfies the following constraints for $m\ne 0$.
	\begin{gather}\label{eq:linop_realSH_azim1}
		\rmK_{l_om_ol_im_i}^R =0 \quad\text{whenever}\quad \abs{m_o}\ne\abs{m_i}, \\
		\label{eq:linop_realSH_azim2}
		\rmK_{l_om,l_im}^R = \rmK_{l_o,-m,l_i,-m}^R , \quad\text{and}\quad \rmK_{l_om,l_i,-m}^R = -\rmK_{l_o,-m,l_im}^R.
	\end{gather}
\end{proposition}
\vspace{-10mm}
\begin{proof}
	Since $\rmK_{l_om,l_im'}^R$ is a linear combination of $\rmK_{l_o,\pm m,l_i,\pm m'}^C$ (four combinations of $\pm$ signs), where $\rmK_{mm'}^C$ is the $\left(l_o,m\right)-\left(l_i,m'\right)$-th coefficient of $K$ with respect to complex SH, we get Equation~\eqref{eq:linop_realSH_azim1} from Proposition~\ref{prop:bkgnd_isoBRDF}. Then we only have to check constraints on $\rmK_{l_o,\pm m, l_i, \pm m}$ (four combinations). Note that $\rmK_{l_o0,l_i0}^R=\rmK_{l_o0,l_i0}^C$, we should only care about cases of $m\ne 0$. Without loss of generality, suppose that $m >0$. Rewriting Equation~\eqref{eq:SH_op_coeff_C2R} in a matrix product with the constraint in Proposition~\ref{prop:bkgnd_isoBRDF}, we get:
	\begin{equation}\nonumber
		\begin{split}	
			\begin{bmatrix}
				\rmK_{l_o,+m,l_i,+m}^R & \rmK_{l_o,+m,l_i,-m}^R \\
				\rmK_{l_o,-m,l_i,+m}^R & \rmK_{l_o,+m,l_i,-m}^R
			\end{bmatrix} &= \begin{bmatrix}
				M_{+m,+m}^{C\to R} & M_{+m,-m}^{C\to R} \\
				M_{-m,+m}^{C\to R} & M_{-m,-m}^{C\to R}
			\end{bmatrix}^* \begin{bmatrix}
			\rmK_{l_o,+m,l_i,+m}^C & 0 \\ 0 & \rmK_{l_o, -m, l_i, -m}^C
			\end{bmatrix} \begin{bmatrix}
			M_{+m,+m}^{R \to C} & M_{+m,-m}^{R \to C} \\
			M_{-m,+m}^{R \to C} & M_{-m,-m}^{R \to C}
			\end{bmatrix}^* \\
			&= \frac12\begin{bmatrix}
				1 & \pownu{m} \\ i & -\pownu{m}i
			\end{bmatrix} \begin{bmatrix}
			\rmK_{l_o,+m,l_i,+m}^C & 0 \\ 0 & \rmK_{l_o, -m, l_i, -m}^C
			\end{bmatrix} \begin{bmatrix}
				1 & -i \\ \pownu{m} & \pownu{m}i
			\end{bmatrix} \\
			&= \frac12 \begin{bmatrix}
				\rmK_{l_o,+m,l_i,+m}^C + \rmK_{l_o,-m,l_i,-m}^C & -i\left(\rmK_{l_o,+m,l_i,+m}^C - \rmK_{l_o,-m,l_i,-m}^C\right) \\
				i\left(\rmK_{l_o,+m,l_i,+m}^C - \rmK_{l_o,-m,l_i,-m}^C\right) & \rmK_{l_o,+m,l_i,+m}^C + \rmK_{l_o,-m,l_i,-m}^C
			\end{bmatrix}.
		\end{split}
	\end{equation}
	The right-hand side implies Equation~\eqref{eq:linop_realSH_azim2}.
\end{proof}

\subsection{Spherical Convolution}
\label{sec:spherical-convolution}

\subsubsection{Senses to define convolution}
Before investigating spherical convolution, let's review about convolution on planar (Euclidean) domains, $\R^n$. First, the convolution of two functions $k$ and $f\in\calF\left(\R^n, \bbK\right)$ ($\bbK=\R$ or $\C$) is defined by $k*f\left(\bfx\right) = \int_{\R^n}{k\left(\bfx-\bfx'\right)f\left(\bfx'\right)\rmd \bfx'}$. While it is a binary operation of functions in $\calF\left(\R^n, \C\right)$ into the same function space $\calF\left(\R^n, \C\right)$ yet, this property no more holds for spherical domains. To extend the definition of convolution to spherical domains, consider a linear operator $K\in\calL\left(\calF\left(\R^n, \C\right), \calF\left(\R^n, \C\right)\right)$ defined by $K\left[f\right]=k*f$. Then we observe an important property that $K$ is a translation equivariant linear operator, i.e., it commutes an arbitrary translation. Conversely, if a translation equivariant linear operator $K$ is given first, then there exists some function $k\in\calF\left(\R^n,\C\right)$ such that $K\left[f\right]=k*f$ under the assumption of the existence of an operator kernel of $K$.

\subsubsection{Spherical convolution}
Spherical convolution is defined as a binary operation of an azimuthally symmetric spherical function \NEW{$k:\left[0,\pi\right]\to\R$} and a spherical function $f\in\calF\left(\Sspv,\R\right)$ that does not need to have any symmetry. \NEW{Note that azimuthal symmetry of spherical functions, not operators, is discussed in Section~\ref{sec:bkgnd_ZH}.}
\begin{definition}{Spherical convolution}{sph_conv}
	$k$ and $f\in\calF\left(\Sspv,\bbK\right)$ ($\bbK=\R$ or $\C$) are spherical functions. Suppose that $k$ has azimuthal symmetry. Then spherical convolution of $k$ and $f$ is defined as follows:
	\begin{equation}\label{eq:sph_conv}
		k*f\left(\homega\right) = \int_{\Sspv}{ k\left(\cos^{-1}\homega\cdot\homega'\right)f\left(\homega'\right) \rmd \homega' %
		}.
	\end{equation}
	
	In this operation, $k$ is called the \emph{convolution kernel}.
\end{definition}
Due to the azimuthal symmetry of $k$, Equation~\eqref{eq:sph_conv} can be rewritten in several forms using the following property:
\begin{equation}\label{eq:sph_conv2}
	k\left(\cos^{-1}\homega\cdot\homega'\right) = k\left(\vec R_{\homega\to \hat z_g} \homega'\right) = k\left(\vec R_{\homega' \to\hat z_g}\homega\right),
\end{equation}
\begin{proof}
	\NEW{
	Recall that the inner product is preserved under rotation, as written in Equation~\eqref{eq:prem_inner_rot}. Then we get
	\begin{equation}\nonumber
		\begin{split}
			k\left(\cos^{-1}\homega\cdot\homega'\right) \underarrowref{=}{\kern-2emEq.~\eqref{eq:prem_inner_rot}\kern-2em}&\, k\left(\cos^{-1}\left(\vec R_{\homega\to \hat z_g}\homega\right)\cdot\left(\vec R_{\homega\to \hat z_g}\homega'\right)\right) = k\left(\cos^{-1}\hat z_g\cdot\left(\vec R_{\homega\to \hat z_g}\homega'\right)\right) \\
			\underarrowref{=}{\kern-2emEq.~\eqref{eq:bkgnd_azimforms_s2z}\kern-2em}&\,k\left(\vec R_{\homega\to \hat z_g} \homega'\right).
		\end{split}
	\end{equation}
	Then the remaining term can also be obtained in the same way.}
\end{proof}
where in the first term, $k$ is written as a function of a single real value of zenith angle, and in the second and third terms, $\vec R_{\hat a\to \hat b}$ denotes any rotation in $\SOgroupv$ such that $\vec R_{\hat a\to \hat b}\hat a = \hat b$. Note that the second and third terms are well-defined independent of choices of such rotations due to the symmetry of $k$. Note that we can rewrite Equation~\eqref{eq:sph_conv} in two other forms as follows:
\begin{align} \label{eq:sph_conv_form1}
	k*f &= \int_{\Sspv}{f\left(\homega '\right) \vec R_{\hat z_g \to \homega',\calF}\left[k\right]\rmd \homega'}, \\ \label{eq:sph_conv_form2}
	k*f\left(\homega\right) &= \lrangle{\vec R_{\hat z_g \to\homega, \calF}\left[k^*\right], f }_{\calF}.
\end{align}
While Equation~\eqref{eq:sph_conv_form1} views the convolution as a linear combination of rotated kernel, Equation~\eqref{eq:sph_conv_form2} views a single point at the operation result as an inner product of the kernel $k$ and the operand function $f$. When approximating such integral operations on a discrete point set of the domain $\Sspv$, we can consider each function $k$ and $f$ as numeric vectors whose indices indicate each point on $\Sspv$, and the convolution operation can be considered as a matrix related to $k$. Then Equation~\eqref{eq:sph_conv_form1} can be considered as a linear combination of column vectors of the matrix of $k$, while Equation~\eqref{eq:sph_conv_form2} does as the inner product of a row vector the matrix of $k$ and the vector of $f$. We call these views \emph{column view of convolution} and \emph{row view of convolution}, respectively. While in scalar spherical convolution, the format of the kernel $k$ in the two views seem straightforwardly equivalent, excepting just complex conjugation, in polarized spherical convolution, which will be introduced in a later section, the kernel will be defined slightly differently depending on each view. In that section, we will focus on the column view in Equation~\eqref{eq:sph_conv_form1}, which is related to the view of convolution as a linear operator, which will be introduced now.

Rather than viewing the convolution as a binary operation on spherical functions, it can be considered as a special case of linear operation on spherical functions with fixing the kernel. The following key property of spherical convolution as a linear operator explains why spherical convolution is defined in the above way.

\begin{proposition}{Spherical convolution and rotation equivariance}{sph_conv_roteqv}
	Suppose that a linear operator $K_\calF\in \calL\left(\calF\left(\Sspv,\bbK\right), \calF\left(\Sspv,\bbK\right)\right)$ on spherical functions is rotation equivariant, i.e., $K_\calF\left[\vec R_\calF\left[f\right]\right]=\vec R_\calF\left[K_\calF\left[f\right]\right]$ for any $f\in\calF\left(\Sspv,\bbK\right)$, and has an operator kernel $K\in\calF\left(\Sspv\times \Sspv,\bbK\right)$.
	Then the linear operator $K_\calF$ is characterized by a spherical convolution with a function $k: \left[0,\pi\right] \to \bbK$, i.e., $T\left[f\right] = k*f$. Here, the kernel is obtained as:
	\begin{equation}
		k\left(\theta\right)\coloneqq K\left(\homega, \homega'\right) \text{ for any }\homega,\homega'\in\Sspv \text{ with } \homega\cdot\homega' = \cos\theta.
	\end{equation}
	Moreover,
	\begin{equation} \label{eq:sph_conv_roteqv_delta}
		k = K_\calF\left[\delta\left(\homega,\hat z_g\right)\right].
	\end{equation}
	
	Conversely, convolution $k*f$ is rotation equivariant for $f$.
\end{proposition}
\vspace{-10mm}
\begin{proof}
	For any function $f$,
	\begin{equation}
		K_\calF\left[f\right]\left(\homega_0\right) = \int_{\Sspv}{K\left(\homega_i, \homega_o\right)f\left(\homega_i\right)\rmd\homega_i} = \int_{\Sspv}{k\left(\theta\right)f\left(\homega_i\right)\rmd \homega_i},
	\end{equation}
	where $\cos \theta = \homega_i\cdot \homega_o$. Then it is equivalent to Definition~\ref{def:sph_conv}.
	
	For Equation~\eqref{eq:sph_conv_roteqv_delta},
	\begin{equation}
		K_\calF\left[\delta\left(\homega,\hat z_g\right)\right]\left(\homega_o\right) = \int_{\Sspv}{K\left(\homega_i, \homega_o\right)\delta\left(\homega_i, \hat z_g\right)\rmd\homega_i} = K\left(\hat z_g, \homega_o\right) = k\left(\theta_o\right),
	\end{equation}
	where $\homega_o=\homega_{\mathrm{sph}}\left(\theta_o, \phi_i\right)$.
\end{proof}
\vspace{-4mm}
Here, 
\NEW{it is worth noting not to confuse the operator kernel and the convolution kernel. The rotation equivariant linear operator $K_{\calF}$ is characterized by the \emph{operator kernel} $K$, and at the same time by the \emph{convolution kernel} $k$, where $K\left(\homega,\homega'\right)=k\left(\cos^{-1}\homega\cdot\homega'\right)$ holds.} When handling with rotation equivariant operators, some formulae require distinction of the two types of kernels.

\subsubsection{Convolution in spherical harmonics}
As Fourier transform (both continuous and discrete versions) reduces convolution into the simpler pointwise product in the Euclidean domain, spherical harmonics can reduce the integral formula of spherical convolution in Equation~\eqref{eq:sph_conv} into the following formula for coefficient vectors, which is almost an element-wise product.

\begin{proposition}{Spherical convolution theorem: convolution in SH coefficients}{sph_conv_theorem}
	Denote SH coefficients of an azimuthally symmetric spherical function $k\in\calF\left(\Sspv,\bbK\right)$ by $\rmk_{l0}$ and SH coefficients of a spherical function $f\in\calF\left(\Sspv,\bbK\right)$ by $\rmf_{lm}\coloneqq\lrangle{Y_{lm}, f}_{\calF}$. Then, the SH coefficient of the convolution $k*f$ can be evaluated as follows:
	\begin{equation}\label{eq:sph_conv_SH}
		\lrangle{Y_{lm},k*f}_{\calF} = \sqrt\frac{4\pi}{2l+1}\rmk_{l0}\rmf_{lm}.
	\end{equation}
\end{proposition}
\vspace{-10mm}
\begin{proof}
	We refer to \cite{driscoll1994computing}.
\end{proof}

\NEW{Considering convolution with a fixed kernel as a linear operator, the above fact can be rewritten in terms of a coefficient matrix.}
\begin{proposition}{Spherical convolution theorem: linear operator form}{sph_conv_theorem_linop}
	\NEW{A rotation equivariant linear operator $K_\calF\in\calL\left(\calF\left(\Sspv,\bbK\right), \calF\left(\Sspv,\bbK\right)\right)$ is characterized by the convolution kernel $k\in\calF\left(\Sspv,\bbK\right)$. Denote the SH coefficients of $k$ by $\rmk_{l0}\coloneqq \lrangle{Y_{l0}, k}_\calF$. Then the SH coefficients of $K_\calF$, denoted by $\rmK_{l_om_o,l_im_i}\coloneqq \lrangle{Y_{l_om_o}, K_\calF\left[Y_{l_im_i}\right]}_\calF$, are evaluated as follows.
	\begin{equation}
		\rmK_{l_om_o,l_im_i} = \delta_{l_ol_i}\delta_{m_om_i} \sqrt\frac{4\pi}{2l+1}\rmk_{l0}.
	\end{equation}
	}
\end{proposition}
\NEW{Imagine that the element-wise product of two vectors with a fixed left operand is equivalent to the matrix-vector product with a diagonal matrix.

In Section~\ref{sec:theory-conv} later, we will derive our polarized spherical convolution theorem using the new basis as a generalization of Proposition~\ref{prop:sph_conv_theorem_linop}.}

\subsection{Reflection operator in SH}
\label{sec:sh-reflection}
In the context of rendering, we sometimes need a reflection operator which flips $\homega\in\Sspv$ with respect to a given axis.

We call a transform $T: \Sspv\to \Sspv$ as the reflection operator along $\hat z$, if
\begin{equation}
	T\left(\frF\left[\omega_1, \omega_2, \omega_3\right]^T\right) = T\left(\frF\left[\omega_1, \omega_2, -\omega_3\right]^T\right),
\end{equation}
for any $\homega=\frF\left[\omega_1, \omega_2, \omega_3\right]^T\in\Sspv\subset \Rspv$ and $\frF\in\FRsp$ such that $\extColz{\frF}=\hat z$. Note that it is well-defined independent of choice of the frame $\frF$. Note that it is self inversion and it acts on $\calF\left(\Sspv, \C\right)$ as a linear operator as follows:
\begin{equation}
	T_\calF\left[f\right]\left(\homega\right) = f\left(T\left(\homega\right)\right) = f\left(T^{-1}\left(\homega\right)\right), \quad \forall f\in\calF\left(\Sspv,\C\right).
\end{equation}
Then, its SH coefficients can be obtained as follows:
\begin{equation}
	\lrangle{Y_{lm}, T_{\calF}\left[Y_{l'm'}\right]} = \int_{\bbS^2}{Y_{lm}^*\left(\theta,\phi\right)Y_{l'm'}\left(\pi-\theta,\phi\right)\rmd\omega} = \delta_{ll'}\delta_{mm'} \left(-1\right)^{l+m}.
\end{equation}

\clearpage

%% file: equation_table.tex
\newcommand{\ct}[1]{{\color{blue}{#1}}} %

\setlength\tabcolsep{4pt}
\begin{table*}[htpb]%
\rotatebox{90}{
\begin{minipage}{1.0\textheight}
	\caption{\label{tb:elem_set_desc} 
		Summary of our fundamental building blocks for frequency domain theory of polarized rendering, compared with conventional scalar rendering. All the quantities (3rd-7th rows) and operations (8th-10th rows) have their frequency domain formulations. $l$ is the maximum level of spherical harmonics basis. \NEW{Each cell contains references both for the main paper (denoted by ``M-'') and this supplemental document (denoted by ``S-'').}
		}
	\vspace{-3mm}
	\small
	\begin{tabular}{
			P{0.001\linewidth} M{0.01\linewidth}
			? M{0.14\linewidth} | M{0.157\linewidth} | M{0.045\linewidth}
			? M{0.188\linewidth} | M{0.16\linewidth} | M{0.06\linewidth}
		}
		\thickhline
		&
		& \multicolumn{3}{c?}{\textbf{Scalar} rendering} & \multicolumn{3}{c}{\textbf{Polarized} rendering} \\
		
		\cline{3-8}
		&
		&\multirow{2}{*}{Angular domain}
		&\multicolumn{1}{c}{Freq. domain} &
		& Angular~domain
		&\multicolumn{1}{c}{Freq. domain} & \\
		
		\cline{5-5} \cline{8-8}
		&
		&& \footnotesize{(M-Sec.\ref{sec:background_sh}, S-Sec.\ref{sec:bkgnd_SH})} & \#\,coeff. & \footnotesize{(M-Sec.\ref{sec:background_polar}, S-Sec.\ref{sec:background_polar})} & \footnotesize{(M-Sec.\ref{sec:our_theory}, S-Sec.\ref{sec:our_theory})} & \#\,coeff. \\
		
		\thickhline
		\multicolumn{2}{l?}{\textbf{Radiance}}
		& Scalar radiance	& --	& --
		& Stokes vector	& -	& --\\
		\hline
		\multicolumn{2}{l?}{\textbf{Environment map}}
		& spherical function \linebreak (scalar field)	& SH coeff. vector \footnotesize{(M-Eq.\,\eqref{eq:bkgnd_sh_coeff}, S-Eq.\,\eqref{eq:bkgnd_SH_coeff})}
		& \footnotesize{$O\left(l^2\right)$}
		& Stokes vector field \linebreak \footnotesize{(M-Sec.\ref{sec:stokes-on-sphere}, S-Sec.\ref{sec:scalar_vector_fields})}	& PSH coeff. vector \footnotesize{(M-Eq.\,\eqref{eq:theory_PSH_lincomb})}
		& \footnotesize{$O\left(4 l^2\right)$}\\
		\hline
		\multicolumn{2}{l?}{\textbf{BRDF}}
		& BRDF 			& SH coeff. matrix \footnotesize{(M-Eq.\,\eqref{eq:bkgnd_SHcoeff_operator}, S-Eq.\,\eqref{eq:bkgnd_SH_coeffmat})}
		& \footnotesize{$O\left(l^4\right)$}
		& Mueller pBRDF	& PSH coeff. matrix \footnotesize{(M-Eq.\,\eqref{eq:theory_psh_coeff_linop})}
		& \footnotesize{$O\left(4\times 4 l^4\right)$}\\
		
		\cline{2-8}
		\rule{0pt}{1em} %
		& \multicolumn{1}{|l?}{$+$\,Axial symmetry}
		& isotropic BRDF & \, $+$\,Sparsity \, \linebreak \footnotesize{(M-Eq.\,\eqref{eq:bkgnd_SHcoeff_isobrdf}, S-Prop.\ref{prop:bkgnd_isoBRDF})}
		& \footnotesize{$O\left(l^3\right)$}
		& isotropic pBRDF & \, $+$\,Sparsity \, \linebreak \footnotesize{(M-Eq.\,\eqref{eq:theory_iso2coeff})}
		& \footnotesize{$O\left(4\times 4 l^3\right)$} \\
		
		\cline{2-8}
		& \multicolumn{1}{|l?}{$+$\,Radial symmetry}
		& convolution kernel \linebreak \footnotesize{(S-Prop.\ref{prop:sph_conv_roteqv})} & \, $+$\,Sparsity \, \linebreak \footnotesize{(M-Eq.\,\eqref{eq:bkgnd_SH_sph_conv})}
		& \footnotesize{$O(l)$}
		& polarized convolution kernel \linebreak \footnotesize{(M-Eq.\,\eqref{eq:theory_convkernel_geo2num}, S-Prop.\,\ref{prop:theory_conv_kernel})}
		& $+$\,Sparsity \linebreak \footnotesize{(M-Eq.\,\eqref{eq:theory_conv_coeff}, {S-Eqs.\,(\ref{eq:convp0_convcoeff},\ref{eq:conv0p_convcoeff},\ref{eq:convpp_convcoeff},\ref{eq:convpp_convcoeffc})})}
		& \footnotesize{$O\left(4\times 4 l\right)$} \\
		
		\thickhline
		\multicolumn{2}{l?}{\textbf{Rotation}}
		& \footnotesize{(M-Eq.\,\eqref{eq:bkgnd_rot_sph_func}, S-Eq.\,\eqref{eq:rotation_on_scalar_field})}
		& Real Wigner-D function \linebreak \footnotesize{(M-Eq.\,\eqref{eq:bkgnd_real_wignerD}, S-Eq.\,\eqref{eq:real_wignerD})}
		& \footnotesize{$O\left(l^3\right)$}
		& \footnotesize{(M-Eq.\,\eqref{eq:theory_rot_STKfield}, S-Def.\ref{def:field-stk-rotation})}
		& Real \& complex Wigner-D function \footnotesize{(M-Eq.\,\eqref{eq:theory_PSH_rotmat}, S-Prop.\ref{prop:theory-rot})}
		& \footnotesize{$O\left(l^3\right)$} \\
		\hline
		\multicolumn{2}{l?}{\textbf{Light interaction}}
		& rendering equation 			& matrix-vector product \linebreak \footnotesize{(M-Eq.\,\eqref{eq:bkgnd_SHcoeff_matmul})}
		& \footnotesize{$O\left(l^4\right)$}
		& polarized rendering equation	& matrix-vector product \linebreak \footnotesize{(M-Eq.\,\eqref{eq:theory-PSH-matmul})}
		& \footnotesize{$O\left(4\times 4 l^4\right)$}\\

		\cline{2-8}
		& \multicolumn{1}{|l?}{$+$\,Radial symmetry}
		& convolution \linebreak \footnotesize{(M-Eq.\,\eqref{eq:bkgnd_sph_conv}, S-Def.\ref{def:sph_conv})}
		& SH convolution \footnotesize{(M-Eq.\,\eqref{eq:bkgnd_SH_sph_conv}, S-Prop.\ref{prop:sph_conv_theorem})}
		& \footnotesize{$O(l^2)$}
		& polarized convolution \footnotesize{(M-Eq.\,\eqref{eq:theory_conv_def})}
		& PSH convolution \linebreak\footnotesize{(M-Eq.\,\eqref{eq:theory_conv_to}, S-Eqs.\,(\ref{eq:convp0_operator},\ref{eq:conv0p_operator},\ref{eq:convpp_operator}))}
		& \footnotesize{$O\left(4\times 4 l^2\right)$} \\
		
		\hline
		
		\multicolumn{2}{l?}{\textbf{Visibility mask}}
		& Point-wise product & Triple product \footnotesize{(M-Eq.\,\eqref{eq:theory_tp_s2s}, S-Eq.\,\eqref{eq:theory_tp_s2s})} & --
		& Point-wise product & Triple product \linebreak \footnotesize{(M-Eq.\,\eqref{eq:theory_tp_v2v}, S-Eq.\,\eqref{eq:theory_tp_v2v})} & --  \\
		\thickhline
	\end{tabular}

\end{minipage}}
\end{table*}

%% file: background2-supp.tex
\section{Background: Polarization and Mueller Calculus}
\label{sec:background_polar}

Here, we introduce the theoretical background of polarization in Mueller calculus. \NEW{Section~\ref{sec:background_polar_intro} gives brief introduction for novice readers who are not familiar with Mueller calculus formulation. Section~\ref{sec:background_polar_formal} provides a reformulation of it in a more rigorous manner to construct a solid theory of our polarized SH in later sections. Section~\ref{sec:background_polar_formal} is aimed at dedicated readers who are familiar with rigorous mathematics.} While Mueller calculus and its formal definition using equivalence classes already exist, this section contains our novel usage of terminology which distinguishes \emph{Stokes vectors} and \emph{Stokes component vectors} and notations $\left[\cdot\right]_{\frF}$ and $\left[\cdot\right]^{\frF}$.

\subsection{\NEW{Introduction to Mueller Calculus}}
\label{sec:background_polar_intro}

To take polarization into account, several intensity-related quantities, including radiance and BSDF, should be reformulated. The polarized intensity of rays is usually described by Jones calculus, which includes phase information of electromagnetic waves, or Mueller calculus, which includes unpolarized intensity due to incoherent light. Following recent works in computer graphics~\cite{baek2018simultaneous, baek2020image, hwang2022sparse} we focus on Mueller calculus.
	
	Suppose that there is a polarized ray and a local frame $\frF=\left[\hat x,\hat y,\hat z\right]$, where $\hat z$ is equal to the propagation direction of the ray. Then the polarized intensity of the ray is characterized by the four Stokes parameters $\bfs=\left[s_0,s_1,s_2,s_3\right]^T$. Here, each component $s_0$ to $s_3$ indicates total intensity, linear polarization in horizontal/vertical direction, linear polarization of diagonal/anti-diagonal direction, and circular polarization, respectively. We refer interested readers to \citet{collett2005field} for a more physical foundation of polarization and Mueller calculus.
	
	While Stokes parameters have linearity so that Stokes parameters obtained under multiple incoherent light sources are equal to the addition of Stokes parameters obtained under each individual source, they have an important property that makes them %
	different from scalars and even vectors. %

\begin{figure}[htpb]%
	\centering
	\vspace{-2mm}
	\includegraphics[width=\columnwidth]{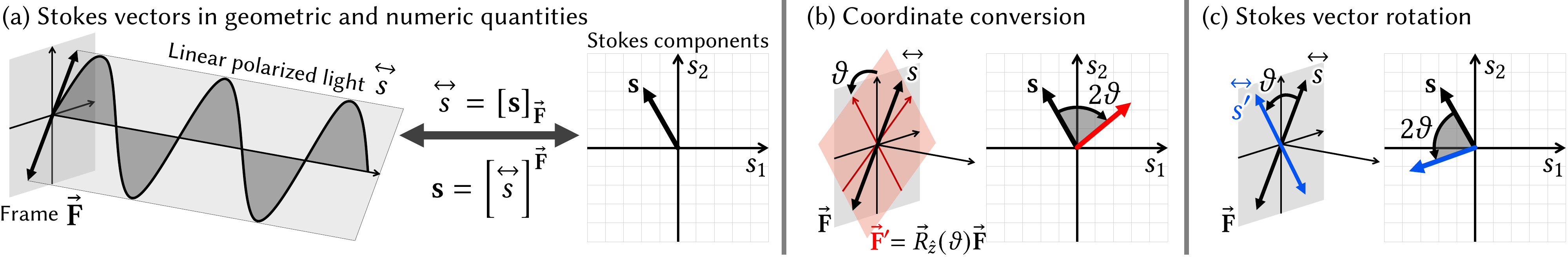}%
	\vspace{-2mm}
	\caption[]{\label{fig:bkgnd_stokes_rotation_wide}
		\NEW{(a) We distinguish a Stokes vector $\dvs$ as geometric quantities and its Stokes component vector $\bfs$ as numeric quantities. (b) Under coordinates conversion, the Stokes component vectors rotate twice while the Stokes vector $\dvs$ does not change. (c) We also define the rotation of the Stokes vector itself.}}
	\vspace{-4mm}		
\end{figure}

When taking another local frame $\frF' = \vec R_{\hat z}\left(\vartheta\right)\frF$, obtained by rotating $\frF$ by $\vartheta$ along its $z$ axis, the Stokes parameters with respect to the new frame $\frF'$ is evaluated as
\begin{equation} \label{eq:stk_coord_convert}
	\bfs' = \bfC_{\frF\to \frF'}\bfs =  \begin{bmatrix}
		1&0&0&0 \\
		0&\cos2\vartheta&\sin2\vartheta&0 \\
		0&-\sin2\vartheta&\cos2\vartheta&0 \\
		0&0&0&1
	\end{bmatrix} \bfs.
\end{equation}
We can observe here that $s_0$ and $s_3$ behave as \emph{scalar}s, which are measured independent of local frames.
On the other hand, $s_1$ and $s_2$ are neither scalars nor coordinates of an ordinary vector, which must have $\vartheta$ rather than $2\vartheta$ in Equation~\eqref{eq:stk_coord_convert}.
This twice rotation property of $s_1$ and $s_2$ under coordinate conversion will be dealt with as \emph{spin-2 functions} in %
Section~\ref{sec:theory-SWSH}.
Figure~\ref{fig:bkgnd_stokes_rotation_wide}(a) visualizes it where the two-sided arrow in the left indicates the actual oscillation direction of polarized ray and the right plot shows $s_1$ and $s_2$ values of it under \NEW{a local frame. Figure~\ref{fig:bkgnd_stokes_rotation_wide}(b) also visualizes coordinate conversion of a fixed ray.}

\subsection{Formal Definitions of Mueller Calculus}
\label{sec:background_polar_formal}

Following \citet{mojzik2016bi}, the \emph{Stokes space} and \emph{Stokes vectors} can be formally defined using equivalence classes. Here, we distinguish \emph{spin-2 Stokes vectors}, which consists of $s_1$ and $s_2$ linear polarization components and \emph{full Stokes vectors} (or simply \emph{Stokes vectors}) to build our polarized SH theory, which requires separating linear operators (Mueller matrix or transforms) into spin-0 and spin-2 parts.
\vspace{-1mm}
\begin{definition}{Spin-2 Stokes spaces}{bkgnd-stokes-space}
	For any $\homega\in\Sspv$, the \emph{Spin-2 Stokes space with respect to $\homega$}, denoted by $\STKsp{\homega}^2$ is defined as follows.
	\begin{equation}
		\STKsp{\homega}^2\coloneqq \left\{\left[\left(\bfs, \frF\right)\right]_{\sim}\mid \bfs\in\R^2, \frF\in\FRspo{\homega}\right\}
	\end{equation}
	Here, $\left[\cdot\right]_\sim$ denotes an an equivalence class with respect to a relation $\sim$ on a pair of a numeric vector in $\R^2$ and a frame in $\FRspo{\homega}$ defined as:
	\begin{equation}
		\left(\bfs, \frF\right)\sim \left(\bft, \frG\right) \text{ if and only if } \bft=\bfR_{2}\left(-2\vartheta\right)\bfs , \quad \forall\bfs, \bft\in\R^2, \frF, \frG\in\FRspo{\homega},
	\end{equation}
	where $\vartheta$ is uniquely determined to satisfy $\frG = \frF \bfR_z\left(\vartheta\right)$ up to $+2n\pi$.
\end{definition}
\vspace{-2mm}
Note that our main paper writes as $\frG = \vec R_{\hat z}\left(\vartheta\right)\frF$, where $\hat z = \frF$, to avoid introducing notations for numeric rotations. These are equivalent due to a relationship discussed in Section~\ref{sec:bkgnd_geo_num}. Now we introduce Stokes vectors, which are geometric quantities, and Stokes components, which are numeric ones, and notations to convert them to each other.
\vspace{-1mm}
\begin{definition}{Spin-2 Stokes vectors and spin-2 Stokes component vectors}{bkgnd-s2-stk}
	Using notations Definition~\ref{def:bkgnd-stokes-space}, we denote $\left[\bfs\right]_{\frF}\coloneqq\left[\left(\bfs, \frF\right)\right]_\sim\in \STKsp{\homega}^2$, which called a \emph{spin-2 Stokes vector of a ray along $\homega$}. $\bfs$ is called the \emph{spin-2 Stokes component vector of $\left[\bfs\right]_{\frF}$ with respect to $\frF$}. Conversely, for any $\dvs\in\STKsp{\homega}^2$, $\left[\dvs\right]^{\frF}$ is defined as some $\bfs'\in\R^2$ which satisfies $\dvs = \left[\bfs'\right]_{\frF}$. Note that it is well-defined, independent of the choice of a frame\footnote{Our $\left[\cdot\right]_{\frF}$ and $\left[\cdot\right]^{\frF}$ notations are slightly inspired from a convention in Riemannian geometry, where coordinates $v^i$ which depends on an observer can be converted to an invariant quantity $v^i \bfe_i$ by attaching the subscripted quantity $\bfe_i$, which indicates a basis for the local tangent space.}.
\end{definition}
\vspace{-2mm}
Now, full Stokes vectors can be defined similarly or just by taking the direct sum of scalars and spin-2 Stokes vectors.
\vspace{-1mm}
\begin{definition}{(Full) Stokes spaces}{}
	For any $\homega\in\Sspv$, the \emph{full Stokes space} (or \emph{Stokes space}, simply) \emph{with respect to} $\homega$, denoted by $\STKsp{\homega}^4$ (or $\STKsp{\homega}$) is defined by two ways, equivalently.
	\begin{enumerate}
		\item $\STKsp{\homega}^4 \coloneqq \R \oplus \STKsp{\homega}^2 \oplus \R$
		\item $\STKsp{\homega}^4\coloneqq \left\{\left[\left(\bfs, \frF\right)\right]_{\sim}\mid \bfs\in\R^4, \frF\in\FRspo{\homega}\right\}$, where $\left(\bfs, \frF\right)\sim \left(\bft, \frG\right)$ if and only if $\bft=\bfC_{\frF\to\frG}\bfs$.
	\end{enumerate}
	Here, $\bfC_{\frF\to\frG}$ is defined using $\vartheta$ such that $\frG=\frF\bfR_z\left(\vartheta\right)$ as follows.
	\begin{equation}\label{eq:polar-coord-conv}
		\bfC_{\frF\to\frG} \coloneqq \begin{bmatrix}
			1&0&0&0\\
			0 & \cos2\vartheta & \sin2\vartheta & 0 \\
			0 & -\sin2\vartheta & \cos2\vartheta & 0 \\
			0 & 0 & 0 & 1
		\end{bmatrix}.
	\end{equation}
\end{definition}
\vspace{-1mm}
Here, we sometimes denote the matrix in the right-hand side of Equation~\eqref{eq:polar-coord-conv} as $\bfR_{1:2}\left(-2\vartheta\right)$, which indicates embed $\bfR_2$ into a $4\times 4$ matrix (with index based on zero) at indices 1 and 2.

We also define the \emph{(entire) spin-2 Stokes space} as $\calS^2\coloneqq \sqcup_{\homega\in\Sspv}\STKsp{\homega}^2$ and the \emph{(entire) Stokes space} as $\calS^4\coloneqq \sqcup_{\homega\in\Sspv}\STKsp{\homega}^4$, where $\sqcup$ indicates disjoint union\footnote{For readers who are not familiar to disjoint union, it can be just considered as union.}.
(Full) Stokes vectors, (full) Stokes components, and $\left[\cdot\right]_{\frF}$ and $\left[\cdot\right]_{\frF}$ notations from Definition~\ref{def:bkgnd-s2-stk} can be redefined for full Stokes spaces similarly. Note that for $\dvs_2 \coloneqq \left[\left[s_1,s_2\right]^T\right]_{\frF}\in\STKsp{\homega}^2$ and $\dvs_4\coloneqq \left[\left[s_0, s_1,s_2, s_3\right]^T\right]_{\frF}\in\STKsp{\homega}^4$, we use notations for theirs relationship as $\dvs_4 = s_0 \oplus \dvs_2 \oplus s_3$ or $\dvs_4 = \left(s_0, \dvs_2, s_3\right)$.

We define operations on (spin-2) Stokes vectors, which are well-defined independent of the choice of a frame $\frF\in\FRspo{\homega}$ below.
\vspace{-1mm}
\begin{definition}{Stokes vector operations}{}
	For $\dvs$ and $\dvt\in\STKsp{\homega}^{\left\{2,4\right\}}$,
	\begin{enumerate}
		\item Linear combination: for any $a,b\in \R$, $a\dvs + a\dvt \coloneqq \left[a\left[\dvs\right]^{\frF} + a\left[\dvt\right]^{\frF}\right]_{\frF}$ for any $\frF\in\FRspo{\homega}$.
		\item Inner product: $\lrangle{\dvs, \dvt}_{\STKsp{\homega}^{\left\{2,4\right\}}} \coloneqq \left[\dvs\right]^{\frF} \cdot \left[\dvt\right]^{\frF}$ (or denoted as simply $\lrangle{\cdot, \cdot}_{\calS}$, or explicitly $\lrangle{\cdot, \cdot}_{\calS|\R}$, etc.)
		\item Rotation: for any $\vec R\in\SOgroupv$, $\vec R_{\calS}\in\calL\left(\calS,\calS\right)$ is defined as $\vec R_{\calS}\dvs = \left[\left[\dvs\right]^{\frF}\right]_{\vec R \frF}$.
	\end{enumerate}
	
	When $\dvs$ and $\dvt\in\STKsp{\homega}^2$, the following is additionally defined.
	\begin{enumerate}
		\item Complex scalar multiplication: for any $z\in\C$, $z\dvs \coloneqq\left[\R^2\left(z\C\left(\left[\dvs\right]^{\frF}\right)\right)\right]_{\frF}$
		\item Inner product over scalar $\C$: $\lrangle{\dvs, \dvt}_{\STKsp{\homega}^2|\C} \coloneqq \C\left(\left[\dvs\right]^{\frF}\right)^* \C\left(\left[\dvt\right]^{\frF}\right)\in\C$ (or denoted simply $\lrangle{\cdot,\cdot}_{\calS|\C}$).
	\end{enumerate}
\end{definition}
\vspace{-1mm}
Here, $\C: \R^2\to \C$ and $\R^2:\C\to \R^2$ denote the canonical conversions between $\R^2$ and $\C$. In addition, note that we sometimes denotes $\left[z\right]_{\frF}\coloneqq\left[\left[\Re z, \Im z\right]^T\right]_{\frF}\in\STKsp{\extColz{\frF}}^2$  for a complex number $z\in \C$.

We observe that $\STKsp{\homega}^4$ is an inner product space over $\R$, while $\STKsp{\homega}^2$ can be handled as an inner product space over both $\R$ or $\C$. Two inner products satisfy the relationship described in Proposition~\ref{prop:linalg-linsp-rc}.

\begin{figure}[tbp]
	\centering
	\includegraphics[width=0.7\columnwidth]{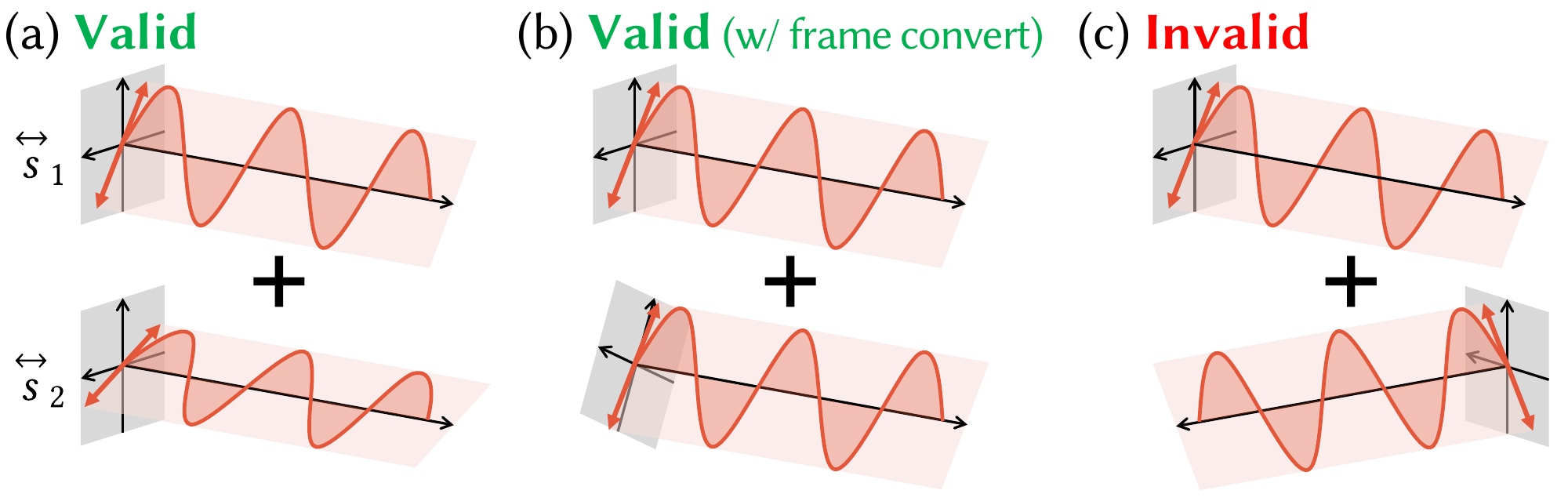}%
	\vspace{-3mm}
	\caption[]{\label{fig:bkgnd_stokes_sum}
		Addition between two Stokes vectors $\dvs_1$ and $\dvs_2$.
		(a) If two Stokes component vectors have the same frame, we can perform addition directly.
		(b) If two Stokes component vectors have different frames but on the same Stokes space, addition can be performed with frame conversion. 
		(c) If two Stokes vectors belong to different Stokes spaces (different ray directions), addition cannot be defined.
	}
	\vspace{-2mm}		
\end{figure}

Not only just a vector space, linear operators (transforms) also have to be formulated in Mueller calculus.
\vspace{-1mm}
\begin{definition}{Mueller transform space}{polar-mueller-space}
	The \emph{(full) Mueller space with respect to $\homega_i$ and $\homega_o\in\Sspv$}, denoted by $\MUEsp{\homega_i}{\homega_o}^4$ (or $\MUEsp{\homega_i}{\homega_o}$ simply), and \emph{spin 2-to-2 Mueller space with respect to $\homega_i$ and $\homega_o$}, denoted by $\MUEsp{\homega_i}{\homega_o}^2$ are defined as follows, equivalently.
	\begin{enumerate}
		\item $\MUEsp{\homega_i}{\homega_o}^{\left\{2,4\right\}}\coloneqq \calL\left(\STKsp{\homega_i}^{\left\{2,4\right\}}, \STKsp{\homega}^{\left\{2,4\right\}}\right)$, respectively.
		\item $\MUEsp{\homega_i}{\homega_o}^{\left\{2,4\right\}} \coloneqq \left\{\left[\left(\bfM, \frF_i, \frF_o\right)\right]_{\sim} \mid \bfM\in \R^{\left\{2\times 2,4\times 4\right\}}, \frF_i\in\FRspo{\homega}, \frF_o\in\FRspo{\homega_o}\right\}$, where $\left(\bfM, \frF_i, \frF_o\right)\sim \left(\bfN, \frG_i, \frG_o\right)$ if and only if
		\begin{equation}
			\begin{split}
				\bfN &= \bfR_2\left(-2\vartheta_o\right) \bfM \bfR_2\left(2\vartheta_i\right), \text{ for }\MUEsp{\homega_i}{\homega_o}^2, \\
				\bfN &= \bfC_{\frF_o\to\frG_o} \bfM \bfC_{\frF_i\to\frF_i}^{-1}, \text{ for }\MUEsp{\homega_i}{\homega_o}^4,
			\end{split}
		\end{equation} 
	\end{enumerate}
	where $\frG_i=\bfF_i\bfR_z\left(\vartheta_i\right)$, $\frG_o=\bfF_o\bfR_z\left(\vartheta_o\right)$, and $\bfC$ from Equation~\eqref{eq:polar-coord-conv}.
\end{definition}
\NEW{Similar to Stokes spaces, we can define the \emph{(entire) Mueller space} as $\calM^{\left\{2,4\right\}}\coloneqq \sqcup_{\homega_i,\homega_o\in\Sspv}\MUEsp{\homega_i}{\homega_o}^{\left\{2,4\right\}}$ in both senses of spin-2 and full.} As a full Stokes vector contains a spin-2 Stokes vector as its subpart, a full Mueller transform $\dvM\in\calM^4$ contains a spin 2-to-2 Mueller transform as its subpart, which is denoted by $\dvM\left[1{:}2,1{:}2\right]\in\calM^2$. Note that separately taking a single index 1 or 2 for $\dvM$ is illegal since it yields a frame-dependent quantity. We also define \emph{Mueller matrices}, numeric quantities measured from Mueller transforms.

\vspace{-1mm}
\begin{definition}{Mueller transforms and Mueller matrices}{}
	Using notations Definition~\ref{def:polar-mueller-space}, we denote $\left[\bfM\right]_{\frF_i\to\frF_o}\coloneqq\left[\left(\bfM, \frF_i, \frF_o\right)\right]_\sim\in \MUEsp{\homega_i}{\homega_o}^{\left\{2,4\right\}}$, which called a \emph{Mueller transform from a ray along $\homega_i$ to one along $\homega_o$}. $\bfM$ is called the \emph{Mueller matrix of $\left[\bfM\right]_{\frF_i\to\frF_o}$ with respect to $\frF_i$ and $\frF_o$}. Conversely, for any $\dvM\in\MUEsp{\homega_i}{\homega_o}^{\left\{2,4\right\}}$, $\left[\dvM\right]^{\frF_i\to\frF_o}$ is defined as some $\bfM'\in\R^{\left\{2\times 2,4\times 4\right\}}$ which satisfies $\dvM = \left[\bfM'\right]_{\frF_i\to\frF_o}$. Note that it is well-defined, independent of the choice of frames.
\end{definition}
\vspace{-2mm}

Since a Mueller space is a space of linear maps, linear combination and product between two Mueller transforms in the same space is naturally defined. For a rotation $\vec R\in\SOgroupv$, $\vec R_{\calM}:\calL\left(\calM, \calM\right)$ is defined as:
\begin{equation}
	\vec R_{\calM}\left[\dvM\right] = \vec R_{\calS} \dvM \vec R_{\calS}^{-1},
\end{equation}
where the right-hand side consists of the product of Mueller transforms by considering $\vec R_\calS$ as a Mueller transform. Also note that the coordinate conversion matrix for Stokes vectors can be rewritten as:
\begin{equation}
	\bfC_{\frF\to\frG} = \left[\dvI\right]^{\frF\to\frG},
\end{equation}
where $\dvI$ indicates the identity Mueller transform.

\clearpage

%% file: stokesfield-supp.tex
\section{Analysis on Stokes Vector Fields}
\label{sec:stokes-vector-fields}

\NEW{Sections~\ref{sec:scalar_vector_fields} to~\ref{sec:naive-SH} will provide descriptions for analysis on Stokes vector fields to help understand why naively applying conventional scalar SH to rendering with polarized lights fails. It will support the fact that scalar SH suffers from a singularity problem for Stokes vector fields, and the singularity problem violates rotation invariance.	

In addition, Sections~\ref{sec:stokes-vector-fields-operations} and~\ref{sec:naive-SH} provide some formal techniques that will be used for the proofs in Section~\ref{sec:our_theory}.}

\subsection{Preliminaries: Continuity of Scalar and Tangent Vector Fields}
\label{sec:scalar_vector_fields}

For better intuition, we first introduce scalar and tangent vector fields, which are simpler types than Stokes vector fields. Observing the difference between Stokes vector fields and the simpler types of fields may help understand the challenges of Stokes vector fields. 

\mparagraph{Scalar fields}
Continuity of a (scalar-valued) spherical function, or scalar field, $f\colon\Sspv\to\R$ or $\C$ is well defined when considering $\Sspv$ as a smooth surface embedded in $\Rspv$. However, it is often more convenient to test the continuity of the spherical function written in spherical coordinates, $f\left(\theta,\phi\right)$. The $f:\Sspv\to\C$ is continuous if and only if its spherical coordinates parameterization $f\left(\theta,\phi\right)$\footnote{For rigorous mathematics we need another symbol rather than $f$, which is defined on the sphere, but we use the symbol for better intuition.} is continuous on $\left[0,\pi\right]\times\left[0,2\pi\right]$ and the following conditions hold.
\begin{equation}
	f\left(0,\phi_1\right)=f\left(0,\phi_2\right), \quad f\left(\pi,\phi_1\right)=f\left(\pi,\phi_2\right), \quad f\left(\theta,0\right)=f\left(\theta,2\pi\right), \quad \forall \phi_1,\phi_2\in\left[0,2\pi\right] \text{ and } \forall\theta\in\left[0,\pi\right] .
\end{equation}
Analogously, the continuity of spherical stokes-valued functions can be tested in the $\left[0,\pi\right]\times \left[0,2\pi\right]$ parameterization domain in the later section, but it has different constraints from the above.

\mparagraph{Tangent vector fields}
Before dealing with Stokes-value spherical functions such as Stokes vectors as a function of propagation directions, we will first explain tangent vector fields on the sphere to show the analogy and difference between them.

A tangent vector field on the sphere $\vec f\colon\Sspv\to \cup_{\homega\in\Sspv}T_{\homega}\Sspv$ is a function defined on the sphere $\Sspv$ of which each value at $\homega\in\Sspv$ takes value from $\vec f\left(\homega\right)\in T_{\homega}\Sspv$, where $T_{\homega}\Sspv$ denotes the tangent plane of $\Sspv$ at $\homega$ defined by $T_{\homega} \Sspv \coloneqq \left\{v\in\Rspv \mid \homega\cdot v = 0\right\}$.

As examples to help intuition of tangent vector fields, one can imagine a tangent vector field on the sphere as a wind velocity map on the earth or the gradient vector field of an omnidirectional image obtained by a fish-eye lens.

\mparagraph{Representation under a coordinates system}
Since a tangent vector field on the sphere takes a value from a different tangent plane at each point $\homega$, representing the tangent vector field is more complicated than scalar fields. One common way is to use frame fields. A \textit{frame field} on $\Sspv$, $\frF\left(\homega\right)$, is defined as a function maps (almost everywhere) each point $\homega\in\Sspv$ to a frame $\frF\left(\homega\right)\in\FRspo{\homega}$, which has $\homega$ as the third axis, i.e., $\frF\left(\homega\right)\left[:,3\right]=\homega$. Note that frame fields are usually required to be continuous except at a zero-measure singularity (usually two points). Then a tangent vector field $\vec f\colon\bbS^2\to \cup_{\homega\in\bbS^2}T_{\homega}\bbS^2$ can be represented as:
$$
\vec f\left(\homega\right) = a\left(\homega\right) \frF\left(\homega\right)\left[:,1\right] + b\left(\homega\right) \frF\left(\homega\right)\left[:,2\right],
$$
for some scalar-valued spherical functions $a$ and $b$. A usual way to select the $\theta\phi$ frame field is introduced in Equation~\eqref{eq:bkgnd_tp_moving_frame}, which is aligned to the spherical coordinates. Recall the formulae in more detail; it can be written as follows:
\begin{equation}\label{eq:field-tp-frame}
	\begin{split}
		\frF_{\theta\phi}\left(\theta,\phi\right) &\coloneqq \left[\hat\theta,\hat\phi,\homega\right] \text{,}\\
 \text{where ~ ~}	\hat \theta &\coloneqq \mathrm{normalize}\left(\pfrac{\homega_{\mathrm{sph}}}{\theta}\right) = \frF_g\left[\cos\theta\cos\phi, \cos\theta\sin\phi, -\sin\theta\right]^T , \\
	\hat \phi &\coloneqq \mathrm{normalize}\left(\pfrac{\homega_{\mathrm{sph}}}{\phi}\right) = \frF_g \left[-\sin\phi, \cos\phi, 0\right]^T .
	\end{split}
\end{equation}

Here $\frF_g$ indicates the global (world) frame, and $\homega_{\mathrm{sph}}$ indicates the spherical coordinate system specified by the global frame $\frF_g$ \NEW{as defined by Equation~\eqref{eq:sph_coord}}.

\mparagraph{Continuity of tangent vector fields}
\NEW{Unlike} scalar fields, coordinate systems and frame fields raise discontinuity, which does not contain the original structure of the sphere $\bbS^2$, only testing the continuity of $a\left(\homega\right)$ and $b\left(\homega\right)$ is not enough to test the continuity of the vector field $\vec f$. 
\NEW{The continuity of $\vec f$ is rewritten in terms of $a$ and $b$ as follows:}
\begin{gather}\nonumber
	a\text{ and }b\text{ are continuous on }\Sspv-S_{\frF}, \\ \nonumber
	\forall \homega_s \in S_{\frF}, \quad 
	\lim_{\homega\to\homega_s}a\left(\homega\right) \frF\left(\homega\right)\left[:,1\right] + b\left(\homega\right) \frF\left(\homega\right)\left[:,2\right] \text{ converges}.
\end{gather}
\NEW{where $S_{\frF}\subset \Sspv$ denotes the set of singularities of the frame $\frF$.} Note that every frame field has singularities due to the Hairy ball theorem. For symbolic or numerical evaluation, the above must be reformulated into a coordinate system, usually a spherical one. We observe that the simplest case to describe this constraint occurs when singularities of the frame field are a subset of discontinuity of the coordinate system, for instance, $\theta=0$ or $\pi$ and $\phi=0$ or $2\pi$ for the spherical coordinates. In this context, we investigate continuity conditions of several types of spherical functions in terms of spherical coordinates and $\theta\phi$ frame field.

While $\frF_{\theta\phi}\left(\homega\right)$ at $\homega=\pm\hat z_g$ is considered to be \emph{not defined}, it is more useful to consider that $\frF_{\theta\phi}\left(0 \text{ or }\pi,\phi\right)$ is defined depending on $\phi$  by directly substituting $\theta$ to Equation~\eqref{eq:field-tp-frame} as follows:

\begin{equation}
	\frF_{\theta\phi}\left(0,\phi\right) = \frF_g\begin{bmatrix}
		\cos\phi & -\sin\phi & 0 \\
		\sin\phi & \cos\phi & 0 \\
		0 & 0 & 1
	\end{bmatrix} = \frF_g \bfR_{zy}\left(\phi,0\right), \quad \frF_{\theta\phi}\left(\pi,\phi\right) = \frF_g\begin{bmatrix}
	-\cos\phi & -\sin\phi & 0 \\
	-\sin\phi & \cos\phi & 0 \\
	0 & 0 & -1
	\end{bmatrix} = \frF_g \bfR_{zy}\left(\phi,\pi\right).
\end{equation}

Denoting $\bff \coloneqq \left[a, b\right]^T$ under $\frF_{\theta\phi}$, the tangent vector field $\vec f$ is continuous if and only if $\bff\left(\theta,\phi\right)$ is continuous on $\left[0,\pi\right]\times\left[0,2\pi\right]$ and:

\begin{multline}
	\bff\left(0,\phi_2\right) = \bfR_2\left(-\phi_2 + \phi_1\right)\bff\left(0,\phi_1\right),\quad
	\bff\left(\pi,\phi_2\right) = \bfR_2\left(+\phi_2 - \phi_1\right)\bff\left(\pi,\phi_1\right),\quad
	\bff\left(\theta,0\right)  = \bff\left(\theta, 2\pi\right), \\
	\text{for any } \phi_1,\phi_2\in\left[0,2\pi\right] \text{ and } \theta\in\left[0,\pi\right],
\end{multline}
which has different conditions from scalar fields.

\subsection{Continuity of Stokes Vector Fields}
\label{sec:continuity-stokes-vector-field}
\mparagraph{Stokes vector fields on the sphere}
Now, we can consider applying the advantages of spherical harmonics on spherical functions to polarized intensity. Then, we should first look into the spherical functions of Stokes vectors (or Stokes vector fields on the sphere). \NEW{Different from the case of scalar radiance, but similar to tangent vector fields}, a Stokes vector field $\dvf\in \calF\left(\Sspv,\STKsp{\homega}\right)$
\footnote{Rigorously, it should be written as $\left\{ \dvf\colon \Sspv\to \cup_{\homega\in\Sspv}\STKsp{\homega} \mid \forall\homega\in\Sspv, \dvf\left(\homega\right)\in \STKsp{\homega} \right\}$. But we write as the main text for the sake of simplicity and better intuition.}
has also the challenge that it evaluates given directions $\homega\in\Sspv$ into values from different Stokes spaces $\dvf\left(\homega\right)\in\STKsp{\homega}$.

\mparagraph{Representation under a coordinates system}
Similar to tangent vector fields, we can represent a Stokes field into four components of scalar fields, but this cannot be done directly by applying local frames as linear operators on vectors. We must use the Stokes component conversion defined in Definition~\ref{def:bkgnd-s2-stk}. Then, we can rewrite the Stokes vector field $\dvf$ as follows:

\begin{equation}
	\left[\dvf\left(\homega\right)\right]^{\frF_{\theta\phi}\left(\homega\right)} = \begin{bmatrix}
		f_0\left(\theta,\phi\right) & f_1\left(\theta,\phi\right) & f_2 \left(\theta,\phi\right) & f_3 \left(\theta,\phi\right)
	\end{bmatrix}^T.
\end{equation}
The continuity of $\dvf$ can be represented in terms of each component $f_0$, $\cdots$, $f_3$, and it yields different constraints at the singularities $\pm\hat z_g$ from both scalar and tangent vector fields. By denoting $\bff\coloneqq \left[f_1,f_2\right]^T$,
	
\begin{multline}\label{eq:stk-field-conti}
	\bff\left(0,\phi_2\right) = \bfR_2\left(-2\left(\phi_2 - \phi_1\right)\right)\bff\left(0,\phi_1\right),\quad
	\bff\left(\pi,\phi_2\right) = \bfR_2\left(2\left(\phi_2 - \phi_1\right)\right)\bff\left(\pi,\phi_1\right),\quad
	\bff\left(\theta,0\right)  = \bff\left(\theta, 2\pi\right), \\
	\text{for any } \phi_1,\phi_2\in\left[0,2\pi\right] \text{ and } \theta\in\left[0,\pi\right],
\end{multline}
while $f_0$ and $f_3$ components are conventional scalar fields. Note that the first two constraints of $f_1$ and $f_2$ appear twice the components' rotation. From such different conditions, representing a Stokes vector field using a continuous scalar or tangent vector field yields a discontinuous Stokes vector field, which implies each type of field should have different types of continuous basis functions.

\begin{figure}[tbp]
	\centering
	\includegraphics[width=\columnwidth]{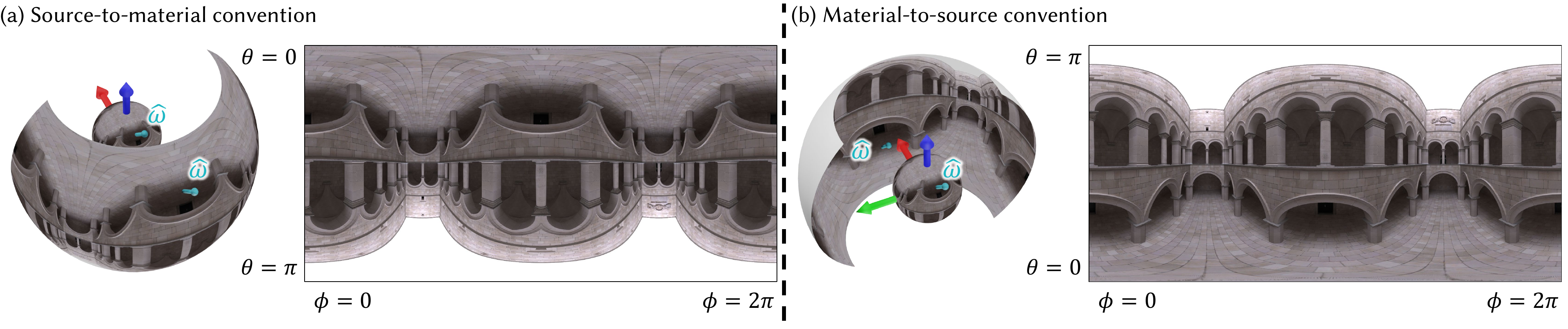}%
	\vspace{-3mm}
	\caption[]{\label{fig:envmap-convention}
		We can unwrap an image on a sphere into an equirectangular image in two ways depending on a spherical point in the domain: (a) a ray propagation direction or (b) a light vector that points to a light source from a material (or observer). In our main paper, Main. Figure~\ref{fig:bkgnd_continuity_stokes_field} follows the convention in (a) since it describes the general properties of spherical functions. The others Main Figures.~\ref{fig:bkgnd_penvmap_visualize},~\ref{fig:bkgnd_stokes_rotation2},~\ref{fig:theory_penvmap_rotation},~\ref{fig:theory_conv_validation}, and~\ref{fig:theory_rendering_pipeline} follows the convention in (b) for better intuition since they describe environment map images. Note that $\theta$ and $\phi$ in our equations always indicate spherical coordinates of ray propagation directions so that the top row of the equirectangular image in (b) indicates $\theta=\pi$ while the one in (a) indicates $\theta = 0$.	}
	\vspace{-2mm}		
\end{figure}

\subsection{\NEW{Stokes Vector Fields Operations}}
\label{sec:stokes-vector-fields-operations}
\NEW{To discuss bases for Stokes vector fields, we should define several operations on Stokes vector fields. It can be done by generalizing scalar field operations in Section~\ref{sec:bkgnd_SH}, based on Stokes vectors operations in Section~\ref{sec:background_polar}.
We define the inner product and rotations of Stokes vector fields as follows.

\begin{definition}{Inner product of Stokes vector fields}{field-stk-inner}
	For Stokes vector fields $\dvf$, $\dvg:\Sspv\to \STKsp{\homega}$, the inner product of them is defined as follows.
	\begin{equation}
		\lrangle{\dvf, \dvg}_{\calF\left(\Sspv, \STKsp{\homega}\right)} \coloneqq \int_{\Sspv}{\lrangle{\dvf\left(\homega\right), \dvg\left(\homega\right)}_\calS \rmd\homega}.
	\end{equation}
\end{definition}
}
\vspace{-3mm}
\begin{definition}{Rotation of Stokes vector fields}{field-stk-rotation}
	For $\vec R\in\SOgroupv$, it can acts as $\vec R_{\calF}\in\calL\left(\calF\left(\Sspv,\STKsp{\homega}\right), \calF\left(\Sspv,\STKsp{\homega}\right)\right)$, a linear operator on Stokes vector fields as follows.
	\begin{equation} \label{eq:field-stk-rotation}
		\vec R_{\calF}\left[\dvf\right]\left(\homega\right) = \vec R_{\calS}\left[\dvf\left(\vec R^{-1}\homega\right)\right], \quad \forall \dvf:\Sspv\to \STKsp{\homega}.
	\end{equation}
\end{definition}
\vspace{-2mm}
\NEW{Note that the inner product in Definition~\ref{def:field-stk-inner} is often written as simply $\lrangle{\dvf, \dvg}_{\calF}$. The rotation defined in Definition~\ref{def:field-stk-rotation} is illustrated in Figure~\ref{fig:bkgnd_sphfunc_rotation}(b).}
\subsection{\NEWJ{Scalar SH to Stokes Vector Fields}}
\label{sec:naive-SH}

Now, we will show two problems when using scalar SH to Stokes vectors: singularity and violation of rotation invariance.
\subsubsection{Singularity}
We first focus on the continuity condition for Stokes vector fields.
Concretely, we can try to na{\"i}vely apply the scalar SH on each component $f_0...f_3$ of the Stokes vector field with respect to the $\theta\phi$-frame field $\frF_{\theta\phi}\left(\homega\right)$ as
\begin{equation}\label{eq:naive_SH}
	\dvY_{lm0}^{\left(\mathrm{naive}\right)} \coloneqq\begin{bmatrix}
		Y_{lm}\left(\homega\right) \\
		0 \\
		0 \\
		0
	\end{bmatrix}_{\frF_{\theta\phi}\left(\homega\right)}\!\!\!\!\!,\cdots,\  \dvY_{lm3}^{\left(\mathrm{naive}\right)} \coloneqq\begin{bmatrix}
		0 \\
		0 \\
		0 \\
		Y_{lm}\left(\homega\right)
	\end{bmatrix}_{\frF_{\theta\phi}\left(\homega\right)},
\end{equation}
which is considered as a basis, where $0\le \abs{m}\le l$.
However, scalar SH satisfy
\begin{equation}\label{eq:SH_at_sing}
	\begin{split}
		Y_{l0}\left(0,\phi\right) &= \text{const.} \ne 0, \\
		Y_{l0}\left(\pi,\phi\right) &= \text{const.} \ne 0,
	\end{split}
\end{equation}
and those conditions never satisfy the continuity condition of spin-2 functions in Equation~\eqref{eq:stk-field-conti}.
Thus, even if a continuous Stokes vector field $\dvf$ is given, its finite projection on the basis in Equation~\eqref{eq:naive_SH} up to $l\le l_\mathrm{max}$ is always discontinuous at $\pm \hat z_g$. 
This is a fundamentally different feature from how the scalar SH behaved on scalar fields, which always converts finite coefficients to continuous functions and has a smoothing role.

\subsubsection{Rotation invariance violation}
The singularity issue of basis functions is not only the presence of singularity itself but also the effects of the continuity of the basis function, which is a necessary condition for rotation invariance.

Note that $\dvY_{l'01}^{\left(\mathrm{naive}\right)}\!\!\!\brahomega$ is discontinuous at $\homega=\pm\hat z_g$.
So when rotating it by $\vec R=\vec R_{y_g}\left(\frac\pi 2\right)$, then the rotated basis $\vec R_{\calF}\left[\dvY_{l'01}^{\left(\mathrm{naive}\right)}\right]$ is discontinuous at $\homega=\pm \hat x_g$.
Thus, when decomposing it into a linear combination of the original basis $\dvY_{l'mp}^{\left(\mathrm{naive}\right)}$, which is always continuous at $\homega=\pm \hat x_g$, the linear combination must be an infinite sum to make such discontinuity since the finite sum of continuous functions is always continuous. Generally, it can be written as a coefficient matrix of the rotation as
\begin{equation}\label{eq:naive_SH_rotmat}
	\lrangle{\dvY_{lmp}^{\left(\mathrm{naive}\right)}, \vec R_\calF \left[\dvY_{l'm'p'}^{\left(\mathrm{naive}\right)}\right]}_\calF \ne 0, \text{ for } l\ne l',
\end{equation}
where an inner product of two Stokes vector fields is %
\NEW{defined in Definition~\ref{def:field-stk-inner}}

Recall that the rotation invariance of SH for scalar fields is represented as a block diagonal coefficient matrix in Equation~\eqref{eq:rotinv_sh}.
However, Equation~\eqref{eq:naive_SH_rotmat} implies that the elements of the coefficient matrix at $l\ne l'$ are nonzero.
This means SH for the Stokes vector field does not yield a block diagonal and violates the rotation invariance.
For further validation related to rotation invariance, refer to Section~\ref{sec:theory_rotinv}\NEW{ in the main paper}.

\subsection{Rotation Form of Stokes Vector Fields} \label{sec:stokes-vector-field-rot}
Rather than unwrapping Stokes vector fields into spherical coordinates, the following formulation is sometimes useful in deriving our theory.
\vspace{-0.5mm}
\begin{definition}{Rotation form of Stokes vector fields}{field-stk-rotform}
	Given a global frame $\frF_g$, for a spin-2 Stokes vector field $\dvf:\Sspv\to \STKsp{\homega}$, its rotation form $f:\SOgroupv\to\C$ is defined as follows.
	\begin{equation}
		f\left(\vec R\right) = \left[\dvf\left(\vec R\hat z_g\right)\right]^{\vec R\frF_g}.
	\end{equation}
\end{definition}
\vspace{-0.5mm}
Note that full Stokes vector fields can be similarly redefined as a function with codomain $\R^4$ rather than $\C$. Note that the following property is converse.
\vspace{-0.5mm}
\begin{proposition}{Stokes vector fields from rotation forms}{field-stk-from-rotform}
	A function $f:\SOgroupv\to\C$ can be a rotation form of a spin-2 Stokes vector field if and only if
	\begin{equation}
		f\left(\vec R \vec R{\hat z_g}\left(\psi\right)\right) = e^{-2i\psi} f\left(\vec R\right).
	\end{equation}
\end{proposition}
\vspace{-0.5mm}
Note that it comes from the continuity condition of Stokes vector fields. For a function $f:\SOgroupv\to\R^4$, the condition to be equivalent to a full Stokes vector can be obtained by substituting $e^{2i\psi}$ by the $\bfC_{\frF\to\frG}$ matrix.

\subsection{Mueller Transform Fields}
Similar to Stokes vector fields, we can also define a Mueller transform field as a function $\dvM:\Sspv\times\Sspv\to \MUEsp{\homega_i}{\homega_o}$ which satisfies $\dvM\left(\homega_i,\homega_o\right)\in\MUEsp{\homega_i}{\homega_o}$.

We define the rotation of a Mueller transform field as follows.
\vspace{-0.5mm}
\begin{definition}{Rotation of Mueller transform fields}{}
	For $\vec R\in\SOgroupv$, it can acts as $\vec R_{\calF}\in\calL\left(\calF\left(\Sspv\times \Sspv,\calM\right), \calF\left(\Sspv\times \Sspv,\calM\right)\right)$, a linear operator on Mueller transform fields as follows.
	\begin{equation}
		\vec R_{\calF}\left[\dvM\right]\left(\homega_i,\homega_o\right) = \vec R_{\calM}\left[\dvM\left(\vec R^{-1}\homega_i, \vec R^{-1}\homega_o\right)\right], \quad \forall \dvf:\Sspv\to \STKsp{\homega}.
	\end{equation}
\end{definition}
\vspace{-0.5mm}
Note that it can be understood as a pBRDF obtained by rotating the material in a rendering context.

Mueller transform fields are more discussed in later Section~\ref{sec:theory_mueller}.

\clearpage

%% file: theory-supp.tex
\section{Polarized Spherical Harmonics for Stokes Vector Field}
\label{sec:our_theory}

\subsection{Spin-weighted Spherical Harmonics}
\label{sec:theory-SWSH}

Note that our definition of spin-weighted functions and SWSH may take a slightly different formulation than other literature, but still equivalent. We chose our formulation for convenience to derive our PSH theory.
\begin{definition}{Spin-weight $s$ functions}{theory-spin-func}
	Given a global frame $\frF_g$, $f:\SOgroupv\left(3\right)\to \C$ (or $f:\FRsp \to \C$) is called a \textit{spin-weight} $s$ \textit{function} (or \emph{spin-}$s$ \emph{function}, simply) if:
	\begin{equation}
		f\left(\vec R \vec R_{\hat z_g}\left(\psi\right)\right) = e^{-is\psi}f\left(\vec R\right)
		\text{ for any }\vec R\in \SOgroupv, \psi\in\R.
	\end{equation}
	Equivalently, it can also be defined as $f : \Sspv\to \bigcup_{\homega\in\Sspv}\left(\C\times \frF_{\homega}\right) / \sim$
	\begin{equation}\nonumber
		f\left(\homega\right) \in \left(\C\times \frF_{\homega}\right) / \sim, \text{ where } \left(z_1, \frF_1\right)\sim \left(z_2 , \frF_2\right) \text{ if and only if } \frF_2 = \frF_1\bfR_z\left(\psi\right) \text{ for some }\vartheta\text{ and }  z_2 = e^{-is\psi} z_1 .
	\end{equation}
\end{definition}
The condition also can be represented as:
\begin{equation}
	f\left(\cos\psi \hat x - \sin\psi\hat y, \sin\psi\hat x+\cos\psi\hat y, \hat z\right)
	= e^{is\psi}f\left(\hat x, \hat y, \hat z\right),
\end{equation}
by considering $f$ as a function on $\FRsp$. Note that the definition of spin-$s$ functions does not depend on the choice of global frame $\frF_g$.
An important property is that there is a natural correspondence between spin-2 functions and Stokes-valued spherical functions by considering $f\left(\hat x,\hat y, \hat z\right)$ as $s_1+i s_2$ where $s_1$ and $s_2$ are linear Stokes parameter for a ray along $\hat z$ with respect to the frame $\left[\hat x,\hat y, \hat z\right]$. One also observes that spin-0 and spin-1 functions are equivalent to the sphere's scalar and tangent vector fields, respectively.

Taking equivalent but slightly different orders to derive SWSH, we define SWSH as follows.

\begin{definition}{Spin-weighted spherical harmonics}{theory-swsh}
	The spin-weighted spherical harmonics with spin $s$, order $l$, and degree $m$ is a spin-$s$ function defined as follows:
	\begin{equation}
		{}_s Y_{lm}\left(\vec R\right) = \left(-1\right)^s\sqrt\frac{2l+1}{4\pi} D_{m,-s}^{l*}\left(\vec R\right).
	\end{equation}
\end{definition}
Note that due to Proposition~\ref{prop:wignerD_identity}(6), SWSH becomes an orthonormal basis for spin-$s$ functions, with a differential measure on $\Sspv$ following the definition through equivalence classes described in Definition~\ref{def:theory-spin-func}.

\begin{proposition}{Spin-2 spherical harmonics in Stokes vector fields}{theory-s2sh-s2stk}
	Defining a spin-2 Stokes vector field $\dvY_{lm}\left(\homega\right)\coloneqq\left[\sY{lm}\left(\vec R\right)\right]_{\vec R\frF_g}$, it becomes the well defined orthonormal basis for Stokes vectors fields, over scalar $\C$.
\end{proposition}
See also the rotation form of Stokes vectors fields discussed in Definition~\ref{def:field-stk-rotform} and Proposition~\ref{prop:field-stk-from-rotform}. Then a representation under the $\theta\phi$ frame field $\frF_{\theta\phi}$, which is introduced in the main paper, is defined as follows:
\begin{equation}\label{eq:theory-s2sh-tp}
	\sY{lm}\left(\theta,\phi\right)\coloneqq \left[\dvY_{lm}\left(\theta,\phi\right)\right]^{\frF_{\theta\phi}\left(\theta,\phi\right)} .
\end{equation}
Note that our main paper introduces the function in Equation~\eqref{eq:theory-s2sh-tp} first and then derives the formulation in Proposition~\ref{prop:theory-s2sh-s2stk} later to start from numerically measurable quantity, which is regarded more practical.

\subsection{Converting Between $\R^2$ and $\C$}
\label{sec:theory_R2_and_C}

we defines symbols to convert $\C$ and $\R^2$ or $\R^{2\times2}$.
\begin{align}\label{eq:R2_to_C}
	\C \left(\begin{bmatrix} x \\ y\end{bmatrix}\right) &\coloneqq x+yi \in \C, \\
	\label{eq:C_to_R2}
	\R^2\left(x+yi\right) &\coloneqq \begin{bmatrix} x \\ y\end{bmatrix} \in \R^2 ,\\ 
	\label{eq:C_to_R22}
	\R^{2\times 2}\left(x+ yi\right) & \coloneqq \begin{bmatrix} x & -y \\ y & x\end{bmatrix} \in \R^{2\times 2}.
\end{align}
Then we get:
\begin{equation}\label{eq:chain_compR2}
	\R^2\left(z_1\right)^T\R^{2\times 2}\left(z_2\right)\cdots \R^{2\times 2}\left(z_{n-1}\right)\R^2\left(z_n\right)
	= \Re\lrangle{z_1, z_2\cdots z_n}_\C = \Re\left(z_1^* z_2 \cdots z_n\right) .
\end{equation}

\mparagraph{Complex pair separation}
We observe that Equations~\eqref{eq:R2_to_C} and~\eqref{eq:C_to_R2} are the inverses of each other, but the function in Equation~\eqref{eq:C_to_R22} has not the inverse since it is not surjective. However, we found that any $2\times 2$ real matrix $\bfM=\begin{bmatrix}
	m_{11} & m_{12} \\ m_{21} & m_{22}
\end{bmatrix}$ can be represented by two complex numbers as follows:
\begin{equation}\label{eq:theory-comppair-sep}
	\begin{split}
		\bfM &= \R^{2\times 2} \left(\C_\mathrm{iso}\left(\bfM\right)\right) + \R^{2\times 2} \left(\C_\mathrm{conj}\left(\bfM\right)\right) \bfJ , \\
		\text{where ~ ~} \Ciso\left(\bfM \right)&\coloneqq \frac{m_{11} + m_{22}}{2} + \frac{m_{21} - m_{12}}{2}i, \\
		\Cconj\left(\bfM\right) &\coloneqq \frac{m_{11} - m_{22}}{2} + \frac{m_{21} + m_{12}}{2}i, \\
		\bfJ &\coloneqq \begin{bmatrix}
			1&0\\0&-1
		\end{bmatrix}.
	\end{split}
\end{equation}
We call $\Ciso\left(\bfM\right)$ and $\Cconj\left(\bfM\right)$ the \emph{isomorphic part} and the \emph{conjugation part of} $\bfM$, respectively.

The matrix $\bfJ$ acts on all right complex representations as complex conjugation, i.e.,
\begin{equation} \label{eq:J_conj}
	\bfJ \R^{2}\left(z\right)= \R^2\left(z^*\right).
\end{equation}
In general,
\begin{multline}\label{eq:J_conj_right}
	\R^2\left(z_1\right)^T\R^{2\times 2}\left(z_2\right)\cdots \R^{2\times 2}\left(z_{d}\right)\bfJ \R^{2\times 2}\left(z_{d+1}\right)\cdots \R^{2\times 2}\left(z_{n-1}\right) \R^2\left(z_n\right)
	\\
	=\R^2\left(z_1\right)^T\R^{2\times 2}\left(z_2\cdots z_{d}\right)\bfJ  \R^2\left(z_{d+1}\cdots z_n\right) 	=\Re\lrangle{z_1, z_2\cdots z_d z_{d+1}^*\cdots z_n^*}_\C
\end{multline}
Please be careful that this fact cannot be reduced to a product of $\bfJ$ and a single $2\times 2$ matrix, i.e.,
\begin{equation}\label{eq:theory-compair-caution}
	 \bfJ \R^{2\times 2}\left(z\right) \ne \R^{2\times 2}\left(z^*\right), %
\end{equation}
since $\bfJ$ cannot be $\R^{2\times 2}\left(z\right)$ for some $z\in \C$. Thus, we observe that Equation~\eqref{eq:J_conj_right} should be obtained by contracting $\R^{2\times 2}\left(z_{d+1}\right)\cdots \R^{2\times 2}\left(z_{n-1}\right) \R^2\left(z_n\right)=\R^2\left(z_{d+1}\cdots z_n\right)$ first, and followed by applying Equation~\eqref{eq:J_conj}

\mparagraph{Complex indexing formulae}
Due to the complexity of our derivation, such as viewing a function space as a linear space over scalar both $\R$ or $\C$, The following conversion equations will be useful. We call them complex indexing formulae.
\begin{align}
	\label{eq:ipow_to_R2}
	\bfMat\left[\Re\left(i^{1-p} z\right) \mid p=1,2\right] &= \begin{bmatrix}
		\Re z \\ \Im z
	\end{bmatrix} = \R^2\left(z\right) ,\\
	\label{eq:ipow_to_R2conj}
	\bfMat\left[\Re\left(i^{p-1} z\right) \mid p=1,2\right] &= \begin{bmatrix}
		\Re z^* \\ \Im z^*
	\end{bmatrix} = \R^2\left(z^*\right) ,\\
	\label{eq:ipow_to_R2x2}
	\bfMat\left[\Re\left(i^{p_i-p_o} z\right)\mid p_o,p_i = 1,2\right] &= \begin{bmatrix}
		\Re z & -\Im z \\ \Im z & \Re z
	\end{bmatrix} = \R^{2\times 2}\left(z\right) ,\\
	\label{eq:ipow_to_R2x2J}
	\bfMat\left[\Re\left(i^{2-p_i-p_o} z\right)\mid p_o,p_i = 1,2\right] &= \begin{bmatrix}
		\Re z & \Im z \\ \Im z & -\Re z
	\end{bmatrix} = \R^{2\times 2}\left(z\right) \bfJ .
\end{align}

\subsection{Polarized Spherical Harmonics}
\label{sec:theory-PSH}
\subsubsection{Discussion on real coefficient formulation}
\label{sec:theory-PSH_RC}

Note that we already discussed the necessity of our real coefficient formulation for spin-2 components for our PSH in the main paper in terms of complex pair separation, which is described both in the main paper and this document through Equations~\eqref{eq:theory-comppair-sep} to~\eqref{eq:theory-compair-caution}. Now, we discuss our real coefficient formulation for spin-0 components.
We now have two choices when we fix spin-2 coefficients as $\R^2$. Using a basis $\delta_{p0}Y_{lm}^R\oplus \left(\delta_{p1}\dvY_{lm1} + \delta_{p2}\dvY_{lm2}\right)\oplus \delta_{p3}Y_{lm}^R$ and coefficients in $\R^4$, or using a basis $\delta_{p0}Y_{lm}^C\oplus \left(\delta_{p1}\dvY_{lm1} + \delta_{p2}\dvY_{lm2}\right)\oplus \delta_{p3}Y_{lm}^C$ and coefficients in $\C\oplus\R^2\oplus \C$. While the former one, which will be selected our polarized spherical harmonics basis in Proposition~\ref{prop:theory-PSH-def}, clearly implies that it encodes general $\R$-linear operators on $\calF\left(\Sspv,\STKsp{\homega}\right)$ into $4\times 4$ real matrices of coefficients for fixed $l$ and $m$ indices, the later one cannot well define coefficient matrices. First, $\R$-linear operators on $\C\oplus\R^2\oplus \C$ belong to $\left(\R^2 \oplus \R^2 \oplus \R^2\right)^2$, which requires $2\times 2$ real coefficients for fixed $l$ and $m$ indices to represent operators from $s_0$ components to $s_0$ components. It contains too much redundant information to describe real-valued data from the original angular domain. It is not even compatible with conventional formulation where SH encodes linear operators on scalar fields to a coefficient simply in $\R$ or $\C$ for fixed $l$ and $m$ indices. As another choice, if one tries to define a coefficient matrix with mixed entry types, $\C$ and $\R$, we cannot define it as closed under matrix multiplication. If taking a product of two such matrices, complex values in spin 0-to-0 submatrices make spin 0-to-2 and 2-to-0 submatrices become complex. Finally, they make spin 2-to-2 submatrices become complex when multiplying another matrix again. It yields a contradiction.

\subsubsection{Polarized spherical harmonics}

As discussed in the previous section, we define our polarized SH by combining spin-0 and spin-2 SH with real coefficient formulation.
\begin{proposition}{Polarized spherical harmonics}{theory-PSH-def}
	With an index set
	\begin{equation}
		I_\mathrm{PSH} = \left\{\left(l,m,p\right)\in \Z^2 \mid \left|m\right|\le l, 0\le p < 4, \text{ and [if }p=1,2\text{ then }l\ge 2\text{]}\right\},
	\end{equation}
	$\dvY_{lmp}$'s are an orthonormal basis for the linear space of Stokes vector fields $\left\{f:\Sspv\to\calS_{\homega}\right\}$ over the scalar $\R$, where
	\begin{equation}
		\dvY_{lm0}\left(\homega\right) = \begin{bmatrix}
			Y_{lm}^R\brahomega \\0\\0\\0
		\end{bmatrix}_{\frtp\left(\homega\right)} \!\!\!\!\!\!\!\!\!\!\!\!\!\!,
		\dvY_{lm1}\left(\homega\right) = \begin{bmatrix}
			0\\ \Re \left[_2Y_{lm}\brahomega\right] \\ \Im \left[_2Y_{lm}\brahomega\right] \\0
		\end{bmatrix}_{\frtp\left(\homega\right)} \!\!\!\!\!\!\!\!\!\!\!\!\!\!,
		\dvY_{lm2}\left(\homega\right) = 
		\begin{bmatrix}
			0\\ -\Im \left[_2Y_{lm}\brahomega\right] \\ \Re \left[_2Y_{lm}\brahomega\right] \\0
		\end{bmatrix}_{\frtp\left(\homega\right)} \!\!\!\!\!\!\!\!\!\!\!\!\!\!,
		\dvY_{lm3}\left(\homega\right) = \begin{bmatrix}
			0 \\ 0 \\ 0 \\ Y_{lm}^R\brahomega
		\end{bmatrix}_{\frtp\left(\homega\right)}\!\!\!\!\!\!\!\!\!\!\!\!\!\!.
	\end{equation}
\end{proposition}

Note that it can be rewritten as $\dvY_{lmp} = \delta_{p0}Y_{lm}^R\left(\homega\right)\oplus\left(\delta_{p1}\dvY_{lm1}+\delta_{p2}\dvY_{lm2}\right)\oplus \delta_{p3}Y_{lm}^R$. Also note that taking spin-2 Stokes vector from $\dvY_{lm1}$ and $\dvY_{lm2}$, $\dvY_{lm}=\dvY_{lm1}$ and $i\dvY_{lm}=\dvY_{lm2}$. The following formula is useful to derive our linear operator formulations through a few equations rather than enumerating each indices $p_i$ and $p_o$. Using Equation~\eqref{eq:ipow_to_R2}, for $p=1,2$,
\begin{equation}
	\dvY_{lmp} = 0 \oplus \left[i^{p-1}\,\sY{lm}\right]_{\frF_{\theta\phi}} \oplus 0.
\end{equation}

\subsection{Rotation of Polarized Spherical Harmonics}
\label{sec:theory_rotinv}
\NEW{Here, we provide \NEW{the statement describing the PSH coefficient matrices of rotations on Stokes vector fields and its proof.}
	}

\begin{proposition}{Rotation coefficients of PSH}{theory-rot}
	The coefficient matrices of a rotation transform $\vec R\in\SOgroupv$ acting on the function space of Stokes vector fields, $\vec R_\calF$, is evaluated as follows.
\begin{equation}
	\begin{split}
	\mathbf{Mat}&\left[\lrangle{\dvY_{l_om_op_o}, \vec R_\calF \left[\dvY_{l_im_ip_i}\right]}_\calF \mid p_0,p_i=0,\cdots,3\right] \\
	&= \delta_{l_il_o}\begin{bmatrix}
			D_{m_om_i}^{l,R}\left(\vec R\right) & 0 & 0 & 0 \\
			0 & \Re D_{m_om_i}^{l,C}\left(\vec R\right) & -\Im D_{m_om_i}^{l,C}\left(\vec R\right) & 0 \\
			0 & \Im D_{m_om_i}^{l,C}\left(\vec R\right) & \Re D_{m_om_i}^{l,C}\left(\vec R\right) & 0 \\
			0 & 0 & 0 & D_{m_om_i}^{l,R}\left(\vec R\right)
	\end{bmatrix} \\
	&= \delta_{l_il_o} \begin{bmatrix}
		D_{m_om_i}^{l,R}\left(\vec R\right) & \mathbf{0}_{1\times 2} & 0  \\
		\mathbf{0}_{2\times 1} & \R^{2\times2}\left(D_{m_om_i}^{l,C}\left(\vec R\right)\right) & \mathbf{0}_{2\times 1} \\
		0 & \mathbf{0}_{1\times 2} & D_{m_om_i}^{l,R}\left(\vec R\right)
	\end{bmatrix}.
	\end{split}
\end{equation}
\end{proposition}

\vspace{-10mm}
\begin{proof}
	Relation between spin-weighted spherical harmonics and Wigner-D matrices:
\begin{equation}
	D_{m,-s}^l\left(\phi,\theta,\psi\right) = \left(-1\right)^s\sqrt{\frac{4\pi}{2l+1}}{}_sY_{lm}^*\left(\theta,\phi\right)e^{si\psi}.
\end{equation}

\noindent Using rotation matrices and $\hat z$:
\begin{equation}
	D_{m,-s}^l\left(\vec R\right) = \left(-1\right)^s\sqrt{\frac{4\pi}{2l+1}}{}_sY_{lm}^*\left(\vec R\hat z\right)e^{si\gamma_{zyz}\left(\vec R\right)},
\end{equation}
where $\gamma_{zyz}\left(\vec R\right)$ indicates an angle $\gamma$ such that $\vec R = \vec R_{\hat z_g \hat y_g \hat z_g}\left(\alpha,\beta,\gamma\right)$.

Rotated basis can be evaluated as:
\begin{equation}
		\left[\left(\vec R_\calF\dvY_{lmp}\right)\left(\vec R'\hat z\right)\right]^{\vec R'\frF} = \left[\dvY_{lmp}\left(\vec R^{-1}\vec R'\hat z\right)\right]^{\vec R^{-1}\vec R'\frF} 
		= \sqrt\frac{2l+1}{4\pi} \R^2\left[i^{p-1}D_{m,-2}^{l*}\left(\vec R^{-1}\vec R'\right)\right]
\end{equation}

\begin{equation}
	\begin{split}
		\lrangle{\dvY_{l'm'p'}, \vec R_\calF\left[\dvY_{lmp}\right]} &= \frac{\sqrt{\left(2l+1\right)\left(2l'+1\right)}}{8\pi^2} \Re\int_{SO\left(3\right)}{
			i^{p-p'}D_{m',-2}^{l'}\left(S\right)D_{m,-2}^{l*}\left(R^{-1}S\right)
			\rmd S} \\
		&= \delta_{ll'}\delta_{mm'}\Re\left[i^{p-p'}D_{mm'}^{l*}\left(R^{-1}\right)\right]
		= \delta_{ll'}\delta_{mm'}\Re\left[i^{p-p'}D_{m'm}^{l}\left(R\right)\right] \\
		&= \delta_{ll'}\delta_{mm'} \left(\R^{2\times 2}\circ D_{m'm}^l\left(R\right)\right)_{p'p} .
	\end{split}
\end{equation}

See also \citet{boyle2013angular}.
\end{proof}
\subsection{Linear Operators (pBRDF, Radiance Transfer)}
\label{sec:theory_mueller}

A linear operator on Stokes fields is characterized as a function of two directions into Mueller spaces.
\begin{definition}{Linear operators and kernels}{}
	Suppose there is a Mueller transform field $\dvK \colon \bbS^2\times \bbS^2 \to \calM_{\homega_i\to\homega_o}$. The \emph{linear operator of the kernel $\dvK$},  denoted by $\dvK_\calF\in \calL\left(\calF\left(\Sspv,\STKsp{\homega_i}\right), \calF\left(\Sspv,\STKsp{\homega_o}\right)\right)$, is defined as follows:
	\begin{equation}
		\forall \dvs\in \calF\left(\Sspv,\STKsp{\homega}\right),\quad \dvK_\calF\left[\dvs\right]\left(\homega_i\right) = \int_{\Sspv}{ \dvK\left(\homega_i,\homega_o\right)\dvs\left(\homega_i\right) \rmd \homega_i}.
	\end{equation}
	If a linear operator $\dvK_\calF$ is given first, a Mueller field $\dvK$ satisfying the above equation is called the \emph{kernel} of the operator $\dvK_\calF$.
\end{definition}

A linear operator on Stokes fields can also be written as a function of two rotation transforms, similar to rotation forms for Stokes vector fields.
\begin{definition}{Rotation form of a Mueller transform field}{}
	The \emph{rotation form of the Mueller transform field} $\dvK \colon \bbS^2\times \bbS^2 \to \calM_{\homega_i\to\homega_o}$ (or the \emph{rotation form of the operator} $\dvK_\calF$) is defined as:
	\begin{equation}\label{eq:theory_mueller_rotform}
		\begin{split}
		\tilde\bfK\colon \SOgroupv\times\SOgroupv\to \R^{4\times 4}, \\
		\tilde K\left(\vec R_i,\vec R_o\right) = \left[\dvK\left(\vec R_i \hat z,\vec R_o \hat z\right)\right]^{\vec R_i \frF_e\to\vec R_o \frF_e}.
		\end{split}
	\end{equation}
	Conversely, when a function $\tilde\bfK\colon \SOgroupv\times\SOgroupv\to \R^{4\times 4}$ is given, it can be the rotation form of a Mueller transform field if and only if it satisfies the following constraints:
	\begin{equation}\label{eq:theory_muelrot_constrain}
		\tilde K\left(\vec R_i \vec R_z\left(\psi_1\right), \vec R_o \vec R_z\left(\psi_2\right)\right) = \Rns{-2\psi_2} \tilde K\left(\vec R_i,\vec R_o\right) \Rns{2\psi_1} .
	\end{equation}
\end{definition}

We found that applying a linear operator to a Stokes vector field can be done in rotation forms of the Mueller transform field and the Stokes vector field.

\begin{proposition}{Applying linear operator in rotation forms}{theory-linop-apply-rotform}
	$\dvf:\Sspv\to\STKsp{\homega}$ and $\dvK:\Sspv\times\Sspv\to \MUEsp{\homega_i}{\homega_o}$ are a Stokes vector field and Mueller transform field, respectively. The rotation forms of $\dvf$ and $\dvK$ are denoted by and $\bfM$, respectively. Then the rotation form of $\dvK_{\calF}\left[\dvf\right]$ can be evaluated as follows.
	\begin{equation}\nonumber
		\bfg\left(\vec R_o\right) = \frac1{2\pi}\int_{\SOgroupv}{\bfK\left(\vec R_i, \vec R_o\right) \bff\left(\vec R_i\right) \rmd \vec R_i},
	\end{equation}
	where $\bfg$ denotes the rotation form of the resulting Stokes vector field.
\end{proposition}
\vspace{-10mm}
\begin{proof}
	By definition $\bfg$ is obtained as:
	\begin{equation}\nonumber
		\bfg\left(\vec R_o\right) = \left[\dvK_{\calF}\left[\dvf\right]\left(\vec R_o\hat z_g\right)\right]^{\vec R_o \frF_g}. 
	\end{equation}
	The term inside $\left[\cdot\right]$ can be obtained as follows using the integral conversion in Equation~\eqref{eq:inttech_sph2rot}:
	\begin{equation}\nonumber
		\int_{\Sspv}{\dvK\left(\homega_i, \vec R_o \hat z_g\right) \dvf\left(\homega_i\right)\rmd \homega_i} = \frac1{2\pi}\int_{\SOgroupv}{\dvK\left(\vec R_i\hat z_g, \vec R_o \hat z_g\right) \dvf\left(\vec R_i\hat z_g\right)\rmd\vec R_i}.
	\end{equation}
	Note that the integrand on the right hand side is $\left[ \bfK\left(\vec R_i, \vec R_o\right) \bff\left(\vec R_i\right) \right]_{\vec R_o\frF_g}$. Substituting all equations into the first one,
	\begin{equation}
		\bfg\left(\vec R_o\right) = \frac1{2\pi}\int_{\SOgroupv}{\bfK\left(\vec R_i, \vec R_o\right) \bff\left(\vec R_i\right) \rmd \vec R_i},
	\end{equation}
	which also yields
	\begin{equation}
		\dvg\left(\vec R_o \hat z_g\right) = \frac 1{2\pi}\left[\int_{\SOgroupv}{\bfK\left(\vec R_i, \vec R_o\right) \bff\left(\vec R_i\right) \rmd \vec R_i}\right]_{\vec R_o \frF_g}.
	\end{equation}
\end{proof}

\begin{definition}{Complex form of a Mueller transform field}{theory-muel-compform}
	The \emph{complex form of} a Mueller transform field $\dvK \colon \bbS^2\times \bbS^2 \to \calM_{\homega_i\to\homega_o}$ is defined as ten functions $\tilde K_{0|3,0|3}\colon\Sspv\times \Sspv\to\R$ and $\tilde K_{0|3,\bfp},\tilde K_{\bfp,0|3}, \tilde K_{\bfp\bfp\mathrm{i}}, \tilde K_{\bfp\bfp\mathrm{c}}\colon\Sspv\times \Sspv\to\C$ which satisfy:
	\begin{equation}
		\left[\dvK\left(\homega_i,\homega_o\right)\right]^{\frtp\left(\homega_i\right)\to\frtp\left(\homega_o\right)} = \begin{bmatrix}
			\tilde K_{00} & \R^2\left(\tilde K_{0\bfp}\right)^T &  \tilde K_{03} \\
			\R^2\left(\tilde K_{\bfp0}\right) & \R^{2\times 2}\left(\tilde K_{\bfp\bfp\mathrm{i}}\right) + \R^{2\times 2}\left(\tilde K_{\bfp\bfp\mathrm{c}}\right)\mathbf{J} & \R^2\left(\tilde K_{\bfp3}\right) \\
			\tilde K_{30} & \R^2\left(\tilde K_{3\bfp}\right)^T &  \tilde K_{33}
		\end{bmatrix}.
	\end{equation}
	Note that we omit the variables $\left(\homega_i,\homega_o\right)$ for each $\tilde K$ component for simplicity.
\end{definition}

In Definition~\ref{def:theory-muel-compform}, each component should satisfy the following quantities and functions on the rotation group, which satisfy the following can be complex forms of a Mueller transform field, conversely.
\begin{align}
	\tilde K_{0|3,p}\left(\vec R_i \vec R_z\left(\psi\right), \vec R_o\right) &= \tilde K_{0|3,p}\left(\vec R_i , \vec R_o\right)e^{-2\psi i} \\
	\tilde K_{p,0|3}\left(\vec R_i , \vec R_o \vec R_z\left(\psi\right)\right) &= \tilde K_{p,0|3}\left(\vec R_i , \vec R_o\right)e^{-2\psi i} \\
	\tilde K_{ppa}\left(\vec R_i \vec R_z\left(\psi_1\right) , \vec R_o \vec R_z\left(\psi_2\right)\right) &= \tilde K_{ppa}\left(\vec R_i , \vec R_o\right)e^{-2\left(\psi_2-\psi_1\right) i} \\
	\tilde K_{ppb}\left(\vec R_i \vec R_z\left(\psi_1\right) , \vec R_o \vec R_z\left(\psi_2\right)\right) &= \tilde K_{ppa}\left(\vec R_i , \vec R_o\right)e^{-2\left(\psi_2+\psi_1\right) i}
\end{align}

The coefficient matrix of a linear operator on Stokes vector fields can be defined and evaluated by 16 integral formulae obtained by directly extending Proposition~\ref{prop:linalg-coeff-linop}. However, we found that they can be evaluated with fewer formulae using the complex form of the Mueller transform field.

\begin{proposition}{Coefficient matrix using the complex form of a Mueller field}{}
	The polarized spherical harmonics coefficients $\rmM_{l_om_op_p,l_im_ip_i}\coloneqq\lrangle{\dvY_{l_om_op_p}, \dvM_\calF\left[\dvY_{l_im_ip_i}\right]}$ of a linear operator with the kernel $\dvM \colon \bbS^2\times \bbS^2 \to \calM_{\homega_i\to\homega_o}$ is evaluated using the complex form of $\dvM$ as follows: 
	\begin{equation}
		\begin{split}
			\rmM_{l_om_o0|3,l_im_i0|3} &= \int_{\Sspv\times \Sspv}{
				Y_{l_om_o}^{R}\left(\homega_o\right)\tilde M_{0|3,0|3}\left(\homega_i,\homega_o\right)Y_{l_im_i}^R\left(\homega_i\right)
			\rmd \homega_i\rmd \homega_o}, \\
			\begin{bmatrix}
				\rmM_{l_om_o0|3,l_im_i1} & \rmM_{l_om_o0|3,l_im_i2}
			\end{bmatrix} &= \R^2\left(\int_{\Sspv\times \Sspv}{
				Y_{l_om_o}^{R}\left(\homega_o\right)\tilde M_{0|3,\bfp}\left(\homega_i,\homega_o\right){}\sY{l_im_i}^*\left(\homega_i\right)
				\rmd \homega_i\rmd \homega_o}\right)^T, \\
			\begin{bmatrix}
				\rmM_{l_om_o1,l_im_i0|3} \\ \rmM_{l_om_o2,l_im_i0|3}
			\end{bmatrix} &= \R^2\left(\int_{\Sspv\times \Sspv}{
				\sY{l_om_o}^{*}\left(\homega_o\right)\tilde M_{\bfp,0|3}\left(\homega_i,\homega_o\right)Y_{l_im_i}^R\left(\homega_i\right)
				\rmd \homega_i\rmd \homega_o}\right), \\
			\begin{bmatrix}
				\rmM_{l_om_o1,l_im_i1} & \rmM_{l_om_o1,l_im_i2} \\
				\rmM_{l_om_o2,l_im_i1} & \rmM_{l_om_o2,l_im_i2}
			\end{bmatrix} &= \R^{2\times 2}\left(\int_{\Sspv\times \Sspv}{
			\sY{l_om_o}^{*}\left(\homega_o\right)\tilde M_{\bfp\bfp\mathrm{i}}\left(\homega_i,\homega_o\right)\sY{l_im_i}\left(\homega_i\right)
			\rmd \homega_i\rmd \homega_o}\right) \\
			&+ \R^{2\times 2}\left(\int_{\Sspv\times \Sspv}{
				\sY{l_om_o}^{*}\left(\homega_o\right)\tilde M_{\bfp\bfp\mathrm{c}}\left(\homega_i,\homega_o\right)\sY{l_im_i}^*\left(\homega_i\right)
				\rmd \homega_i\rmd \homega_o}\right) \bfJ.
		\end{split}
	\end{equation}
\end{proposition}
\vspace{-10mm}
\begin{proof}
	Proof can be done by using the definition of the complex forms. Here, we clarify derivation steps for spin 2-to-2 components, $p_i,p_o=1, 2$, which additionally utilize our complex pair separation and complex indexing formulae in Equations~\eqref{eq:ipow_to_R2x2} and~\eqref{eq:ipow_to_R2x2J} as
	\begin{equation}
		\label{eq:pbrdf_coeff_deriv}
		\begin{split}
			\rmf\left[l_im_ip_i, l_om_op_o\right] &= \lrangle{\dvY_{l_om_op_o},\dvM\dvY_{l_im_ip_i}} \\
			&= \Re\int_{S^2\times S^2}{ i^{1-p_o}{}_2Y_{l_om_o}^*\brao \left[ i^{p_i-1}\tilde M_{\ppi}\,{}_2Y_{l_im_i} + i^{1-p_i} \tilde M_{\ppi}\,{}_2Y_{l_im_i}^* \right] \rmd\homega_i\rmd\homega_o}  \\
			&= \R^{2\times2}\left[ \int_{S^2\times S^2}{ \tilde M_{\ppi}\braio{}_2Y_{l_om_o}^*\brao {}_2Y_{l_im_i}\brai \rmd\homega_i\rmd\homega_o} \right] \\
			&+ \R^{2\times2}\left[ \int_{S^2\times S^2}{ \tilde M_{\ppc}\braio{}_2Y_{l_om_o}^*\brao {}_2Y_{l_im_i}^*\brai \rmd\homega_i\rmd\homega_o} \right] \bfJ .
		\end{split}
	\end{equation}
\end{proof}

\subsection{Reflection Operator}
\label{sec:theory-refl}

To adapt Section~\ref{sec:sh-reflection} to our PSH, we first define a reflection operator $T_\calS\in\calL\left(\calS,\calS\right)$ with respect to $\hat z_g$ as follows.
\begin{equation}
	\left[T_\calS\left(\dvs\right)\right]^{\vec R_{zyz}\left(\alpha,\beta,\gamma\right)\frF_g}
	=\left( \left[\dvs\right]^{\vec R_{zyz}\left(\alpha,\pi -\beta,0|\pi-\gamma\right)\frF_g}\right)^* .
\end{equation}
Note that it can be understood by flipping the double-sided arrow, which visualizes a Stokes vector. It is also equivalent to perfect mirror reflection by the dielectric material of infinite index of refraction.

It can also act on Stokes vector fields, and its PSH coefficients are obtained in the following steps. Using ZYZ Euler angles for rotations,

\begin{equation}
	\lrangle{\dvY_{lm},T_\calF\left[\dvY_{l'm'}\right]} = \int_{\bbS^2}{
		\sY{lm}^*\left(\alpha,\beta,\gamma\right) \sY{l'm'}^*\left(\alpha,\pi-\beta,0|\pi-\gamma\right)
		\rmd\homega}.
\end{equation}
Note that it is constant for $\gamma$. For Wigner-D form:
\begin{equation}
	\sqrt\frac{\left(2l+1\right)\left(2l'+1\right)}{16\pi^2}\frac1{2\pi}\int_{SO\left(3\right)}{
		D_{m,-2}^{l}\left(\alpha,\beta,\gamma\right) D_{m',-2}^{l'}\left(\alpha,\pi-\beta,0|\pi-\gamma\right)
		\rmd \vec R} .
\end{equation}
By symmetry of small-D $d_{mm'}^l$ and Wigner-D,
\begin{equation}
	D_{m',-2}^{l'}\left(\alpha,\pi-\beta,0|\pi-\gamma\right) = \left(-1\right)^{l'+m'}D_{m',2}^{l'}\left(\alpha,\beta,\gamma\right) = \left(-1\right)^{l'}D_{-m',-2}^{l'*}\left(\alpha,\beta,\gamma\right) .
\end{equation}
Substituting the above, the orthogonality of Wigner D-functions yields:
\begin{equation}
	\lrangle{\dvY_{lm},T_\calF\left[\dvY_{l'm'}\right]} =\left(-1\right)^l \delta_{ll'}\delta_{m,-m'}.
\end{equation}

\subsection{Triple Product of SWSH}
\label{sec:theory-tp}

There are special symbols to represent the SWSH triple product. Spin-0 SH can be written as:
\begin{equation}\label{eq:theory_tp_s2s}
	\int_{\Sspv}{Y_{l_1m_1}^*Y_{l_2m_2}Y_{l_3m_3}\rmd\homega} = \\ \left(-1\right)^{m_1}\sqrt\frac{\left(2l_1+1\right)\left(2l_2+1\right)\left(2l_3+1\right)}{4\pi}\begin{pmatrix}
		l_1 & l_2 & l_3 \\
		-m_1 & m_2 & m_3
	\end{pmatrix} \begin{pmatrix}
		l_1 & l_2 & l_3 \\
		0 & 0 & 0
	\end{pmatrix},
\end{equation}
where the symbol $\begin{pmatrix}
	l_1 & l_2 & l_3 \\
	-m_1 & m_2 & m_3
\end{pmatrix}$ is called a \emph{Wigner 3-j symbol}. Then the triple product of two spin-2 SH and one spin-0 SH, which is a spin-SH coefficient of scalar multiplication of another spin-2 SH by a spin-0 SH, is written as:
\begin{equation}\label{eq:theory_tp_v2v}
	\int_{\Sspv}{\sY{l_1m_1}^*Y_{l_2m_2}\sY{l_3m_3}\rmd\homega} = \\ \left(-1\right)^{m_1}\sqrt\frac{\left(2l_1+1\right)\left(2l_2+1\right)\left(2l_3+1\right)}{4\pi}\begin{pmatrix}
		l_1 & l_2 & l_3 \\
		-m_1 & m_2 & m_3
	\end{pmatrix} \begin{pmatrix}
		l_1 & l_2 & l_3 \\
		-2 & 0 & 2
	\end{pmatrix}.
\end{equation}

We do not explicitly introduce what Wigner 3-j symbols are and how we can compute them. However, note that the above two equations have the same kind of special symbols, which only depend on integer indices. Since existing PRT methods have been used to spin-0 triple product, we can also compute spin-2 SH from their implementation.

\subsection{Convolution on Stokes Vectors Fields}
\label{sec:theory-conv}
In this section, we derive polarized spherical convolution as a rotation equivariant linear operator on Stokes vector fields.

\begin{definition}{Rotation equivariant opeartor}{}
	A linear operator $K_{\calF}\in\calL\left(\calF\left(\Sspv,\calS\right), \calF\left(\Sspv,\calS\right)\right)$ on Stokes vector fields called \emph{rotation equivariant} if $\vec R_{\calF} \left(K_{\calF}\left[\dvf\right]\right) = K_{\calF}\left[\vec R_{\calF} \dvf\right]$ holds for any $\vec R\in\SOgroupv$ and $\dvf:\Sspv\to\STKsp{\homega}$.
\end{definition}
If such an operator has a kernel, Mueller transform field, then rotation equivariant can be stated as follows.
\begin{proposition}{Rotation equivariant for operator kernel}{}
	Suppose there is a Mueller transform field $\dvK \colon \bbS^2\times \bbS^2 \to \calM_{\homega_i\to\homega_o}$. The linear operator of the kernel $\dvK$, $\dvK_\calF$ is rotation equivariant if and only if $\vec R_{\calF} \left[\dvK \right]=\dvK$ for any $\vec R\in\SOgroupv$.
\end{proposition}
The above condition $\vec R_{\calF} \left[\dvK \right]=\dvK$ can be written using the rotation form $\bfK:\SOgroupv\times\SOgroupv\to\R^{4\times 4}$ of the Mueller transform as follows:
\begin{equation}
	\bfK\left(\vec R \vec R_i, \vec R\vec R_o\right) = \bfK\left( \vec R_i, \vec R_o\right), \quad \forall \vec R,\vec R_i,\vec R_o\in\SOgroupv.
\end{equation}

Then, we finally obtain a minimal form of the rotation equivariant (operator) kernel. It can be considered an extension that a rotation equivariant operator on scalar fields has been characterized by a simple azimuthal symmetric scalar field. However, we have more information in the codomain of the polarized convolution kernel.
\begin{proposition}{Minimal form of a rotation equivariant operator kernel}{theory_conv_kernel}
	A Mueller transform field $\dvK \colon \bbS^2\times \bbS^2 \to \calM_{\homega_i\to\homega_o}$ which is a kernel of rotation equivariant linear operator can be characterized by a Mueller transform function of a single angle as $\bfK\left(\vec I, \vec R_{\hat y_g}\left(\beta\right)\right)$, where $\bfK$ denotes the rotation form of $\dvK$.
\end{proposition}
\vspace{-10mm}
\begin{proof}
	Using rotation equivariance $\bfK\left(\vec R \vec R_i, \vec R\vec R_o\right) = \bfK\left( \vec R_i, \vec R_o\right)$, we get:
	\begin{equation} \nonumber
		\bfK\left(\vec R_i, \vec R_o\right) = \bfK\left(\vec I, \vec R_i^{-1}\vec R_o\right).
	\end{equation}
	With ZYZ Euler angle $\vec R_{\hat z_g \hat y_g \hat z_g}\left(\alpha,\beta,\gamma\right) = \vec R_i^{-1}\vec R_o$ and the constraint of rotation forms of Mueller transform fields,
	\begin{equation} \label{eq:theory-conv-kernel-rotform3}
		\bfK\left(\vec I, \vec R_i^{-1}\vec R_o\right) = \bfR_{1:2}\left(-2\gamma\right)\bfK\left(\vec I, \vec R_{\hat y_g}\left(\beta\right)\right)\bfR_{1:2}\left(-2\alpha\right).
	\end{equation}
\end{proof}

Note that $\bfK\left(\vec I,\vec R_{\hat y_g}\left(\beta\right)\right)\eqqcolon \bfk\left(\beta\right)\in\R^{4\times 4}$ is the polarized convolution kernel which also introduced in our main paper. Equation~\eqref{eq:theory-conv-kernel-rotform3} yields its constraints by substituting $\beta=0$ and $\beta=\pi$ and using $\vec R_{zyz}\left(\alpha,0,\gamma\right) = \vec R_z\left(\alpha + \gamma\right)$ and $\vec R_{zyz}\left(\alpha , \pi, \gamma\right) = \vec R_{zy} \left(\alpha-\gamma,\pi\right)$:
\begin{equation}
	\begin{split}
		\bfk\left(0\right) = \bfR_{1:2}\left(\psi\right)\bfk\left(0\right)\bfR_{1:2}\left(-\psi\right), \\
		\bfk\left(\pi\right) = \bfR_{1:2}\left(\psi\right)\bfk\left(\pi\right)\bfR_{1:2}\left(\psi\right), \\
	\end{split}
\end{equation}
for any $\psi$. A particular corollary of it is that the isomorphic and conjugation parts of spin 2-to-2 submatrix of $\bfk$ become zero at $\theta=\pi$ and $\theta=0$, respectively. These constraints are highly related to each subspace of PSH bases for each submatrix of convolution kernels.

\subsection{Convolution in Polarized Spherical Harmonics}
\label{sec:theory-conv-SH}

Note that the following lemma is useful. It comes from Wigner D-function identities.
\begin{lemma}\label{lem:wignerD_integral_3rot}
	For any indices $l_i$, $m_i$, and $m'_i$ for $i=1,2$ in the valid range and rotation transforms $\vec S,\vec T\in\SOgroupv$,
	\begin{equation}
		\begin{split}
			\int_{\SOgroupv}{ D_{m_1 m'_1}^{l_1}\left(\vec R\vec S\right)D_{m_2 m'_2}^{l_2*}\left(\vec R\vec T\right) \rmd \mu\left(\vec R\right)}
			&= \frac{8\pi^2}{2l_1+1}\delta_{\left(l_1m_1\right)\left(l_2m_2\right)} D_{m'_1m'_2}^{l_1*}\left(\vec S^{-1}\vec T\right) \\
			&= \frac{8\pi^2}{2l_1+1}\delta_{\left(l_1m_1\right)\left(l_2m_2\right)} D_{m'_2m'_1}^{l_1}\left(\vec T^{-1}\vec S\right) .
		\end{split} 
	\end{equation}
\end{lemma}

\paragraph{Spin 0-to-2}
Starting from the definition, an entry of the coefficient matrix of a rotation equivariant linear operator on Stokes vector fields $\dvK_{\calF}$ is obtained as follows. Note that in this section, $\tilde K$ denotes the complex form of the Mueller transform $\dvK$, and $\lrangle{\cdot, \cdot}_{\C}$ denotes the inner product on the Stokes space over scalar $\C$.
\begin{multline}\label{eq:convp0_org}
	\mathrm{K}_{l_om_op_o,l_im_i0} =\lrangle{\dvY_{l_om_op_o}, \dvK_\calF\left[Y_{l_im_i}^R\right]} = \iint_{\Sspv\times \Sspv }{ \lrangle{\dvY_{l_om_op_o}\left(\homega_o\right), \dvK_{\bfp 0}\left(\homega_i,\homega_o\right)Y_{l_im_i}^R\left(\homega_i\right)}_\calS \rmd\homega_i \rmd \homega_o}= \\
	= \int_0^\pi{\!\!\!\!\int_{\SOgroupv}{ \!\! \left[\dvY_{l_om_o p_o}\left(\vec R\vec R_y\left(\frac\theta2\right)\hat z\right)\right]^{\vec R \vec R_y\left(\frac\theta2\right)\frF_g} \!\!\cdot  \left[\dvK_{\bfp0}\left(\vec R\vec R_y\left(-\frac\theta2\right)\hat z,\vec R\vec R_y\left(\frac\theta2\right)\hat z\right)Y_{l_im_i}^R\left(\vec R\vec R_y\left(-\frac\theta2\right)\hat z\right)\right]^{\vec R \vec R_y\left(\frac\theta2\right)\frF_g}\!\!\rmd \vec R}\sin \theta \rmd \theta}.
\end{multline}
Using the relation between spin-weighted spherical harmonics and Wigner-D functions in Definition~\ref{def:theory-swsh},

\begin{equation}\label{eq:convp0_Rtheta}
	\begin{split}
		\mathrm{Eq.}~\eqref{eq:convp0_org}&= B_{l_il_o}\Re\left[i^{1-p_o}\int_0^\pi{\!\!\!\!\int_{\SOgroupv}{\!\!  D_{m_o,-2}^{l_o}\left(\vec R\vec R_y\left(\frac\theta2\right)\right) \tilde K_{\bfp0}\left(\vec R\vec R_y\left(-\frac\theta2\right),\vec R\vec R_y\left(\frac\theta2\right)\right) D_{m_i0}^{l_i,R}\left(\vec R\vec R_y\left(-\frac\theta2\right)\right)\rmd \vec R}\sin\theta\rmd\theta}\right] \\
		&\eqqcolon B_{l_il_o}\Re \left[i^{1-p_o} I_1 \right] ,
	\end{split}
\end{equation}
where $B_{l_il_o}\coloneqq \frac{\sqrt{\left(2l_i+1\right)\left(2l_i+1\right)}}{4\pi}$, and we are denoting the integral term as $I_1$ for simplicity of later steps. In the integrand of $I_1$, rotation equivariance of $\tilde K_{\bfp0}$ yields:
\begin{equation}
	\tilde K_{\bfp0}\left(\vec R\vec R_y\left(-\frac\theta2\right),\vec R\vec R_y\left(\frac\theta2\right)\right) = \tilde K_{\bfp0}\left(\vec I, \vec R_y\left(\theta\right)\right).
\end{equation}
We observe that it is independent of $\vec R$ so that it can go outside of the integral over $\vec R$. Then, using the relation between real and complex Wigner-D functions in Equation~\eqref{eq:wignerD_RC}, $I_1$ becomes:
\begin{equation}
	I_1 = \int_0^\pi{\tilde K_{\bfp0}\left(\vec I,\vec R_y\left(\theta\right)\right)\int_{\SOgroupv}{ D_{m_o,-2}^{l_o}\left(\vec R \vec R_y\left(\frac\theta2\right)\right) \sum_{m_c=\pm m_i}\left(M_{m_i m_c}^{C\to R}\right)^* D_{m_c0}^{l_i}\left(\vec R \vec R_y\left(-\frac\theta2\right)\right) \rmd \vec R} \sin\theta\rmd\theta}.
\end{equation}
Then we can use Lemma~\ref{lem:wignerD_integral_3rot} to the inner integral with a symmetry of Wigner-D functions $D_{m_c 0}^{l_i}=\left(-1\right)^{m_c}D_{-m_c0}^{l_i,*}$.
\begin{equation} \label{eq:convp0_wignerD_final}
	I_1 =\delta_{l_il_o} \frac{2\pi}{B_{l_il_i}}\int_0^\pi{\tilde K_{\bfp0}\left(\vec I,\vec R_y\left(\theta\right)\right) \sum_{m_c=\pm m_i} \delta_{m_o,-m_c}\left(-1\right)^{m_c} \left(M_{m_i m_c}^{C\to R}\right)^* D_{0,-2}^{l_i}\left(\vec R_y\left(\theta\right)\right) \sin\theta\rmd\theta}.
\end{equation}
Here, we observe that $I_1=0$ if $\abs{m_i}=\abs{m_o}$, terms containing $m_i$, $m_o$, and $m_s$, denoted by $\bfU^{\bfp0}$, is evaluated as follows.

\begin{multline}\label{eq:convp0_phase} %
	\bfU^{\bfp0} \coloneqq \bfMat\left[\sum_{m_c=\pm m_i}\!\!\! \delta_{m_o,-m_c}\left(-1\right)^{m_c} \left(M_{m_i m_c}^{C\to R}\right)^* \mid m_o,m_i=+\abs{m},-\abs{m}\right] \\= \left(-1\right)^{m} \begin{bmatrix}
		0&1\\1&0
	\end{bmatrix} \frac1{\sqrt2}\begin{bmatrix}
		1 & i \\
		\left(-1\right)^{m} & - \left(-1\right)^{m}i
	\end{bmatrix} = \frac1{\sqrt{2}} \begin{bmatrix}
		1 & -i \\
		\left(-1\right)^{m} & \left(-1\right)^{m}i 
	\end{bmatrix} .
\end{multline}
Then, combining Equations~\eqref{eq:convp0_wignerD_final} and \eqref{eq:convp0_phase} and converting the Wigner-D to a spin-weighted spherical harmonics conversely, we get
\begin{equation}\label{eq:convp0_I1_s2SH}
	I_1 = \delta_{\left(l_i\abs{m_i}\right)\left(l_o\abs{m_o}\right)} U_{m_om_i}^{\bfp0} \frac{2\pi}{B_{l_il_il_i}} \int_0^{2\pi}{\lrangle{\dvY_{l0}\left(\theta,0\right),\dvK_{\bfp0}\left(\hat z_g, \homega_{\mathrm{sph}}\left(\theta,0\right)\right) }_\C\sin\theta\rmd\theta} .
\end{equation}
We observe here that, similar to conventional convolution through scalar spherical harmonics, the only degree of freedom comes from the order $l$. Thus, we can define the \emph{convolution coefficient} of the scalar-to-Stokes part of $\dvK$ as follows.
\begin{equation} \label{eq:convp0_convcoeff}
	\mathrm{k}_{l,\bfp0} = 2\pi \int_0^\pi{ \lrangle{\dvY_{l0p}\left(\theta,0\right),\dvK_{\bfp0}\left(\hat z_g, \homega_{\mathrm{sph}}\left(\theta,0\right)\right) }_\C \sin\theta\rmd \theta} \in \C,
\end{equation}
which can be considered as an inner product over the entire $\Sspv$ as scalar SH convolution is. Note that we are defining the convolution coefficient $\rmk_{l,\bfp0}$ as a complex number so that we will take $\Re$ in later steps.

Now we finally get the coefficient of the linear operator by combining Equations~\eqref{eq:convp0_Rtheta}, \eqref{eq:convp0_I1_s2SH}, and \eqref{eq:convp0_convcoeff}.

\begin{equation} \label{eq:convp0_operator}
	\rmK_{l_om_op_o,l_im_i0}  =\delta_{\left(l_i\abs{m_i}\right)\left(l_o\abs{m_o}\right)} \sqrt\frac{4\pi}{2l_i+1} \Re\left(i^{1-p_o} U_{m_om_i}^{\bfp0} \mathrm{k}_{l,\bfp0}\right),
\end{equation}
where $p_o=1,2$.

\paragraph{Spin 2-to-0}
We can follow similar steps to scalar-Stokes components. The coefficient of the linear operator is:
\begin{multline}\label{eq:conv0p_org}
	\mathrm{K}_{l_om_o0,l_im_ip_i} =\lrangle{Y_{l_om_o}^R, \dvK_\calF\left[\dvY_{l_im_ip_i}\right]} = \iint_{\Sspv\times \Sspv }{ Y_{l_om_o}^R\left(\homega_o\right) \lrangle{\dvK_{0\bfp}\left(\homega_i,\homega_o\right),\dvY_{l_im_ip_i}\left(\homega_i\right)}_{\calS} \rmd\homega_i \rmd \homega_o}= \\
	= \int_0^\pi{\!\!\!\!\int_{\SOgroupv}{ \!\! Y_{l_om_o}^R \left(\vec R\vec R_y\left(\frac\theta2\right)\hat z\right)   \left[\dvK_{0\bfp}\left(\vec R\vec R_y\left(-\frac\theta2\right)\hat z,\vec R\vec R_y\left(\frac\theta2\right)\hat z\right)\right]^{\vec R \vec R_y\left(-\frac\theta2\right)\frF_g} \!\!\cdot \left[\dvY_{l_im_i p_i}\left(\vec R\vec R_y\left(-\frac\theta2\right)\hat z\right)\right]^{\vec R \vec R_y\left(-\frac\theta2\right)\frF_g}\!\!\rmd \vec R}\sin \theta \rmd \theta} \\
	=B_{l_il_o}\Re\left[i^{p_i-1} \int_0^\pi{\!\!\!\!\int_{\SOgroupv}{\!\!  D_{m_o,0}^{l_o,R}\left(\vec R\vec R_y\left(\frac\theta2\right)\right) \tilde K_{0\bfp}\left(\vec I,\vec R_y\left(\theta\right)\right) D_{m_i,-2}^{l_i,*}\left(\vec R\vec R_y\left(-\frac\theta2\right)\right)\rmd \vec R}\sin\theta\rmd\theta} \right] \eqqcolon B_{l_il_o}\Re\left[p^{p_i - 1}I_2\right].
\end{multline}
Here, we denote the integral term by $I_2$. Using the relation between real and complex Wigner-D functions in Equation~\eqref{eq:wignerD_RC} followed by Lemma~\ref{lem:wignerD_integral_3rot},
\begin{equation} \label{eq:conv0p_Rtheta}
	\begin{split}
		I_2 &= \int_0^\pi{\tilde K_{0\bfp}\left(\vec I,\vec R_y\left(\theta\right)\right)\int_{\SOgroupv}{ \sum_{m_c=\pm m_o}\left(M_{m_o m_c}^{C\to R}\right)^*D_{m_c,0}^{l_o}\left(\vec R \vec R_y\left(\frac\theta2\right)\right)  D_{m_i,-2}^{l_i,*}\left(\vec R \vec R_y\left(-\frac\theta2\right)\right) \rmd \vec R} \sin\theta\rmd\theta}  \\
		&= \delta_{l_il_o} \frac{2\pi}{B_{l_il_i}}\int_0^\pi{\tilde K_{0\bfp}\left(\vec I,\vec R_y\left(\theta\right)\right) \sum_{m_c=\pm m_o} \delta_{m_cm_i} \left(M_{m_o m_c}^{C\to R}\right)^* D_{-2,0}^{l_i}\left(\vec R_y\left(\theta\right)\right) \sin\theta\rmd\theta}.
	\end{split}
\end{equation}

On the right-hand side, we can reduce the terms containing $m_c$, denoted by $\bfU^{0\bfp}$ as follows:
\begin{multline}\label{eq:conv0p_phase} %
	\bfU^{0\bfp} \coloneqq \bfMat\left[\sum_{m_c=\pm m_o}\!\!\! \delta_{m_cm_i}\left(M_{m_o m_c}^{C\to R}\right)^* \mid m_o,m_i=+\abs{m},-\abs{m}\right] \\= \bfMat\left[\left(M_{m_o m_i}^{C\to R}\right)^* \mid m_o,m_i=+\abs{m},-\abs{m}\right] = \frac1{\sqrt{2}} \begin{bmatrix}
		1 & \left(-1\right)^{m} \\
		i & -\left(-1\right)^{m}i
	\end{bmatrix}.
\end{multline}
Combining Equations~\eqref{eq:conv0p_Rtheta} and \eqref{eq:conv0p_phase} yields:
\begin{equation}
	I_2 = \delta_{\left(l_i\abs{m_i}\right)\left(l_o\abs{m_o}\right)}U_{m_om_i}^{0\bfp} \frac{2\pi}{B_{l_il_il_i}}\int_0^\pi{Y_{l,-2}^{C,*}\left(\theta,0\right)\lrangle{\dvK_{0\bfp}\left(\hat z_g, \homega_{\mathrm{sph}}\left(\theta,0\right)\right), \begin{bmatrix}
		1 \\ 0
	\end{bmatrix}_{\frF_g}}_\C \sin\theta\rmd\theta}.
\end{equation}
Now we can finally define the convolution coefficient of the Stokes-to-scalar part of $\dvK$, denoted by $\rmk_{l,0\bfp}$, and obtain the coefficient of a linear operator in terms of $\rmk_{l,0\bfp}$.
\begin{align} \label{eq:conv0p_convcoeff}
	\mathrm{k}_{l,0\bfp} &= 2\pi \int_0^\pi{Y_{l,-2}^{C,*}\left(\theta,0\right)\lrangle{\dvK_{0\bfp}\left(\hat z_g, \homega_{\mathrm{sph}}\left(\theta,0\right)\right), \begin{bmatrix}
		1 \\ 0
	\end{bmatrix}_{\frF_g}}_\C \sin\theta\rmd\theta} \in \C, \\
	\label{eq:conv0p_operator}
	\rmK_{l_om_o0,l_im_ip_i}  &=\delta_{\left(l_i\abs{m_i}\right)\left(l_o\abs{m_o}\right)} \sqrt\frac{4\pi}{2l_i+1} \Re\left(i^{p_i-1} U_{m_om_i}^{0\bfp} \mathrm{k}_{l,0\bfp}\right),
\end{align}
where $p_o=1,2$.

\paragraph{Spin 2-to-2}
The coefficient of the linear operator is:

\begin{align}\label{eq:convpp_org}
	\mathrm{K}_{l_om_op_o,l_im_ip_i} &=\lrangle{\dvY_{l_om_op_o}, \dvK_\calF\left[\dvY_{l_im_ip_i}\right]} = \iint_{\Sspv\times \Sspv }{  \lrangle{\dvY_{l_om_op_o}\left(\homega_o\right),\dvK_{\bfp\bfp}\left(\homega_i,\homega_o\right)\left[\dvY_{l_im_ip_i}\left(\homega_i\right)\right]}_{\calS} \rmd\homega_i \rmd \homega_o} \\
	&= \int_0^\pi{\!\!\!\!\int_{\SOgroupv}{ \!\! \left[\dvY_{l_om_op_o} \left(\vec R_o\hat z\right)\right]^{\frF_o} \!\!\cdot   \left[\dvK_{\bfp\bfp}\left(\vec R_i\hat z,\vec R_o\hat z\right)\right]^{\frF_i\to\frF_o}  \left[\dvY_{l_im_i p_i}\left(\vec R_i\hat z\right)\right]^{\frF_o}\!\!\rmd \vec R}\sin \theta \rmd \theta},
\end{align}
where $\vec R_i\coloneqq \vec R \vec R_y\left(-\frac{\theta}{2}\right)$, $\vec R_o\coloneqq \vec R \vec R_y\left(\frac{\theta}{2}\right)$, $\frF_i \coloneqq \vec R_i \frF_g$, and $\frF_o \coloneqq \vec R_o \frF_g$. It can be rewritten in terms of Wigner-D functions as:
\begin{equation}\label{eq:convpp_wigner} %
	\rmK_{l_om_op_o,l_im_ip_i}= B_{l_il_o}\int_0^\pi{\!\!\!\!\int_{\SOgroupv}{ \!\!  \R^2\left(i^{p_o-1}D_{m_o,-2}^{l_o,*}\left(\vec R_o\right)\right)^T \tilde\bfK_{\bfp\bfp}\left(\vec R_i,\vec R_o\right) \R^2 \left(i^{p_i-1} D_{m_i,-2}^{l_i,*}\left(\vec R_i\right)\right) \rmd \vec R}\sin \theta \rmd \theta}.
\end{equation}
Here, we need an additional step that was not needed for scalar-to-Stokes and Stokes-to-scalar terms. Note that $2\times 2$ matrix $\tilde\bfK$ can be decompose into two terms $\tilde \bfK_{\bfp\bfp}=\R^{2\times 2}\left(\tilde K_{\ppi}\right)+ \R^{2\times 2}\left(\tilde K_{\ppc}\right) \bfJ$. Then right two terms in the integral in Equation~\eqref{eq:convpp_wigner} become as follows by Equation~\eqref{eq:J_conj_right}:
\begin{equation}
	\tilde\bfK_{\bfp\bfp}\left(\vec R_i,\vec R_o\right) \R^2 \left(i^{p_i-1} D_{m_i,-2}^{l_i,*}\left(\vec R_i\right)\right) = \R^2\left( i^{p_i-1} \tilde K_{\ppi}\left(\vec R_i,\vec R_o\right)D_{m_i,-2}^{l_i,*}\left(\vec R_i\right) + i^{1-p_i}\tilde K_{\ppc}\left(\vec R_i,\vec R_o\right) D_{m_i,-2}^{l_i}\left(\vec R_i\right) \right).
\end{equation}
Substituting this results into Equation~\eqref{eq:convpp_wigner},
\begin{equation} \label{eq:convpp_wigner3} %
	\begin{split}
		\rmK_{l_om_op_o,l_im_ip_i}&= B_{l_il_o} \Re\left[i^{p_i-p_o} \int_0^\pi{\!\!\!\!\int_{\SOgroupv}{ \!\!  D_{m_o,-2}^{l_o}\left(\vec R_o\right) \tilde K_{\ppi}\left(\vec R_i,\vec R_o\right) D_{m_i,-2}^{l_i,*}\left(\vec R_i\right) \rmd \vec R}\sin \theta \rmd \theta} \right] \\
		&+  B_{l_il_o} \Re\left[i^{2-p_i-p_o} \int_0^\pi{\!\!\!\!\int_{\SOgroupv}{ \!\!  D_{m_o,-2}^{l_o}\left(\vec R_o\right) \tilde K_{\ppc}\left(\vec R_i,\vec R_o\right) D_{m_i,-2}^{l_i}\left(\vec R_i\right) \rmd \vec R}\sin \theta \rmd \theta} \right] \\
		&\eqqcolon B_{l_il_o}\Re\left[i^{p_i-p_o}I_3 + i^{2-p_i-p_o}I_4\right] .
	\end{split}
\end{equation}
Here, we denote two integral terms by $I_3$ and $I_4$, respectively. First, $I_3$ can be evaluated similarly to previous components.
\begin{align}
	I_3 &=\delta_{\left(l_im_i\right)\left(l_om_o\right)} \frac{2\pi}{B_{l_il_i}} \int_0^\pi{\tilde K_{\ppi}\left(\vec I,\vec R_y\left(\theta\right)\right) D_{-2,-2}^{l_i}\left(\vec R_y\left(\theta\right)\right) \sin\theta\rmd\theta} \\
	&= \delta_{\left(l_im_i\right)\left(l_om_o\right)} \frac{2\pi}{B_{l_il_il_i}}\int_0^\pi{\lrangle{\prescript{}{2}{Y}_{l_i,-2}\left(\theta,0\right), \tilde K_{\ppi}\left(\vec I,\vec R_y\left(\theta\right)\right) }_\C \sin\theta\rmd\theta}.
\end{align}

Similarly, $I_4$ is:
\begin{align}
	I_4 &=\delta_{\left(l_i,-m_i\right)\left(l_om_o\right)}\left(-1\right)^{m_i} \frac{2\pi}{B_{l_il_i}} \int_0^\pi{\tilde K_{\ppc}\left(\vec I,\vec R_y\left(\theta\right)\right) D_{2,-2}^{l_i}\left(\vec R_y\left(\theta\right)\right) \sin\theta\rmd\theta} \\
	&= \delta_{\left(l_i,-m_i\right)\left(l_om_o\right)} \left(-1\right)^{m_i} \frac{2\pi}{B_{l_il_il_i}}\int_0^\pi{\lrangle{\prescript{}{2}{Y}_{l_i2}\left(\theta,0\right), \tilde K_{\ppc}\left(\vec I,\vec R_y\left(\theta\right)\right) }_\C \sin\theta\rmd\theta}.
\end{align}

Finally, the convolution coefficient of the Stokes-to-Stokes part of $\dvK$ can be defined as two complex numbers $\rmk_{l,\bfp\bfp a}$ and $\rmk_{l,\bfp\bfp b}$, and the coefficient of linear operator can be written in terms of $\rmk_{l,\bfp\bfp a}$ and $\rmk_{l,\bfp\bfp b}$.
\begin{align} \label{eq:convpp_convcoeff}
	\mathrm{k}_{l,\ppi} &=  2\pi \int_0^\pi{\lrangle{\prescript{}{2}{Y}_{l_i,-2}\left(\theta,0\right), \tilde K_{\ppi}\left(\vec I,\vec R_y\left(\theta\right)\right) }_\C \sin\theta\rmd\theta} \in\C, \\
	\label{eq:convpp_convcoeffc}
	\mathrm{k}_{l,\ppc} &=  2\pi \int_0^\pi{\lrangle{\prescript{}{2}{Y}_{l_i,2}\left(\theta,0\right), \tilde K_{\ppc}\left(\vec I,\vec R_y\left(\theta\right)\right) }_\C \sin\theta\rmd\theta} \in\C, \\
	\label{eq:convpp_operator}
	\rmK_{l_om_op_o,l_im_ip_i}  &=\delta_{l_il_o} \sqrt\frac{4\pi}{2l_i+1} \Re\left( \delta_{m_im_o} i^{p_i-p_o}  \mathrm{k}_{l,\ppi} + \delta_{-m_im_o} \left(-1\right)^{m_i}i^{2-p_i-p_o}  \mathrm{k}_{l,\ppc} \right),
\end{align}
where $p_o=1,2$. Note that the final equation can be rewritten using Equations~\eqref{eq:ipow_to_R2x2} and \eqref{eq:ipow_to_R2x2J}:
\begin{equation}\label{eq:convpp_convcoeff2}
	\bfMat\left[\rmK_{l_om_op_o,l_im_ip_i} \mid p_o,p_i\right]  =\delta_{l_il_o} \sqrt\frac{4\pi}{2l_i+1} \left( \delta_{m_im_o} \R^{2\times2}\left(\mathrm{k}_{l,\ppi}\right) + \delta_{-m_im_o} \left(-1\right)^{m_i}\R^{2\times2}\left(\mathrm{k}_{l,\ppc}\right)\bfJ \right).
\end{equation}
We observe that this expression is natural since $\tilde K_{\ppi}$ and $\tilde K_{\ppc}$ was decomposed from $\tilde \bfK_{\bfp\bfp} = \R^{2\times2}\left( \tilde K_{\ppi} \right) + \R^{2\times2}\left( \tilde K_{\ppc} \right)\bfJ$.

\clearpage

%% file: application1-supp.tex
\section{Results and Discussion}
\label{sec:result-discussion}

\subsection{\NEW{Results for Precomputed Polarized Radiance Transfer}}

\begin{figure*}[tp]
	\centering
	\vspace{-4mm}	
	\includegraphics[width=\linewidth]{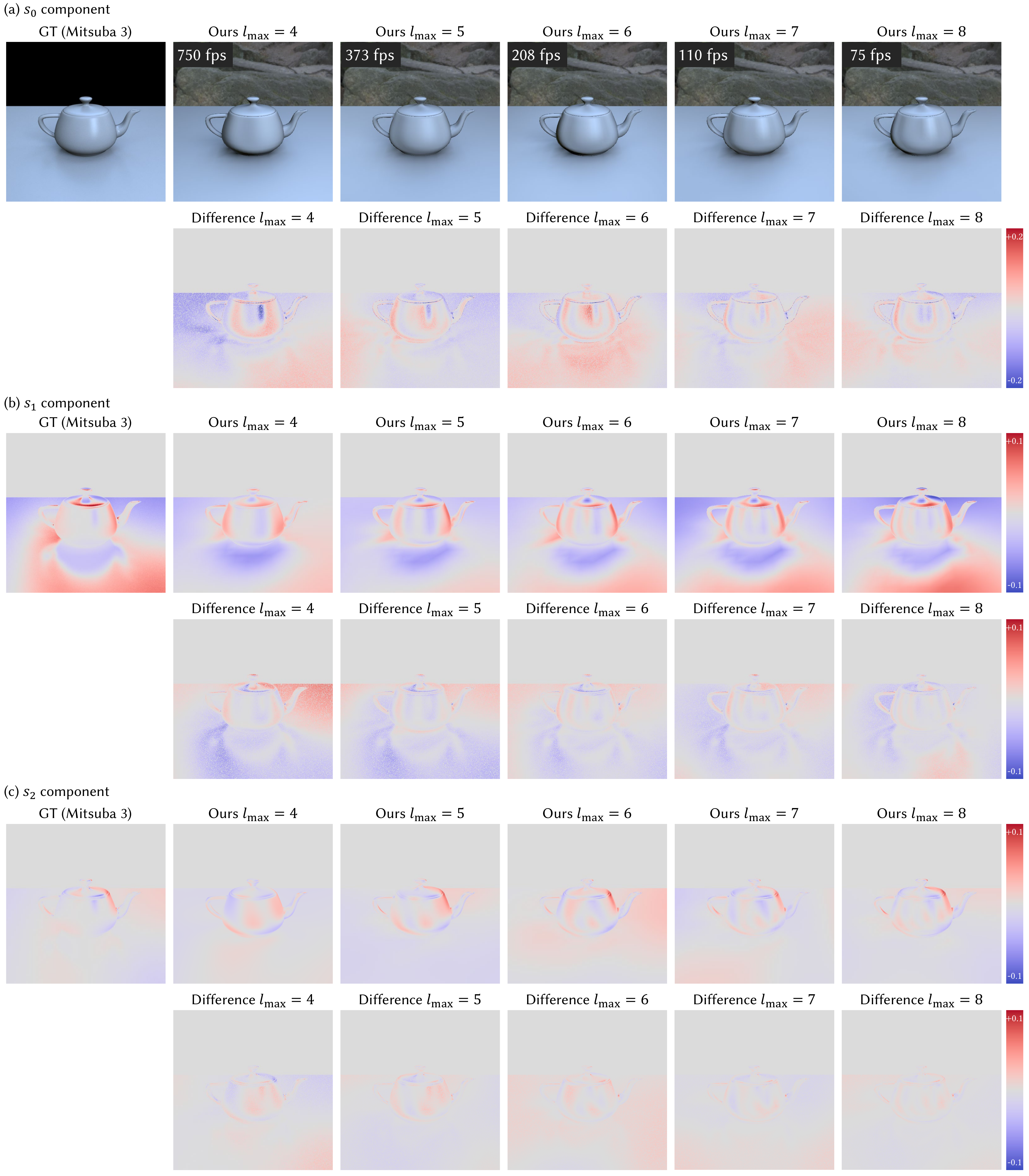}%
	\vspace{-4mm}
	\caption[]{\label{fig:result_GT_comparison}
		\NEW{Rendered images for Figure~\ref{fig:result_GT_comparison_plot} in the main paper. We validate our real-time polarized rendering with shadowed radiance transfer compared with Mitsuba 3 GT ray tracer. (a) to (c) shows $s_0$, $s_1$, and $s_2$ Stokes components of polarized images, respectively. We observe our results get closer to GT results as the cut-off frequency $l_{\mathrm{max}}$ increases.
		Note that differences in this figure and Figure~\ref{fig:result_GT_comparison_plot} in the main paper were computed only at pixels where the object exists.}
	}
	\vspace{-2mm}		
\end{figure*}

\NEW{
\mparagraph{Scene specification}
We provide technical details of the scene setups throughout the main paper (Figures~\ref{fig:teaser},~\ref{fig:result_ablations}~\ref{fig:result_ablations},~\ref{fig:app_prt_result},~\ref{fig:theory_rendering_pipeline}, and~\ref{fig:result_validprojection}) and this supplemental document (Figure~\ref{fig:result_GT_comparison}) in Table~\ref{tb:result_info}. The reported numbers of vertices include 3D models themselves and ground planes. Note that  While (1) lighting (environment map), (2) radiance transfer matrix of pBRDF and shadow, and (3) high-order convolution approximation of pBRDF are encoded in PSH coefficients, each single coefficient contains trichromatic RGB values, refer to 12 bytes (4x3 bytes float). Each scene uses two materials. Note that while transfer matrices differ for each vertex, convolution coefficients for high-order pBRDF are shared by all vertices of the same material due to rotation equivariance.

}

\NEW{\mparagraph{Validation against GT}
Here we provide rendered images of Mitsuba 3 GT render and our PPRT method for each cut-off frequency $l_{\mathrm{max}}$, which are discussed in Figure~\ref{fig:result_GT_comparison_plot} in the main paper Section~\ref{sec:pprt}. The resulting images and difference maps are shown in Figure~\ref{fig:result_GT_comparison}.  Since Mitsuba 3 does not support polarized environment map emitters, we are using an unpolarized environment map for this scene. In addition, \citet{baek2020image}'s data-based pBRDFs are only supported by multispectral variants of Mitsuba 3, while our implementation is based on conventional RGB rendering by projecting multi-channel \citet{baek2020image} pBRDF into RGB in advance. Instead, for this quantitative validation, we conducted this scene with an analytic pBRDF model \citet{baek2018simultaneous}.
Specific configurations of this scene are also reported in Table~\ref{tb:result_info}.
}

\begin{table*}[htpb]\small
	\caption{\label{tb:result_info} 
		\NEW{The scene setups specification throughout the main paper and this supplemental document. For several scenes which do not use high-order convolution approximation we are not reporting numbers of such coefficients.}}
	\vspace{-3mm}
	\begin{tabular}{
			m{0.12\linewidth} | M{0.09\linewidth} 
			| M{0.12\linewidth} | M{0.12\linewidth} | M{0.15\linewidth}
			| M{0.15\linewidth} | M{0.05\linewidth}
		}
		\thickhline
		\multicolumn{2}{l|}{Scene} & \# of vertices & Lighting coeff. & Radiance transfer matrix (per vertex) & Convolution coeff. (per material) & FPS \\
		\hline
		
		\multicolumn{2}{l|}{Main Fig.~\ref{fig:teaser}} & 21,087 & 300 & 5,625 & 45 & 100 \\
		\hline
		
		\multirow{2}{*}{Main Fig.~\ref{fig:result_ablations}(a)} & {Rows 1 \& 2} & \multirow{2}{*}{10,115} & 75 & 5,625 & -- & 475\\
		& Row 3 &  & 300 & 5,625 & 45 & 306\\
		\hline

		\multirow{2}{*}{Main Fig.~\ref{fig:result_ablations}(b)} & {Rows 1 \& 2} & \multirow{2}{*}{20,545} & 75 & 5,625 & -- & 162\\
		& Row 3 &  & 300 & 5,625 & 45 & 102\\
		\hline
		
		\multirow{3}{*}{Main Fig.~\ref{fig:app_prt_result}} & {(b)} & \multirow{3}{*}{10,115} & 75 & 5,625 & -- & 480\\
		& (c) &  & 108 & 11,664 & -- & 210\\
		& (d) &  & 300 & 5,625 & 45 & 308\\
		\hline
		
		\multicolumn{2}{l|}{Main Figs.~\ref{fig:theory_rendering_pipeline},~\ref{fig:result_validprojection}} & 19,944 & 300 & 5,625 & 45 & 111\\
		\hline
		
		\multirow{5}{*}{Fig.~\ref{fig:result_GT_comparison}} & {$l_{\mathrm{max}}=4$} & \multirow{5}{*}{3,482} & 75 & 5,625 & -- & 750 \\
		& $l_{\mathrm{max}}=5$ &  & 108 & 11,664 & -- & 373\\
		& $l_{\mathrm{max}}=6$ &  & 147 & 21,609 & -- & 208\\
		& $l_{\mathrm{max}}=7$ &  & 192 & 36,864 & -- & 110\\
		& $l_{\mathrm{max}}=8$ &  & 243 & 59,049 & -- & 75\\
		\hline
		
		\thickhline
	\end{tabular}
\end{table*}
\subsection{\NEW{Discussion on SWSH Formulations in Previous Work}}
\label{sec:discussion-SWSH}

\NEW{
\mparagraph{Definitions of spin-weight spherical harmonics}
For interested readers, we briefly review the formulations of SWSH in previous work here. When SWSH were originally introduced by \citet{newman1966note}, they were defined using a special kind of differential operators, spin raising and lowering operators $\eth$ and $\bar\eth$. Then \citet{newman1966note} defined SWSH in the spherical coordinates $\left(\theta, \phi\right)$, and dependency of local frames is regarded implicitly.  \citet{goldberg1967spin} found a relationship between SWSH and Wigner D-functions. Our description of SWSH in Definition~\ref{def:theory-swsh} is based on this relationship to make the frame dependency clear rather than implicit. Note that we do not cover what $\eth$ and $\bar\eth$ operators are.

\mparagraph{Spin-weight $s=-2$ spherical harmonics}
\citet{zaldarriaga1997all,ng1999correlation} used both spin $+2$ and $-2$ SH to handle the correlation of Stokes vector fields, but the necessity of two types of special functions for describing a single type of quantity, Stokes vectors, has been somewhat counterintuitive. While the occurrence of complex conjugation in several equations in this work ($\sY{l_im_i}^*$ and $\tilde \rmf_{l,-mp}^*$ in Equations~\eqref{eq:theory_mueller_coeff_V2Vc} and~\eqref{eq:theory_conv_to2} in the main paper, respectively) can be considered to correspond to the spin $-2$ coefficients in \citet{zaldarriaga1997all,ng1999correlation}, we do not need to introduce spin $-2$ SH in our paper. Instead, the complex conjugation is explained not as a property of special functions such as spin $\pm 2$ SH, but by the \emph{complex pair separation} of Mueller transform (Equation~\eqref{eq:theory_mat2comp_property} in the main paper and Equation~\eqref{eq:theory-comppair-sep} in this supplemental document), which is defined in the angular domain without regarding any basis function.
}

\clearpage